\documentclass[10pt,journal,compsoc]{IEEEtran}

\usepackage{makecell}
\usepackage{amsmath,amsfonts}
\usepackage{algorithmic}
\usepackage{array}
\usepackage[caption=false,font=footnotesize,labelfont=sf,textfont=sf]{subfig}
\usepackage{textcomp}
\usepackage{stfloats}
\usepackage{url}
\usepackage{verbatim}
\usepackage[most]{tcolorbox}
\usepackage{graphicx}
\usepackage{cite}
\usepackage{balance}
\usepackage{subcaption}
\usepackage[table]{xcolor}
\usepackage{caption}
\usepackage[table]{xcolor}

\newcommand{\negcell}[1]{\cellcolor{red!18}#1}
\newcommand{\poscell}[1]{\cellcolor{green!18}#1}
\usepackage{booktabs}
\usepackage{multirow}
\usepackage{makecell}
\usepackage{float}
\usepackage{tikz}
\usepackage{ragged2e}
\usetikzlibrary{fit}
\usepackage{enumitem}
\usepackage{amsthm}
\newtheorem{definition}{Definition}
\usepackage{acronym}
\usepackage{hyperref}
\usepackage{comment}
\usepackage{cleveref}
\usepackage{prisma-flow-diagram}
\usepackage{listings}
\usepackage{comment}
\definecolor{OliveGreen}{cmyk}{0.64,0,0.95,0.40}
\definecolor{lightlightgray}{gray}{0.93}
\definecolor{phdcolor}{RGB}{230, 242, 255}
\definecolor{masccolor}{RGB}{255, 248, 220}
\definecolor{undergradcolor}{RGB}{245, 245, 245}
\definecolor{postdoccolor}{RGB}{240, 255, 240}
\definecolor{phddefense}{RGB}{220, 230, 241}
\definecolor{proposalexam}{RGB}{255, 242, 204}
\definecolor{mascdefense}{RGB}{234, 241, 221}
\definecolor{oralexam}{RGB}{255, 219, 219}

\crefname{equation}{Eq.}{Eqs.}
\Crefname{equation}{Equation}{Equations}

\newcounter{myalgorithm}
\renewcommand{\themyalgorithm}{\arabic{myalgorithm}}
\begin{document}

\title{Image Augmentation as Test Generation for Deep Learning-Based Image Retrieval Systems}

\author{
Yehan~De~Silva,
Anirudh~Sridhar,
Armin~Lotfy,
Nafiseh~Kahani,
Yvan~Labiche,
Ziyu~Wang,
Frank~Ouyang,
Clare~Carty,
and~Azalia~Shamsaei%
\IEEEcompsocitemizethanks{
\IEEEcompsocthanksitem
Y. De Silva, A. Sridhar, A. Lotfy, N. Kahani, and Y. Labiche
are with the Department of Systems and Computer Engineering,
Carleton University, Ottawa, ON, Canada.\\
{\raggedright
E-mail: [YehandeSilva, AnirudhSridhar]@cmail.carleton.ca,
[arminlotfy, nafisehahani, yvanlabiche]@cunet.carleton.ca.
\par}
\IEEEcompsocthanksitem
Z. Wang, F. Ouyang, C. Carty, and A. Shamsaei
are with March Networks, Ottawa, ON, Canada.\\
{\raggedright E-mail: [zwang,
fouyang,
ccarty,
ashamsaei]@marchnetworks.com.\par}
%
}%
}








\IEEEtitleabstractindextext{%
\begin{abstract}
\justifying
Ensuring the reliability of deep learning-based image retrieval systems is a significant software engineering challenge, especially in safety-critical and high-stakes deployment scenarios.  This paper presents a dual contribution: (1) a systematic literature review of augmentation and generation techniques used in articles published between 2020 and 2025, which resulted in the identification of 50 such techniques which we organized into a ten-category taxonomy, and (2) a large-scale empirical study that evaluates these techniques as candidate test generators for embedding-based image retrieval systems. Augmented images are embedded using Amazon Titan and OpenCLIP, and evaluated across four analytical dimensions: (1) embedding-space similarity, (2) embedding uncertainty measured via four estimators (dispersion, pairwise distance, Mahalanobis distance, and ensemble agreement), (3) semantic realism scored by LLaVA, and (4) retrieval failure rate. Experiments are performed on three datasets: CIFAR-10, ImageNet-1K, and a dataset from an industrial partner (March Networks), {utilizing both the Amazon Titan Multimodal Embeddings G1 model and the OpenCLIP ViT-H/14 model pretrained on the LAION-2B dataset.
} Across all evaluated datasets and embedding models, and under the single severity level tested for each technique, weather simulation and {SaSPA} are the image augmentation/generation techniques that produce the highest embedding uncertainty and failure rates while maintaining a favorable balance between performance stability, visual realism, and augmentation effectiveness. The results we discuss are configuration-specific and may shift under milder or stronger perturbation settings. In contrast, GAN-based augmentation techniques are among the lowest in realism, indicating the presence of synthetic artifacts and perceptual inconsistencies that reduce their suitability to produce realistic test inputs.  Overall, our findings provide practical guidelines for selecting augmentation techniques that maximize test diversity while preserving realistic image characteristics, thereby enabling the construction of more comprehensive and effective test suites for image retrieval systems while reducing the cost of manual data labeling through the use of metamorphic testing.

\end{abstract}

}

\maketitle

\IEEEdisplaynontitleabstractindextext
\IEEEpeerreviewmaketitle

\section{Introduction}

AI-based image search and retrieval systems are increasingly deployed in production environments where input conditions are far from ideal. In surveillance and security applications, cameras are mounted at varying angles, overhead, side-mounted, or oblique, and must operate under challenging environmental conditions including low light, rain, fog, and glare \cite{pan2024solving}. In industrial inspection and medical imaging, images may be degraded by sensor noise, motion blur, compression artifacts, or resolution loss \cite{gordo2016deep, cao2020unifying}. In all of these settings, the retrieval system is expected to return semantically correct results regardless of the quality or viewing conditions of the query image. Ensuring this property through manual testing is prohibitively expensive. The combinatorial space of environmental factors, camera perspectives, object dynamics, and image quality degradations is too large to enumerate by hand. An automated strategy is therefore needed to generate large and diverse query inputs; however, automatically judging whether the retrieved results are correct raises an oracle problem. Metamorphic testing addresses this by defining an expected relation between related test inputs and their outputs, removing the need for a manually labelled oracle \cite{chen2018metamorphic}.

A scalable way to generate such tests is \emph{image augmentation} --- controlled image modifications that we refer to as augmentation techniques (or simply augmentations) throughout this manuscript. Applying transformations such as lighting changes, weather effects, noise, viewpoint shifts, partial occlusion, motion blur, and adversarial perturbations to a query image and submitting the result to the retrieval system probes its robustness across realistic deployment conditions without requiring manual labeling. This instantiates the metamorphic relation used in this study: because a semantics-preserving transformation should not change an image's category, the system should retrieve images of the same category for both the original and the augmented query, and any deviation is a detected fault. Not every transformation is a valid test generator, however. A useful augmentation must be strong enough to expose weaknesses yet preserve the image's semantic category. If it destroys that category, a resulting failure cannot be attributed to the model, whereas a transformation too weak to change anything yields only trivial tests.

Identifying which augmentation techniques are the most challenging to an AI-based system requires more than simply counting failures. Different techniques cause difficulty in different ways, and understanding \emph{why} a model fails is as important as knowing \emph{that} it fails. We therefore evaluate each technique --- both the validity of the follow-up inputs it produces and its fault-exposure power --- through four complementary metrics:
\begin{itemize}
    \item \textbf{Retrieval failure count}: how often the model retrieves an image from the wrong category for an augmented query, directly quantifying how frequently each technique causes a misclassification. Since a semantics-preserving technique should yield an image of the same category as the original, any retrieved image from a different category is a violation of the metamorphic relation and hence a detected fault.
    \item \textbf{Embedding similarity}: the cosine similarity between the original and augmented image embeddings, revealing how much each augmentation displaces the model's internal representation of the image.
    \item \textbf{Embedding uncertainty}: the stability of the model's representations under augmented inputs, assessed through four estimators: dispersion relative to the class centroid, pairwise distance between original and augmented embeddings, Mahalanobis distance from the embedding distribution, and agreement across an ensemble of k-nearest neighbor (KNN) classifiers. High uncertainty indicates that the augmented images occupy ambiguous regions of the embedding space, exactly the conditions under which misclassification is most likely.
    \item \textbf{Semantic realism}: the visual plausibility of the augmented images, scored by LLaVA \cite{liu2023visual}, a multimodal large language model (LLM), against nine perceptual criteria including lighting consistency, texture integrity, and semantic coherence. This distinguishes augmentations that cause failures because they represent realistic hard cases from those that cause failures only because they produce unnatural, artifact-ridden images with no correspondence to real deployment scenarios.
\end{itemize}

\noindent Together, these metrics provide a complete profile of each augmentation technique from a testing perspective: how hard it is for the model, how the technique causes difficulty, and whether that difficulty reflects realistic deployment scenarios. These metrics directly operationalize the research questions defined in Section~\ref{sec:evaluation_questions}.

Despite the wide variety of augmentation techniques available, there is no systematic study that (a) catalogues the full landscape across all major categories and (b) evaluates them through this multi-dimensional testing lens to identify which augmentations are most challenging for embedding-based retrieval models. Existing surveys \cite{shorten2019survey, MUMUNI2022100258} focus on training-time accuracy improvements and do not study augmentation in the context of automated robustness testing, embedding-space analysis, or semantic realism assessment. Consequently, it remains unclear which augmentation techniques preserve the semantic category while producing sufficiently challenging queries for retrieval evaluation. Fixed corruption benchmarks such as ImageNet-C \cite{hendrycks2019benchmarking} evaluate classifier robustness against a small set of predefined corruptions applied to a single model, whereas we catalogue the full augmentation landscape and assess each technique as a test generator for embedding-based retrieval, across both open and closed models.

This gap motivated the present work. We make two complementary contributions. First, we conduct a systematic literature review (SLR) of image augmentation techniques published between 2020 and 2025, identifying 56 relevant papers from an initial pool of over 88 candidates and organizing the techniques they describe into a ten-category taxonomy spanning geometric transforms, photometric adjustments, noise injection, selection-based techniques, filtering operations, self-mixing, sample mixing, generative adversarial network (GAN)-based synthesis, diffusion-based generation, and adversarial perturbation. Second, we conduct a large-scale empirical study that applies each catalogued technique as a source of challenging test inputs for two embedding-based retrieval models, Amazon Titan and OpenCLIP, and evaluates it across the four metrics above on three datasets: CIFAR-10, ImageNet-1K, and a dataset from our industry partner, March Networks.  We report a ranked comparison of all evaluated techniques across the four metrics, enabling practitioners to identify which augmentations offer the best trade-off between fault-exposure power and semantic realism for their specific deployment context. 


The remainder of this manuscript is structured as follows. Section~\ref{sec:systematic_literature_review} describes the SLR protocol and the resulting taxonomy of augmentation techniques. Section~\ref{sec:background} provides background for the remainder of the manuscript. Section~\ref{sec:augmentation_techniques} presents all identified augmentation techniques organized by category. Section~\ref{sec:empirical_study_design} presents the empirical study design, embedding models, datasets, and evaluation metrics. Section~\ref{sec:results} reports the experimental findings. Section~\ref{sec:threats} discusses threats to validity. Section~\ref{sec:conclusion} concludes the paper.

\section{Systematic Literature Review}
\label{sec:systematic_literature_review}

The standard systematic review process we followed was established by Kitchenham et al. \cite{kitchenham2009systematic} and is structured into three main phases: defining the review objective and search criteria; conducting the review by searching scholarly databases and reference lists; and reporting on the review by combining the findings into a single source. The objective of this SLR is to identify and catalogue the full landscape of image {image augmentation techniques, including generative augmentation techniques} 
techniques published between 2020 and 2025. The resulting catalogue (Section~\ref{sec:augmentation_techniques}) forms the input to the empirical study in Section~\ref{sec:empirical_study_design}.

\subsection{Search Strategy}
\label{sec:search_strategy}

We focus on high-quality academic publications sourced from reputable databases, including Google Scholar, IEEE Xplore, ACM Digital Library, SpringerLink, ScienceDirect, and Scopus. These sources were chosen for their broad and reliable coverage of peer-reviewed journals and top conference proceedings in the fields of computer vision and image retrieval. We also manually reviewed proceedings from three conferences chosen for their relevance: CVPR (Conference on Computer Vision and Pattern Recognition), ICCV (International Conference on Computer Vision), and NeurIPS (Neural Information Processing Systems).

We constructed the search query as a conjunction of main terms, incorporating synonyms and alternative spellings as OR-alternatives, and applied it to titles and abstracts:
\begin{quote}
(``image augmentation'' OR ``data augmentation'' OR ``image transformation'') AND\\
(``image retrieval'' OR ``image classification'' OR ``image recognition'' OR ``object detection'') AND\\
(``deep learning'' OR ``convolutional neural network'' OR ``neural network'' OR ``vision transformer'')
\end{quote}
We applied field restrictions to titles, abstracts, and keywords where supported by the database interface. We applied the same logical structure uniformly across all  databases to ensure consistency and reproducibility.

\subsection{Inclusion/Exclusion Criteria}

To ensure the relevance and quality of the studies selected for this review, we defined clear inclusion and exclusion criteria prior to the identification process. Only studies that included a quantitative evaluation of the impact of augmentation on model performance were considered to ensure the analysis was based on reproducible and measurable outcomes. Both journal articles and conference proceedings were included, provided that they presented sufficient experimental details, including dataset descriptions, model configurations, and evaluation metrics. Preprints were also considered when they included comprehensive methodological sections with replicable results.

We excluded studies with no empirical evaluation of the proposed augmentation techniques, that lacked adequate methodological information, such as missing performance metrics or unclear model descriptions, or if the content was not related to image augmentation in the deep learning context. We removed duplicates and non-English publications, as well as conference proceedings entries.

Furthermore, the review was limited to studies published within the last five years (January 2020 -- September 2025). This boundary is motivated by two landmark developments that define the modern augmentation landscape: Denoising Diffusion Probabilistic Models (DDPM) \cite{ho2020denoising}, published in 2020, made diffusion-based image synthesis practically viable for the first time, and Contrastive Language--Image Pretraining (CLIP) \cite{radford2021learning}, published in 2021, established the embedding paradigm that underpins the retrieval models evaluated in this study. Techniques predating these advances are less directly relevant to the research questions, which focus on augmentation in the context of deep embedding-based retrieval. To ensure complete coverage of all techniques within this window, all techniques identified in the 56 included papers were implemented in a unified evaluation framework, allowing direct comparison across categories under identical experimental conditions. A small number of foundational methods that predate this window (e.g., MixUp \cite{zhang2018mixupempiricalriskminimization}, CutMix, and Cutout) were nonetheless retained and evaluated because later techniques build directly on them; aside from these, excluding pre-2020 work remains a potential threat to completeness.

\subsection{Systematic Review Process}
The systematic review process was conducted following the PRISMA 2020 guidelines \cite{page2021prisma} to improve the transparency, completeness, and reproducibility of the review. 


Querying the databases led us to identify 88 studies. We identified 16 duplicates or conference proceedings, which we removed. Studies that passed these preliminary phases were subjected to a full-text review, during which each article was assessed against the predefined inclusion/exclusion criteria. 

The review process was conducted by three individuals between June and September 2025. To efficiently manage the review process, it was subdivided into two categories based on the publication date. The first subset included articles published in the years 2020 to 2022, while the other subset included articles published in the years 2023 to 2025. Any conflict in inclusion decisions were resolved through discussion among all three reviewers until consensus was reached. At the end of the process, 56 papers were included in the literature review. These 56 papers collectively span ten augmentation categories and provide the complete set of techniques evaluated in the empirical study (Section~\ref{sec:empirical_study_design}).

\section{Related Works and Background} \label{sec:background}

This section reviews relevant articles and delineates the technological principles that support the augmentation methods and assessment framework elaborated in later sections.
Several studies have investigated the impact of data augmentation techniques on model training and representation quality. Cui et al. \cite{3780338.3780785} analyzes the influence of different augmentation strategies on self-supervised contrastive learning and characterizes their effects through a mathematical error-bound formulation. Similarly, Huang et al. \cite{10.1007/978-3-031-19821-2_3} systematically evaluates multiple augmentation techniques to quantify the sensitivity of different self-supervised learning models to augmentation choices. The studies in \cite{purushwalkam2020demystifying,tian2020makes} further examine the role of augmentation in contrastive self-supervised learning and demonstrate that the selection of augmentation transformations can substantially affect the quality of learned representations. Lee et al. \cite{lee2021improving} investigates the effect of different augmentation techniques on representation-learning models, while Lai et al. \cite{lai2025enhancing} evaluates augmentation techniques within unsupervised sentence-embedding frameworks based on LLMs, with particular emphasis on data diversity and robustness to noise. In addition, Feng et al. \cite{feng2024geometry} studies the impact of textual data augmentation techniques on learned representations. 

This study aims to address two research deficiencies by assessing severe augmentation methods and datasets, as well as the absence of integrating other complementing measures. These two elements are crucial for delivering a comprehensive assessment of augmentation efficacy, encompassing facets that have been inadequately explored in earlier research. Moreover, the use of multiple datasets with diverse characteristics enables the analysis to capture augmentation behavior under different data distributions and application contexts, thereby supporting a broader assessment of the obtained findings. Several studies have investigated the impact of data augmentation techniques on model training and representation quality. In the remainder of this section, we discuss the terminology and background required for this work.

\textbf{Image Representation} Digital images are matrices of \textit{pixels}, where each pixel stores red, green, and blue (RGB) channel values in $[0,255]$; color may also be expressed in other spaces such as HSV \cite{7756307}. Photometric augmentations modify individual pixel values, whereas geometric augmentations rearrange their spatial positions.

\textbf{Image Transformations} An \textit{affine transformation} is a spatial mapping that preserves collinearity \cite{weisstein2004affine}, ensuring that parallel lines remain parallel after the transformation \cite{martin1982affine}. Flipping, rotating, shearing, and translating are all special cases of affine transformations and form the basis of geometric augmentations. \textit{Histogram Equalization (HE)} is a contrast-enhancement technique that redistributes pixel intensities to produce a roughly uniform histogram, making full use of the available brightness range \cite{gonzalez2009digital}.

\textbf{Deep Learning Architectures} Convolutional Neural Networks (CNNs) \cite{krizhevsky2012imagenet} and Vision Transformers (ViTs) \cite{han2022survey} are the two dominant image encoders; ViTs apply self-attention over image patches to capture long-range dependencies. {CNNs extract hierarchical spatial features through convolutional operations. While CNNs construct visual representations from local patterns to higher-level abstractions using predefined spatial inductive biases, and map them into embedding vectors. The patch embeddings are compared through the self-attention mechanism, which captures their contextual relationships in the learned feature space while positional embeddings preserve spatial information. The resulting representations are then refined by feed-forward networks and propagated through successive Transformer layers to extract increasingly high-level visual features.}


\textbf{Generative Models} A \textit{Generative Adversarial Network (GAN)} trains a generator and a discriminator in competition: applied to images, the generator synthesizes images from random noise while the discriminator tries to distinguish them from real images. At equilibrium the generator produces images the discriminator can no longer reliably tell from real data \cite{goodfellow2014generative}. \textit{Contrastive Language--Image Pretraining (CLIP)} trains aligned image and text encoders so that semantically similar visual and textual inputs are positioned closer together in a shared embedding space \cite{che2023enhancing}. \textit{OpenCLIP} is an open-source reproduction of CLIP trained on the LAION-400M and LAION-2B datasets rather than the proprietary WebImageText corpus used by OpenAI \cite{ilharco2021openclip}. \textit{Denoising Diffusion Probabilistic Models (DDPM)} generate images through a forward process that gradually adds Gaussian noise and a learned reverse process that iteratively removes it \cite{ho2020denoising}. \textit{Denoising Diffusion Implicit Models (DDIM)} extend DDPMs with a non-Markovian diffusion process that enables faster and more computationally efficient generation \cite{song2020denoising}.

\textbf{Domain Translation} {Domain-translation methods are augmentation techniques that transform images from one visual domain to another while preserving their semantic content and spatial structure. A domain refers to a collection of images that share common visual characteristics, such as illumination, season, texture, or artistic style. These techniques modify such domain-specific attributes while retaining the essential content of the original image. Style transfer changes visual properties such as texture, color, or artistic appearance while preserving the depicted objects and overall scene composition. Object transfiguration alters the appearance or category-specific attributes of an object while maintaining its pose and surrounding context. Season transfer modifies environmental conditions, such as converting a summer scene into a winter scene, without changing the underlying scene layout. In unpaired image-to-image translation, cycle-consistency constrains an image translated to the target domain and subsequently mapped back to the source domain to remain consistent with the original image \cite{zhu2017unpaired}.}
Motivated by these findings, the present study evaluates augmentation techniques across multiple datasets and assessment criteria. The evaluation incorporates several complementary metrics to provide a multidimensional characterization of augmentation effectiveness, including aspects that have received limited attention in previous studies. Moreover, the use of multiple datasets with diverse characteristics enables the analysis to capture augmentation behavior under different data distributions and application contexts, thereby supporting a broader assessment of the obtained findings.
Several studies have investigated the impact of data augmentation techniques on model training and representation quality.  
In the remainder of this section, we discuss the terminology and background required for this work.

\textbf{Image Representation} Digital images are matrices of \textit{pixels}, where each pixel stores red, green, and blue (RGB) channel values in $[0,255]$; color may also be expressed in other spaces such as HSV \cite{7756307}. Photometric augmentations modify individual pixel values, whereas geometric augmentations rearrange their spatial positions.

\textbf{Image Transformations} An \textit{affine transformation} is a spatial mapping that preserves collinearity \cite{weisstein2004affine}, ensuring that parallel lines remain parallel after the transformation \cite{martin1982affine}. Flipping, rotating, shearing, and translating are all special cases of affine transformations and form the basis of geometric augmentations. \textit{Histogram Equalization (HE)} is a contrast-enhancement technique that redistributes pixel intensities to produce a roughly uniform histogram, making full use of the available brightness range \cite{gonzalez2009digital}.

\textbf{Deep Learning Architectures} Convolutional Neural Networks (CNNs) \cite{krizhevsky2012imagenet} and Vision Transformers (ViTs) \cite{han2022survey} are the two dominant image encoders; ViTs apply self-attention over image patches to capture long-range dependencies. {CNNs extract hierarchical spatial features through convolutional operations. While CNNs construct visual representations from local patterns to higher-level abstractions using predefined spatial inductive biases, and map them into embedding vectors. The patch embeddings are compared through the self-attention mechanism, which captures their contextual relationships in the learned feature space while positional embeddings preserve spatial information. The resulting representations are then refined by feed-forward networks and propagated through successive Transformer layers to extract increasingly high-level visual features.}


\textbf{Generative Models} A \textit{Generative Adversarial Network (GAN)} trains a generator and a discriminator in competition: applied to images, the generator synthesizes images from random noise while the discriminator tries to distinguish them from real images. At equilibrium the generator produces images the discriminator can no longer reliably tell from real data \cite{goodfellow2014generative}. \textit{Contrastive Language--Image Pretraining (CLIP)} trains aligned image and text encoders so that semantically similar visual and textual inputs are positioned closer together in a shared embedding space \cite{che2023enhancing}. \textit{OpenCLIP} is an open-source reproduction of CLIP trained on the LAION-400M and LAION-2B datasets rather than the proprietary WebImageText corpus used by OpenAI \cite{ilharco2021openclip}. \textit{Denoising Diffusion Probabilistic Models (DDPM)} generate images through a forward process that gradually adds Gaussian noise and a learned reverse process that iteratively removes it \cite{ho2020denoising}. \textit{Denoising Diffusion Implicit Models (DDIM)} extend DDPMs with a non-Markovian diffusion process that enables faster and more computationally efficient generation \cite{song2020denoising}.

\textbf{Domain Translation} {Domain-translation methods are augmentation techniques that transform images from one visual domain to another while preserving their semantic content and spatial structure. A domain refers to a collection of images that share common visual characteristics, such as illumination, season, texture, or artistic style. These techniques modify such domain-specific attributes while retaining the essential content of the original image. Style transfer changes visual properties such as texture, color, or artistic appearance while preserving the depicted objects and overall scene composition. Object transfiguration alters the appearance or category-specific attributes of an object while maintaining its pose and surrounding context. Season transfer modifies environmental conditions, such as converting a summer scene into a winter scene, without changing the underlying scene layout. In unpaired image-to-image translation, cycle-consistency constrains an image translated to the target domain and subsequently mapped back to the source domain to remain consistent with the original image \cite{zhu2017unpaired}.}


\textbf{Embedding-Based Retrieval} Given a query image $q$, an \textit{embedding-based retrieval system} maps $q$ to a vector $\mathbf{v}_q$ in a learned feature space and returns the $k$ database images whose embeddings are closest to $\mathbf{v}_q$ under a distance metric such as cosine similarity or Euclidean distance. Correct retrieval requires that semantically equivalent images, regardless of viewpoint, lighting, or minor corruption, remain close in embedding space. 

\textbf{Metamorphic Testing} \cite{chen2018metamorphic} is a software testing technique that verifies the correctness of a program by checking whether its inputs and outputs satisfy expected relationships (called metamorphic relations) between multiple executions, rather than relying on a known correct output (a test oracle). A semantics-preserving augmentation therefore constitutes a \textit{metamorphic oracle} \cite{chen2018metamorphic}: the top-$k$ results for an augmented query should belong to the same category as the original query, and any deviation constitutes a detectable fault without requiring a human-labelled ground truth.

\section{Augmentation Techniques}
\label{sec:augmentation_techniques}

In the following sections (sections \ref{sec:geometric} to \ref{sec:adversarial}), we discuss all the augmentation techniques identified through the systematic literature in groups as per our taxonomy (Section \ref{subsec:taxonomy}). We also mention their benefits, as discovered during the SLR, when used in deep learning. Table \ref{tab:all_techniques} provides a summary, including where in the 56 primary studies augmentation techniques have been used. Finally, each technique is illustrated by the original source image together with its corresponding augmented output.

\subsection{Taxonomy}
\label{subsec:taxonomy}
\begin{figure*}[!htbp]
  \centering
  \includegraphics[width=\textwidth]{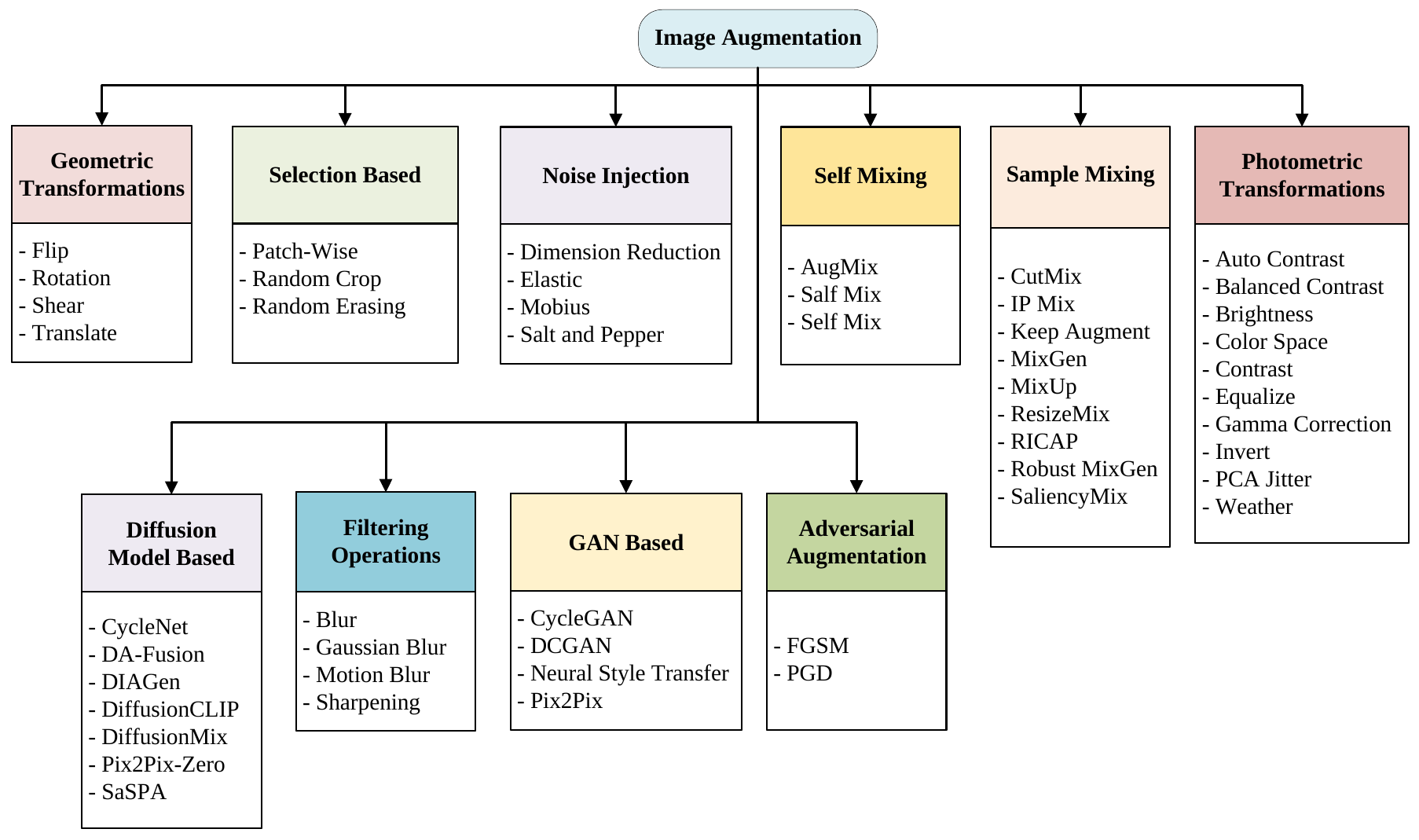} 
  \caption{Taxonomy of image data augmentation techniques.}
  \label{fig:taxonomy}
\end{figure*}

Figure~\ref{fig:taxonomy} presents our classification of image augmentation techniques into ten main categories. The techniques were grouped based on their underlying transformation principles or the specific image features they aim to modify. Note that GAN-based and diffusion-based techniques are presented as separate categories in this taxonomy (reflecting their distinct generative mechanisms), though they are grouped together in some prior surveys. 

Geometric augmentations apply spatial transformations to an image. Photometric augmentations modify the individual pixel values of the image. Selection-based techniques selectively add, remove, or emphasize regions of an image. Filtering applies convolutional filters to blur or sharpen an image. Noise injection introduces random perturbations to simulate real-world corruptions. Mixing-based augmentations combine multiple inputs into a single image where self-mixing operates on different regions of the same image, and sample mixing combines regions from multiple images.  GAN-based augmentation generates entirely new images through adversarial training. Diffusion-based augmentation generates and manipulates images through a learned denoising process, enabling fine-grained text- and image-guided synthesis. Adversarial augmentation crafts worst-case, gradient-guided perturbations that remain perceptually close to the original image. 


\subsection{Geometric Transformations}
\label{sec:geometric}
A geometric augmentation (flipping, rotating, shearing, and translating) alters the spatial configuration of an image by modifying its geometric attributes, such as position, orientation, and aspect ratio \cite{imageaugtech2024image}. They are the most common augmentation techniques after being popularized by Krizhevsky et al. \cite{krizhevsky2012imagenet} which we attribute to their simplicity. 

\begin{figure*}[!htbp]
    \centering
    \subfloat[\footnotesize Original \label{fig:original}]{\includegraphics[width=0.18\textwidth]{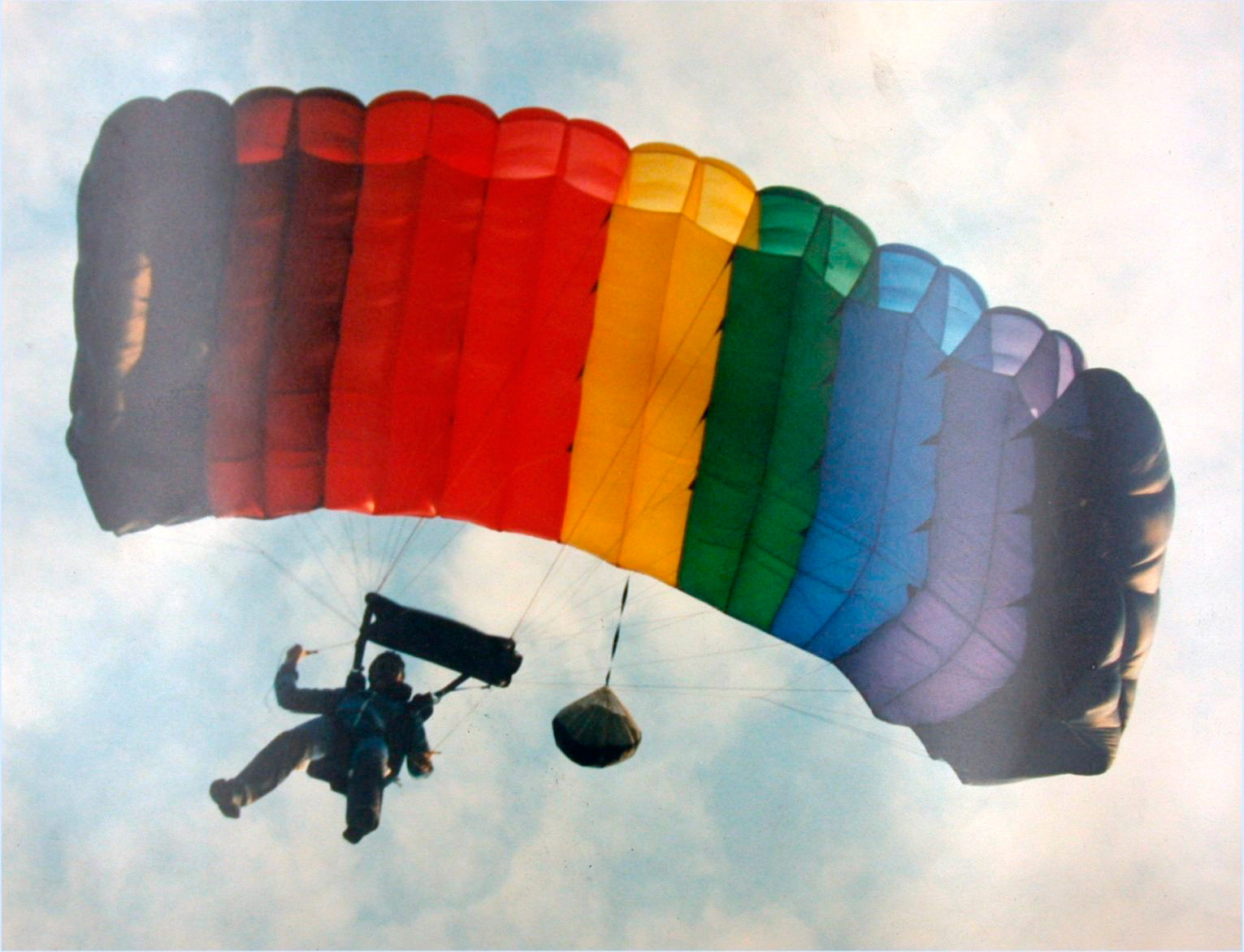}}
    \hfill
    \subfloat[\footnotesize Flip \label{fig:flipped}]{\includegraphics[width=0.18\textwidth]{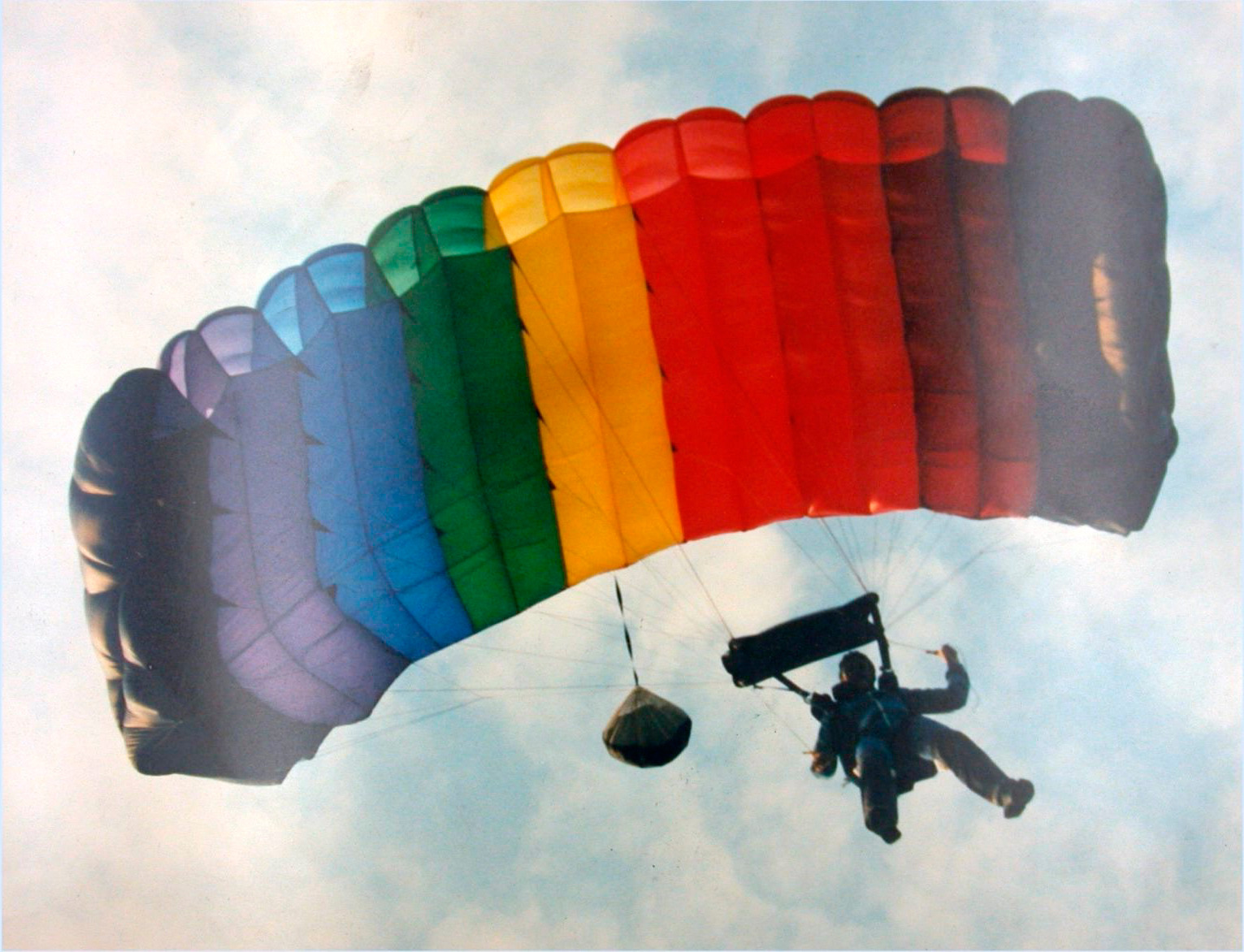}}
    \hfill
    \subfloat[\footnotesize Rotate \label{fig:rotated}]{\includegraphics[width=0.18\textwidth]{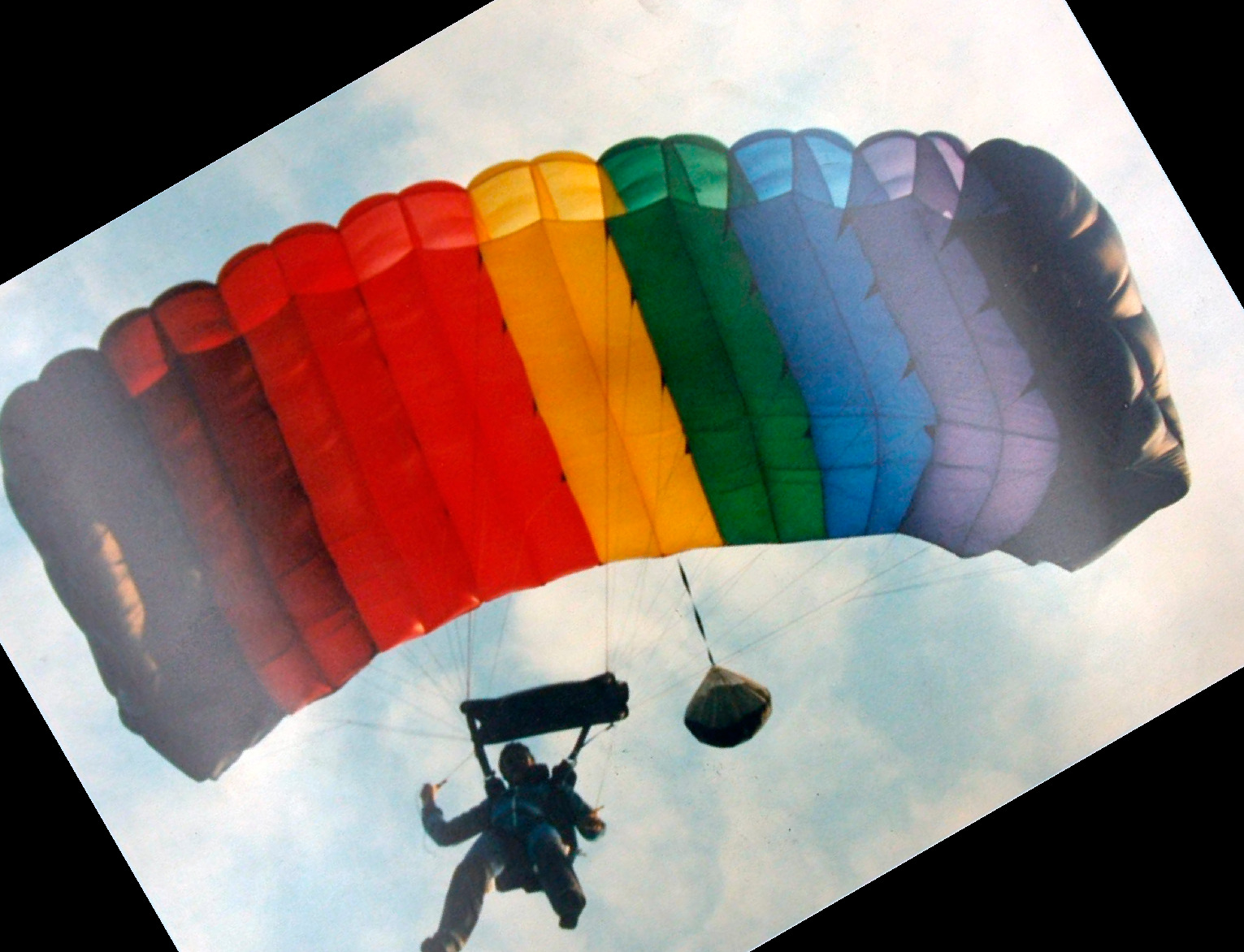}}
    \hfill
    \subfloat[\footnotesize Shear \label{fig:sheared}]{\includegraphics[width=0.18\textwidth]{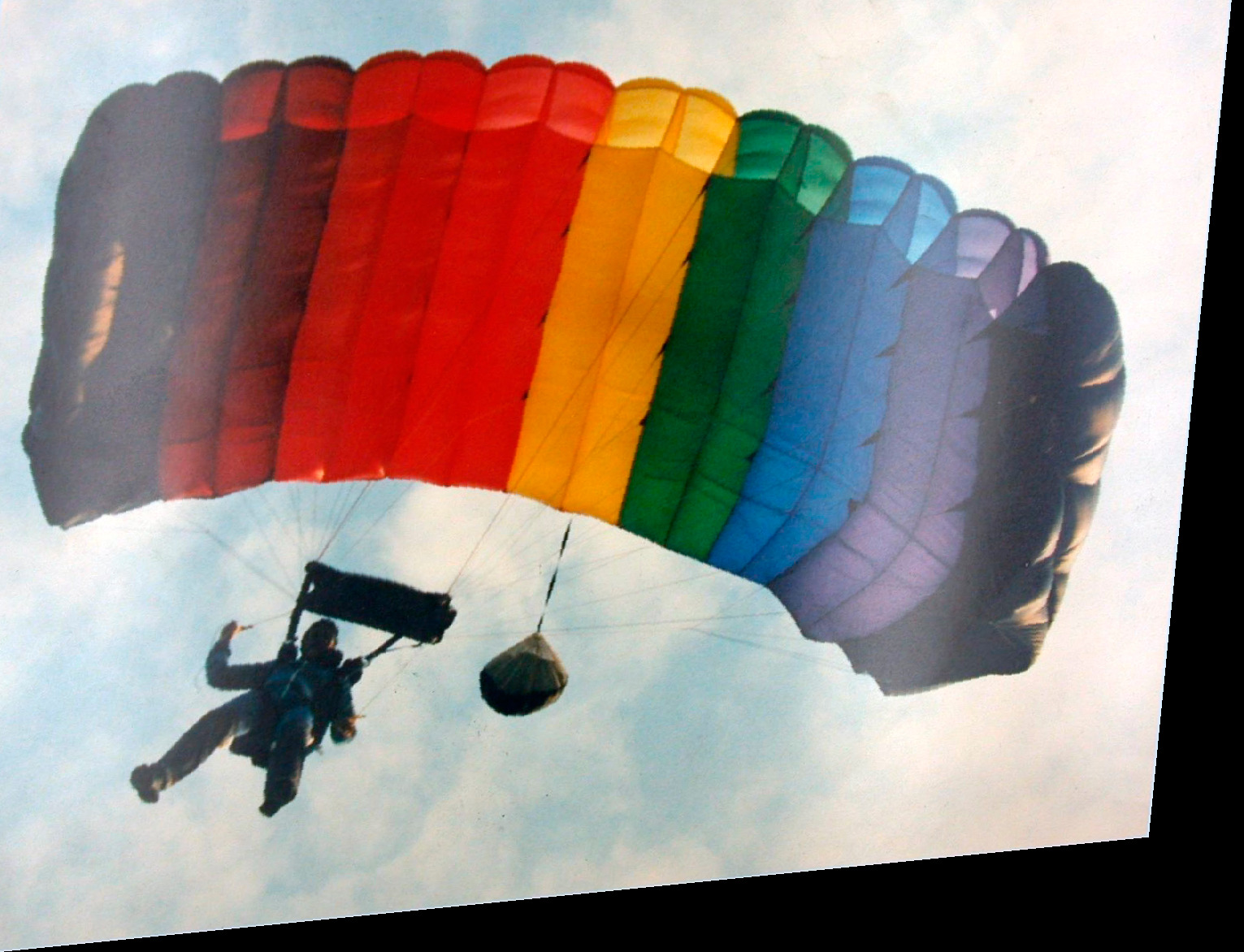}}
    \hfill
    \subfloat[\footnotesize Translate \label{fig:translated}]{\includegraphics[width=0.18\textwidth]{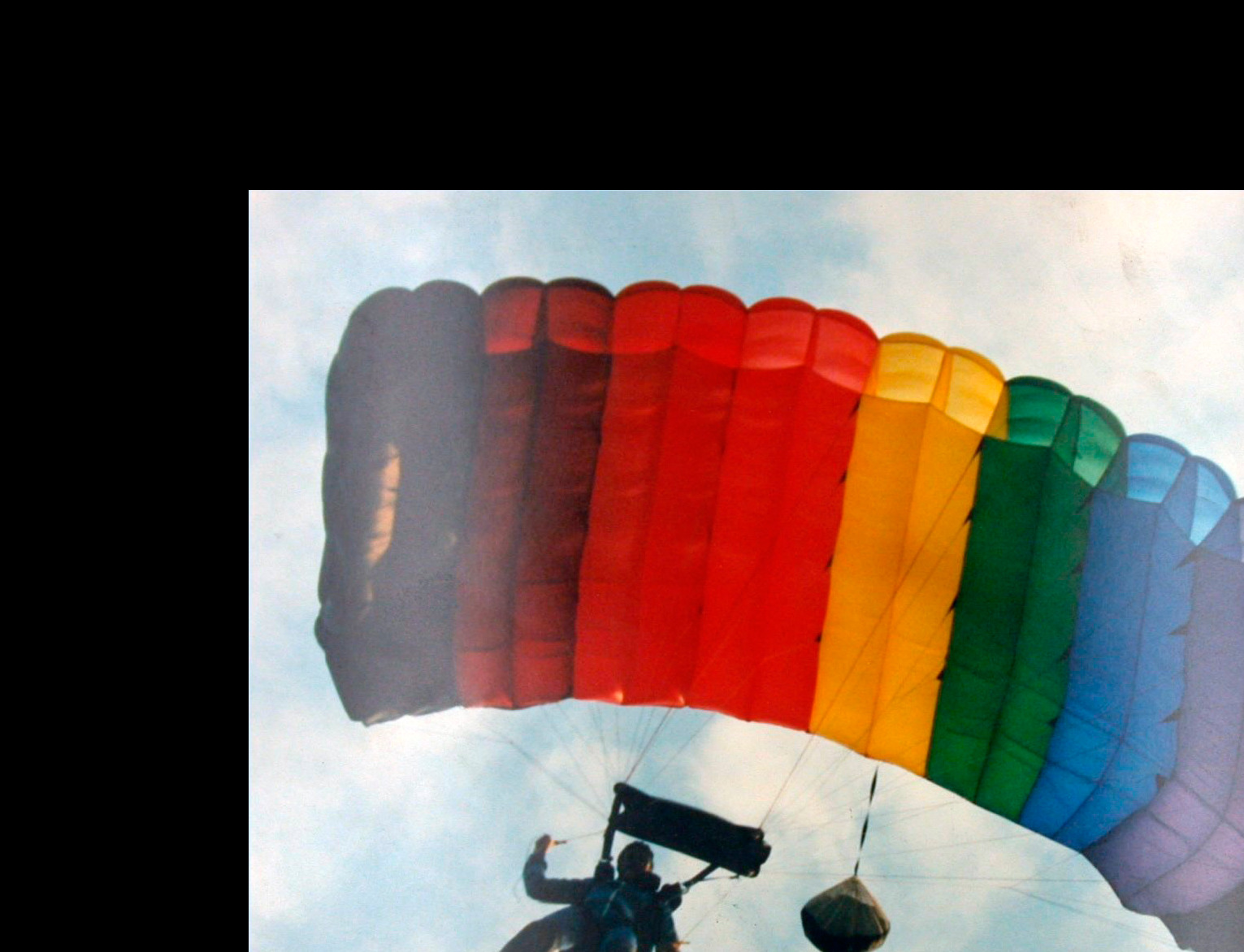}}
    \caption{(a) unmodified, (b) flipped horizontally, (c) rotated 30$^\circ$ counterclockwise, (d) sheared horizontally and vertically with a factor of 0.2 and (e) translated 20\% to the bottom right.}
    \label{fig:geometric_augmentations}
\end{figure*}

Although simple geometric transformations may appear negligible to the human eye, they significantly alter an image's underlying pixel structure, leading models to perceive transformed images as fundamentally different \cite{9596262}. While geometric transformations preserve the original structure of the image, excessive or domain-invalid augmentations can impair image recognition and introduce inaccuracies. Excessive shearing or translating will lead to significant image distortion or move key features out of frame, degrading model performance  \cite{Abdollahi2020}. Furthermore, in datasets where spatial orientation is semantically important, flipping or rotating can lead the model to develop inaccuracies (e.g., \cite{9596262, khosla2020enhancing}).  The visual effect of each technique is illustrated  in Figure \ref{fig:geometric_augmentations}.

\subsection{Photometric Transformations}
\label{sec:photometric}
\begin{figure*}[!htbp]

    \centering
    \subfloat[\footnotesize Auto-Contrast \label{fig:autocontrast}]{\includegraphics[width=0.18\textwidth]{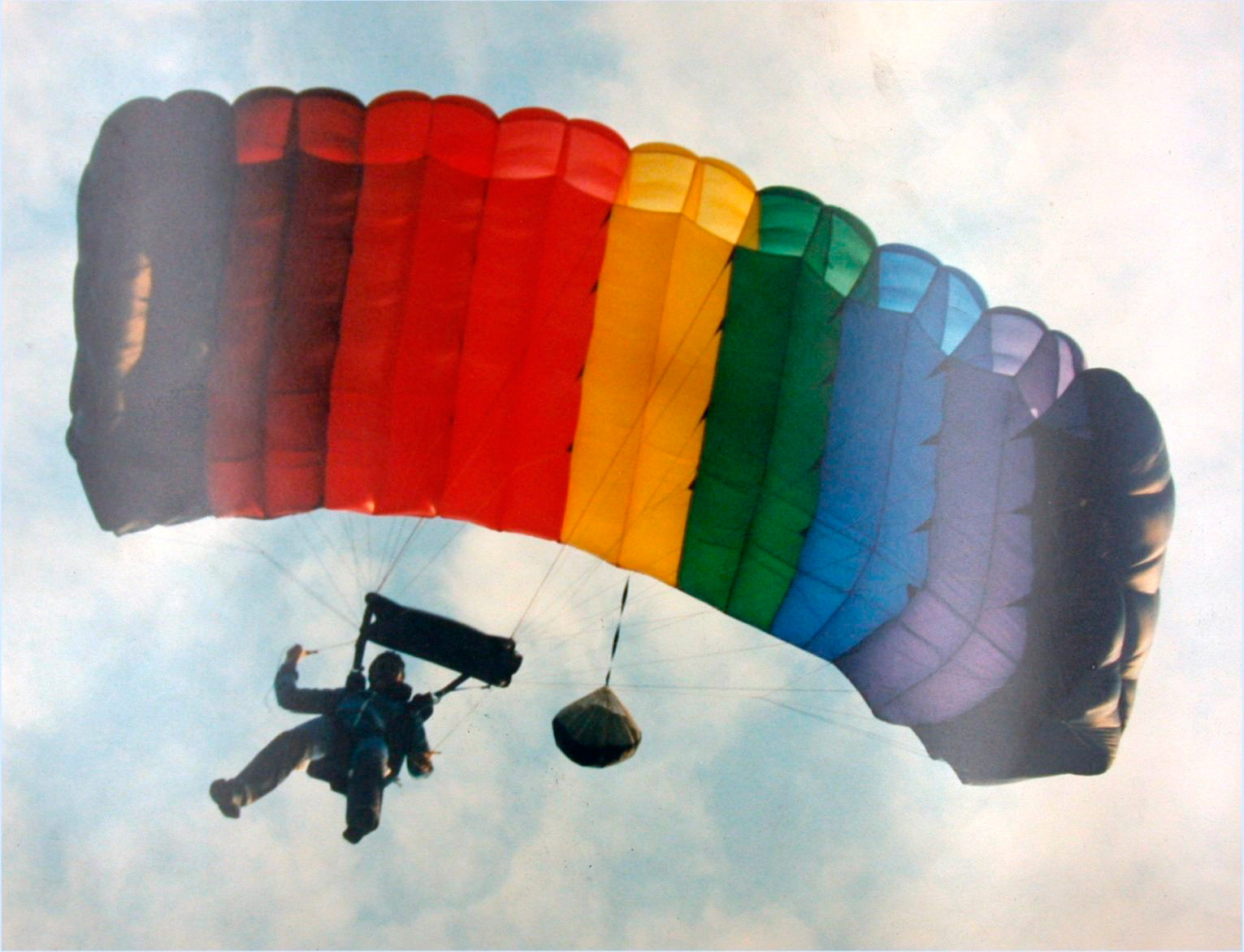}}
    \hfill
    \subfloat[\footnotesize BCET \label{fig:clahe}]{\includegraphics[width=0.18\textwidth]{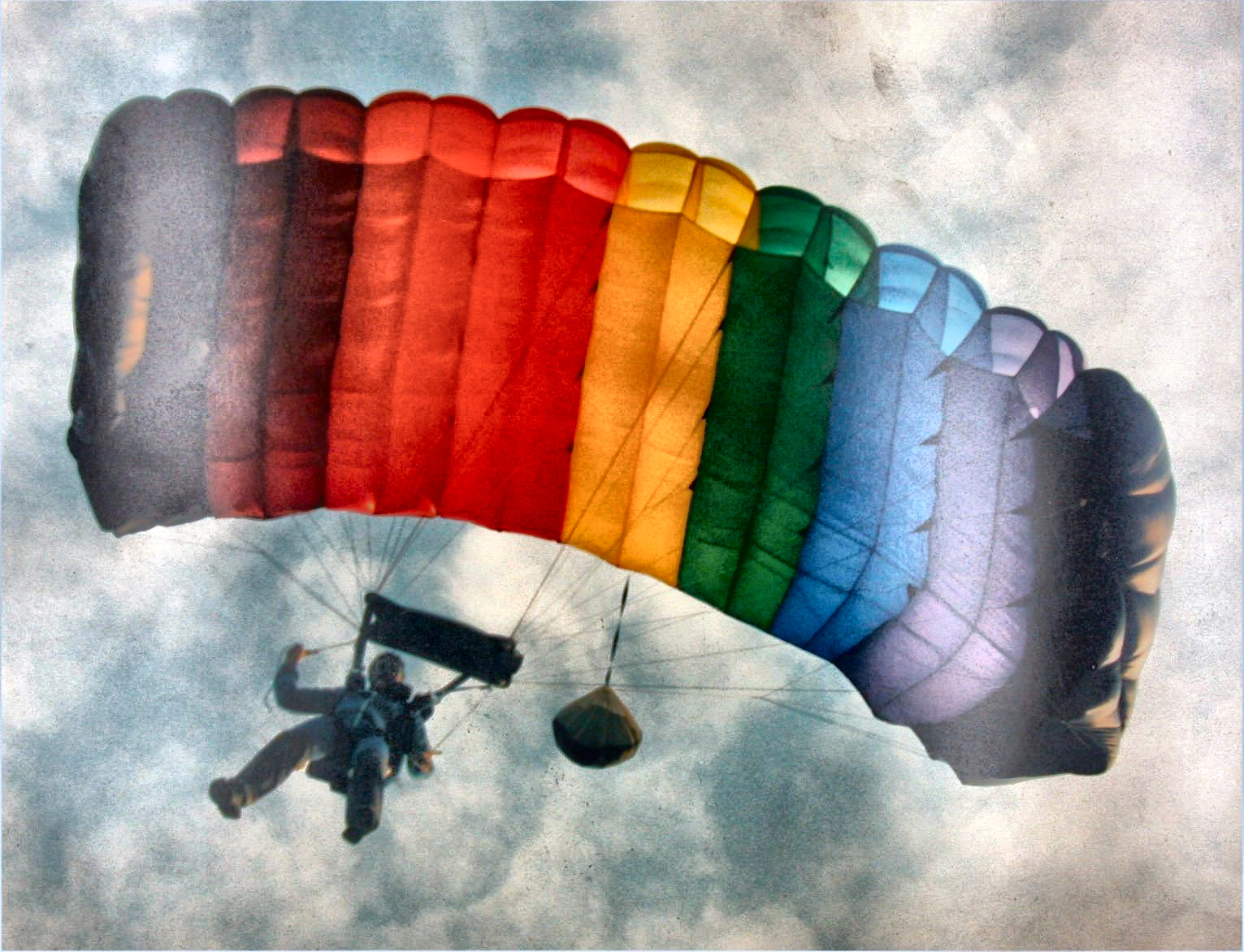}}
    \hfill
    \subfloat[\footnotesize Brightness \label{fig:brightness}]{\includegraphics[width=0.18\textwidth]{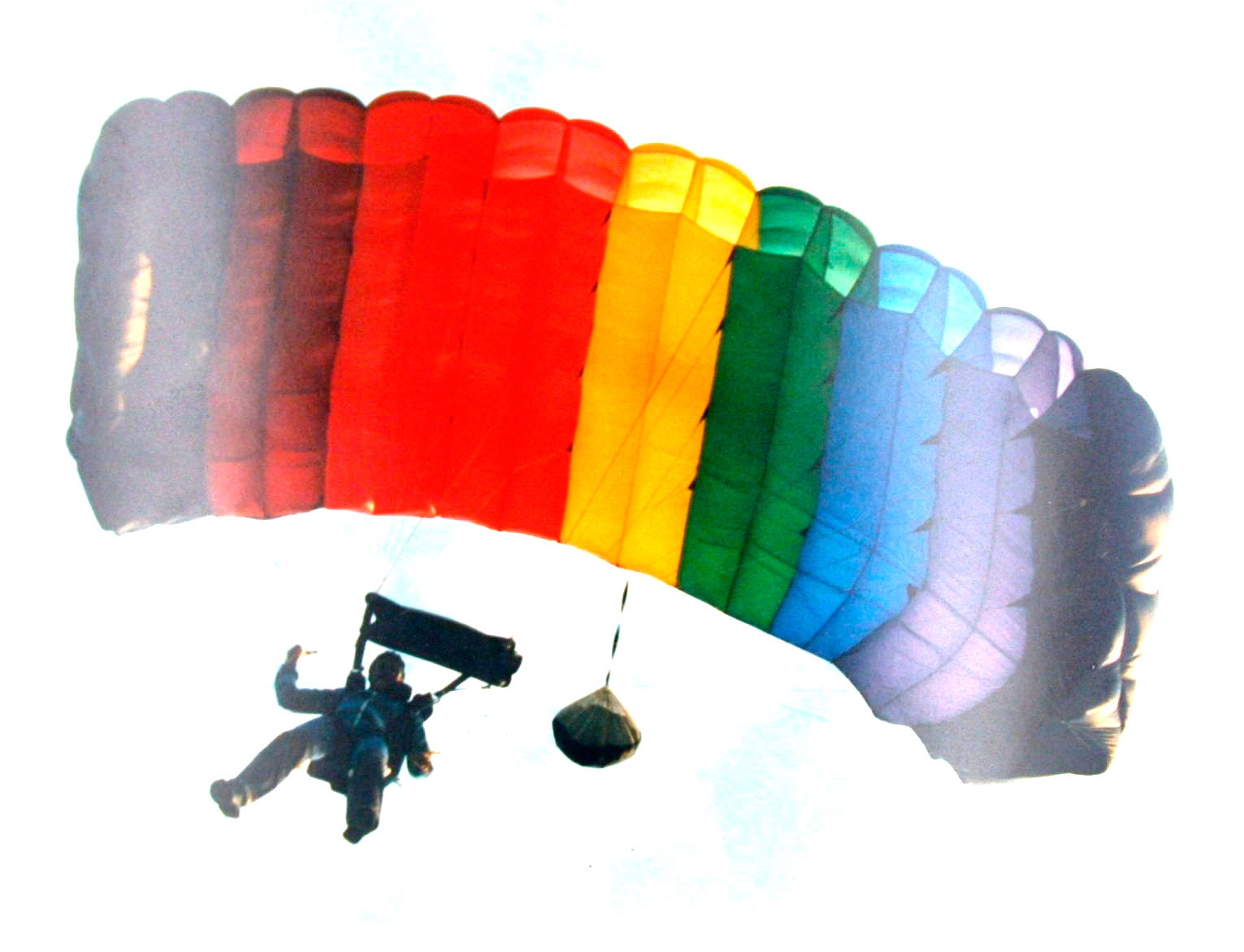}}
    \hfill
    \subfloat[\footnotesize Color Space \label{fig:colorspace}]{\includegraphics[width=0.18\textwidth]{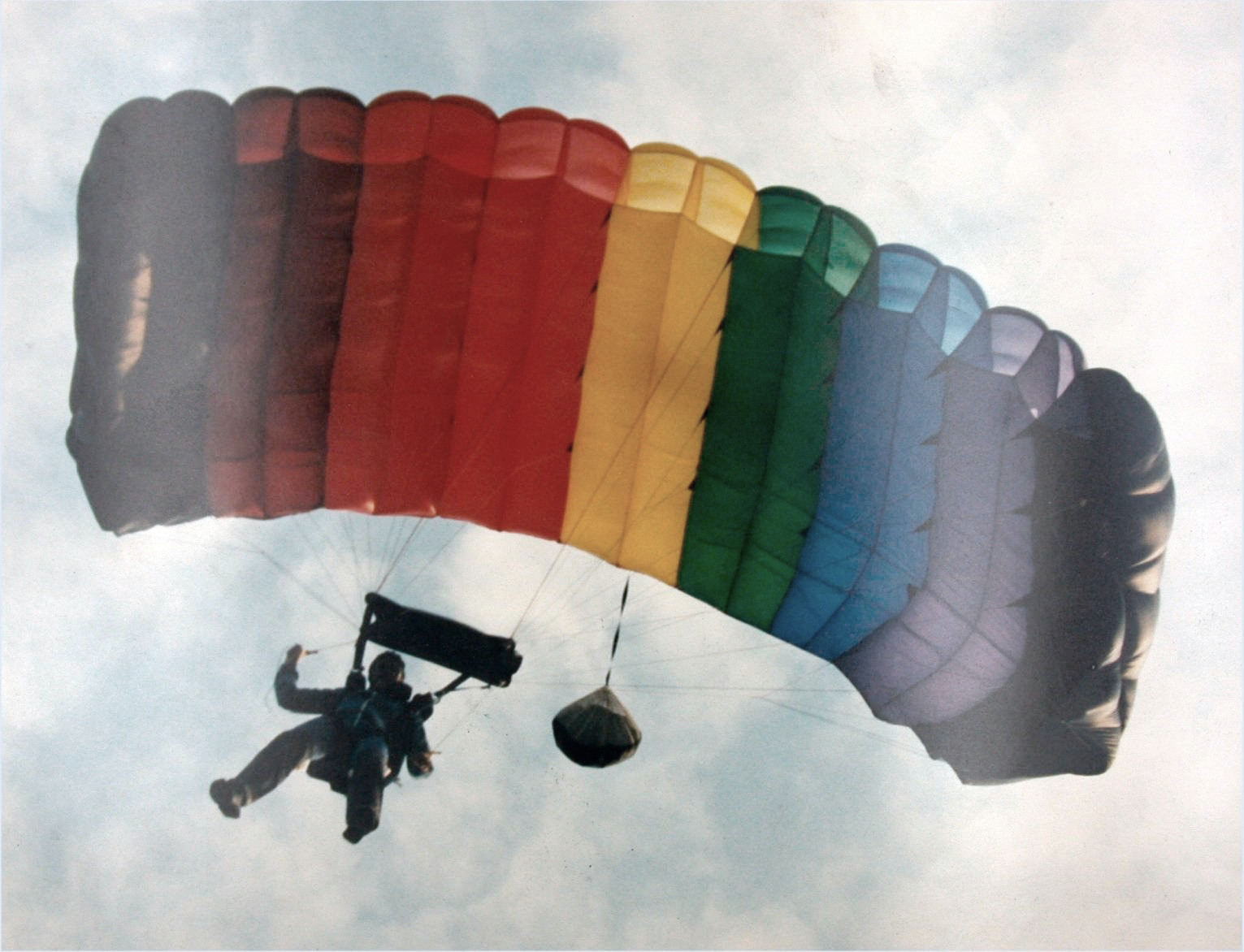}}
    \subfloat[\footnotesize Contrast \label{fig:contrast}]{\includegraphics[width=0.18\textwidth]{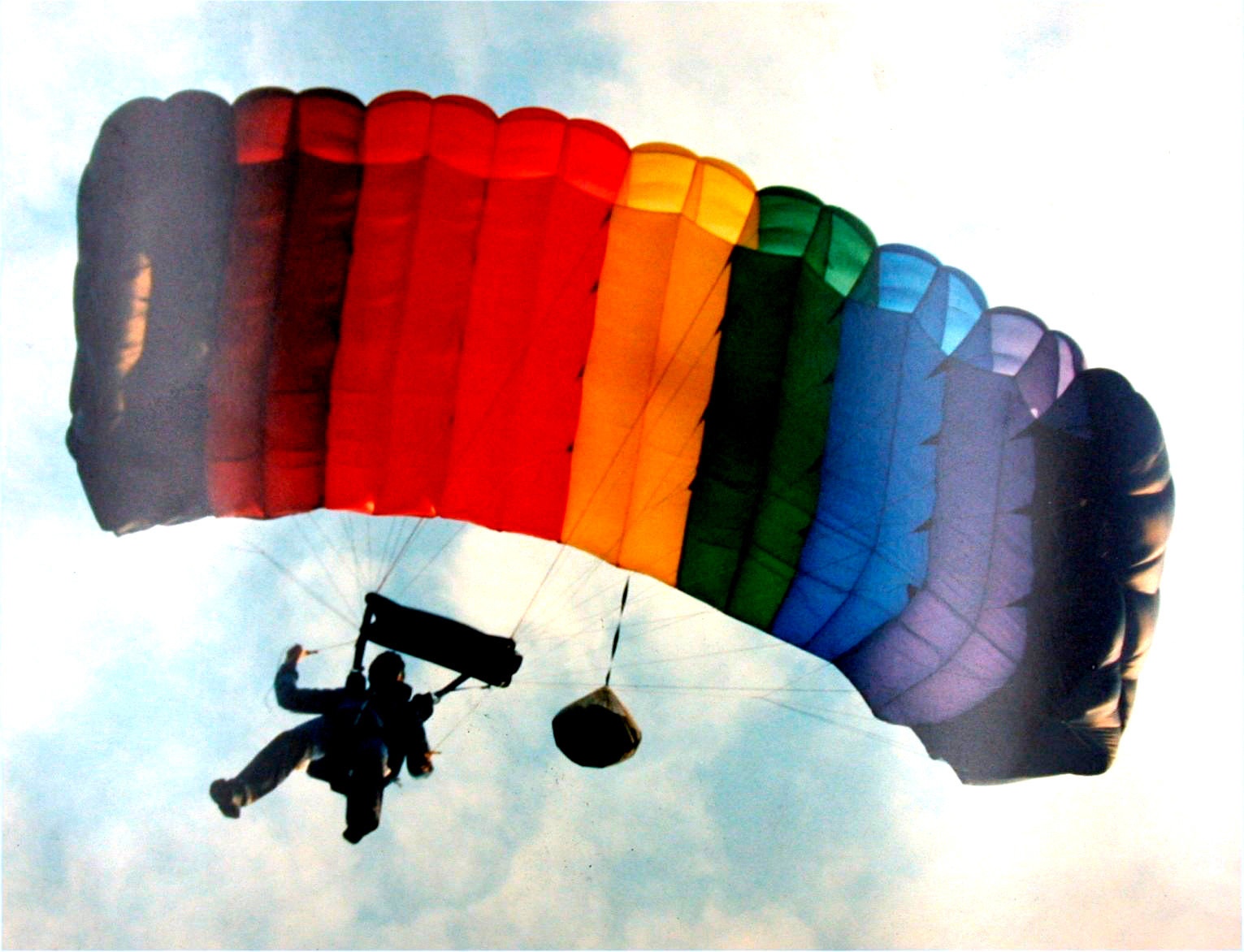}}
    \hfill
    \subfloat[\footnotesize Equalization \label{fig:equalize}]{\includegraphics[width=0.18\textwidth]{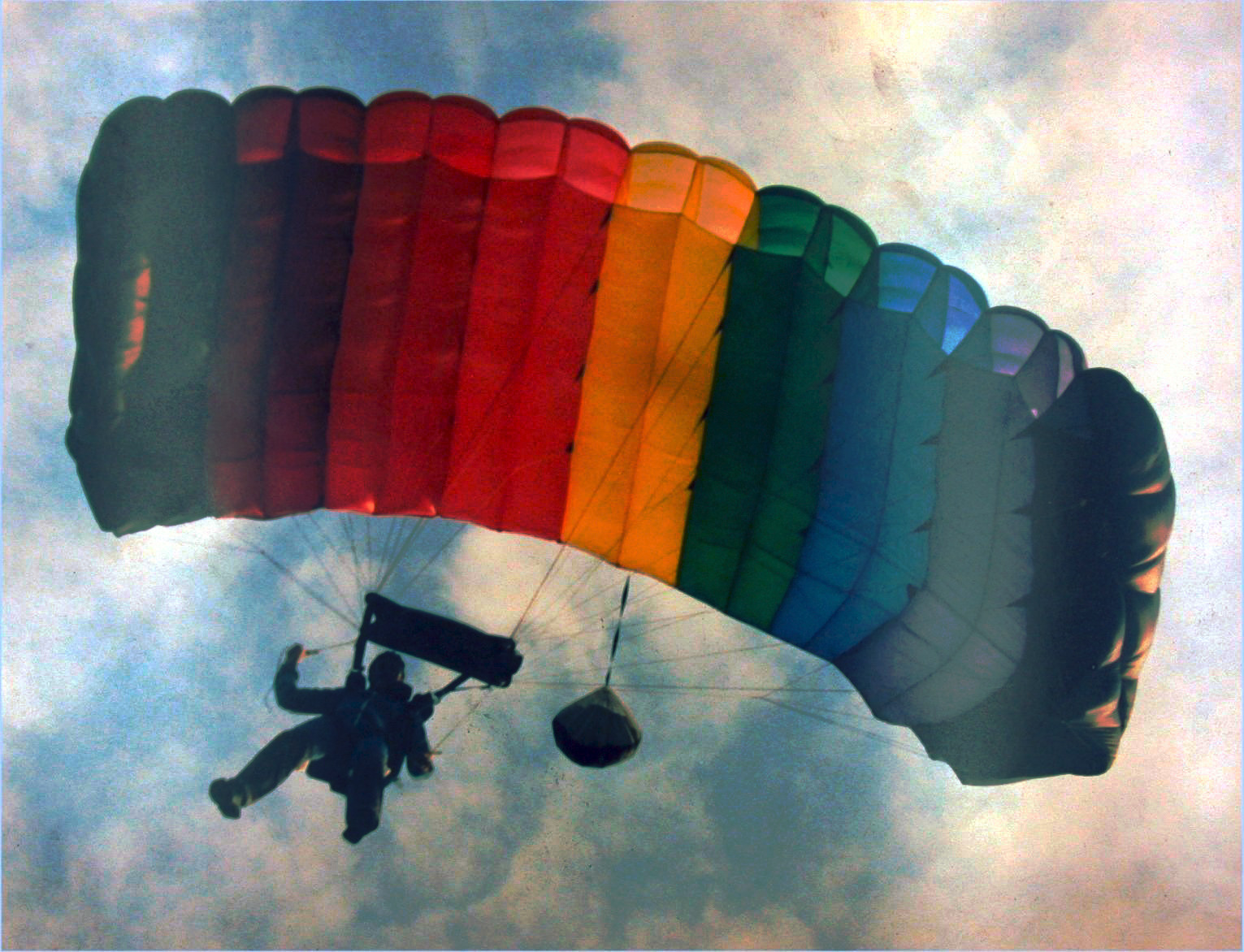}}
    \hfill
    \subfloat[\footnotesize Gamma Correction \label{fig:gamma}]{\includegraphics[width=0.18\textwidth]{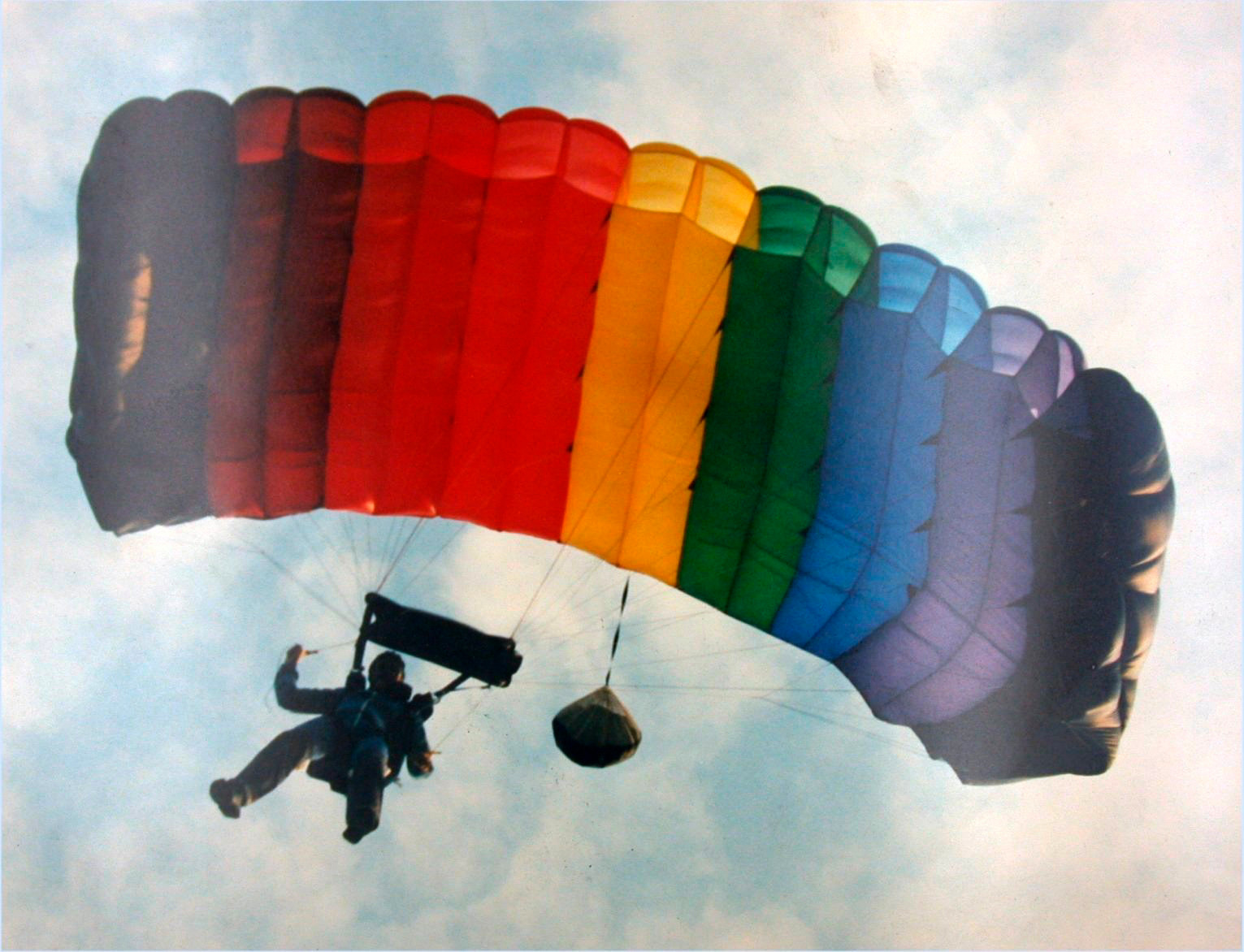}}
    \hfill
    \subfloat[\footnotesize Inversion \label{fig:invert}]{\includegraphics[width=0.18\textwidth]{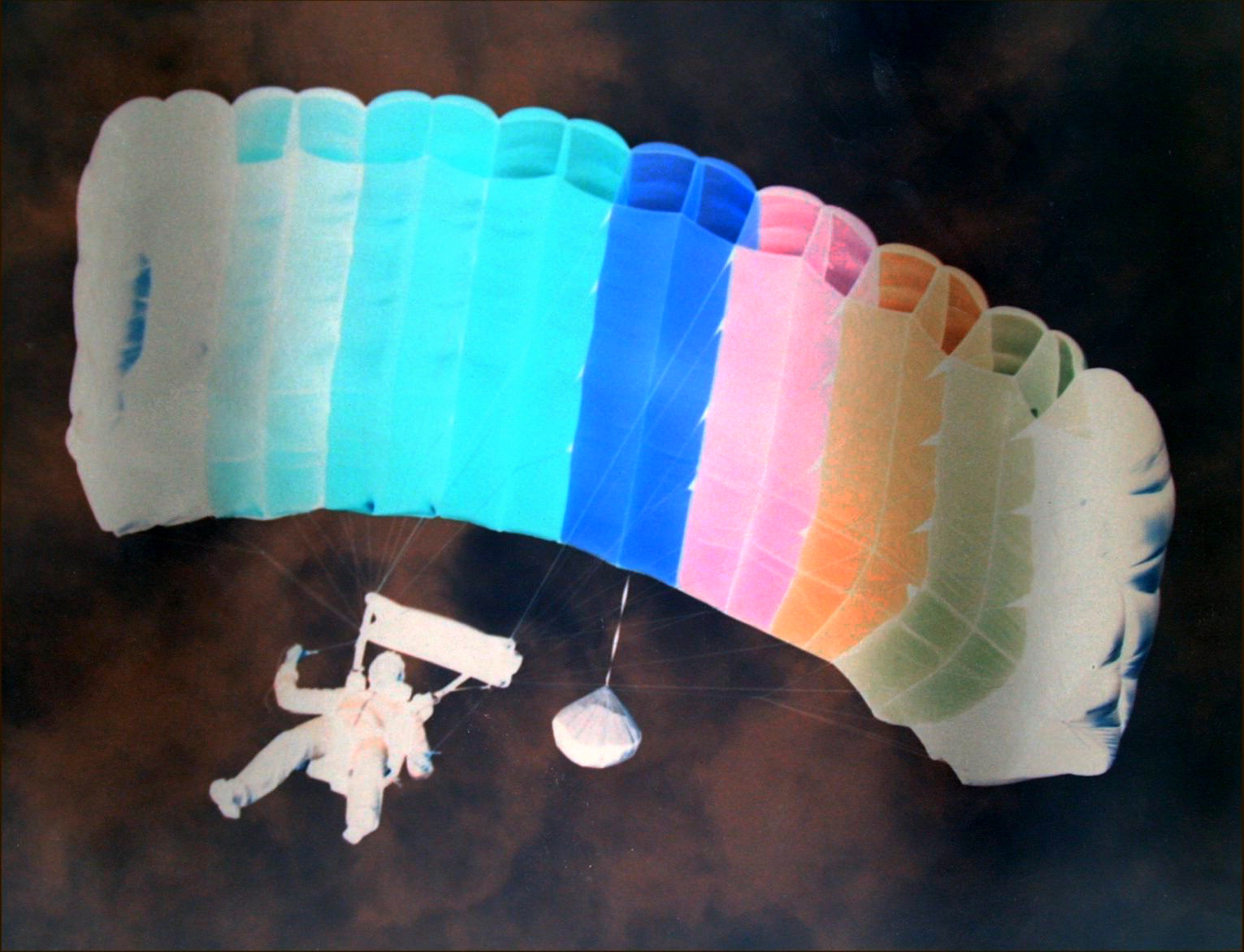}}
    \hfill
    \subfloat[\footnotesize PCA Jitter \label{fig:pcajitter}]{\includegraphics[width=0.18\textwidth]{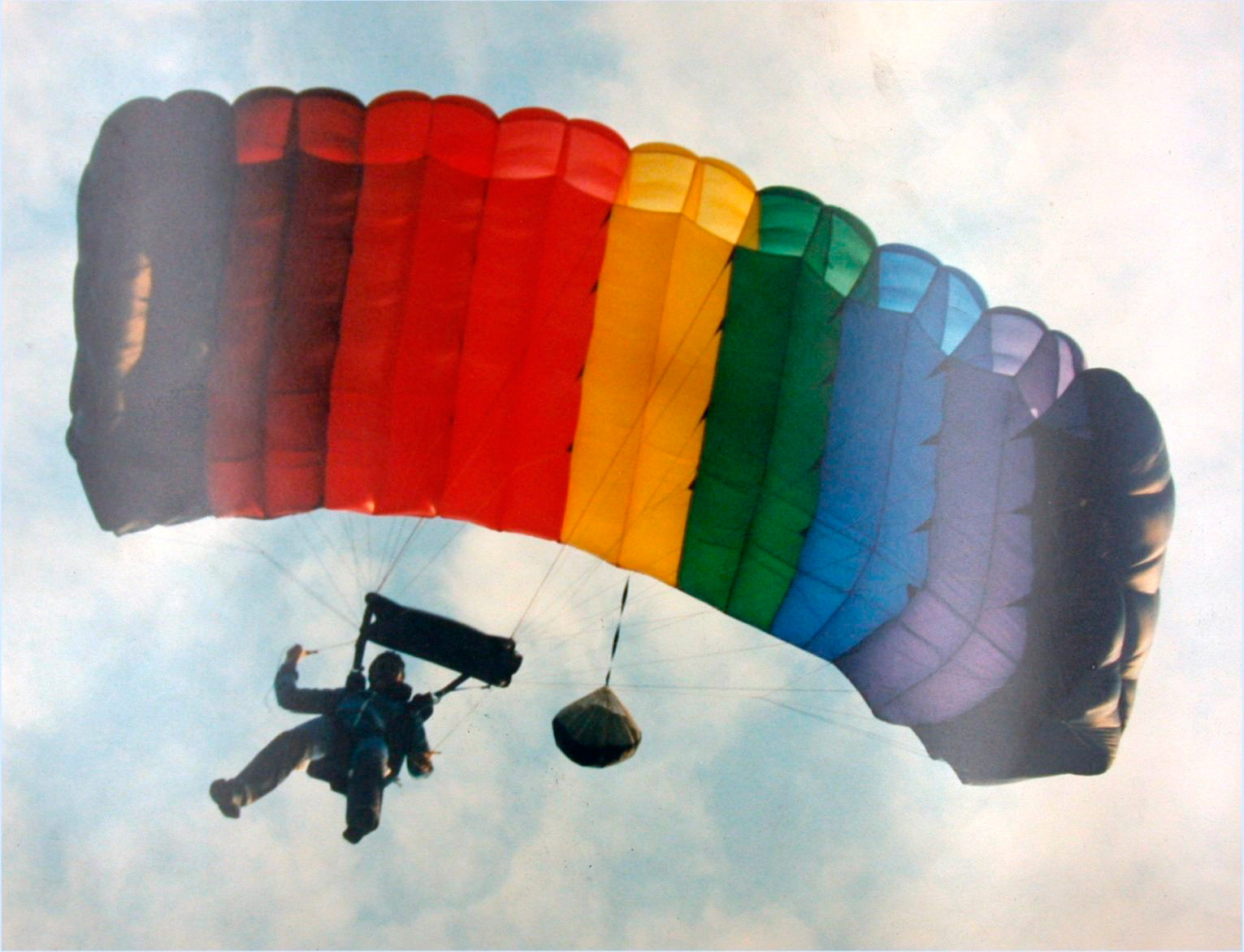}}
    \hfill
    \subfloat[\footnotesize Weather \label{fig:weather_image}]{\includegraphics[width=0.18\textwidth]{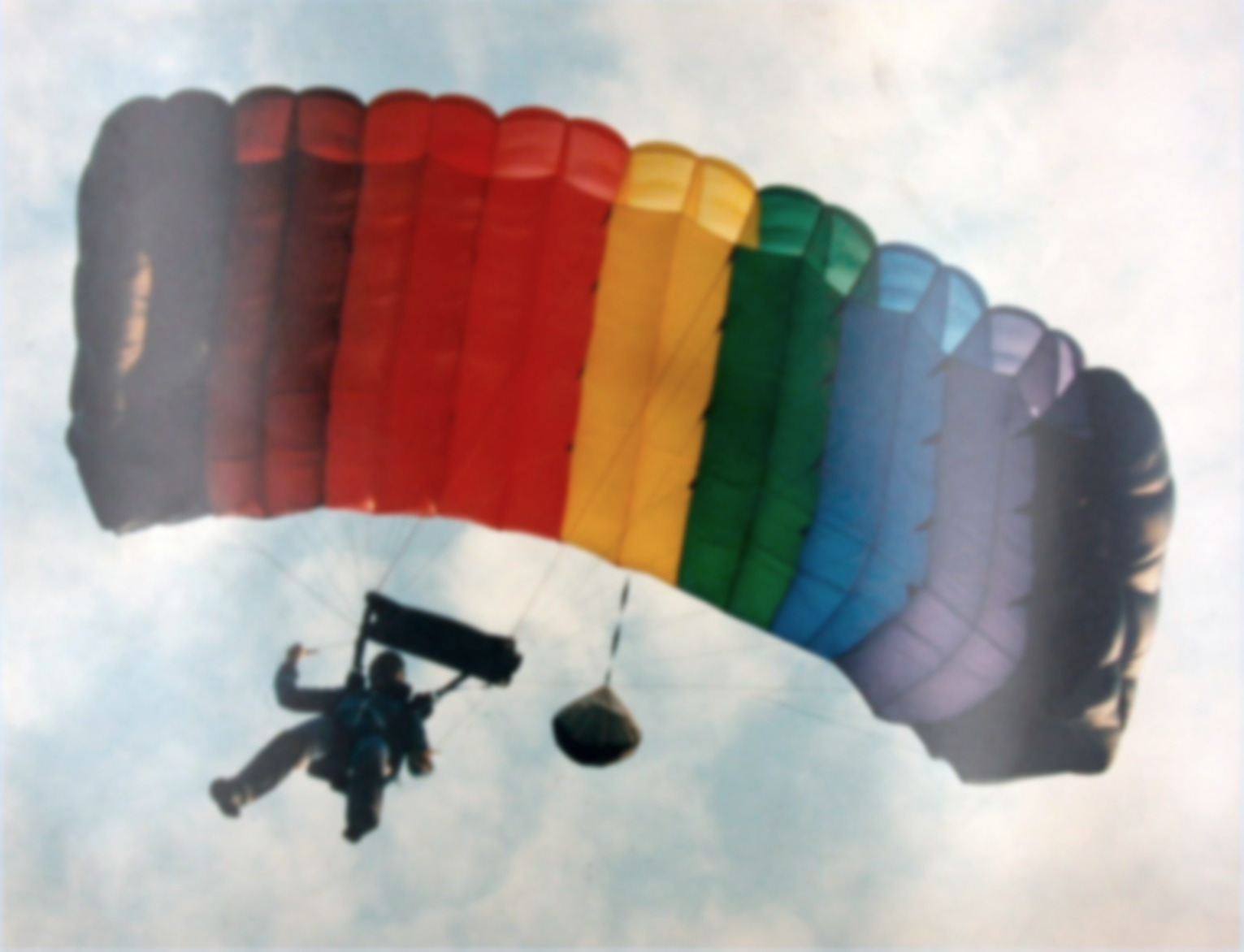}}
    
    \caption{({a) Auto-contrast adjustment. (b) Balanced contrast adjustment. (c) Brightness increase by 50\%. (d) Reduced color saturation by 50\%. (e) Increased contrast by 50\%. (f) Histogram equalization. (g) Gamma correction adjustment. (h) Inverted image. (i) PCA color jitter adjustment. (j) Weather augmentation by 50\%}}
    \label{fig:photometric_augmentation}
\end{figure*}

Photometric augmentation involves altering the visual appearance of an image by modifying its pixel intensity values while keeping its spatial structure intact. These transformations target the radiometric properties of images, such as brightness, contrast, color and illumination by adjusting pixel values across one or more channels to simulate real-world variability in lighting conditions, sensor characteristics or environmental effects. Unlike geometric transformations which affect the positional arrangement of pixels, photometric augmentations modify how objects in the image are perceived in terms of visual tone and color. This makes them particularly effective in tasks where lighting inconsistencies or poor exposure can hinder model performance, such as outdoor object detection, medical imaging or low-light scene recognition. The visual effect of each technique is illustrated in Figure \ref{fig:photometric_augmentation}.

\subsection{Noise Injection}
\label{sec:noise} 
\begin{figure*}[!htbp]
    \centering
    \subfloat[\footnotesize Original \label{fig:original_noise}]{\includegraphics[width=0.18\textwidth]{images/original_image.jpg}}
    \hfill
    \subfloat[\footnotesize Elastic Transformation \label{fig:elastic}]{\includegraphics[width=0.18\textwidth]{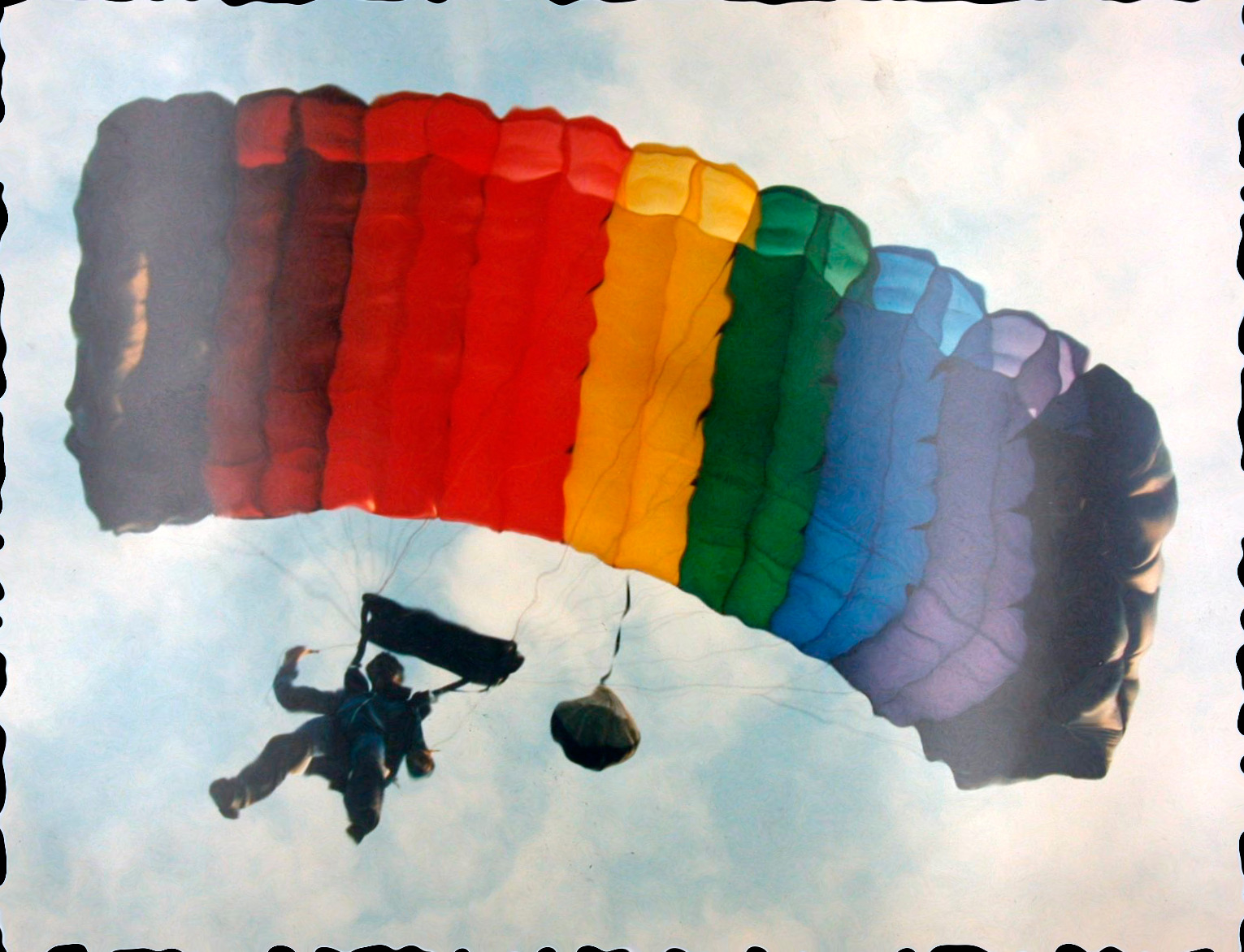}}
    \hfill
    \subfloat[\footnotesize Salt and Pepper Noise \label{fig:salt_pepper}]{\includegraphics[width=0.18\textwidth]{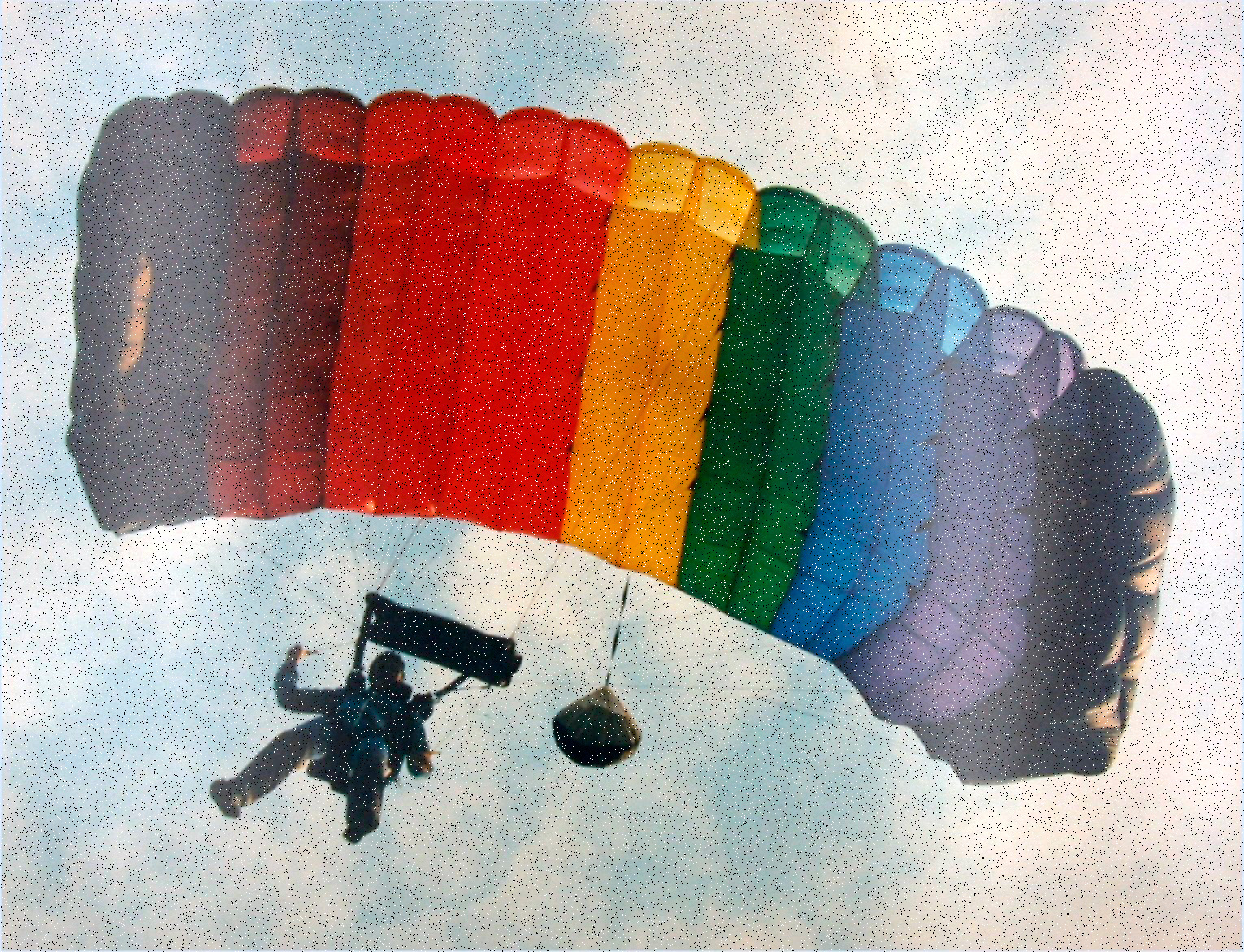}}
    \hfill
    \subfloat[\footnotesize Dimension Reduction \label{fig:feat_space}]{\includegraphics[width=0.18\textwidth]{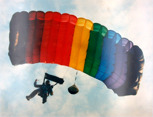}}
    \hfill
    \subfloat[\footnotesize Mobius Transformation \label{fig:mobius}]{\includegraphics[width=0.18\textwidth]{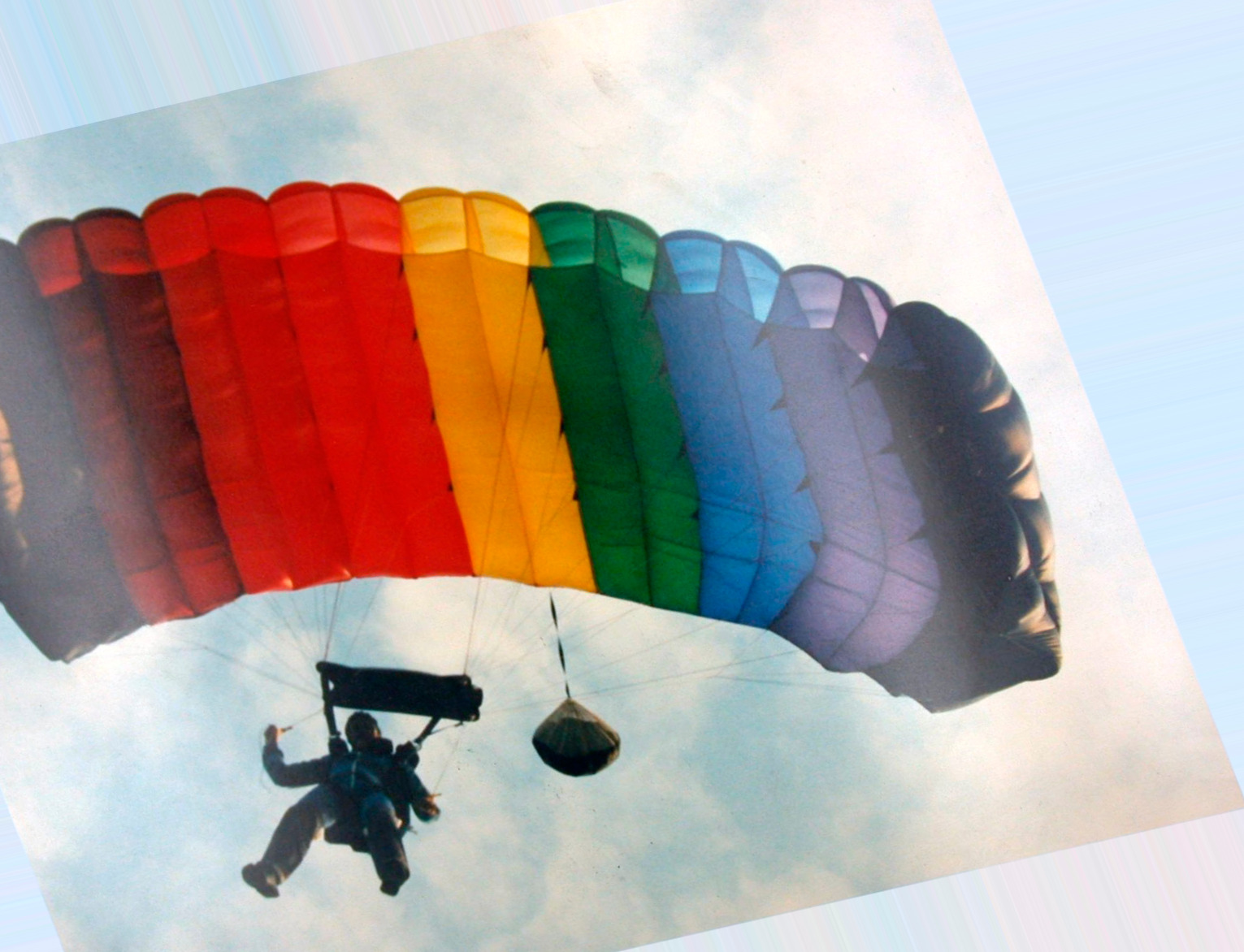}}
    \caption{(a) unmodified, (b) elastically transformed, (c) added 10\% salt and pepper noise to make the image grainier, (d) reduced to 40\% of the original size, (e) Möbius transformation applied.}
    \label{fig:noise_images}
\end{figure*}

Noise injection techniques introduce controlled random perturbations into image features to enhance variability and robustness. By stochastically modifying feature maps, they improve diversity in generated samples while preserving the underlying semantic structure of the image \cite{noh2017regularizing}. The visual effect of each technique is illustrated in  Figure \ref{fig:noise_images}.

\textbf{Elastic Deformation} is a non-linear, smooth distortion applied to images to simulate realistic variations in shape, texture, or structure. Unlike rigid transformations (e.g., rotation or scaling), it preserves local topology while introducing natural-looking deformations, making it particularly useful for tasks where objects exhibit flexibility, such as medical imaging and handwriting recognition \cite{9506328}. 
\textbf{Dimension Reduction} operates in the latent embedding space rather than on pixels. High-dimensional image representations are mapped into lower-dimensional vector embeddings that capture salient visual characteristics such as shape, color distribution, and edge structures. Once projected into this learned feature space, perturbations such as noise injection, interpolation between embeddings, or manifold-aware transformations can be applied directly to the representation, producing augmented samples that are semantically plausible but may not correspond to any single concrete pixel-level image. This is particularly useful for generating diverse training images in unsupervised or few-shot learning settings \cite{devries2017datasetaugmentationfeaturespace}.
\textbf{Möbius Transformation} performs complex transformations and inversions in the pixel space of an image. These transformations enable perspective projection (transforming the perceived distance of objects in an image) and act as structure-preserving coordinate mappings that integrate naturally with other photometric or geometric operations. {Möbius-based augmentation can achieve higher classification accuracy than Cutout , a region-masking augmentation strategy, on standard image-classification benchmark datasets \cite{Zhou_2021}.} 
\textbf{Salt and Pepper Noise}
is a combination of salt noise (the addition of random white pixels) and pepper noise (the addition of random black pixels) to simulate real-world sensor or environmental noise. It is particularly useful for assessing robustness to scenarios with high-amplitude disturbances rather than uniformly distributed noise. Applying this technique makes the model more robust to input perturbations and noise, thereby improving generalization performance \cite{Qiu2025AnIE}

\subsection{Selection Based}
\label{sec:selection}
\begin{figure*}[!htbp]
    \centering
    \subfloat[\footnotesize Original \label{fig:original}]{\includegraphics[width=0.18\textwidth]{images/original_image.jpg}}
    \hfill
    \subfloat[\footnotesize Random Crop \label{fig:cropped}]{\includegraphics[width=0.18\textwidth]{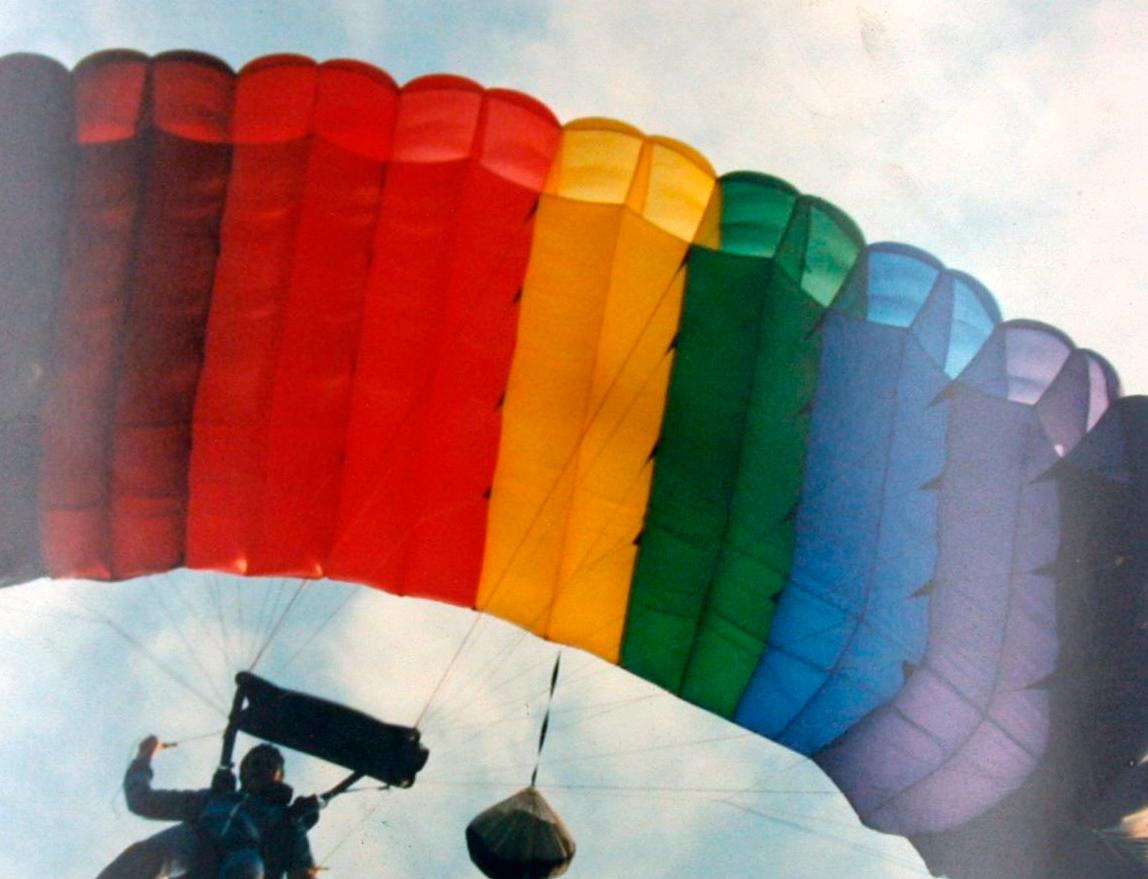}}
    \hfill
    \subfloat[\footnotesize Random Erase \label{fig:random_erased}]{\includegraphics[width=0.18\textwidth]{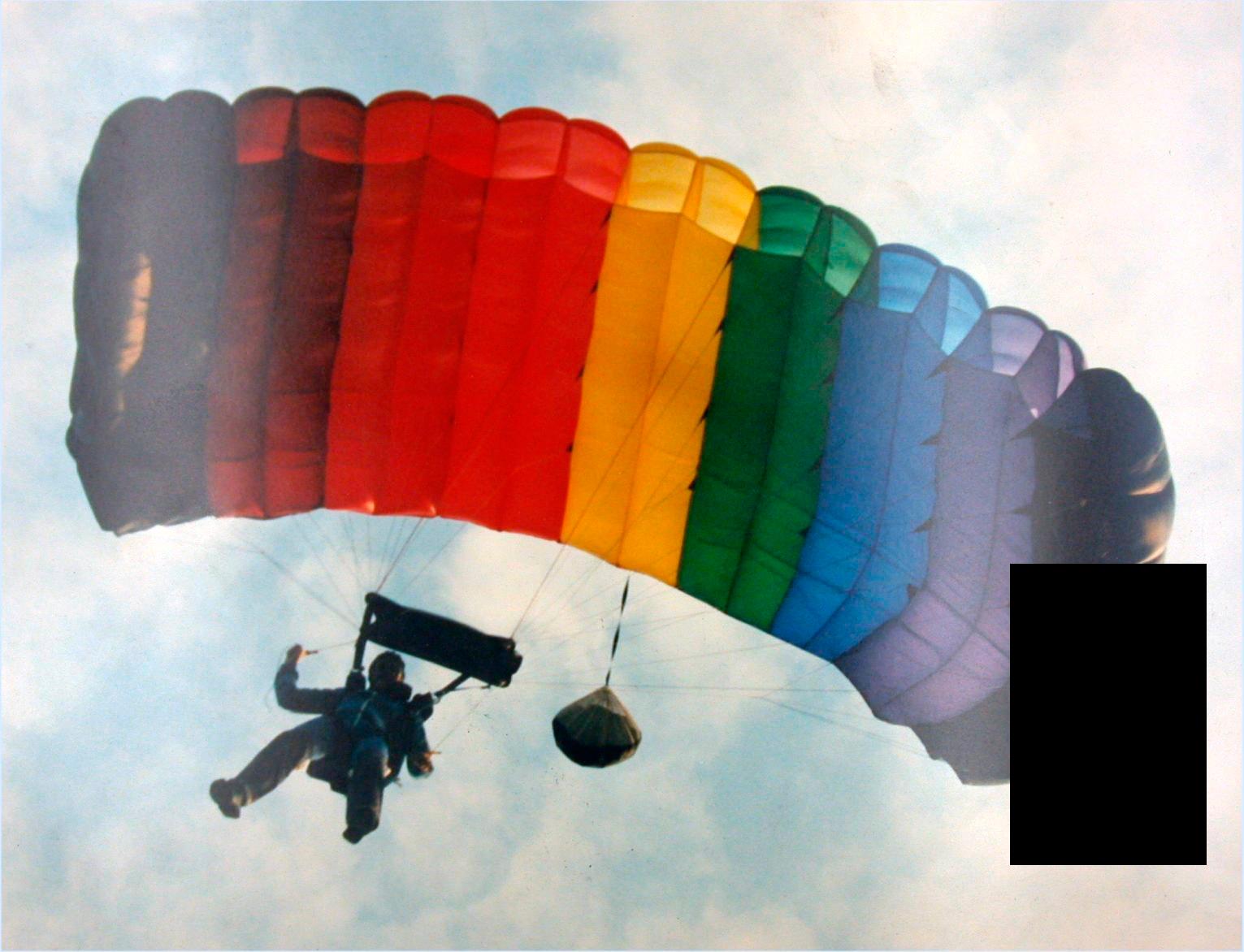}}
    \hfill
    \subfloat[\footnotesize Patch-wise \label{fig:patchwise}]{\includegraphics[width=0.18\textwidth]{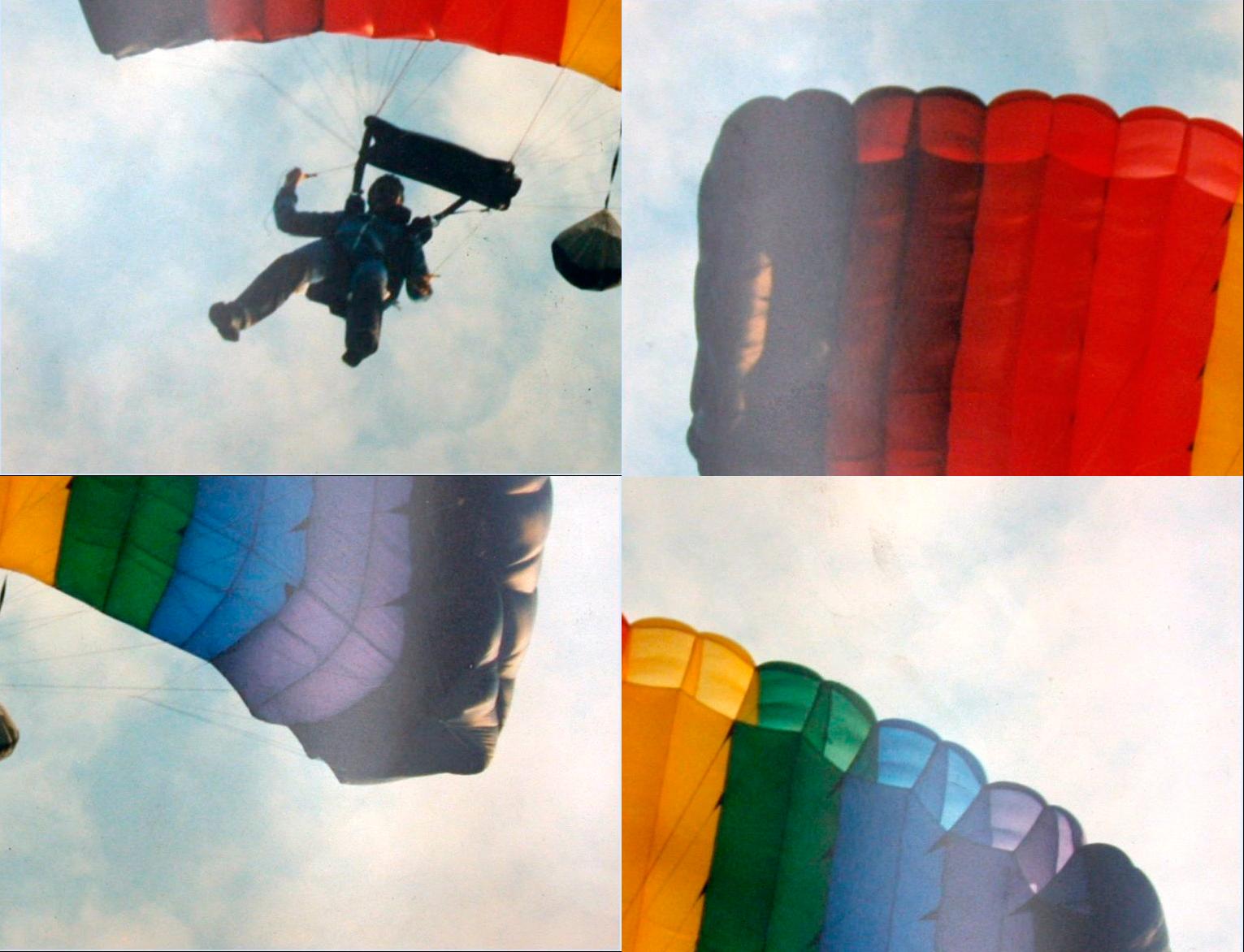}}
    \caption{(a) unmodified, (b) randomly cropped to 75\% of the original image, (c) erased 5\% of the image, (d) shuffled 4 patches.}
    \label{fig:selection_images}
\end{figure*}

Selection augmentations operate on specific regions of an image, typically chosen randomly to minimize human intervention. Most remove or modify a selected region to force the model to rely on the full image rather than localized cues. Because critical object regions may be excised, label-preservation should be verified, either by a domain expert or by saliency-guided region selection \cite{uddin2021saliencymixsaliencyguideddata}.
 The visual effect of each technique is illustrated in Figure \ref{fig:selection_images}.

\subsection{Filtering Operations}
\label{sec:filtering}
Filtering operations work by convolving predefined kernels over local pixel neighborhoods to apply modifications such as smoothing, sharpening, or edge enhancement through convolution, a linear filtering operation. These methods are illustrated in \Cref{fig:noise_images}

\textbf{Blur}
is a parent augmentation to both gaussian and motion blur where a blur effect or image smoothing is applied equally to the entire image. It aims to simulate images taken from an out-of-focus or blurry camera \cite{bang2020image}. 
\textbf{Gaussian Blur}
smooths an image by averaging each pixel with its kernel neighbours using a Gaussian distribution, producing a blurring effect that decreases sharpness and details in the image with a stronger effect in the middle of the image compared to its edges. It has been employed to simulate motion-induced image degradation \cite{bang2020image}. Buddenkotte and Buchert also applied an unnaturally strong Gaussian blur to measure model performance on drastically blurred images \cite{Buddenkotte1463}.
\textbf{Motion Blur} simulates the visual smearing caused by relative motion between the camera and the subject during image capture. It is implemented by convolving the image with a directional linear kernel oriented along a randomly sampled angle $\theta \in [0^\circ, 360^\circ]$. The resulting streak artifact approximates the perceptual degradation seen in images captured during camera shake, panning, or fast-moving object tracking \cite{bang2020image}.
\textbf{Sharpening} enhances the perceived detail of an image by amplifying high-frequency components, which correspond to rapid intensity differences between neighboring pixel values. Techniques such as unsharp masking and high-pass filtering increase local contrast around edges, making boundaries and fine textures more prominent. Unsharp masking sharpens an image by subtracting a blurred version from the original to emphasize edge information, whereas high-pass filtering suppresses low-frequency content while preserving rapid intensity variations, allowing edges and fine structures to be selectively enhanced. Sharpening is often used to simulate over-processed or artificially enhanced images and to evaluate a model’s robustness to increased edge intensity \cite{shorten2019survey}.

\begin{figure*}[!htbp]
    \centering
    \subfloat[\footnotesize Original \label{fig:original_filt}]{\includegraphics[width=0.18\textwidth]{images/original_image.jpg}}
    \hfill
    \subfloat[\footnotesize Blur \label{fig:blur}]{\includegraphics[width=0.18\textwidth]{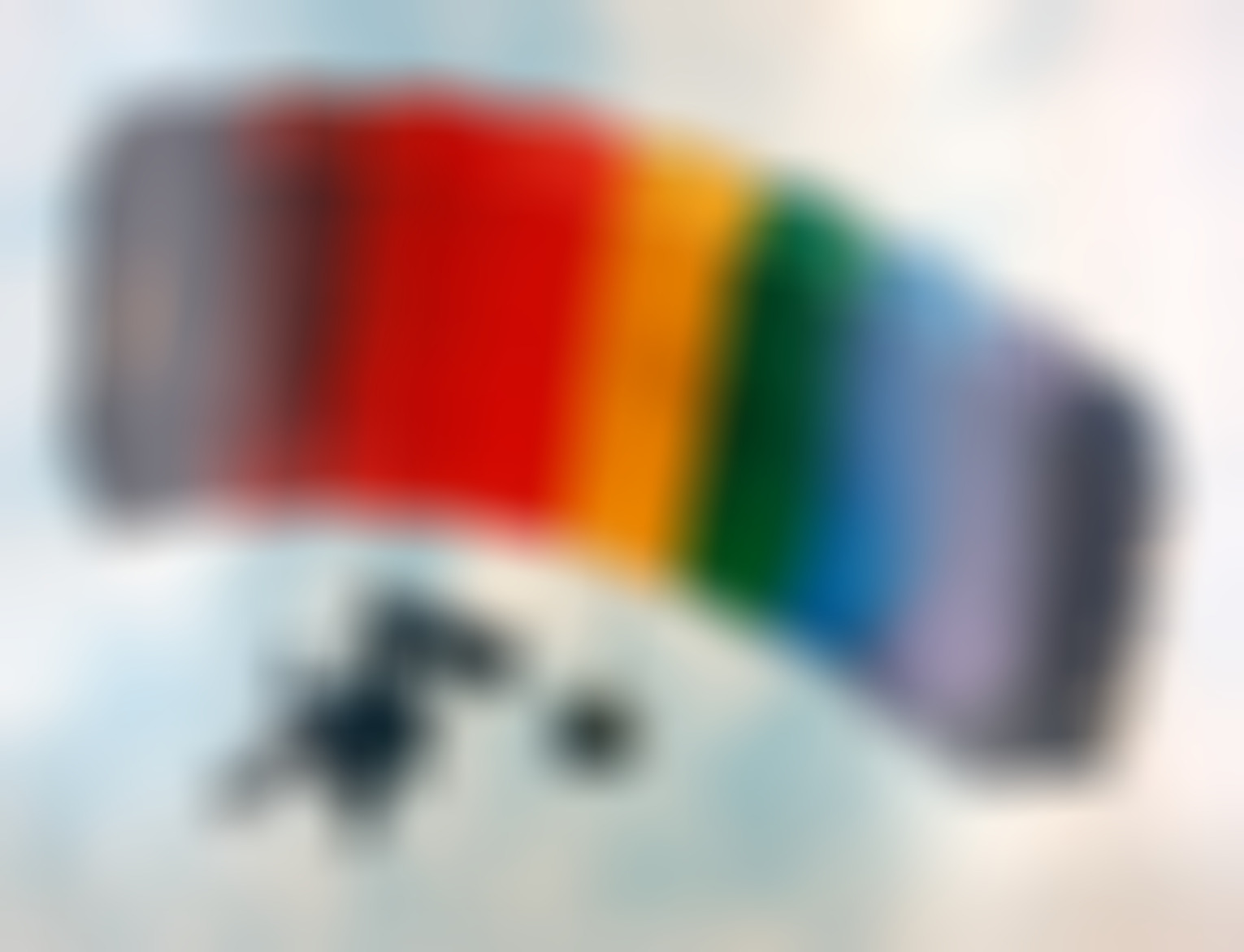}}
    \hfill
    \subfloat[\footnotesize Gaussian Blur \label{fig:gaussian_blur}]{\includegraphics[width=0.18\textwidth]{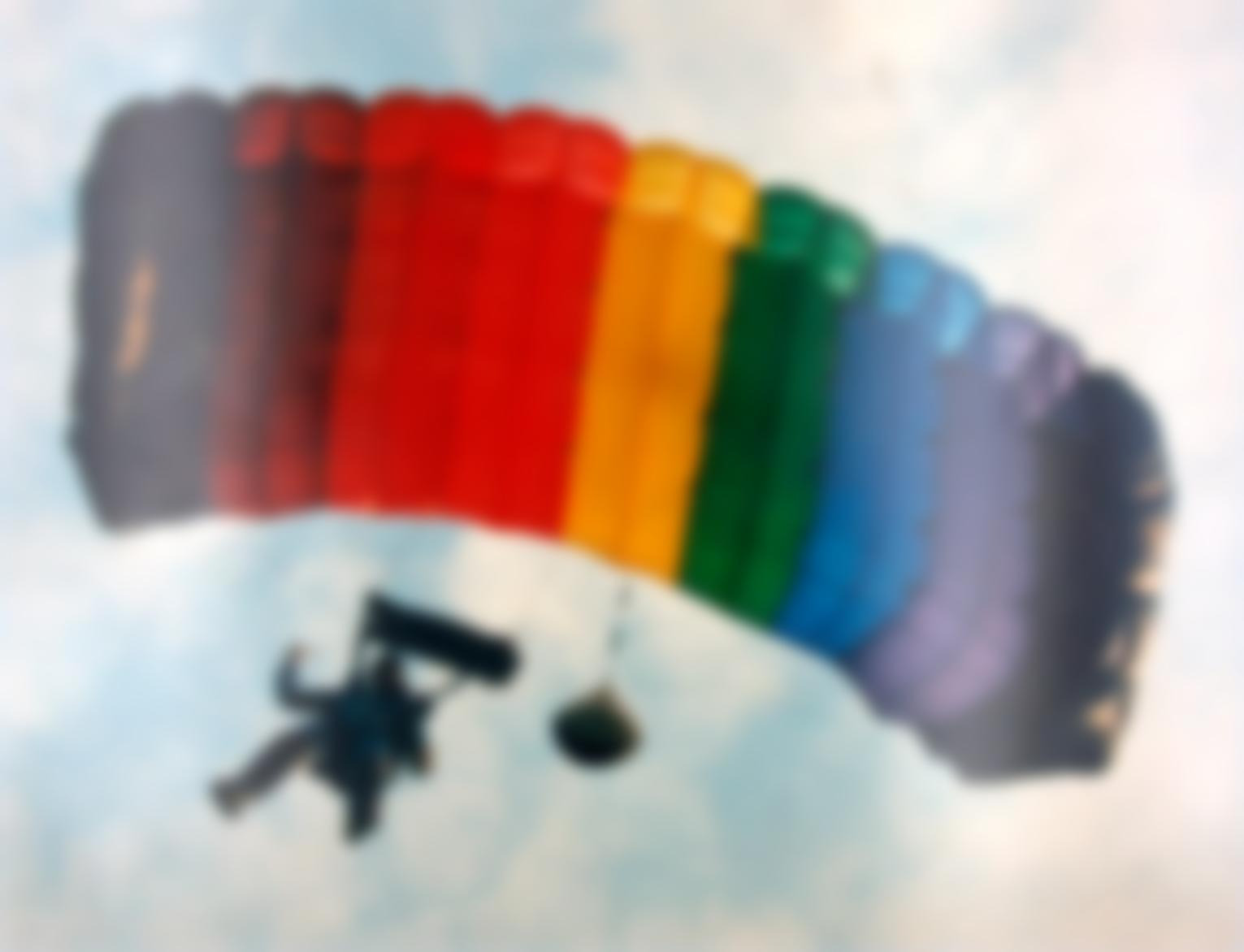}}
    \hfill
    \subfloat[\footnotesize Sharpen \label{fig:sharpen}]{\includegraphics[width=0.18\textwidth]{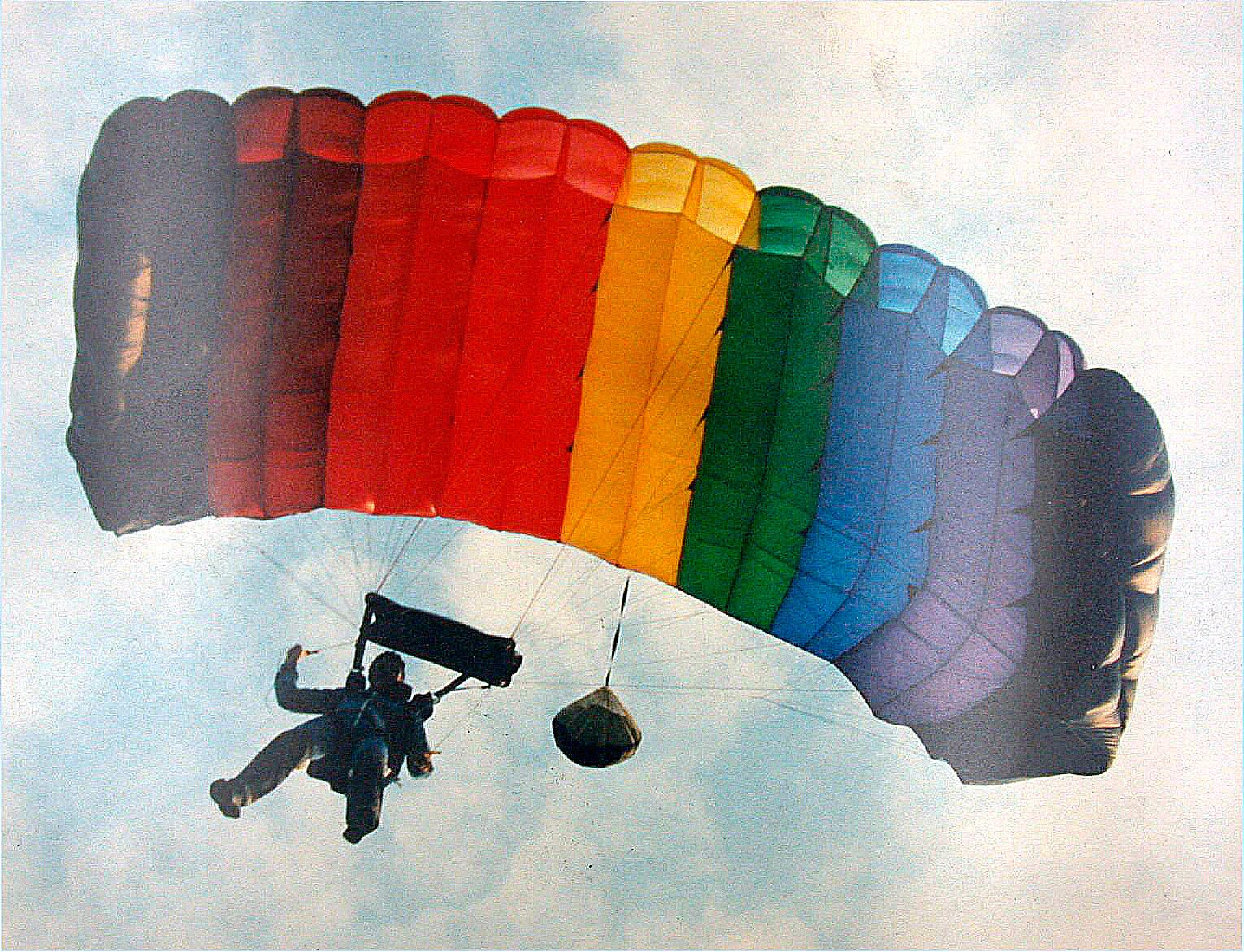}}
    \caption{(a) unmodified, (b) 50\% blur, (c) 100\% Gaussian blur with a kernel size of 5, (d) sharpen by 50\%}
    \label{fig:filtering_augmentations}
\end{figure*}

\subsection{Self Mixing}
\label{sec:selfmix} 
\begin{figure*}[!htbp]
    \centering
    \subfloat[\footnotesize Original \label{fig:original_selfmix}]{\includegraphics[width=0.155\textwidth]{images/original_image.jpg}}
    \hfill
    \subfloat[\footnotesize AugMix (sev.\ 1) \label{fig:augmix1}]{\includegraphics[width=0.155\textwidth]{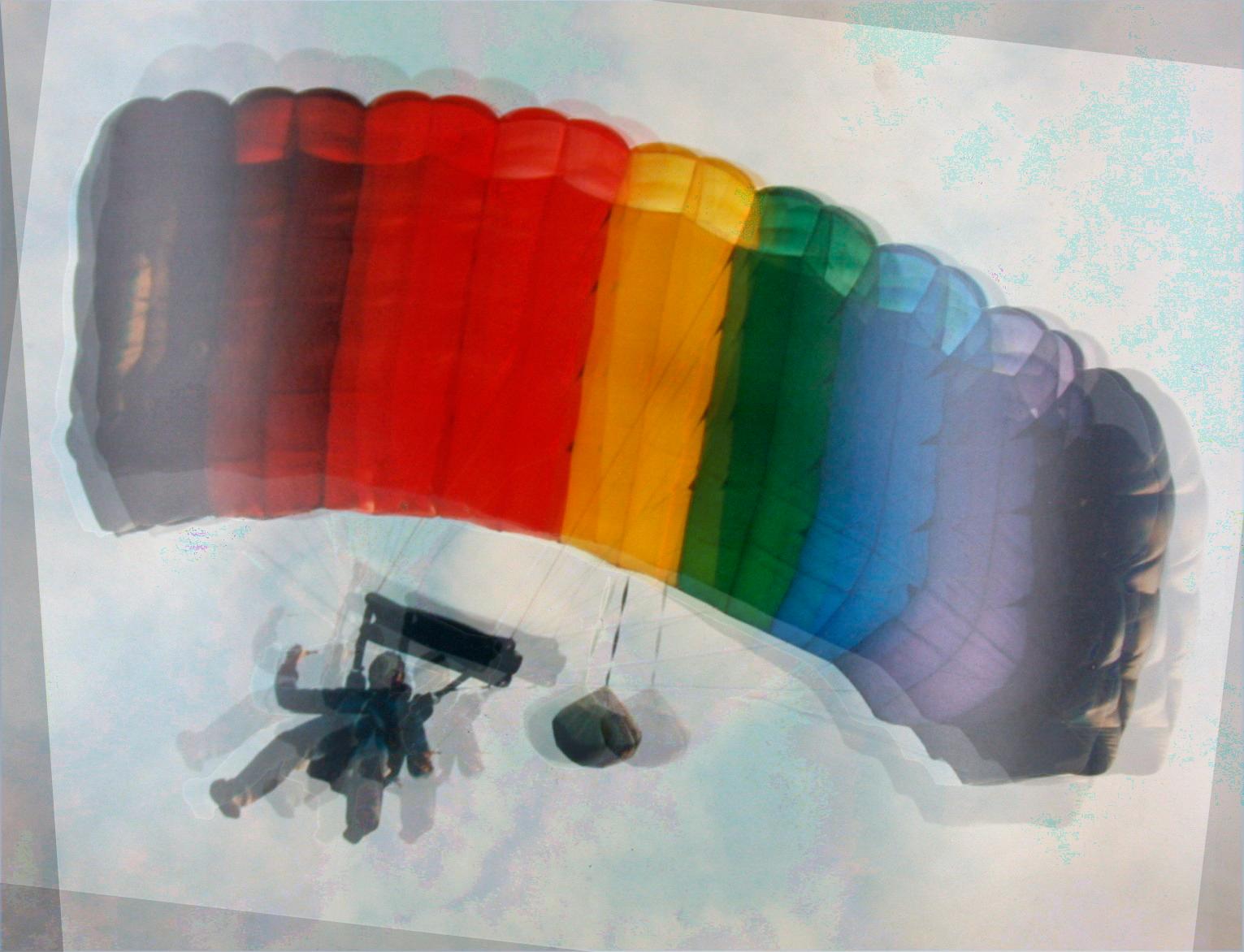}}
    \hfill
    \subfloat[\footnotesize AugMix (sev.\ 3) \label{fig:augmix}]{\includegraphics[width=0.155\textwidth]{images/augmix_image.jpg}}
    \hfill
    \subfloat[\footnotesize Self Mix \label{fig:selfmix}]{\includegraphics[width=0.155\textwidth]{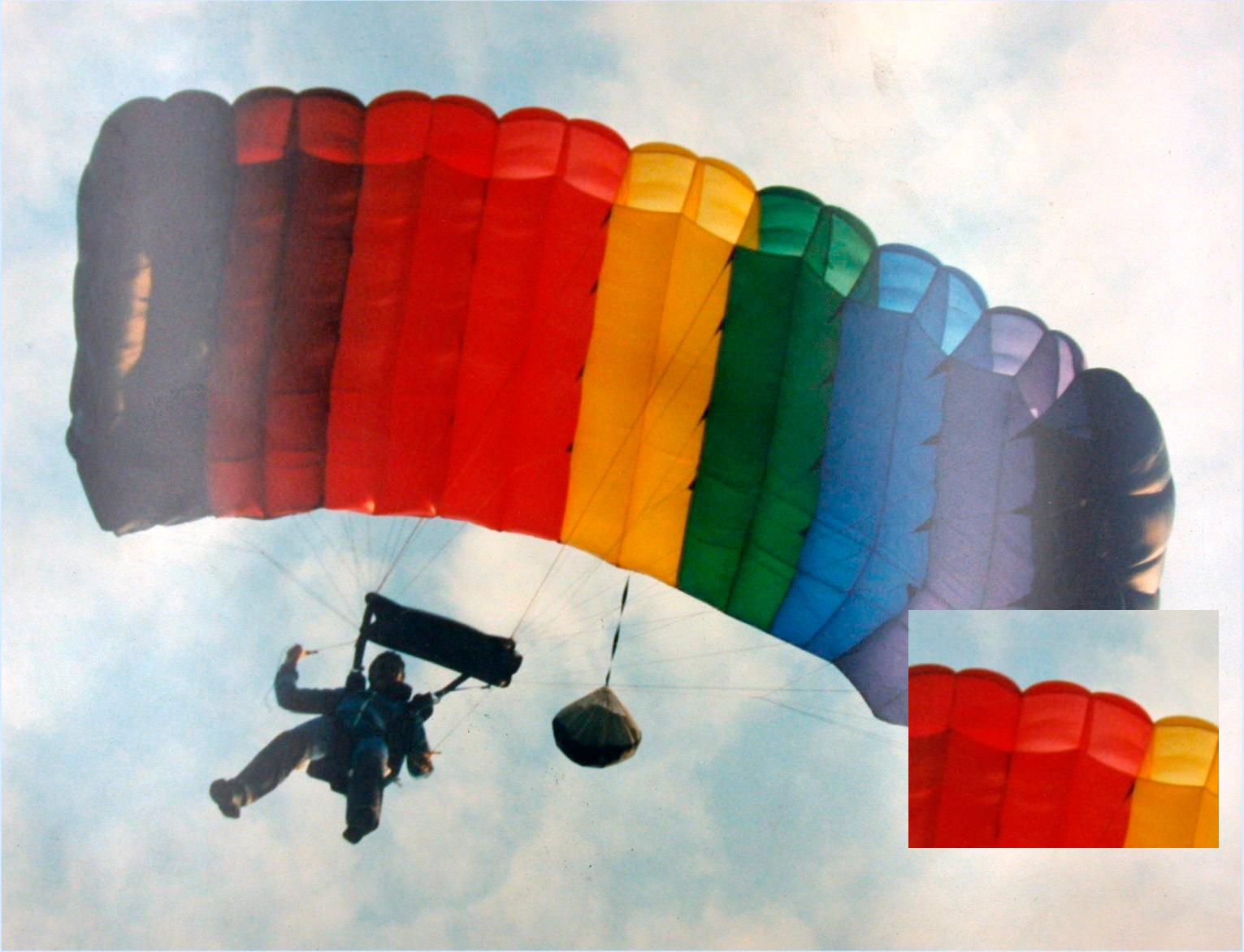}}
    \hfill
    \subfloat[\footnotesize Salf Mix \label{fig:selfmix}]{\includegraphics[width=0.155\textwidth]{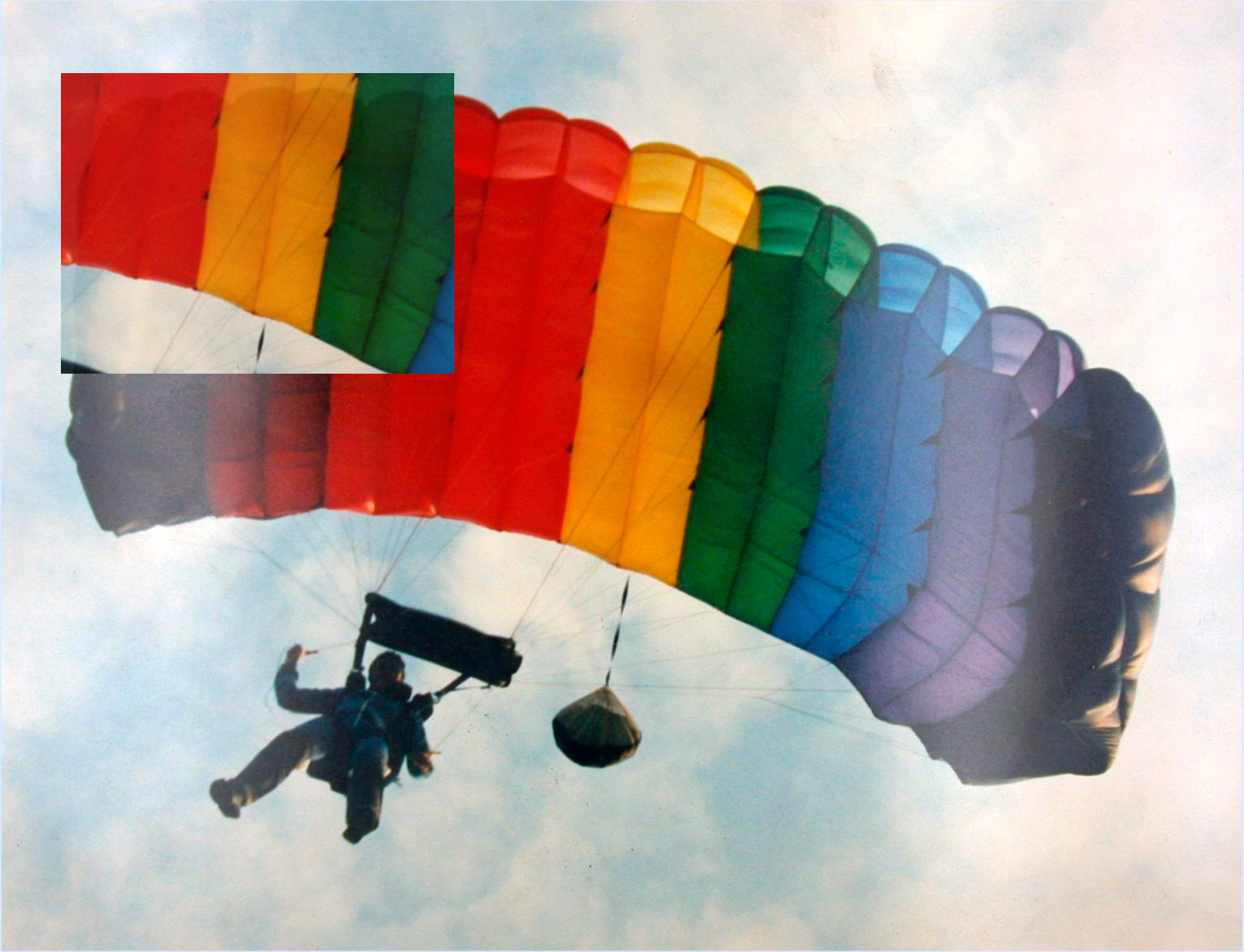}}
    \hfill
    \caption{(a) unmodified; (b) AugMix at severity 1; (c) AugMix at severity 3; (d) Self Mix with patch size 10\%; 
    (e) Salf Mix with patch size 10.}
    \label{fig:selfmix_augmentations}
\end{figure*}

Self mixing involves mixing multiple versions of an image with itself. Either sections of an image may be cutout and patched elsewhere, or multiple augmented versions of the same image may be combined. {The visual effect of each technique is illustrated in Figure  \Cref{fig:selfmix_augmentations}}

\textbf{AugMix} creates multiple augmented versions of the same image that get overlayed with each other. Each chain of augmentations (typically three) are randomly chosen and as a final step, the original image is overlayed as well \cite{hendrycks2020augmixsimpledataprocessing}.
\textbf{Self Mix} cuts a random patch from an image and pastes it into the same image. The goal is to improve generalization ability in a few-shot learning scenario \cite{Seo_2021}.
\textbf{Salf Mix} finds the most and least salient regions of an image, and replaces the least salient with the most salient regions of an image. {This saliency-guided perturbation generates challenging training samples in which some semantic cues are displaced or partially obscured, thereby encouraging the model to learn more robust and generalizable representations.} {Models trained using SalfMix achieve higher image-classification accuracy than models trained using Cutout under the evaluated experimental settings \cite{s21248444}.}


\subsection{Sample Mixing}
\label{sec:samplemix}
\begin{figure*}[!htbp]
    \centering
    \subfloat[\footnotesize Original 1 \label{fig:original_sm}]{\includegraphics[width=0.18\textwidth]{images/original_image.jpg}}
    \hfill
    \subfloat[\footnotesize Original 2 \label{fig:original2}]{\includegraphics[width=0.18\textwidth]{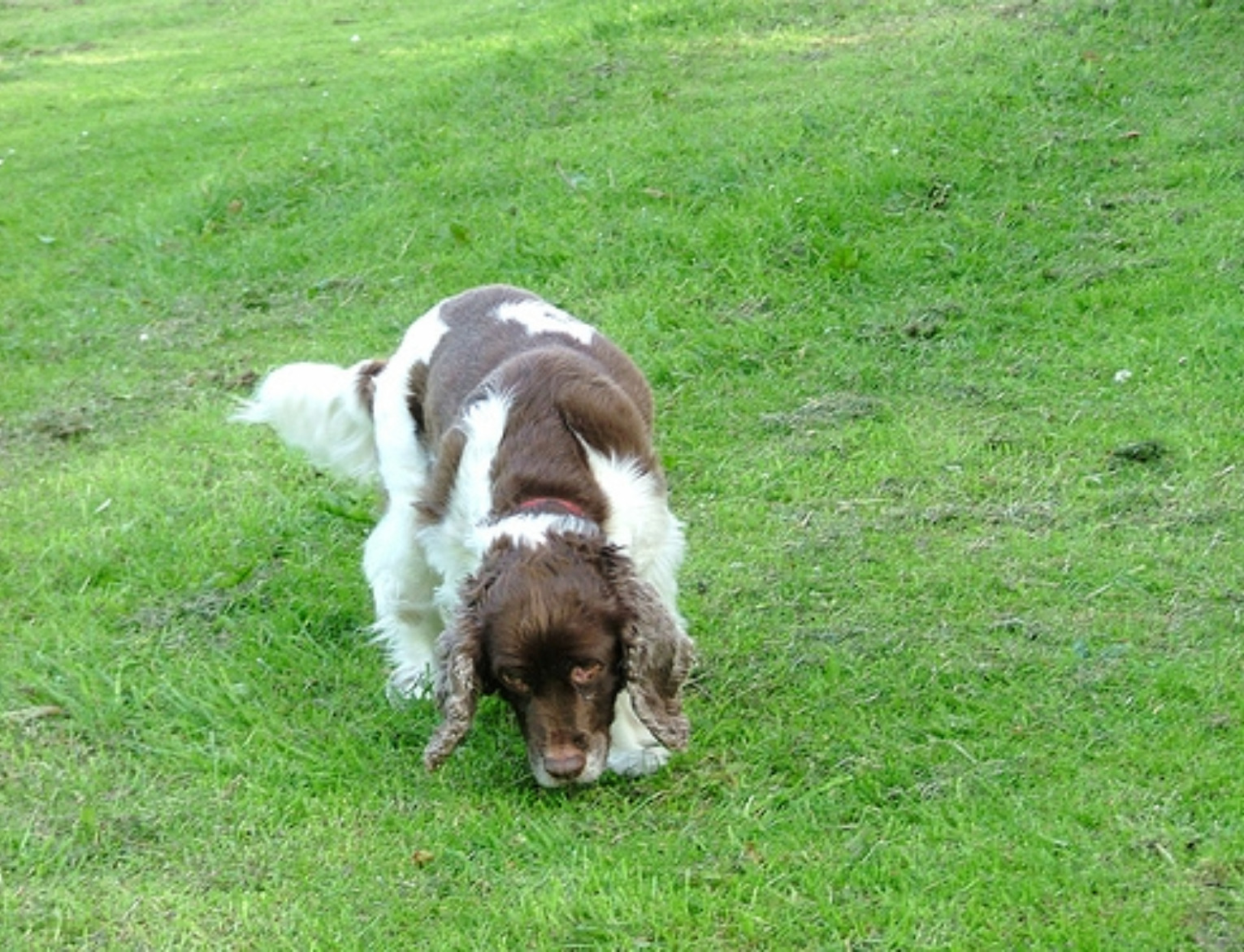}}
    \hfill
    \subfloat[\footnotesize Saliency Mix \label{fig:saliencyMix}]{\includegraphics[width=0.18\textwidth]{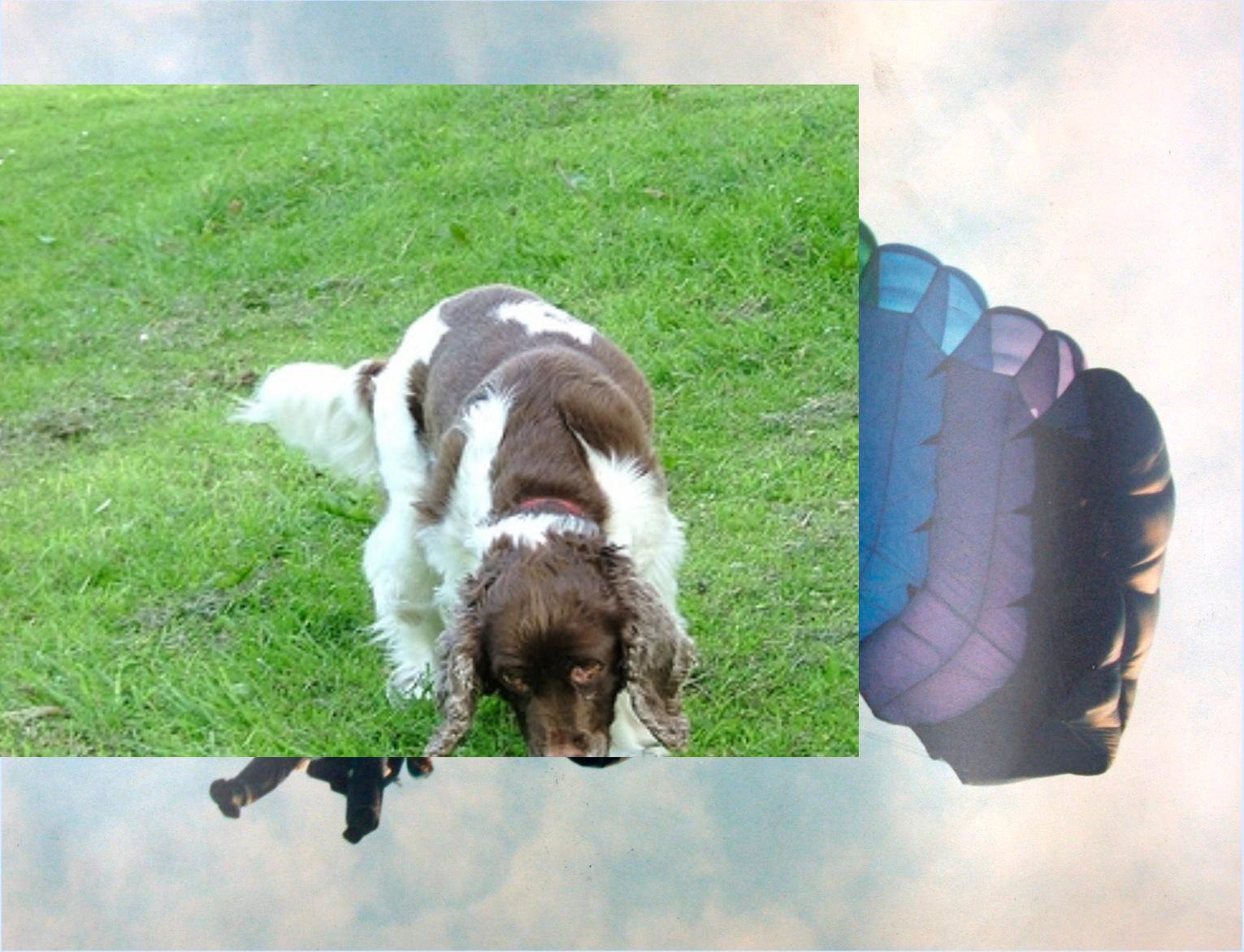}}
    \hfill
    \subfloat[\footnotesize Keep Augment \label{fig:keepAugment}]{\includegraphics[width=0.18\textwidth]{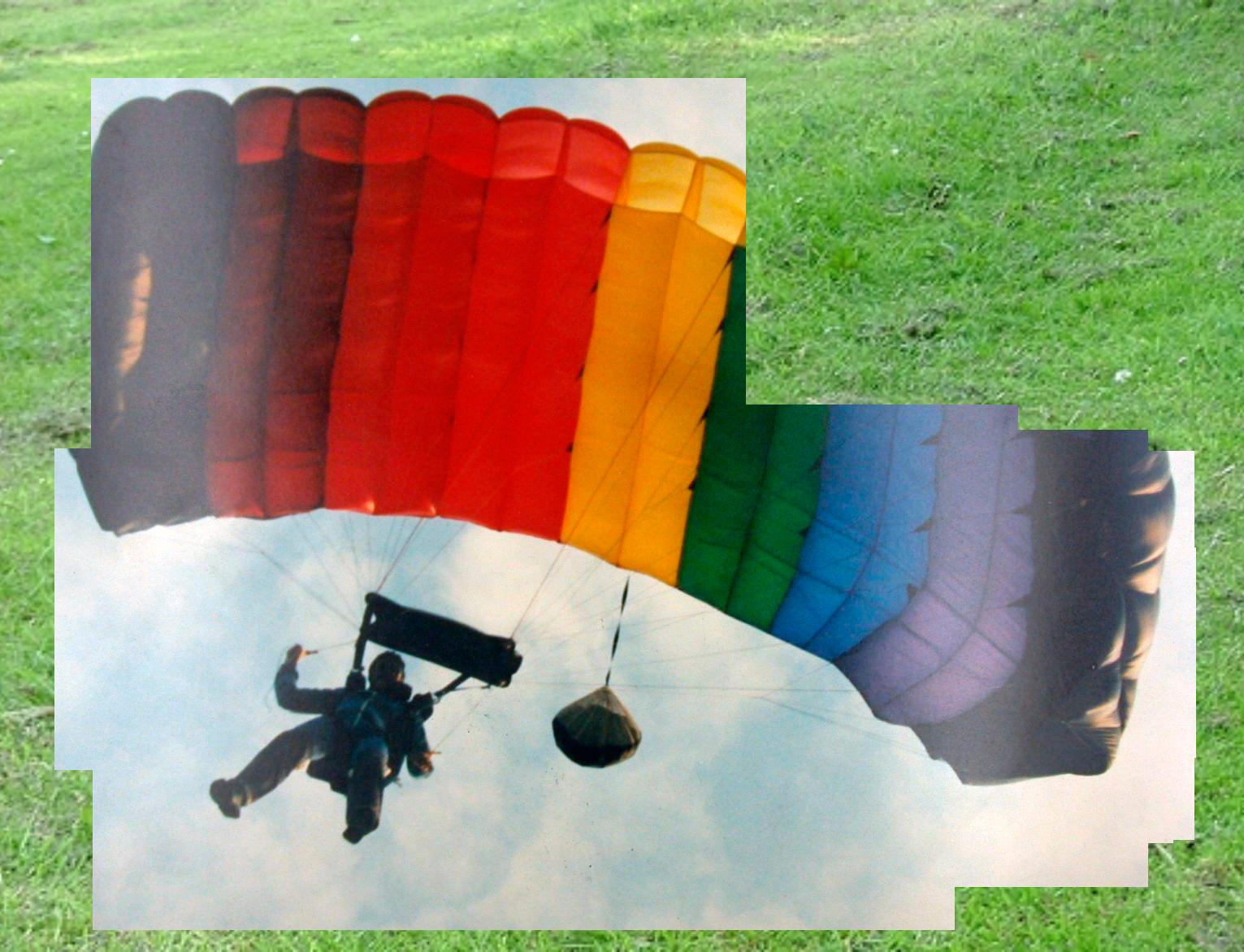}}
    \\[0.5em]
    \subfloat[\footnotesize RICAP \label{fig:ricap}]{\includegraphics[width=0.18\textwidth]{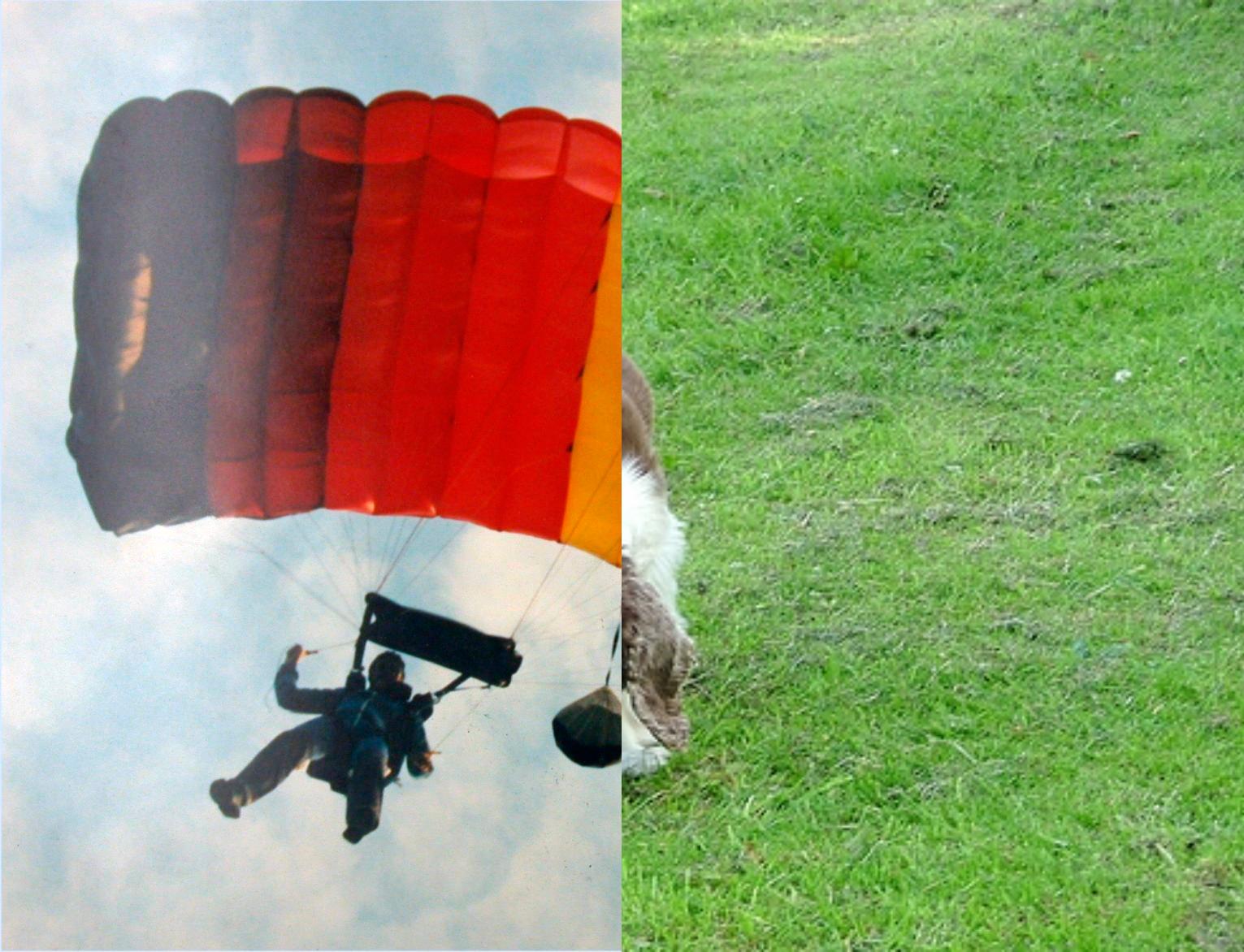}}
    \hfill
    \subfloat[\footnotesize MixGen \label{fig:mixgen}]{\includegraphics[width=0.18\textwidth]{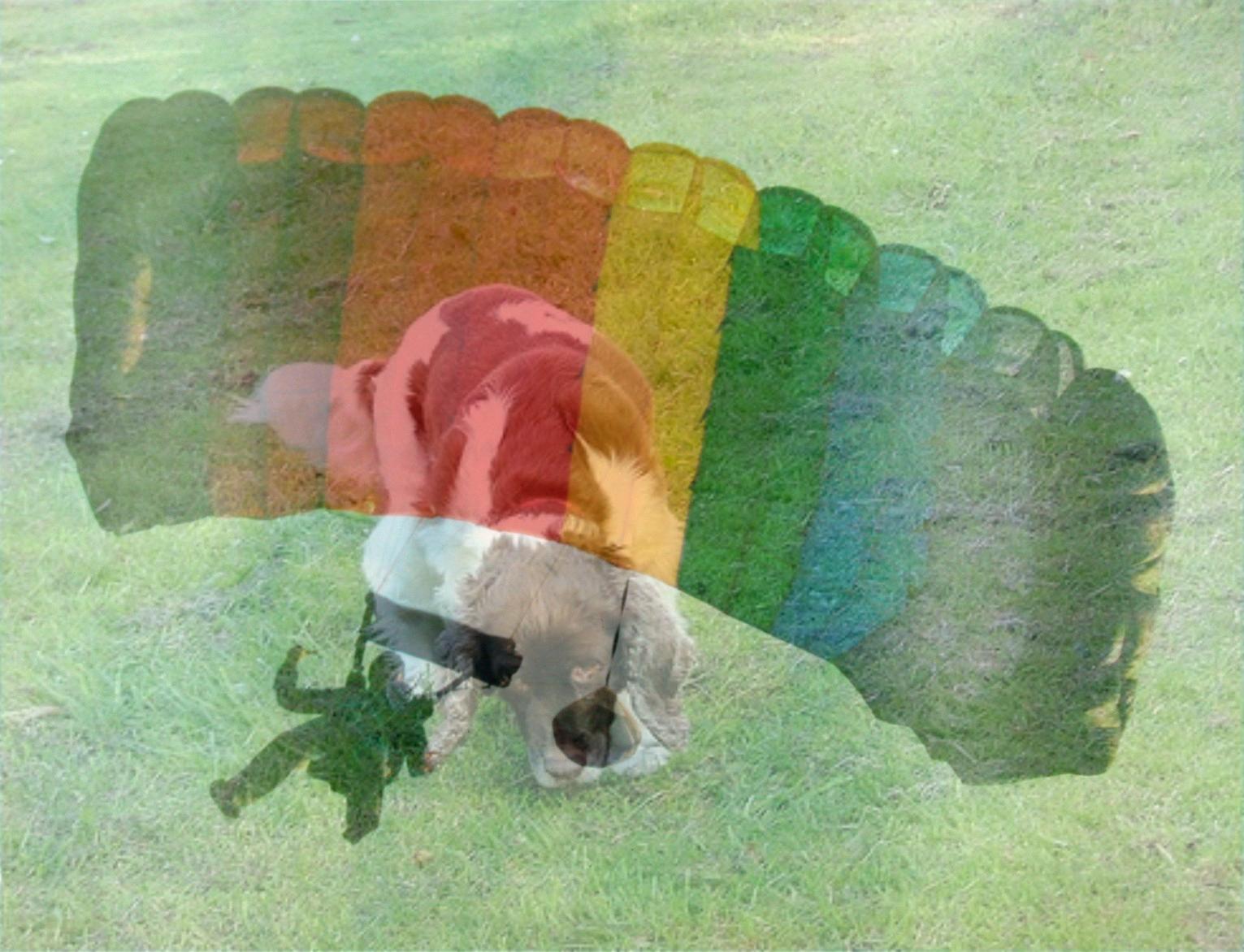}}
    \hfill
    \subfloat[\footnotesize Robust MixGen \label{fig:robustMixgen}]{\includegraphics[width=0.18\textwidth]{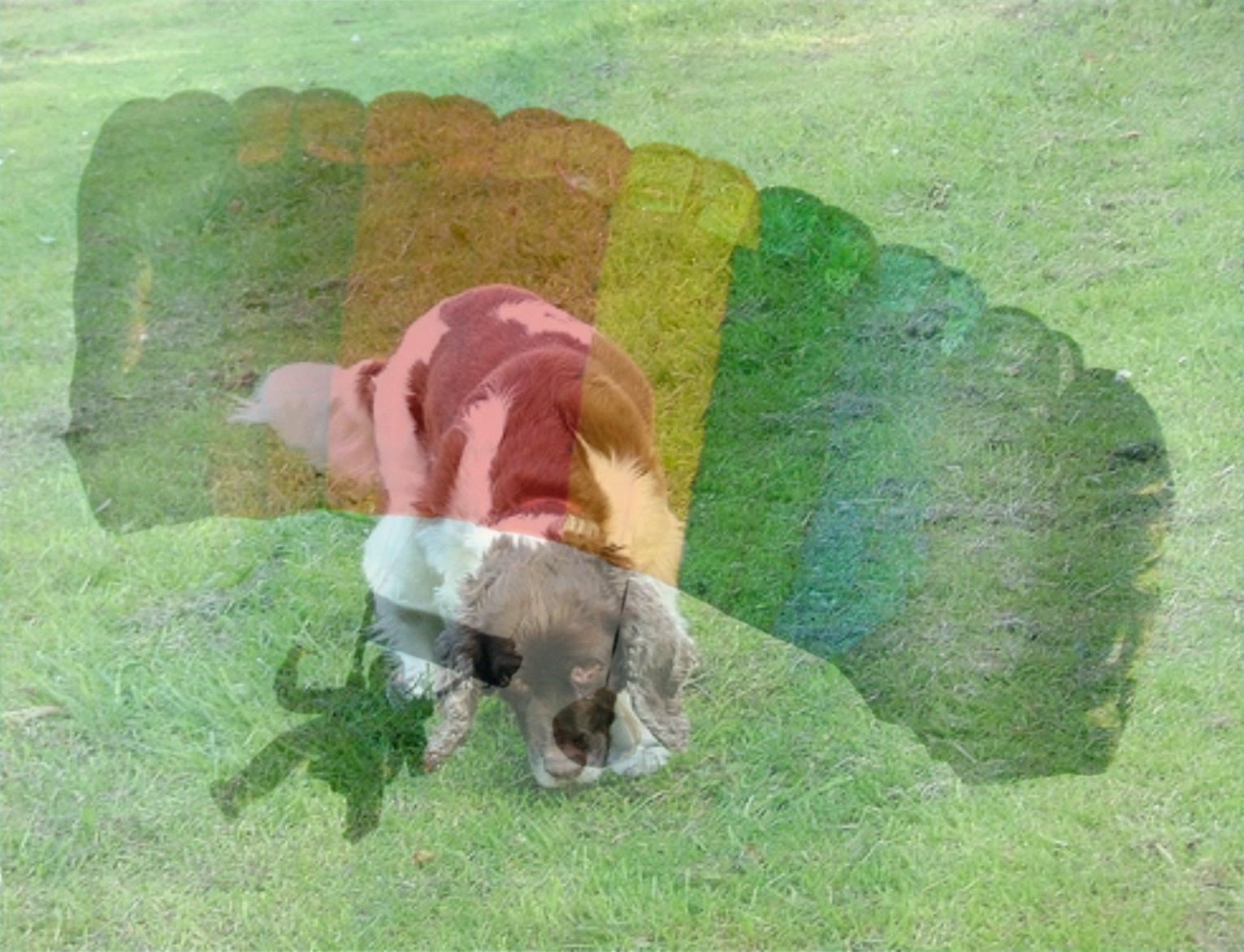}}
    \hfill
    \subfloat[\footnotesize IP Mix \label{fig:ipMix}]{\includegraphics[width=0.18\textwidth]{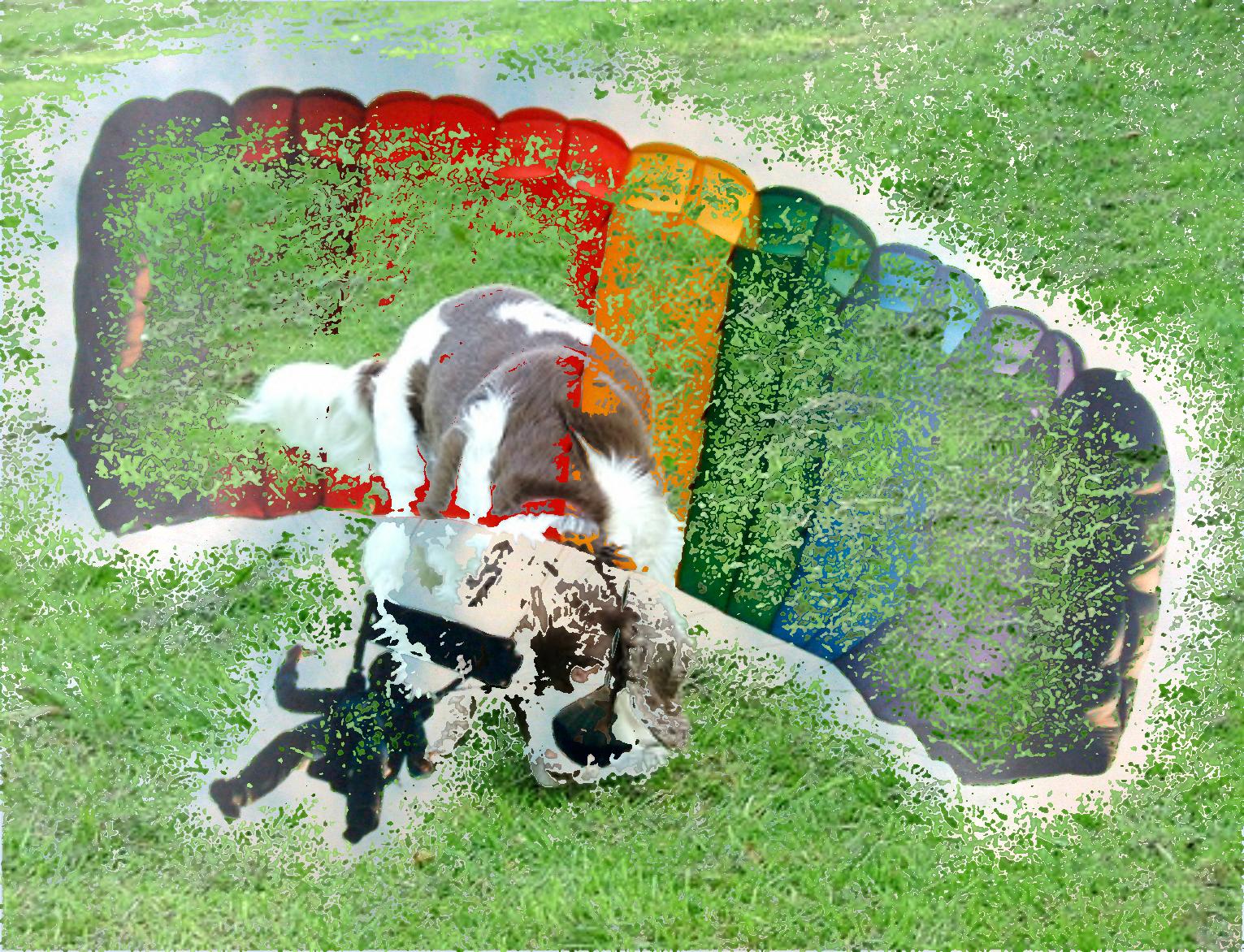}}
    \caption{(a) unmodified 1, (b) unmodified 2, (c) saliency mixed at 50\%, (d) keep augmented with a ratio of 0.5, (e) RICAP vertically, (f) MixGen mixed at 50\%, (g) Robust MixGen mixed at 50\%, (h) IP Mix with a ratio of 0.5.}
    \label{fig:sample_mixing_augmentations}
\end{figure*}

Sample mixing is similar to self-mixing, but differs by blending several images into one new image as shown in \Cref{fig:sample_mixing_augmentations}. Images may originate from the same class or not. In the latter case, the result is represented as a soft label reflecting the weighted contribution of each class.

\textbf{Mix Gen} overlays two input images through linear interpolation \cite{hao2023mixgennewmultimodaldata}. While simple to implement, ignoring the contents and similarity of the two mixed images leads to robustness problems which are addressed with Robust Mix Gen \cite{KIM2025129167}.
\textbf{Robust MixGen} extends MixGen by incorporating similarity-aware blending between input images. First, the two images' feature embeddings are extracted and divided into local areas. Cosine similarity between corresponding regions is calculated and employed as a weighting coefficient during the mixing process. This makes it possible to blend visually consistent areas more strongly while maintaining differences in dissimilar regions. As a result, mixing is applied locally rather than uniformly throughout the image, producing augmented samples that are structurally coherent \cite{KIM2025129167}.
\textbf{MixUp} blends two images pixel-by-pixel via linear interpolation to produce a single composite image and a proportionally blended label. MixUp helps produce smoother decision boundaries and improves generalization \cite{zhang2018mixupempiricalriskminimization}.
\textbf{CutMix} replaces one rectangular region of an image with a patch from another image while combining their weighted labels corresponding to the patch area. CutMix preserves local spatial structure by introducing region-level compositional mixing. This encourages a model to learn from multiple visual contexts while the rest of the regions are maintained \cite{yun2019cutmixregularizationstrategytrain}.
\textbf{ResizeMix} differs from CutMix as it incorporates a patch of the resized source image to the target image rather than substituting another patch for a randomly cropped area. ResizeMix is computationally efficient because it only requires image-resizing operations, as opposed to saliency-guided augmentation approaches that utilize attention or importance maps to determine where to paste content \cite{qin2020resizemixmixingdatapreserved}.
\textbf{SaliencyMix} { is a patch-based augmentation technique that combines two images using a saliency-guided source patch. A saliency map is first generated using the bottom-up method proposed by Montabone and Soto to identify the most visually informative region of the source image. A patch containing this salient region is then extracted and inserted into a target image using a binary mask. By guiding patch selection through saliency estimation, SaliencyMix increases the likelihood that the transferred patch contains informative object content rather than background or other non-discriminative regions. This distinguishes SaliencyMix from CutMix, which selects the source patch randomly \cite{uddin2021saliencymixsaliencyguideddata}.}
\textbf{RICAP} (Random image cropping and patching) randomly crops multiple images and combines the cropped parts to generate a single composite image. RICAP typically selects source images that share the same class label to preserve label identity, since the cropping procedure does not account for the boundaries of semantic objects \cite{Takahashi_2020}.
\textbf{Keep Augment} {operates on a single input image by first applying a base augmentation to generate an intermediate augmented image. A gradient-based saliency map is then computed from the original image to identify the most informative region. The final output is constructed by retaining the most salient patch from the original image while using the corresponding regions from the augmented image for the remaining content. This process preserves discriminative visual information while still introducing sufficient variation for regularization, thereby reducing the risk of removing features that are important for classification}\cite{gong2020keepaugmentsimpleinformationpreservingdata}.
\textbf{IP Mix} {combines an original labeled image with an unlabeled synthetic pattern through image-level, patch-level, and pixel-level mixing operations. Unlike SaliencyMix, which uses a saliency map to identify and transfer an informative region between two labeled images, IPMix applies its mixing operations randomly and does not rely on saliency estimation. These operations introduce structural and textural diversity without incorporating content associated with another semantic class. Consequently, the generated sample retains the label of the original image \cite{huang2023ipmix}.}


\subsection{GAN Based}
\label{sec:gan}
GAN techniques employ adversarial learning to translate images between domains, using a generator–discriminator pair that trains jointly to produce realistic, semantically consistent outputs while preserving scene structure.

\textbf {Deep Convolutional Generative Adversarial Networks (DCGANs)} are a common variant of GANs that replace fully connected layers with convolutional structures, stabilizing training and improving image quality \cite{radford2015dcgan}. For data augmentation, DCGANs generate synthetic samples from random noise. DCGANs have been used to synthesize minority-class samples to address class imbalance in medical-imaging tasks \cite{srivastav2021improved}.
{
\textbf {Neural Style Transfer} \cite{gatys2015neuralalgorithmartisticstyle} uses a pretrained Convolutional Neural Network model (e.g., VGG-19) to record features of the content and style representation of input images. During generation, these recorded features act as targets for an optimization process, where standard gradient descent iteratively updates a random noise image until its features match the targets, thereby transferring the style.}
\textbf {Pix2Pix} is a conditional GAN for supervised image-to-image translation, using a U-Net generator and a PatchGAN discriminator \cite{isola2017image}. 
\textbf {CycleGAN} tackles image-to-image translation without paired training data. It learns two mappings, $G: X \to Y$ and $F: Y \to X$, guided by adversarial discriminators that push generated images toward the target distribution. 
CycleGAN performs effectively on unpaired image translation tasks such as style transfer, object transfiguration, and season transfer \cite{zhu2017unpaired}.

\subsection{Diffusion Model Based}
\label{sec:diffusion}
\begin{figure*}[!htbp]
    \centering

    \subfloat[\footnotesize Original \label{fig:original_gan}]
    {\includegraphics[width=0.18\textwidth]{images/original_image.jpg}}
    \hfill
    \subfloat[\footnotesize DiffusionCLIP \label{fig:diffusionCLIP}]
    {\includegraphics[scale=0.14]{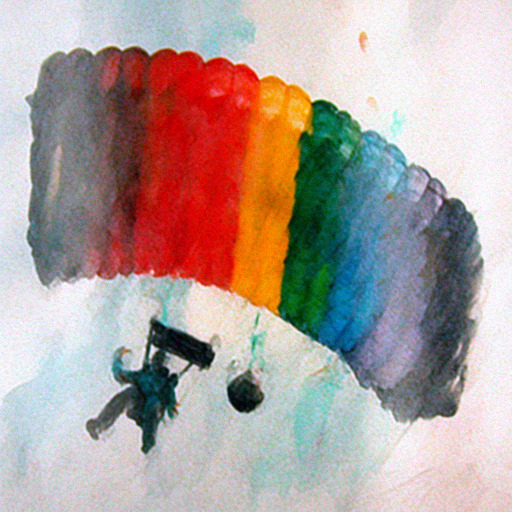}}
    \hfill
    \subfloat[\footnotesize DiffuseMix \label{fig:diffuseMix}]
    {\includegraphics[scale=0.14]{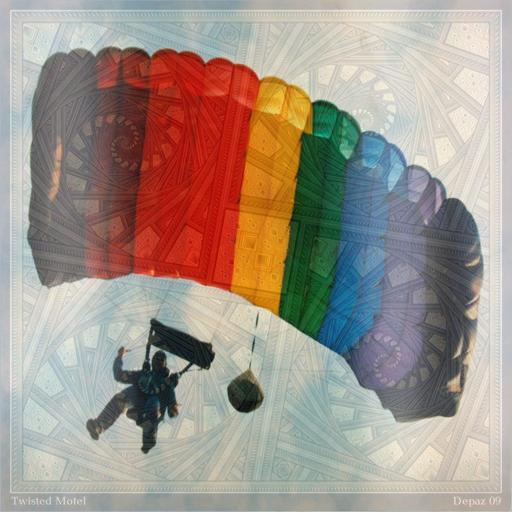}}
    \hfill
    \subfloat[\footnotesize SaSPA \label{fig:saSPA}]
    {\includegraphics[width=0.18\textwidth]{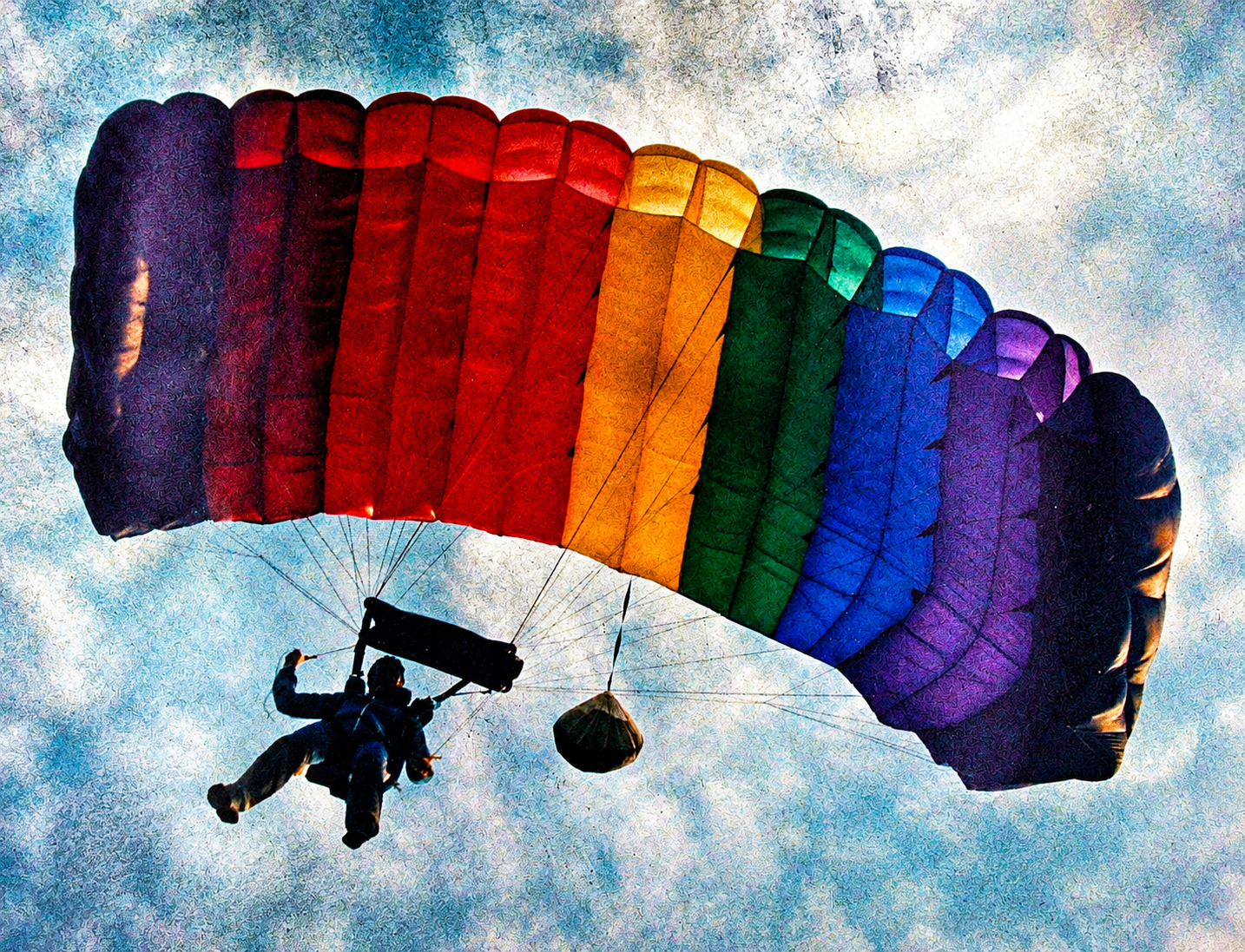}}
    \hfill
    \subfloat[\footnotesize Da-Fusion \label{fig:da-fusion}]
    {\includegraphics[width=0.18\textwidth]{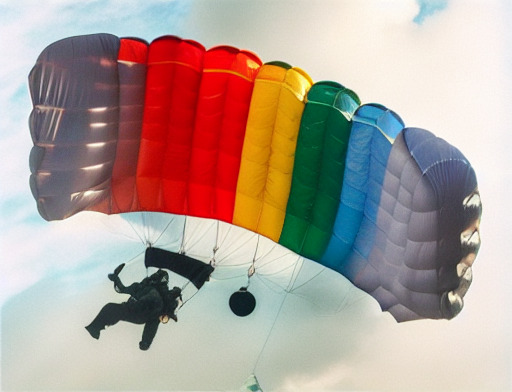}}

    \par\medskip

    \subfloat[\footnotesize DIAGen \label{fig:diaGen}]
    {\includegraphics[width=0.18\textwidth]{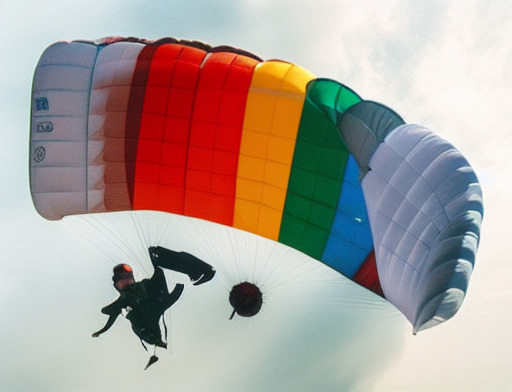}}
    \hfill
    \subfloat[\footnotesize DCGAN \label{fig:dcGAN}]
    {\includegraphics[width=0.18\textwidth]{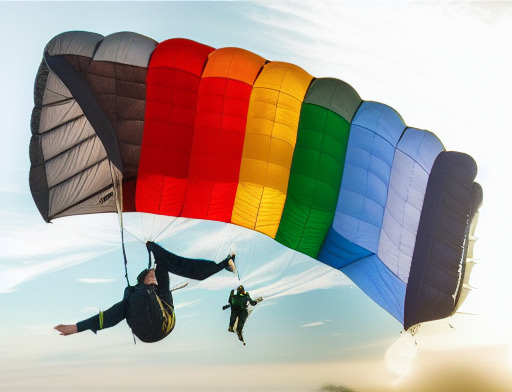}}
    \hfill
    \subfloat[\footnotesize Pix2pix-Zero \label{fig:pix2pixZero}]
    {\includegraphics[width=0.18\textwidth]{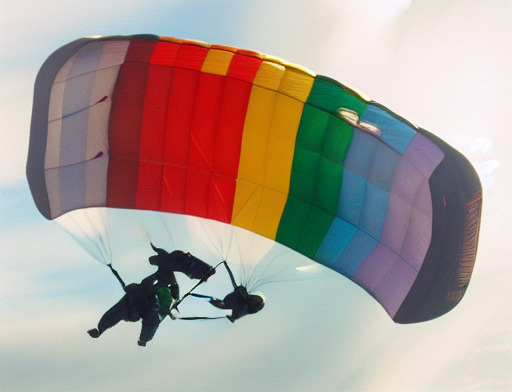}}
    \hfill
    \subfloat[\footnotesize Neural Style Transfer \label{fig:nst}]
    {\includegraphics[width=0.18\textwidth]{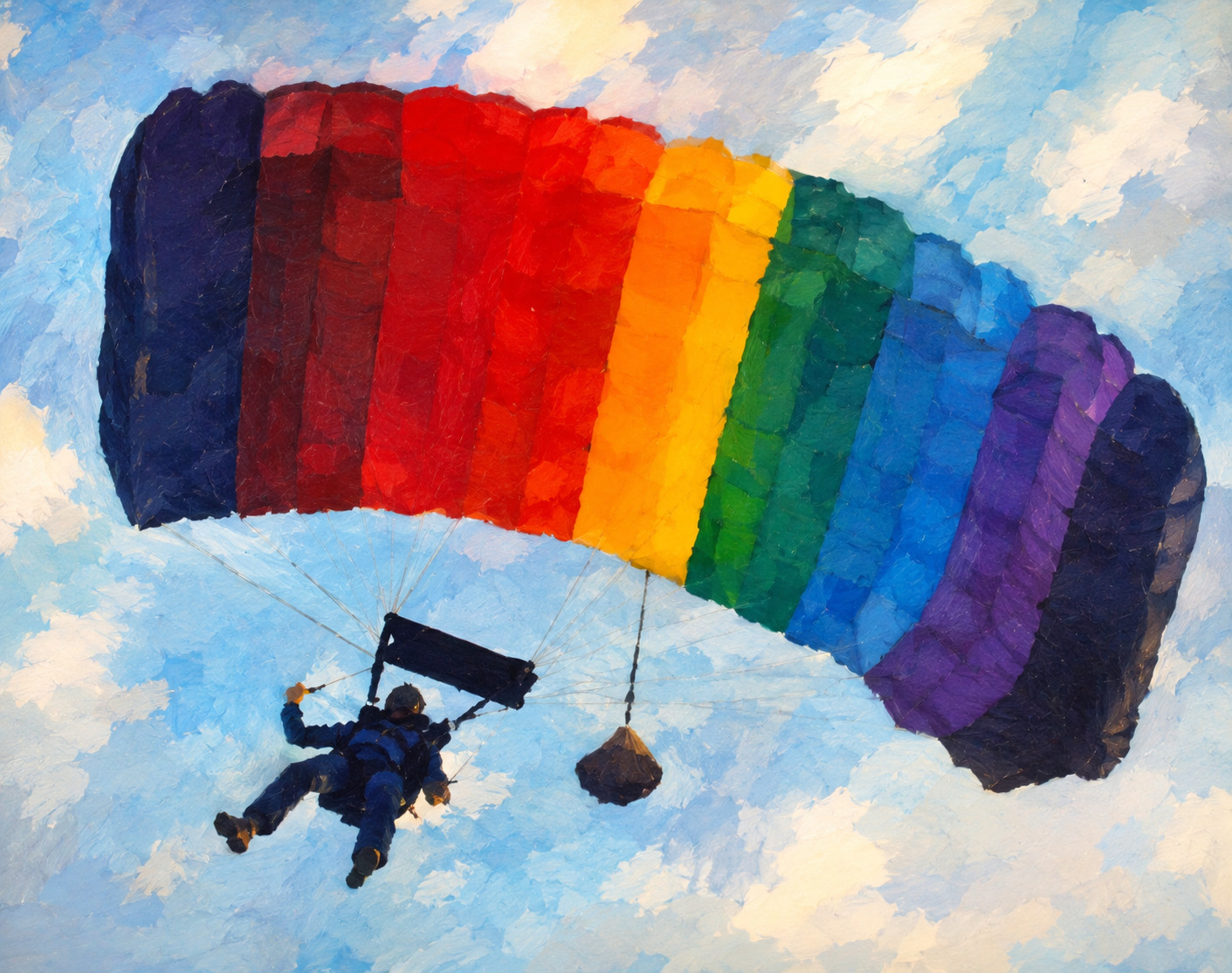}}
    \hfill
    \subfloat[\footnotesize CycleGAN \label{fig:cycleGAN}]
    {\includegraphics[width=0.18\textwidth]{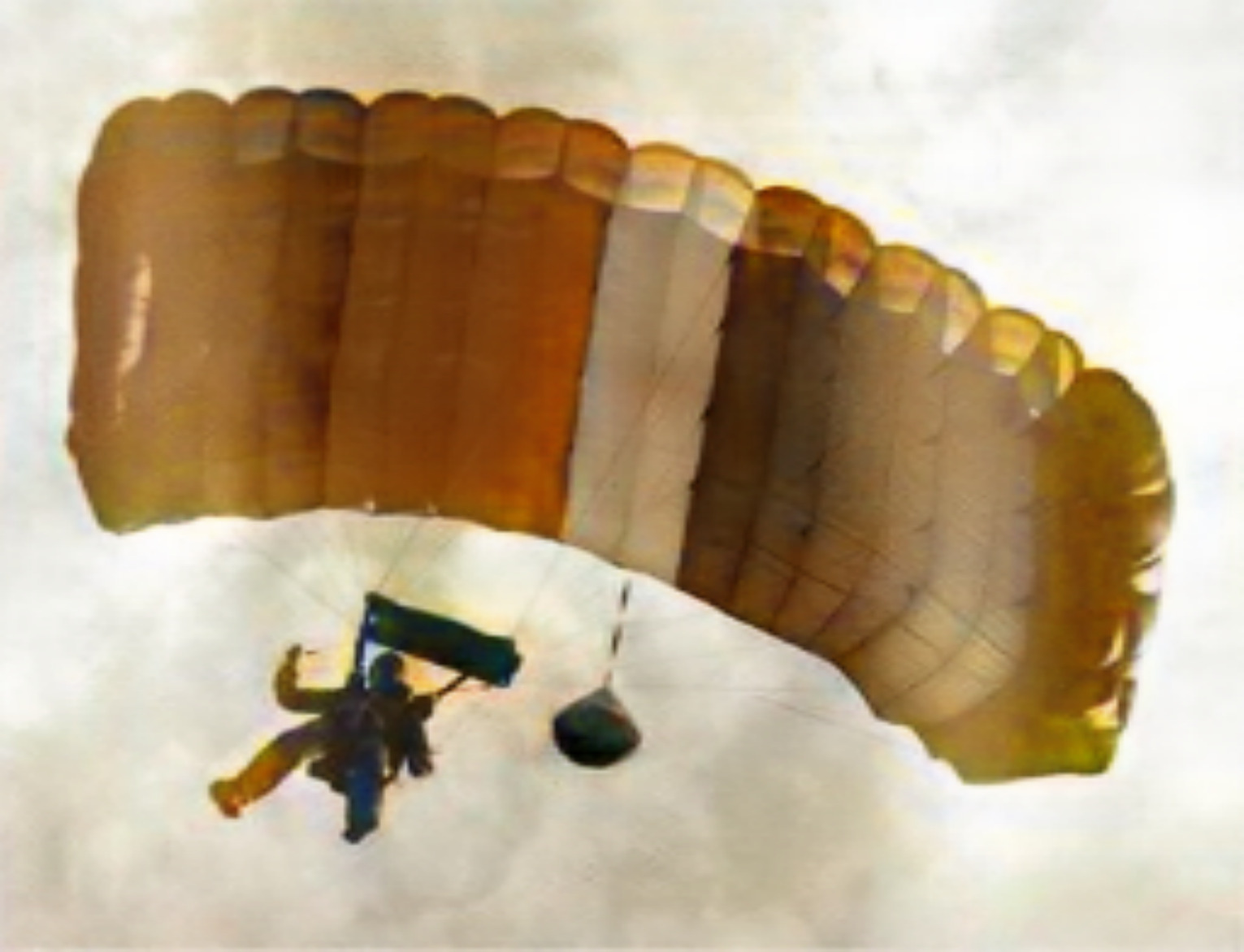}}

    \caption{ (a) unmodified; (b) DiffusionCLIP; (c) DiffuseMix; (d) SaSPA; (e) Da-Fusion; (f) DIAGen; (g)DCGAN; (h) Pix2pix-Zero; (i) Neural Style Transfer; (j) CycleGAN }
    \label{fig:photometric_augmentations}
\end{figure*}

Diffusion models generate images by gradually refining random noise through a learned denoising process \cite{ho2020denoising}. A forward process progressively corrupts an image with noise while the reverse process learns to reconstruct it step by step, enabling precise modification of image structure and style. The specimens produced by these methodologies are illustrated in \Cref{fig:photometric_augmentations}.

\textbf {CycleNet} builds on the principle of cycle consistency introduced in CycleGAN \cite{zhu2017unpaired} by integrating it into text-guided latent diffusion frameworks for unpaired image-to-image translation.
It leverages a pre-trained stable diffusion backbone \cite{rombach2022high} with ControlNet \cite{zhang2023adding} and introduces forward–backward translation cycles as a form of consistency regularization. This feature encourages generated images to retain semantic content and structural fidelity. It also aligns them with the target domain specified by textual prompts \cite{xu2023cyclenet}.  
\textbf {DiffusionCLIP} is a zero-shot modification of an image according to a text prompt, thanks to the Contrastive Language--Image Pretraining (CLIP) framework. The reverse diffusion process is optimized with directional and identity-preserving losses that preserve the original image's visual features while making the necessary change. In addition, DiffusionCLIP allows changing many attributes at once by conditioning the diffusion model on numerous text prompts. It also speeds up inference by cutting down on the number of diffusion sampling steps \cite{kim2022diffusionclip}.
\textbf {Pix2Pix Zero-shot} {is a zero-shot image augmentation technique that modifies an input image according to a specified transformation without requiring additional model training or manually defined editing masks. Given an input image and a textual description of the desired change, Pix2Pix-Zero generates an edited image that reflects the requested modification while preserving the original spatial layout, semantic content, and visual elements unrelated to the edit. The method can be applied to both real and synthetically generated images \cite{parmar2023zero}.}
\textbf {DiffuseMix}
generates a controlled variant of an input image, which is merged with the original image through masked compositing and fractal-based blending operations. This process modifies the appearance while retaining class-relevant semantic content intact. At its core, the method generates a corresponding image using a pre-trained diffusion model guided by filter-style textual prompts. The prompts alter global appearance but avoid structural change. The original image and its generated counterpart are then concatenated with a binary mask to form a hybrid image. A fractal pattern is finally blended with a small weight to add structural variety and reduce overfitting. This preserves salient content while injecting controlled diversity \cite{islam2024diffusemix}. It addresses common failures of mixup techniques that paste across classes.
\textbf {SaSPA}
generates diverse yet class-consistent images for fine-grained visual classification. It addresses the challenge of subtle class distinctions by guiding the diffusion process with edge maps and subject features rather than real images. Building on previous Img2Img \cite{fan2024scaling} and Text2Image \cite{meng2021sdedit} diffusion approaches, {It introduces abstract structural conditioning, which refers to a simplified representation of the original image, including its edge map and subject features, to improve image diversity while preserving class consistency.
} SaSPA enhances semantic variation using prompts from large language models such as GPT-4. It then filters generated images with CLIP and dataset-specific classifiers to remove low-quality samples \cite{michaeli2024advancing}.
\textbf {DA-Fusion}
utilizes pre-trained text-to-image diffusion models to edit images and create new augmented ones. 
DA-Fusion produces semantically significant alterations derived from the acquired knowledge of the pretrained model rather than from manually specified transformations. This makes the visual content more diverse while preserving semantic integrity. By conditioning the generation process on textual prompts, the method produces concept-aware alterations and realistic variations that can be applied even to previously unseen categories \cite{trabucco2023effective}.
\textbf {DIAGen} generates 
{diverse} semantic variations by changing lighting and background while moving beyond simple texture or color alterations. The method adds Gaussian noise to embeddings made using Textual Inversion to make the synthesis process more varied. Class-specific textual cues then lead the diffusion process toward changes in meaning. This yields a wide variety of samples with controlled changes, while keeping the content that is relevant to the class \cite{lingenberg2024diagen, trabucco2023effective}.

\subsection{Adversarial Augmentation}
\label{sec:adversarial}

\begin{figure*}[!htbp]
    \centering
    \subfloat[\footnotesize Original \label{fig:original_selfmix}]{\includegraphics[width=0.155\textwidth]{images/original_image.jpg}}
    \hfill
    \subfloat[\footnotesize FGSM (OpenCLIP) \label{fig:fgsm_open_clip}]{\includegraphics[width=0.155\textwidth]{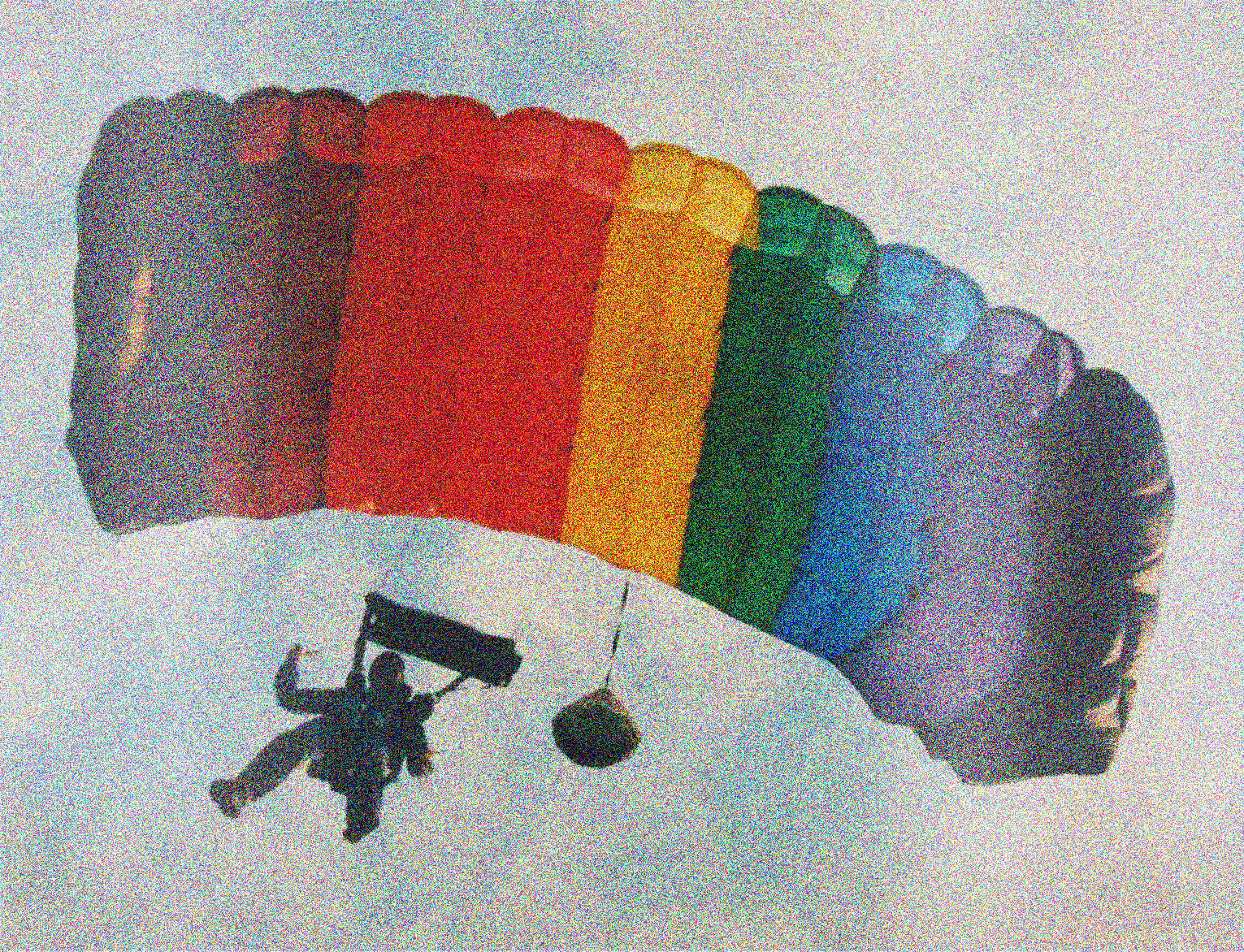}}
    \hfill
    \subfloat[\footnotesize FGSM (Titan) \label{fig:fgsm_titan}]{\includegraphics[width=0.155\textwidth]{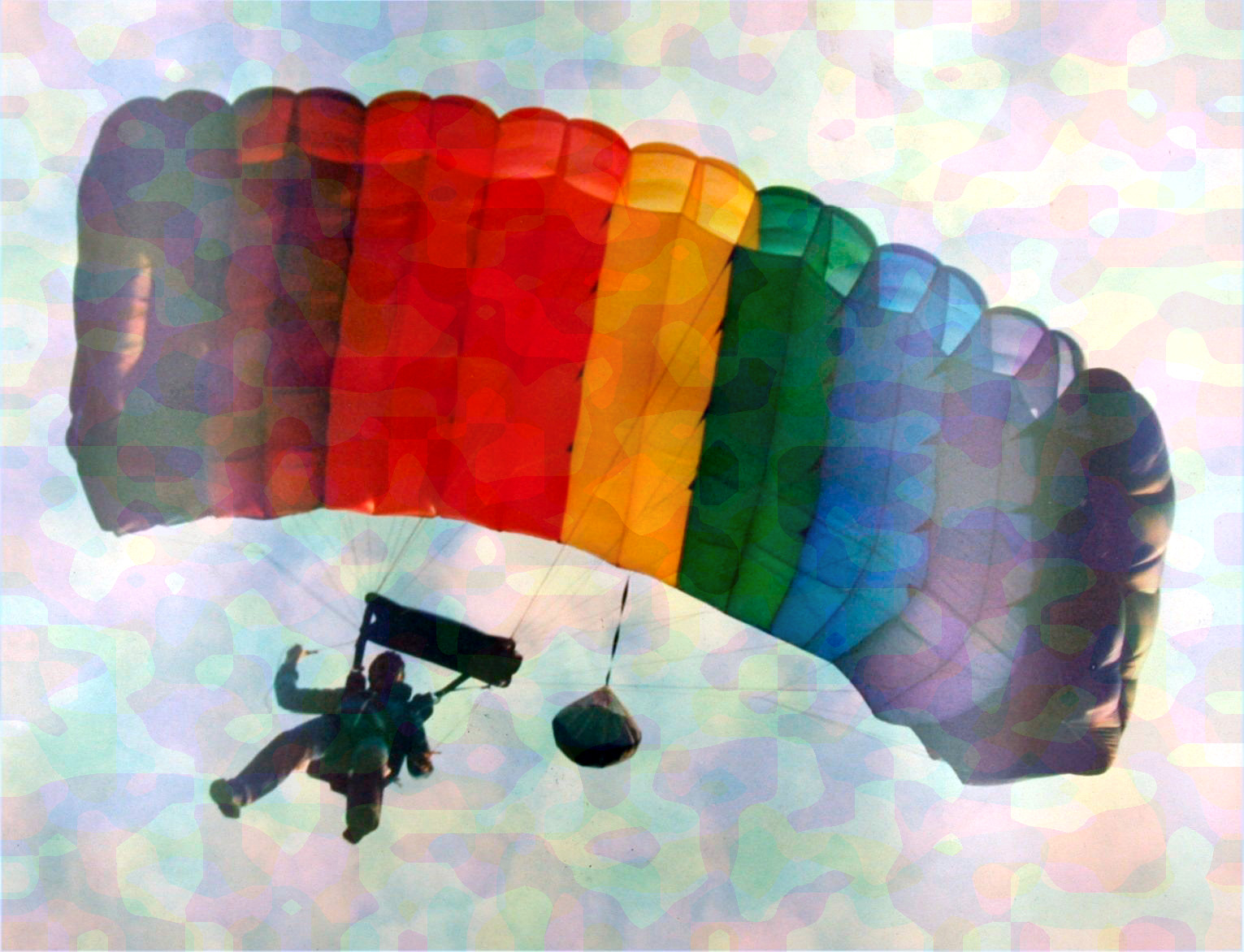}}
    \hfill
    \subfloat[\footnotesize PGD (OpenCLIP) \label{fig:pgd_open_clip}]{\includegraphics[width=0.155\textwidth]{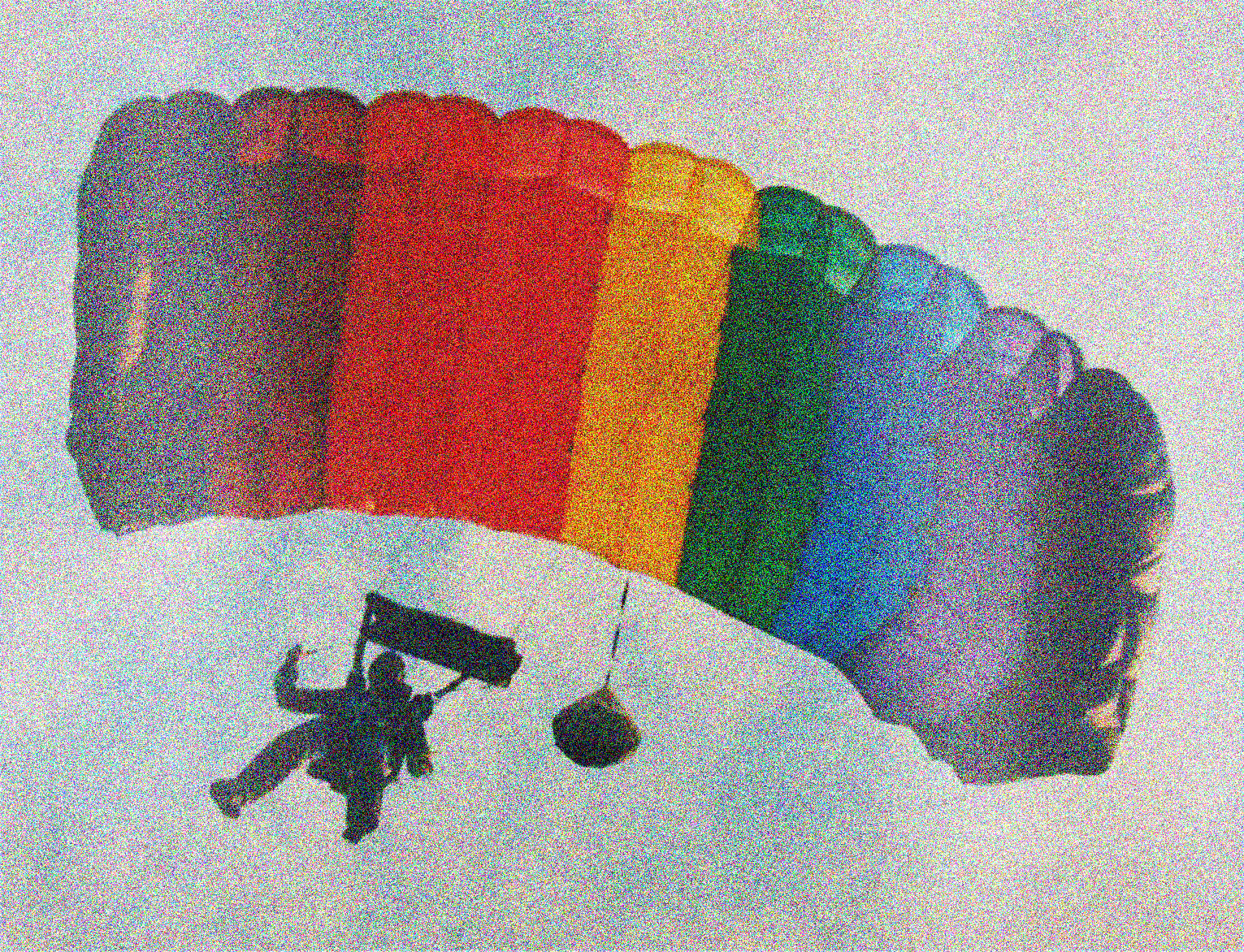}}
    \hfill
    \subfloat[\footnotesize PGD (Titan) \label{fig:pgd_titan}]{\includegraphics[width=0.155\textwidth]{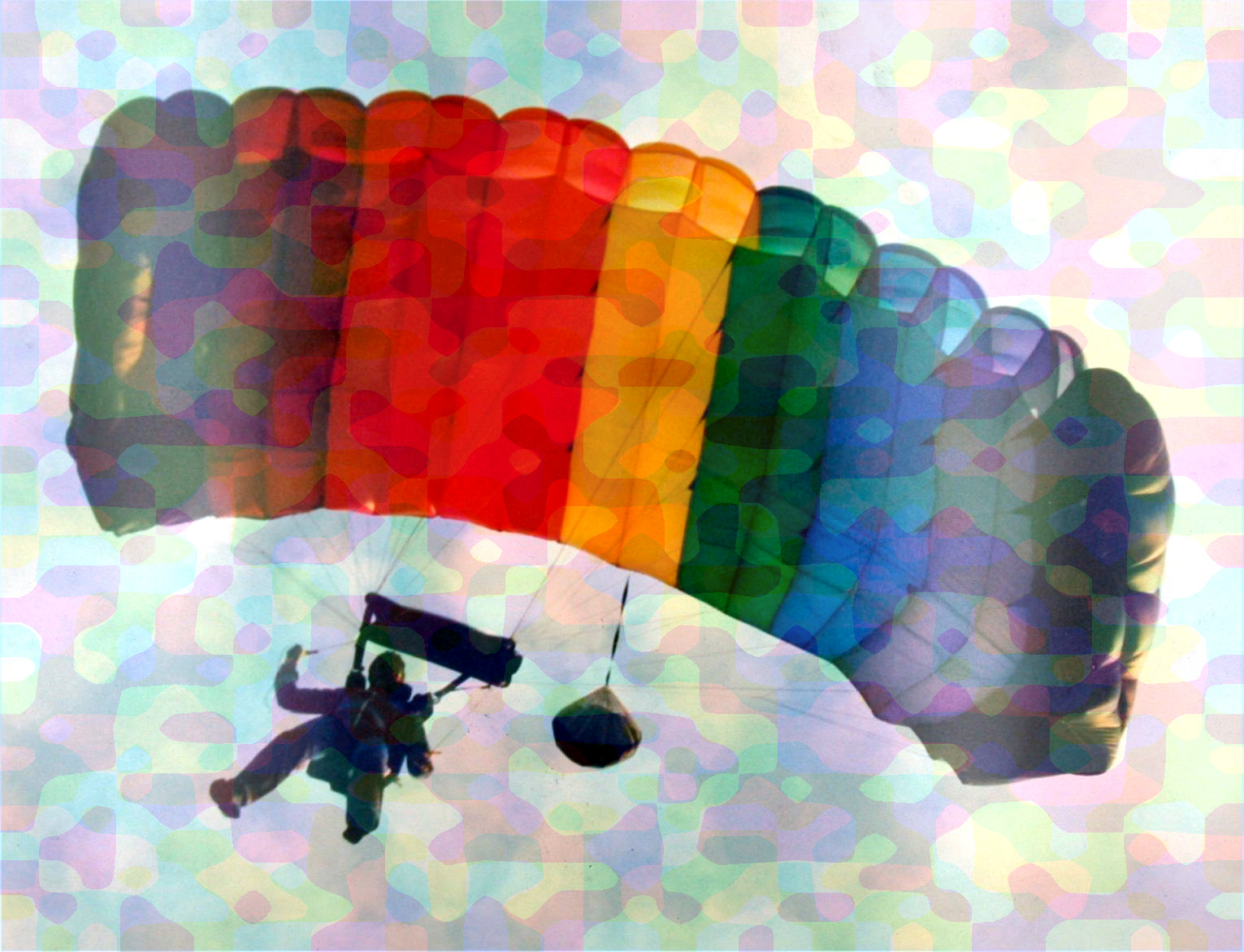}}
    \hfill
    \caption{{(a) unmodified; (b) FGSM on OpenCLIP; (c) FGSM on Titan; (d) PGD on OpenCLIP; (e) PGD on Titan}}
    \label{fig:adversarial_augmentations}
\end{figure*}

Adversarial augmentation treats adversarial examples, inputs crafted to maximally increase model loss while remaining perceptually indistinguishable from originals, as training-time augmentations. By including such inputs in the training dataset, the model is exposed to worst-case perturbations during optimization, improving its resilience to adversarial attacks at inference time \cite{naqvi2023adversarial}. The samples generated by these techniques are depicted in \Cref{fig:adversarial_augmentations}.

\begin{table*}[!htbp]
\centering
\footnotesize
\setlength{\tabcolsep}{4pt}
\renewcommand{\arraystretch}{1.20}
\caption{Summary of Image Augmentation Techniques and Their Descriptions}
\label{tab:all_techniques}
\begin{tabular}{@{}p{0.005\textwidth}p{0.15\textwidth}p{0.80\textwidth}@{}}
\toprule
\textbf{} & \textbf{Technique} & \textbf{Description} \\
\midrule

\multirow{4}{*}{\rotatebox[origin=c]{90}{\textbf{Geometric}}}
& Rotation & Rotates images to vary viewpoint while preserving object structure.
\cite{khosla2020enhancing, elgendi2021effectiveness, chlap2021review, 9596262, 10.3389/fncom.2019.00083, tasci2021voting, nanni2020deep, Abdollahi2020, Elgendi2021} \\
& Flipping & Mirrors the image horizontally or vertically to increase spatial diversity. \cite{khosla2020enhancing, elgendi2021effectiveness, chlap2021review, 9596262, 10.3389/fncom.2019.00083, tasci2021voting, Abdollahi2020, Elgendi2021} \\
& Translation & Shifts the image along horizontal or vertical axes to simulate object displacement and improve spatial generalization. \cite{khosla2020enhancing, chlap2021review, 9596262, tasci2021voting, nanni2020deep, wang2024evaluation, Abdollahi2020, Elgendi2021} \\
& Shearing & Skews the image by shifting one side relative to the other, introducing perspective-like distortions for spatial variability. \cite{elgendi2021effectiveness, chlap2021review, 10.3389/fncom.2019.00083, tasci2021voting, Abdollahi2020, Elgendi2021} \\

\midrule

\multirow{8}{*}{\rotatebox[origin=c]{90}{\textbf{Photometric}}}
& Auto Contrast & Maximizes image contrast by remapping the darkest pixel to black and the brightest to white. 
\cite{ningsih2020improving} \\
& BCET & Balances contrast without altering the overall histogram distribution, preserving mean brightness. \cite{rahman2021exploring} \\
& Brightness &  Brightness uniformly adjusts pixel intensities to simulate lighting changes.\cite{chlap2021review, bang2020image} \\
& Contrast & Contrast alters pixel values relative to the image mean to emphasize local intensity differences. \cite{cirillo2021best, wang2024evaluation} \\
& Color Space & Converts images across color spaces to alter channel-wise intensity distributions. \cite{chlap2021review, buslaev2020albumentations, khosla2020enhancing, wang2024evaluation} \\
& Equalization & Improves global or local contrast by redistributing pixel intensities using HE or CLAHE. \cite{chlap2021review, rahman2021exploring, tasci2021voting} \\
& Gamma Correction & Applies a non-linear pixel transformation to simulate human-perceived changes in real-world lighting.
 \cite{rahman2021exploring} \\
& PCA Jittering & Applies PCA to RGB values and adds noise along principal axes to simulate realistic lighting variations. \cite{buslaev2020albumentations} \\
& Inversion & Inverts pixel intensities to enhance light–dark contrast and improve visibility of structural features.
 \cite{rahman2021exploring} \\
& Weather & Synthetically renders adverse conditions (fog, rain, snow, frost) onto the image.\cite{rahman2021exploring} \\

\midrule

\multirow{4}{*}{\rotatebox[origin=c]{90}{\textbf{Noise}}}
& Elastic Deformation & Applies smooth non-linear distortions to vary shape and texture while preserving local topology.
 \cite{9506328} \\
& Dimension Reduction & Projects high-dimensional image features into compact embeddings that retain salient visual information. \cite{devries2017datasetaugmentationfeaturespace} \\
& Möbius & Applies complex coordinate mappings in pixel space to simulate perspective projection effects. \cite{Zhou_2021} \\
& Salt \& Pepper & Randomly adds white and black pixels to simulate real-world sensor or environmental noise. \cite{Qiu2025AnIE, 10.3389/fpls.2025.1555440} \\

\midrule

\multirow{3}{*}{\rotatebox[origin=c]{90}{\textbf{Selection}}}
& Random Crop & Randomly extracts a sub-region of the image, forcing the model to rely on partial visual information. 
\cite{Takahashi2020} \\
& Random Erase & Removes a random rectangular region and fills it with noise or constant values to simulate occlusion.
 \cite{zhong2020random} \\
& Patch-wise & Modifies local image patches by copying, moving, or replacing them while preserving global structure. \cite{fan2024patch} \\
\addlinespace[2pt]
\midrule

\multirow{4}{*}{\rotatebox[origin=c]{90}{\textbf{Filtering}}}
&Blur & {smooths an image to simulate camera defocus and includes Gaussian and motion blur. \cite{bang2020image}}\\
& Motion Blur & Simulates motion by adding blur around image subjects. \cite{bang2020image} \\
& Gaussian Blur & Smooths an image with a Gaussian kernel to approximate motion-induced degradation.
 \cite{bang2020image, Buddenkotte1463} \\
& Sharpening & Amplifies high-frequency components to enhance perceived edge detail and fine texture. \cite{shorten2019survey} \\

\midrule
\multirow{3}{*}{\rotatebox[origin=c]{90}{\textbf{Self-Mix}}}
& Self Mix & Cuts a random patch from an image and pastes it back into the same image to improve few-shot generalization. \cite{Seo_2021}\\
& Salf Mix & Produces a self-mixed image based on a saliency map of the same image. \cite{s21248444}\\
& AugMix & Overlays multiple augmented versions of the same image, including the original, to improve consistency. 
\cite{hendrycks2020augmixsimpledataprocessing}\\

\midrule
\multirow{9}{*}{\rotatebox[origin=c]{90}{\textbf{Sample Mixing}}}
& MixGen & Overlays two images via linear interpolation to produce a single blended output. 
\cite{hao2023mixgennewmultimodaldata} \\
& Robust MixGen & Extends MixGen with similarity-aware local blending guided by cosine similarity between feature regions. \cite{KIM2025129167} \\
& MixUp & Blends two images through pixel-level interpolation to produce smooth class-boundary representations. \cite{zhang2018mixupempiricalriskminimization} \\
& CutMix & Replaces an image region with another image patch and mixes labels proportionally. \cite{yun2019cutmixregularizationstrategytrain} \\
& ResizeMix & Pastes a resized version of a source image onto a target image without saliency computation. \cite{qin2020resizemixmixingdatapreserved} \\
& SaliencyMix & Uses saliency maps to select semantically important patches for mixing rather than random regions. \cite{uddin2021saliencymixsaliencyguideddata} \\
& RICAP & Randomly crops and combines patches from multiple same-class images to form a composite sample. \cite{Takahashi_2020} \\
& Keep Augment & Applies augmentation while retaining high-saliency regions to preserve class-relevant content.
 \cite{gong2020keepaugmentsimpleinformationpreservingdata} \\
& IP Mix & Preserves high-saliency regions from both input images during mixing to maintain label-relevant visual content. \cite{huang2023ipmix}\\

\midrule

\multirow{4}{*}{\rotatebox[origin=c]{90}{\textbf{GAN}}}
& DCGAN & Generates synthetic samples from random noise using convolutional generator–discriminator networks.
 \cite{ srivastav2021improved} \\ 
& Neural Style Transfer & Separates and recombines content and style of arbitrary images. \cite{gatys2015neuralalgorithmartisticstyle}\\
& Pix2Pix & Performs supervised image-to-image translation using a U-Net generator and PatchGAN discriminator. \cite{isola2017image} \\
& CycleGAN & Performs unpaired image-to-image translation using cycle-consistency loss to avoid mode collapse. \cite{zhu2017unpaired} \\

\midrule

\multirow{7}{*}{\rotatebox[origin=c]{90}{\textbf{Diffusion}}}
& CycleNet & Extends CycleGAN's consistency principle to text-guided latent diffusion for unpaired image translation. 
\cite{xu2023cyclenet} \\
& DiffusionCLIP & Fine-tunes reverse diffusion with CLIP-guided directional losses for zero-shot text-driven image editing.
 \cite{kim2022diffusionclip} \\
& Pix2Pix-zero & Edits real images without per-image prompting via DDIM inversion and CLIP-guided edit directions. \cite{parmar2023zero} \\
& DiffuseMix & Merges a diffusion-generated image variant with the original using masked compositing and fractal blending.
 \cite{islam2024diffusemix} \\
& SaSPA & Guides diffusion with edge maps and subject features to generate diverse, class-consistent images for fine-grained classification, filtering outputs with CLIP and task-specific classifiers. \cite{michaeli2024advancing} \\
& DA-Fusion & Edits real images using text-to-image diffusion priors to generate diverse concept-aware augmented samples. \cite{trabucco2023effective} \\
& DIAGen & Adds Gaussian noise to Textual Inversion embeddings and guides synthesis with class-specific prompts for semantic diversity. \cite{lingenberg2024diagen, trabucco2023effective} \\

\midrule


\addlinespace[1pt]
\multirow{2}{*}{\rotatebox[origin=c]{90}{\textbf{Adversarial}}}
& FGSM & Generates adversarial examples by adding a small $\epsilon$-bounded perturbation in the direction of the input-loss gradient to increase the model's prediction error. 
\cite{kurakin2016adversarial} \\
& PGD &  Generates stronger adversarial examples by iteratively applying small gradient-based perturbations and projecting each step back into the $\epsilon$-ball around the original input.\cite{geisler2024attacking} \\
\addlinespace[3pt]
\bottomrule
\end{tabular}
\end{table*}

\section{Empirical Study Design}
\label{sec:empirical_study_design}

This section describes the design of the empirical study used to evaluate the augmentation techniques catalogued in Section~\ref{sec:augmentation_techniques}. {The objective of this evaluation is to investigate the effect of each image augmentation technique on the retrieval performance of the selected embedding models. Specifically, the evaluation measures the capability of each model to preserve the semantic category of the original image after applying different visual transformations. A robust embedding model is expected to retrieve images from the same semantic category as the original query, even when the input image is modified by augmentation techniques. }The study assesses each technique across four metrics: retrieval failure {count}, embedding-space similarity and diversity, embedding model uncertainty, and semantic realism. The following subsections present the research questions, the embedding models and benchmark datasets used, the experimental setup, and the evaluation metrics including the metamorphic relation that governs correctness.

\subsection{Research Questions}
\label{sec:evaluation_questions}

The empirical study is guided by the following four research questions, each targeting a specific measurable aspect of augmentation behavior. Whereas the SLR (Section~\ref{sec:systematic_literature_review}) catalogues the techniques, these questions drive their experimental evaluation.

\begin{description}
\item[\textbf{RQ1:}] \emph{Which augmentations most often break correct retrieval?}
    How do augmentation techniques compare in their ability to expose retrieval failures, as measured by the rate of metamorphic relation violations and the category-preservation failure {count} across augmented queries? This identifies the transformations with the greatest raw fault-exposure power.

    \item[\textbf{RQ2a:}] \emph{How far does each augmentation move the image in embedding space?}
    To what extent does each augmentation technique alter the embedding-space representation of the original image in terms of cosine similarity and overall embedding diversity? This reveals whether a technique perturbs the model's internal representation strongly or only marginally.

    \item[\textbf{RQ2b:}] \emph{{Do these changes in the embedding space compromise the stability of the model’s learned representations?}}
    How stable are the model's semantic representations under augmentation, as measured by embedding dispersion relative to class centroids, pairwise distance between original and augmented embeddings, Mahalanobis distance from the embedding distribution, and ensemble agreement across diverse KNN classifiers? {This distinguishes tolerable shifts of augmented images within the embedding space from shifts that push them into ambiguous, failure-prone regions.}

    \item[\textbf{RQ3:}] \emph{Are the challenging images still realistic?}
    How semantically realistic and visually plausible are the augmented images produced by each technique, as evaluated by an automated large language model judge? This separates augmentations that expose realistic hard cases from those that fail only because they produce unnatural, artifact-ridden images.


    \item[\textbf{RQ4:}] \emph{Overall, which augmentations make the best test generators?}
    Combining retrieval failure {count} (RQ1), embedding-space similarity and diversity (RQ2a), embedding uncertainty (RQ2b), and semantic realism (RQ3), which augmentation techniques most effectively expose weaknesses in the retrieval model, and under what transformation conditions are these effects most pronounced?

\end{description}

\subsection{Embedding Models}

Two embedding models are employed to encode images into dense vector
representations for similarity-based retrieval evaluation: Amazon Titan
Multimodal Embeddings G1, a commercial model available through AWS Bedrock \cite{aws_titan_2024},
and OpenCLIP ViT-H/14, an open-source implementation of the CLIP framework \cite{ilharco2021openclip}.
Titan was selected because it represents a commercially deployed embedding
service and was recommended by our industry collaborator, while OpenCLIP
provides a reproducible open-source counterpart with publicly available
weights, source code, and preprocessing pipelines
\cite{aws_titan_2024,cherti2023reproducible}. Amazon Titan produces 1024-dimensional multimodal embeddings and uses a contrastive learning-based training strategy that places semantically related
samples closer together in the embedding space and dissimilar samples farther
apart \cite{chen2020simple}. Its managed-cloud deployment also provides a
realistic black-box setting in which model internals are inaccessible.
For OpenCLIP, we use the ViT-H/14 architecture with the
LAION-2B-S32B-b79K pretrained weights. It also produces 1024-dimensional
embeddings and is trained with a contrastive objective, enabling direct
metric-level comparison with Titan while differing in architecture, training
data, and deployment context.

The use of one closed commercial model and one fully open-source model allows
us to assess whether the effects of augmentation generalize across different
embedding systems while maintaining the same output dimensionality.
Alternative models, including SigLIP, DINOv2, and BLIP, were not included
because of differences in their training objectives, modality support, or
retrieval configuration. Google Vertex AI Multimodal Embeddings was also
considered, but its available output dimensions do not include 1024; including
it would therefore introduce embedding dimensionality as an additional
experimental variable.

Both models are evaluated in a zero-shot setting without fine-tuning on the
evaluation datasets. They are applied directly to CIFAR-10 \cite{krizhevsky2009learning}, ImageNet-1K \cite{deng2009imagenet}, and the 
March Networks dataset. This setting is relevant to robustness evaluation
because zero-shot models may be more sensitive to distribution shifts
introduced by augmentation than models fine-tuned on the target domain.

\subsection{Benchmark Datasets}
We used three datasets to evaluate the effectiveness of each augmentation technique: CIFAR-10, ImageNet-1K and a dataset from March Networks, our industry partner. In our analysis of augmentation techniques (Sections \ref{sec:systematic_literature_review} and \ref{sec:augmentation_techniques}), ImageNet-1K emerged as the most utilized dataset, accounting for 18.56\%, followed by CIFAR-100 at 11.34\% and CIFAR-10 at 10.31\%, over a total of 56 {reviewed papers on augmentation techniques}. CIFAR-10 is incorporated because of its widespread application in augmentation research and its manageable size, comprising 60,000 annotated images across ten object categories. Its compact size and simple structure make it appropriate for fast experimentation, algorithm validation, and comparative assessment of augmentation techniques, facilitating effective testing with constrained computational resources. {CIFAR-10 was selected instead of CIFAR-100 because it provides a substantially larger number of samples per class. Although CIFAR-100 contains a greater number of categories, its test set includes only 100 images per class. Given the low image resolution and the visual ambiguity of some samples, this class-wise sample size may be insufficient for drawing reliable conclusions under the adopted experimental configuration. Therefore, CIFAR-10 was used because its larger number of images per class supports more stable and statistically reliable evaluation results.}

In contrast, ImageNet-1K provides a large and varied benchmark for image classification, consisting of more than 1.28 million training images across 1,000 object categories. This dataset encompasses a diverse array of high-resolution images captured under varying conditions, challenging the capabilities of multiple models and rendering it an excellent benchmark for evaluating the efficacy of each augmentation technique. The final dataset (referred to as the MN dataset) is derived from the March Networks database and comprises approximately 20,000 real-world images distributed across 100 categories. It enables us to assess models and augmentation techniques under real-world conditions. Together, these three datasets allow us to assess augmentation effects under varying data distributions, resolutions, and visual domains.

Unlike CIFAR-10 and ImageNet-1K, which have an approximately uniform number of samples per class, the MN dataset is imbalanced. This nonuniform distribution is explicitly considered throughout all stages of the experimental pipeline, including augmentation generation, database construction, image retrieval, failure-rate computation, and statistical analysis. Accounting for the unequal category sizes ensures that results reflect genuine differences in model performance rather than differences in the number of samples available in each category.

\subsection{Experimental Setup}
The Amazon Titan embedding model is deployed on AWS cloud infrastructure and accessed through the Amazon Bedrock API. OpenCLIP runs locally via the \texttt{open\_clip} library on a high-performance workstation running Ubuntu 22.04, equipped with 128~GB of RAM, an NVIDIA L40S GPU, and a 12th-generation Intel Xeon Ice Lake processor. All embedding vectors produced by both models are stored in a Chroma vector database, which supports efficient indexing and similarity-based retrieval.


\subsection{Design of Experiments} \label{sec:design_of_experiments}
{The evaluation is conduct a controlled evaluation. To this end,} each augmentation technique is applied independently to each dataset. For every augmentation technique, the original images and their corresponding augmented versions are embedded separately using Titan and OpenCLIP. This design allows the effect of each augmentation technique to be analyzed individually and enables a direct comparison between the behavior of the two embedding models under the same testing conditions. 

The ImageNet-1K and CIFAR-10 datasets are randomly sampled to reduce bias. For CIFAR-10, 1,000 images are randomly selected for each of the 10 classes, resulting in 10,000 images. For ImageNet-1K, 100 classes are randomly selected, and 300 images are randomly sampled for each selected category, resulting in 30,000 images. A more extensive selection was chosen from ImageNet-1K due to its samples frequently providing higher resolution and increased visual complexity relative to CIFAR-10 {\cite{rufaida2023looking}}. This makes ImageNet-1K apriori more suitable for generating demanding test cases. {However, CIFAR-10 is employed in this study for two main reasons. First, it enables direct comparison with prior studies because, as discussed in the previous section, CIFAR-10 is one of the most widely used image-classification datasets. Second, its low-resolution images provide a suitable setting for evaluating the robustness of the selected models and their ability to maintain reliable performance when processing images with limited visual detail.} The MN dataset consists of 20,000 images. Overall sample sizes were chosen to balance statistical rigor with computational feasibility. {Each augmentation technique is applied once to every original image and evaluated independently using the original images together with only the samples generated by that technique. Thus, the resulting dataset contains twice as many samples as the original dataset, with equal numbers of original and augmented images in each category. This approach ensures complete dataset coverage while limiting computational cost}.

The queries are constructed using the original photos, and the embedding model must preserve the semantic and visual characteristics to ensure that the returned images including the original and augmented images in each query belong to the same category. The number of retrieved images is set equal to the total number of original and augmented images available in the corresponding semantic category. This setting enables the evaluation to determine whether the retrieved results preserve the semantic category of the query image after augmentation. During the testing process, the statistical evaluation metrics are computed to quantify the retrieval behavior of each model under each augmentation condition. ChromaDB is employed to store the embedding vectors and provide efficient access during the retrieval process. The retrieval operation is performed based on cosine similarity, which is implemented within the database. This setup enables a consistent comparison between Titan and OpenCLIP and supports the systematic analysis of how different augmentation techniques affect the semantic consistency and robustness of the embedding models. 

FGSM and PGD are gradient-based adversarial augmentation techniques that require access to the model weights and input gradients. Such information was fully accessible for OpenCLIP, enabling their direct implementation. In contrast, Titan operates as a black-box service and does not expose its internal parameters or gradients. Therefore, Zeroth-Order Optimization was employed to approximate the required gradients and generate adversarial perturbations for Titan \cite{chen2017zoo}. Style-transfer-based augmentation techniques additionally require a separate set of reference style images, as discussed in \cite{Kitov2026StyleTransferDataset}. Pix2Pix is not evaluated because it requires paired source and target images, which are unavailable for the three datasets considered in this study. Instead, CycleGAN is employed as an unpaired image-to-image translation alternative.

To this end, a mixed-effects logistic regression analysis is performed to quantify the impact of each augmentation strategy to the observed failure {count}. {A mixed-effects logistic regression model is appropriate because the retrieval outcome is binary at the retrieved-image level, indicating whether the retrieved image belongs to the expected semantic class. Since all augmentation techniques are evaluated using the same source-query set and retrieval protocol, the augmentation technique can be modeled as a fixed effect. Dataset category is included as a random effect to account for class-specific variability and the dependence among observations belonging to the same class. Unlike comparisons based only on aggregated failure {Counts}, this approach uses all retrieval outcomes, accommodates unequal class sizes, and provides adjusted odds ratios and confidence intervals for estimating the effect of each augmentation technique.}

In order to accomplish this, the retrieved images associated to each query are classified, and a binary response variable is created to signify whether each retrieved image represents a retrieval failure. This concept allows for modeling the failure outcome at the image-retrieval level instead of solely relying on aggregated failure-rate information. The mixed-effects logistic regression model is especially appropriate for this research as it considers both systematic and categorical causes of variance. The augmentation technique, embedding model, and dataset are regarded as fixed effects, and their influences on the failure probability are clearly examined. Conversely, the image category is treated as a random effect to account for category-specific unpredictability, which, as previously mentioned, significantly impacts retrieval performance. This model quantifies each augmentation's effect while accounting for variation across dataset, embedding model, and image category, and is expressed as follows:

\begin{equation}
\begin{aligned}
\mathrm{logit}(p_{ijkl})
&= \beta_0
+ \beta_1\times \mathrm{Augmentation}_{i} \\
&\quad + \beta_2\times \mathrm{Model}_{j} \\
&\quad + \beta_3\times \mathrm{Dataset}_{k}
+ b_l
\end{aligned}
\end{equation}

\noindent here, $p_{ijkl}$ denotes the probability of retrieval failure, and Augmentation, Model, and Dataset are treated as categorical (dummy-coded) factors, $\beta_{0}$ is the intercept, and $\beta_{1}$ and $\beta_{2}$ represent the fixed-effect coefficients associated with the considered explanatory variables. The term $b_{l}$ denotes the random effect, which is introduced to account for variability associated with image categories. The random effects are commonly assumed to follow a normal distribution, as expressed in \Cref{distro}. 
\begin{equation}
\label{distro}
b_l \sim \mathcal{N}(0,\sigma_{\mathrm{category}}^2)
\end{equation}

A single mixed-effects logistic regression model was fitted for each dataset and embedding model, utilizing all augmentation approaches concurrently. The non-augmented condition served as the reference level, and the fixed-effect coefficient for each augmentation approach was evaluated in relation to this baseline. Each model's failure outcome was represented by a binomial response consisting of the count of failed and successfully retrieved photos. The augmentation approach was considered a fixed effect, whereas the image category was modeled as a random intercept to address category-dependent variability in retrieval difficulties. Following model fitting, the fixed-effect coefficient corresponding to each augmentation strategy was obtained from the fitted model. The calculated coefficients, shown on the log-odds scale, yield the relevant odds ratio by exponentiating the coefficient, as seen in \Cref{odd_ratio}. 
\begin{equation}
\label{odd_ratio}
\mathrm{OR}_{\mathrm{aug}} = e^{\beta_{\mathrm{aug}}}
\end{equation}

The resultant odds ratio measures the relative alteration in the probabilities of retrieval failure caused by each augmentation approach in comparison to the baseline condition. An odds ratio over one signifies that the augmentation elevates the likelihood of retrieval failure, while an odds ratio below one denotes that the augmentation diminishes the likelihood of retrieval failure compared to the original, non-augmented state. Note that the baseline condition retrieves fewer images per query than the augmented configurations, which contain both original and augmented images; odds ratios below one are therefore common across the evaluated techniques. Accordingly, augmentations are compared by the relative ordering of their odds ratios rather than by their position with respect to unity. The mixed-effects odds ratios for all augmentation approaches across the three assessed datasets and embedding models are presented in \Cref{tab:failure_rate_mixed_effects}. These results offer a category-adjusted assessment of the impact of each augmentation strategy on retrieval failure, thereby indicating the degree to which each method diminishes or maintains retrieval performance relative to the baseline.

\subsection{Evaluation Metrics}
Each augmentation technique is assessed across four complementary metrics that jointly capture both the \emph{challenge} an augmentation poses to the retrieval system and the \emph{quality} of the inputs it produces. The implemented metrics are enumerated as follows:
\begin{description}
\item{\textbf{Retrieval Failure {Count}}} The retrieval failure {count} is measured using a metamorphic relation (MR), which states that an augmented image should preserve the original semantic category after applying an augmentation technique. If the retrieval system returns images from a different category, the MR is violated and the result is counted as a retrieval failure.

\begin{definition}[Augmentation Metamorphic Relation]
\label{def:mr}
Let $x$ be a source image with ground-truth category $c(x)$, and let $x' = T(x)$ be a follow-up image produced by a semantic-preserving augmentation $T$. Let $\mathcal{R}(q, \mathcal{D}, k)$ denote the set of $k$ images retrieved from database $\mathcal{D}$ in response to query $q$ using an embedding-based retrieval system $f$. Because a single query may return multiple images, the MR requires that \emph{every} retrieved image belongs to the correct category. The MR is satisfied if and only if:
\[
\forall\, x,\; \forall\, r \in \mathcal{R}(f(x'), \mathcal{D}, k),\; c(r) = c(x)
\]
that is, every image returned in response to the augmented query belongs to the same semantic category as the original. Any retrieved image with $c(r) \neq c(x)$ constitutes a detected retrieval fault.
\end{definition}

Each augmentation cataloged in Section~\ref{sec:augmentation_techniques} serves as a candidate generator of the follow-up input $x'$.

\item{\textbf{Embedding-Space Similarity and Diversity}} This metric measures how much each augmentation shifts an image's position in embedding space. For every augmented image, the cosine similarity between its embedding and that of the original is computed (\cref{cosine_sim}). The complement of this similarity, $\Delta_i$, captures the magnitude of the representation change (\cref{diff_equ}). The mean and standard deviation of $\Delta_i$ across all images summarise each technique: higher means indicate larger average perturbations, higher standard deviations greater variability.
\begin{equation}
\label{cosine_sim}
\cos(\mathbf{u}, \mathbf{v}) = \frac{\mathbf{u} \cdot \mathbf{v}}{|\mathbf{u}| , |\mathbf{v}|}
\end{equation}
\begin{equation}
\label{diff_equ}
\Delta_{i} = 1- \cos(\mathbf{x}i, \mathbf{x}{i_{aug}})
\end{equation}
\item{\textbf{Embedding Model Uncertainty}} Uncertainty estimation measures the stability of the model's representations under augmentation. Because neither Titan nor OpenCLIP exposes model internals, all estimators operate solely on output embeddings. Four estimators are used. \emph{Embedding dispersion} is the average cosine distance of the augmented embeddings from their centroid $\boldsymbol{\mu}$ (\cref{centroid},~\cref{discrepancy}); \emph{pairwise distance} is the cosine distance between an original image and its augmented counterpart (\cref{Pairwise}); the \emph{Mahalanobis distance} measures how unusual an augmented embedding is relative to the distribution, accounting for its covariance (\cref{mahalanobis}); and the \emph{ensemble method} aggregates KNN classifiers ($k \in {5,10,15}$, Euclidean and cosine metrics, optional PCA to 100/200 components), where disagreement indicates higher uncertainty.
\begin{equation}
\label{centroid}
\boldsymbol{\mu} = \frac{1}{N} \sum_{i=1}^{N} \mathbf{x}i
\end{equation}
\begin{equation}
\label{discrepancy}
U{\text{disp}} = \frac{1}{N} \sum_{i=1}^{N} \left( 1 - \cos(\mathbf{x}i, \boldsymbol{\mu}) \right)
\end{equation}
\begin{equation}
\label{Pairwise}
U{\text{pair}} = 1 - \cos(f(x),, f(x'))
\end{equation}
\begin{equation}
\label{mahalanobis}
D_M(\mathbf{x}) = \sqrt{(\mathbf{x} - \boldsymbol{\mu})^{T},\mathbf{\sigma}^{-1},(\mathbf{x} - \boldsymbol{\mu})}
\end{equation}

\item{\textbf{Semantic Realism}} This metric evaluates the visual realism of augmented images and determines whether they could plausibly occur in real-world conditions. It is used to assess the practical effectiveness of each augmentation technique, since a useful test image should not only challenge the embedding model but also remain visually realistic.

In addition to image description, the ability of VLMs to extract and interpret visual features makes them suitable candidates for image assessment tasks \cite{chen2024mllm}. Since these models can jointly process visual and textual information, they can be used as evaluators to assess images from different perspectives, such as semantic consistency, visual realism, object preservation, and the presence of visual artifacts \cite{hu2023tifa}. Therefore, VLMs can serve not only as image understanding models but also as automated judges for evaluating the quality and reliability of augmented images. Semantic realism is assessed using LLaVA \cite{liu2023visual}, an open-source multimodal model that pairs a pre-trained vision encoder with a large language model. {The 7-billion-parameter LLaVA model initialized from the `liuhaotian/llava-v1.5-7b` pretrained checkpoint is adopted to provide a practical tradeoff between multimodal inference capability, computational complexity, and resource requirements.} LLaVA acts as an automated semantic judge, scoring each augmented image against nine application-independent criteria that can be adapted to different operational contexts:
\begin{itemize}
    \item Physical plausibility
    \item Lighting consistency
    \item Perspective and camera geometry
    \item Depth of field and focus realism
    \item Texture and material realism
    \item Edge integrity and object boundaries
    \item Noise/compression/sensor characteristics
    \item Semantic coherence of the scene
    \item Local artifacts
\end{itemize}

Each criterion is scored 0 (unrealistic), 1 (partially realistic), or 2 (fully realistic), and scores are averaged to produce a single realism score per image. This study utilizes a one-shot prompting strategy to guarantee that the judge model produces responses in a uniform and predetermined format, as demonstrated in Algorithm~\ref{alg:realism_pseudocode}. A standardized response format is important because a large number of augmented images must be scored automatically, and a uniform structure enables reliable score extraction.

\begin{table}[!htbp]
\centering
\refstepcounter{myalgorithm}
\textbf{Algorithm~\themyalgorithm: Pseudocode for Realism Evaluation Using a VLM}
\label{alg:realism_pseudocode}

\vspace{2pt}
\setlength{\tabcolsep}{3pt}
\renewcommand{\arraystretch}{1.05}
\footnotesize

\begin{tabular}{p{0.07\columnwidth} p{0.86\columnwidth}}
\hline
\textbf{Step} & \textbf{Operation} \\
\hline
\addlinespace[2pt]
1 & Provide the augmented image $I_{\mathrm{aug}}$ to the VLM $M$. \\

2 & Generate a structured prompt for semantic realism assessment. \\

3 & Evaluate $I_{\mathrm{aug}}$ based on nine realism criteria: physical plausibility, lighting consistency, camera geometry, focus realism, texture realism, edge integrity, noise characteristics, semantic coherence, and local artifacts. \\

4 & Assign $r_i \in \{0,1,2\}$ to each criterion, where 0, 1, and 2 denote unrealistic, partially realistic, and realistic conditions, respectively. \\

5 & Compute the total realism score as $R_s = \sum_{i=1}^{9} r_i$. \\

6 & Ensure that the final score satisfies $0 \leq R_s \leq 18$. \\

7 & Return $R_s$ using the predefined output format. \\
\addlinespace[2pt]
\hline
\end{tabular}
\end{table}
\end{description}

The overall ranking is determined using three evaluation criteria: failure {count}, realism, and stability. These criteria are not assigned equal weights because they represent different levels of importance in the evaluation. failure {count} and realism are each assigned a weight of 0.4, as they directly measure the ability of an augmentation technique to expose model weaknesses while preserving visually plausible content. Stability is assigned a lower weight of 0.2 because it reflects the consistency of the observed performance rather than the primary effectiveness of the augmentation.  
The comprehensive ranking of the augmentation taxonomies was derived using the same weighted scoring scheme introduced in the preceding paragraph, in which greater importance was assigned to the failure {count} and realism score than to performance stability. Specifically, the weighted overall score was first computed for each individual augmentation technique by combining the normalized values of the three evaluation criteria. Subsequently, the category-level score was obtained by averaging the overall scores of all augmentation techniques belonging to the corresponding taxonomy. This aggregation procedure enables a systematic comparison of augmentation categories while accounting for the relative effectiveness of their constituent techniques.
\section{Results}
\label{sec:results}
This section reports the experimental findings organized by the four research questions (RQ1--RQ4) defined in Section~\ref{sec:evaluation_questions}. Results are derived from CIFAR-10, ImageNet-1K, and the March Networks dataset and cover both Titan and OpenCLIP embeddings to enable direct comparison across augmentation categories and embedding models. Hyperparameter configurations for all evaluated techniques are summarized in \Cref{tab:hyperparameters} {provided in Appendix}. Hyperparameter values were chosen to produce meaningful and perceptible
modifications for evaluating retrieval robustness.

\subsection{RQ1: Retrieval Failure {Count}}

The MR retrieval failure {count} evaluates the capability of the embedding model to preserve semantic information when the input image is transformed by an augmentation technique. {This metric quantifies the number of retrieved outputs that violate the defined metamorphic relation with respect to the input query.}

The MR violation rate is reported in \Cref{tab:failure_count_std} {in Appendix} for both models across all three datasets. \Cref{tab:failure_count_std} presents the average \emph{number} of failed retrievals together with its standard deviation for each augmentation technique. These are absolute counts rather than rates: because the augmented configuration retrieves twice as many images as the baseline, counts are not directly comparable between the two settings, and the odds ratios in \Cref{tab:failure_rate_mixed_effects} should be used for that comparison. The mean value measures the overall effect of each augmentation on retrieval performance, whereas the standard deviation (STD) describes the range of failure counts among experimental settings. Collectively, these statistics explain the fault-exposure efficacy of each augmentation strategy and the robustness of each embedding model against various augmentation-induced perturbations. {\Cref{tab:failure_rate_mixed_effects} uses color coding to improve the readability and interpretation of the results. Odds ratios greater than one, indicating an increased failure likelihood relative to the baseline, are highlighted in green, whereas odds ratios below one, indicating a decreased failure likelihood relative to the baseline, are highlighted in red.}

The data shown in \Cref{tab:failure_count_std} encompass the failure counts acquired for both the baseline and augmented-image retrieval configurations. Because the number of retrieved images differs between the baseline and augmented settings, direct comparison of the raw number of failures can lead to a misleading interpretation of augmentation effectiveness. For example, in the ImageNet experiments, each baseline query retrieves 300 images, whereas each augmented query retrieves 600 images because the retrieval database contains both original and augmented images. Therefore, a higher absolute number of failed retrieved images may occur simply because the augmented setting contains twice as many retrieved samples, rather than because the augmentation technique is intrinsically more harmful. Accordingly, the failure increase should be interpreted relative to the corresponding baseline condition and the total number of retrieved images.

The findings reported in {Table~\ref{tab:failure_rate_mixed_effects} }indicate that the effectiveness and impact of the evaluated augmentation techniques are strongly dependent on the characteristics of the dataset and the visual properties of the images. ImageNet produces lower failure {counts} because its images are generally clearer, higher-resolution, and more object-centered. In contrast, CIFAR-10 provides fewer fine-grained visual details due to its low resolution, while MN contains more realistic and less ideal visual scenarios. These properties make semantic category preservation more difficult for the embedding models. This behavior is also reflected in the large STD values observed for several augmentation techniques, suggesting that the model response is highly category-dependent. In other words, the evaluated augmentations do not affect all image categories uniformly. For some categories, extracting stable semantic representations is more challenging, whereas other categories remain relatively robust under the same transformation. For example, in CIFAR-10, augmentation techniques such as Translate, Shear, Brightness, Gamma Correction, PCA Jitter, and Color Space produce STD values close to or above 300. This indicates that, although their mean failure counts may appear moderate, their effects vary substantially across categories. Some classes generate a large number of failures, while others remain comparatively stable. Across the three evaluated datasets, the highest STD values are mainly observed for Brightness, Translate, Shear, Color Space, and Gamma Correction. The high variability of Brightness, Color Space, and Gamma Correction can be attributed to their direct modification of illumination, color distribution, and intensity characteristics, which can strongly affect texture- and color-dependent categories. Similarly, Translate and Shear modify the spatial location and geometric structure of objects. Their effect can be severe when objects are small, located near image boundaries, or highly dependent on spatial structure, while the impact may be limited for large and centered objects. Therefore, these augmentations introduce heterogeneous embedding responses across categories, leading to higher variability in failure counts. Conversely, IP-Mix, MixGen, RICAP, KeepAug, and R-MixGen demonstrate the lowest standard deviations, signifying that these approaches preserve the semantic properties of the original images and result in more consistent performance throughout the assessed conditions. This stability is due to their sample-mixing characteristic, wherein new samples are produced by amalgamating information from previous images while preserving the majority of the original semantic content. Consequently, these augmentations introduce regulated variances without significantly altering the fundamental visual representation. This category-level variability is consistent with the broader dataset-level trends observed in the baseline results. Specifically, the embedding models exhibit different failure-count distributions across CIFAR-10, ImageNet, and MN, indicating that the intrinsic visual characteristics of each dataset influence the stability of the extracted representations. Moreover, OpenCLIP generally produces slightly higher failure counts than Titan, suggesting that Titan provides more stable visual-semantic representations under the evaluated conditions. 

The consistent behavior observed across Titan and OpenCLIP indicates that these augmentation techniques expose general vulnerabilities in vision–language embedding models rather than limitations specific to a single architecture. {Nevertheless, the greater sensitivity of OpenCLIP to CutMix, CycleGan, Blur, and Mixup suggests that certain failure modes remain architecture-dependent.} Accordingly, findings supported by both embedding models may be regarded as more broadly generalizable, whereas discrepancies between Titan and OpenCLIP should be interpreted as model-specific sensitivities.

The high failure counts produced by these augmentations primarily arise from their intrinsic ability to disrupt both the visual structure and semantic content of the input images. However, the behavior of sample-mixing techniques demonstrates that a substantial visual modification does not necessarily lead to retrieval failure. In particular, MixGen, RICAP, and KeepAug modify the visual representation while preserving sufficient class-discriminative information to support correct retrieval.

The most effective augmentation techniques are those that simultaneously achieve a high failure odds ratio relative to the baseline (\Cref{tab:failure_rate_mixed_effects}) and a low standard deviation. Because the baseline and augmented configurations retrieve different numbers of images, the odds ratios rather than the raw counts of \Cref{tab:failure_count_std} provide the appropriate basis for this comparison. A high odds ratio indicates a strong capability to challenge or mislead the embedding models, whereas a low standard deviation reflects consistent effectiveness across different datasets, image categories, embedding architectures, and visual conditions. Based on these criteria { Neural Style Transfer, Weather simulation, Dimension Reduction, and Shuffle} are identified as the most effective techniques for exposing retrieval failures. Note, however, that {Neural Style Transfer and Shuffle} achieves this at the cost of low semantic realism (RQ3), so Weather simulation and Dimension Reduction offer the better overall trade-off. Their combination of high failure {counts} and relatively low variability indicates that they consistently expose weaknesses in the evaluated embedding models under diverse experimental conditions.

{\Cref{tab:glmm_ci_variance} {in Appendix} provides additional statistical support for the mixed-effects logistic regression results by reporting the 95\% confidence intervals (CIs) of the estimated odds ratios together with the category-level random-effect variances. 
The reported results indicate that Shuffle, NST, Weather, and Dimension Reduction are among the augmentation techniques with the highest failure odds across the evaluated datasets and embedding models. The random-effect variances further quantify the extent to which these effects vary among image categories.
}

\Cref{fig:heatmap} {in Appendix} presents the 15 image categories associated with the highest failure frequencies across all evaluated datasets and embedding models. A common characteristic of these categories is their high visual complexity, including cluttered backgrounds, multiple objects, fine-grained textures, and substantial intra-class variability.

\begin{tcolorbox}[title=RQ1 Summary, breakable]{The results show that retrieval failures are strongly influenced by both the augmentation technique and the characteristics of the dataset and image category. ImageNet generally produces lower failure counts, whereas the lower resolution of CIFAR-10 and the more complex real-world conditions represented by the MN dataset make semantic category preservation more challenging. The large variation in failure counts across categories further indicates that the effect of an augmentation is highly category-dependent. Across both embedding models, Neural Style Transfer, Weather simulation, Dimension Reduction, and Shuffle exhibit the strongest and most consistent failure-inducing behavior, while techniques such as KeepAug, RICAP, and MixGen produce comparatively few violations. The similar overall trends observed for Titan and OpenCLIP suggest that several augmentation-induced vulnerabilities generalize across embedding models, although OpenCLIP shows greater sensitivity to some techniques, indicating the presence of model-specific failure modes. The behavior of sample-mixing techniques further demonstrates that substantial visual or embedding-level changes do not necessarily lead to retrieval failure when sufficient class-discriminative information is preserved. Finally, the categories with the highest failure frequencies are generally characterized by greater visual complexity, including cluttered backgrounds, multiple objects, fine-grained textures, and high intra-class variability.}\end{tcolorbox}

\begin{table*}[!htbp]
\centering
\caption{Mixed-Effects Logistic Regression Analysis of Failure-Rate Results}
\label{tab:failure_rate_mixed_effects}
\setlength{\tabcolsep}{3pt}
\renewcommand{\arraystretch}{1.05}
\footnotesize

\resizebox{\textwidth}{!}{%
\begin{tabular}{ll cc cc cc c}
\hline
\noalign{\vskip 2pt}

\textbf{Cat.} & \textbf{Aug.}
& \multicolumn{2}{c}{\textbf{CIFAR-10}}
& \multicolumn{2}{c}{\textbf{ImageNet}}
& \multicolumn{2}{c}{\textbf{MN}}
& \textbf{Mean $\Delta$} \\

\cline{3-4}
\cline{5-6}
\cline{7-8}

\noalign{\vskip 2pt}

& & \textbf{Titan} & \textbf{OpenCLIP} & \textbf{Titan} & \textbf{OpenCLIP} & \textbf{Titan} & \textbf{OpenCLIP} & \\

\hline
\noalign{\vskip 2pt}

\multicolumn{9}{c}{\textbf{\textit{OR format: Odds Ratio ($\Delta$\%).}}} \\
\hline
\addlinespace[2pt]

\multirow{4}{*}{\rotatebox[origin=c]{90}{\scalebox{0.90}{\textbf{Geom.}}}} & Rotate & \negcell{0.729($-27.1$)} & \negcell{0.722($-27.8$)} & \negcell{0.171($-82.9$)} & \negcell{0.148($-85.2$)} & \negcell{0.885($-11.47$)} & \negcell{0.102($-89.82$)} & \negcell{$-54.05$} \\
& Flip & \negcell{0.530($-47.0$)} & \negcell{0.526($-47.4$)} & \negcell{0.188($-81.2$)} & \negcell{0.176($-82.4$)} & \negcell{0.970($-3.01$)} & \negcell{0.895($-10.53$)} & \negcell{$-45.26$} \\
& Translate & \negcell{0.255($-74.5$)} & \negcell{0.272($-72.8$)} & \negcell{0.166($-83.4$)} & \negcell{0.152($-84.8$)} & \poscell{1.136($+13.60$)} & \poscell{1.081($+8.13$)} & \negcell{$-48.96$} \\
& Shear & \negcell{0.255($-74.5$)} & \negcell{0.272($-72.8$)} & \negcell{0.180($-82.0$)} & \negcell{0.156($-84.4$)} & \poscell{1.117($+11.75$)} & \poscell{1.089($+8.90$)} & \negcell{$-48.85$} \\

\addlinespace[2pt]
\hline
\addlinespace[2pt]

\multirow{10}{*}{\rotatebox[origin=c]{90}{\scalebox{0.90}{\textbf{Photo.}}}} & Auto Contrast & \negcell{0.253($-74.7$)} & \negcell{0.271($-72.9$)} & \negcell{0.161($-83.9$)} & \negcell{0.146($-85.4$)} & \poscell{1.001($+0.12$)} & \poscell{1.001($+0.10$)} & \negcell{$-52.78$} \\
& BCET & \negcell{0.316($-68.4$)} & \negcell{0.354($-64.6$)} & \negcell{0.160($-84.0$)} & \negcell{0.146($-85.4$)} & \poscell{1.012($+1.20$)} & \poscell{1.032($+3.16$)} & \negcell{$-49.67$} \\
& Brightness & \negcell{0.255($-74.5$)} & \negcell{0.272($-72.8$)} & \negcell{0.170($-83.0$)} & \negcell{0.155($-84.5$)} & \poscell{1.049($+4.93$)} & \poscell{1.019($+1.90$)} & \negcell{$-51.33$} \\
& Contrast & \negcell{0.267($-73.3$)} & \negcell{0.273($-72.7$)} & \negcell{0.162($-83.8$)} & \negcell{0.142($-85.8$)} & \poscell{1.013($+1.31$)} & \poscell{1.035($+3.53$)} & \negcell{$-51.78$} \\
& Color Space & \negcell{0.254($-74.6$)} & \negcell{0.272($-72.8$)} & \negcell{0.169($-83.1$)} & \negcell{0.152($-84.8$)} & \poscell{1.011($+1.07$)} & \poscell{1.002($+0.23$)} & \negcell{$-52.33$} \\
& Equalize & \negcell{0.327($-67.3$)} & \negcell{0.308($-69.2$)} & \negcell{0.171($-82.9$)} & \negcell{0.150($-85.0$)} & \poscell{1.038($+3.83$)} & \poscell{1.081($+8.10$)} & \negcell{$-48.74$} \\
& Gamma & \negcell{0.255($-74.5$)} & \negcell{0.272($-72.8$)} & \negcell{0.160($-84.0$)} & \negcell{0.146($-85.4$)} & \poscell{1.024($+2.36$)} & \poscell{1.024($+2.37$)} & \negcell{$-51.99$} \\
& PCA Jitter & \negcell{0.255($-74.5$)} & \negcell{0.272($-72.8$)} & \negcell{0.161($-83.9$)} & \negcell{0.147($-85.3$)} & \poscell{1.000($+0.01$)} & \negcell{1.000($-0.01$)} & \negcell{$-52.75$} \\
& Invert & \negcell{0.452($-54.8$)} & \negcell{0.415($-58.5$)} & \negcell{0.214($-78.6$)} & \negcell{0.223($-77.7$)} & \poscell{1.241($+24.14$)} & \poscell{1.209($+20.95$)} & \negcell{$-37.41$} \\
& Weather & \negcell{0.947($-5.3$)} & \negcell{0.924($-7.6$)} & \negcell{0.212($-78.8$)} & \negcell{0.215($-78.5$)} & \poscell{1.216($+21.56$)} & \poscell{1.227($+22.70$)} & \negcell{$-20.99$} \\

\addlinespace[2pt]
\hline
\addlinespace[2pt]

\multirow{4}{*}{\rotatebox[origin=c]{90}{\scalebox{0.90}{\textbf{Noise}}}} & Elastic & \negcell{0.255($-74.5$)} & \negcell{0.271($-72.9$)} & \negcell{0.174($-82.6$)} & \negcell{0.160($-84.0$)} & \poscell{1.095($+9.55$)} & \poscell{1.069($+6.88$)} & \negcell{$-49.58$} \\
& S\&P & \negcell{0.432($-56.8$)} & \negcell{0.399($-60.1$)} & \negcell{0.179($-82.1$)} & \negcell{0.168($-83.2$)} & \poscell{1.176($+17.64$)} & \poscell{1.157($+15.69$)} & \negcell{$-41.47$} \\
& Mobius & \negcell{0.255($-74.5$)} & \negcell{0.272($-72.8$)} & \negcell{0.175($-82.5$)} & \negcell{0.149($-85.1$)} & \poscell{1.121($+12.14$)} & \poscell{1.108($+10.76$)} & \negcell{$-48.68$} \\
& Dim. Reduction & \negcell{0.945($-5.5$)} & \negcell{0.926($-7.4$)} & \negcell{0.192($-80.8$)} & \negcell{0.182($-81.8$)} & \poscell{1.197($+19.66$)} & \poscell{1.217($+21.73$)} & \negcell{$-22.34$} \\

\addlinespace[2pt]
\hline
\addlinespace[2pt]

\multirow{3}{*}{\rotatebox[origin=c]{90}{\scalebox{0.86}{\textbf{Selection}}}} & Random Crop & \negcell{0.566($-43.4$)} & \negcell{0.674($-32.6$)} & \negcell{0.197($-80.3$)} & \negcell{0.187($-81.3$)} & \poscell{1.236($+23.58$)} & \poscell{1.197($+19.74$)} & \negcell{$-32.39$} \\
& Erasing & \negcell{0.270($-73.0$)} & \negcell{0.295($-70.5$)} & \negcell{0.160($-84.0$)} & \negcell{0.149($-85.1$)} & \poscell{1.022($+2.20$)} & \poscell{1.010($+1.02$)} & \negcell{$-51.57$} \\
& Shuffle & \negcell{0.618($-38.2$)} & \negcell{0.619($-38.1$)} & \negcell{0.469($-53.1$)} & \negcell{0.420($-58.0$)} & \poscell{1.277($+27.71$)} & \poscell{1.236($+23.55$)} & \negcell{$-22.71$} \\

\addlinespace[2pt]
\hline
\addlinespace[2pt]

\multirow{4}{*}{\rotatebox[origin=c]{90}{\scalebox{0.93}{\textbf{Filter}}}} & M-Blur & \negcell{0.373($-62.7$)} & \negcell{0.363($-63.7$)} & \negcell{0.165($-83.5$)} & \negcell{0.149($-85.1$)} & \poscell{1.025($+2.53$)} & \poscell{1.017($+1.67$)} & \negcell{$-48.47$} \\
& Blur & \negcell{0.565($-43.5$)} & \negcell{0.569($-43.1$)} & \negcell{0.182($-81.8$)} & \negcell{0.161($-83.9$)} & \poscell{1.066($+6.58$)} & \poscell{1.094($+9.43$)} & \negcell{$-39.38$} \\
& Gauss Blur & \negcell{0.611($-38.9$)} & \negcell{0.632($-36.8$)} & \negcell{0.178($-82.2$)} & \negcell{0.155($-84.5$)} & \poscell{1.202($+20.24$)} & \poscell{1.226($+22.60$)} & \negcell{$-33.26$} \\
& Sharpen & \negcell{0.262($-73.8$)} & \negcell{0.284($-71.6$)} & \negcell{0.159($-84.1$)} & \negcell{0.142($-85.8$)} & \poscell{1.002($+0.15$)} & \poscell{1.033($+3.28$)} & \negcell{$-51.99$} \\

\addlinespace[2pt]
\hline
\addlinespace[2pt]

\multirow{3}{*}{\rotatebox[origin=c]{90}{\scalebox{0.90}{\textbf{Self-Mix}}}} & AugMix & \negcell{0.289($-71.1$)} & \negcell{0.294($-70.6$)} & \negcell{0.175($-82.5$)} & \negcell{0.158($-84.2$)} & \poscell{1.069($+6.89$)} & \poscell{1.060($+5.99$)} & \negcell{$-49.25$} \\
& Self Mix & \negcell{0.286($-71.4$)} & \negcell{0.306($-69.4$)} & \negcell{0.162($-83.8$)} & \negcell{0.150($-85.0$)} & \poscell{1.061($+6.11$)} & \poscell{1.034($+3.40$)} & \negcell{$-50.02$} \\
& Salf Mix & \negcell{0.283($-71.7$)} & \negcell{0.304($-69.6$)} & \negcell{0.163($-83.7$)} & \negcell{0.151($-84.9$)} & \poscell{1.051($+5.11$)} & \poscell{1.029($+2.89$)} & \negcell{$-50.31$} \\

\addlinespace[2pt]
\hline
\addlinespace[2pt]

\multirow{9}{*}{\rotatebox[origin=c]{90}{\scalebox{0.90}{\textbf{Sample-Mix}}}} & MixGen & \negcell{0.111($-88.9$)} & \negcell{0.110($-89.0$)} & \negcell{0.050($-95.0$)} & \negcell{0.045($-95.5$)} & \negcell{0.118($-88.20$)} & \negcell{0.089($-91.14$)} & \negcell{$-91.29$} \\
& R-MixGen & \negcell{0.115($-88.5$)} & \negcell{0.130($-87.0$)} & \negcell{0.054($-94.6$)} & \negcell{0.048($-95.2$)} & \negcell{0.119($-88.09$)} & \negcell{0.089($-91.13$)} & \negcell{$-90.75$} \\
& MixUp & \negcell{0.434($-56.6$)} & \negcell{0.497($-50.3$)} & \negcell{0.187($-81.3$)} & \negcell{0.182($-81.8$)} & \poscell{1.176($+17.58$)} & \poscell{1.132($+13.16$)} & \negcell{$-39.87$} \\
& CutMix & \negcell{0.610($-39.0$)} & \negcell{0.726($-27.4$)} & \negcell{0.171($-82.9$)} & \negcell{0.164($-83.6$)} & \poscell{1.174($+17.36$)} & \poscell{1.119($+11.93$)} & \negcell{$-33.93$} \\
& Saliency & \negcell{0.438($-56.2$)} & \negcell{0.459($-54.1$)} & \negcell{0.270($-73.0$)} & \negcell{0.211($-78.9$)} & \poscell{1.210($+21.01$)} & \poscell{1.184($+18.40$)} & \negcell{$-37.14$} \\
& RICAP & \negcell{0.096($-90.4$)} & \negcell{0.096($-90.4$)} & \negcell{0.050($-95.0$)} & \negcell{0.046($-95.4$)} & \negcell{0.118($-88.16$)} & \negcell{0.089($-91.15$)} & \negcell{$-91.75$} \\
& KeepAug & \negcell{0.082($-91.8$)} & \negcell{0.081($-91.9$)} & \negcell{0.035($-96.5$)} & \negcell{0.032($-96.8$)} & \negcell{0.115($-88.52$)} & \negcell{0.087($-91.34$)} & \negcell{$-92.78$} \\
& IP-Mix & \negcell{0.127($-87.3$)} & \negcell{0.116($-88.4$)} & \negcell{0.057($-94.3$)} & \negcell{0.049($-95.1$)} & \negcell{0.118($-88.22$)} & \negcell{0.089($-91.15$)} & \negcell{$-90.75$} \\
& ResizeMix & \negcell{0.451($-54.9$)} & \negcell{0.482($-51.8$)} & \negcell{0.148($-85.2$)} & \negcell{0.140($-86.0$)} & \poscell{1.133($+13.34$)} & \poscell{1.071($+7.07$)} & \negcell{$-42.91$} \\

\addlinespace[2pt]
\hline
\addlinespace[2pt]

\multirow{3}{*}{\rotatebox[origin=c]{90}{\scalebox{0.90}{\textbf{GAN}}}} & DCGAN & \negcell{0.442($-55.8$)} & \negcell{0.470($-53.0$)} & \negcell{0.194($-80.6$)} & \negcell{0.190($-81.0$)} & \poscell{1.183($+18.31$)} & \poscell{1.138($+13.77$)} & \negcell{$-39.72$} \\
& CycleGAN & \negcell{0.541($-45.9$)} & \negcell{0.591($-40.9$)} & \negcell{0.199($-80.1$)} & \negcell{0.178($-82.2$)} & \poscell{1.088($+8.80$)} & \poscell{1.070($+7.01$)} & \negcell{$-38.88$} \\
& NST & \negcell{0.949($-5.1$)} & \negcell{0.927($-7.3$)} & \negcell{0.278($-72.2$)} & \negcell{0.255($-74.5$)} & \poscell{1.279($+27.93$)} & \poscell{1.241($+24.09$)} & \negcell{$-17.81$} \\

\addlinespace[2pt]
\hline
\addlinespace[2pt]

\multirow{6}{*}{\rotatebox[origin=c]{90}{\scalebox{0.90}{\textbf{Diffusion}}}} & DiffusionCLIP & \negcell{0.352($-64.8$)} & \negcell{0.352($-64.8$)} & \negcell{0.176($-82.4$)} & \negcell{0.157($-84.3$)} & \poscell{1.169($+16.86$)} & \poscell{1.188($+18.82$)} & \negcell{$-43.42$} \\
& Pix2Pix-Zero & \negcell{0.680($-32.0$)} & \negcell{0.712($-28.8$)} & \negcell{0.198($-80.2$)} & \negcell{0.186($-81.4$)} & \negcell{0.158($-84.20$)} & \negcell{0.122($-87.81$)} & \negcell{$-65.73$} \\
& DiffuseMix & \negcell{0.942($-5.8$)} & \negcell{0.870($-13.0$)} & \negcell{0.174($-82.6$)} & \negcell{0.177($-82.3$)} & \poscell{1.199($+19.92$)} & \poscell{1.165($+16.51$)} & \negcell{$-24.54$} \\
& SaSPA & \negcell{0.923($-7.7$)} & \negcell{0.696($-30.4$)} & \negcell{0.187($-81.3$)} & \negcell{0.176($-82.4$)} & \poscell{1.251($+25.07$)} & \poscell{1.224($+22.38$)} & \negcell{$-25.73$} \\
& DA-Fusion & \negcell{0.682($-31.8$)} & \negcell{0.713($-28.7$)} & \negcell{0.157($-84.3$)} & \negcell{0.150($-85.0$)} & \poscell{1.211($+21.12$)} & \poscell{1.163($+16.31$)} & \negcell{$-32.06$} \\
& DIAGen & \negcell{0.680($-32.0$)} & \negcell{0.706($-29.4$)} & \negcell{0.181($-81.9$)} & \negcell{0.175($-82.5$)} & \poscell{1.263($+26.30$)} & \poscell{1.228($+22.82$)} & \negcell{$-29.46$} \\

\addlinespace[2pt]
\hline
\addlinespace[2pt]

\multirow{2}{*}{\rotatebox[origin=c]{90}{\scalebox{0.90}{\textbf{Adv.}}}} & FGSM & \negcell{0.506($-49.4$)} & \negcell{0.511($-48.9$)} & \negcell{0.179($-82.1$)} & \negcell{0.167($-83.3$)} & \poscell{1.277($+27.72$)} & \poscell{1.171($+17.13$)} & \negcell{$-36.48$} \\
& PGD & \negcell{0.341($-65.9$)} & \negcell{0.445($-55.5$)} & \negcell{0.164($-83.6$)} & \negcell{0.148($-85.2$)} & \poscell{1.198($+19.84$)} & \poscell{1.133($+13.29$)} & \negcell{$-42.84$} \\

\addlinespace[2pt]
\hline

\end{tabular}%
}
\end{table*}

\subsection{RQ2a: Embedding-Space Similarity and Diversity}

\Cref{tab:embedding_similarity} {provided in Appendix,} presents the average cosine similarity between the embeddings of the original and augmented images.  Shuffle, Weather, and Dimension Reduction exhibit the lowest similarity values, signifying that these techniques generate the most substantial embedding-level discrepancies from the source images. A notable observation is related to IP-Mix, which produces one of the lowest average similarity values. This indicates that IP-Mix causes a substantial shift in the embedding space and strongly changes the representation of the augmented image. However, this large embedding shift does not necessarily result in a high retrieval failure {count}.  This behavior can be explained by the design of IP-Mix ~\cite{huang2023ipmix}, which combines information from input images while preserving important semantic regions. A comparable trend is evident for KeepAug and RICAP. These approaches alter the visual representation; yet, their comparatively modest failure {counts} suggest that they do not adequately mislead the embedding models.
 Conversely, techniques that integrate low similarity with a high failure {count} are more relevant to this investigation, since they both displace the image from its original embedding representation and induce the retrieval system to yield erroneous categories. PCA Jitter, Auto Contrast, and Gamma Correction yield the highest similarity scores, signifying that they induce only little embedding-level alterations. 
Consequently, high similarity and low failure {count} should be seen as indicators of restricted fault-detection capacity rather than robust enhancement efficacy. Table \Cref{tab:embedding_similarity} presents the standard deviation, demonstrating the superior stability of the Titan model, with more examination available in the Appendix.

\begin{figure*}[!htbp]
\centering
\hfill
\subfloat[Failure {count}]{%
\includegraphics[width=1\columnwidth]{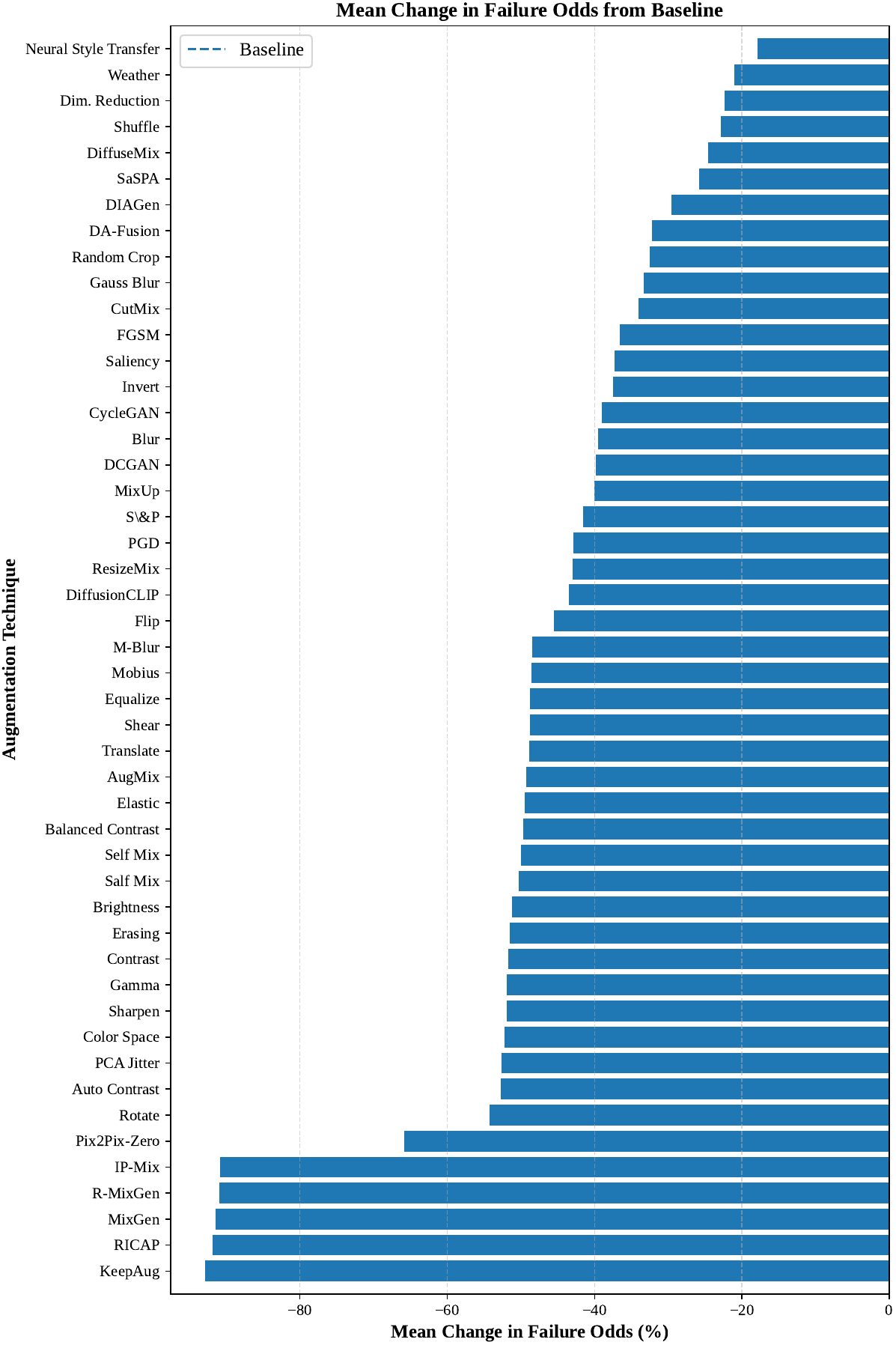}
\label{fig:failure_rate}
}
\subfloat[Embedding similarity]{%
\includegraphics[width=1\columnwidth]{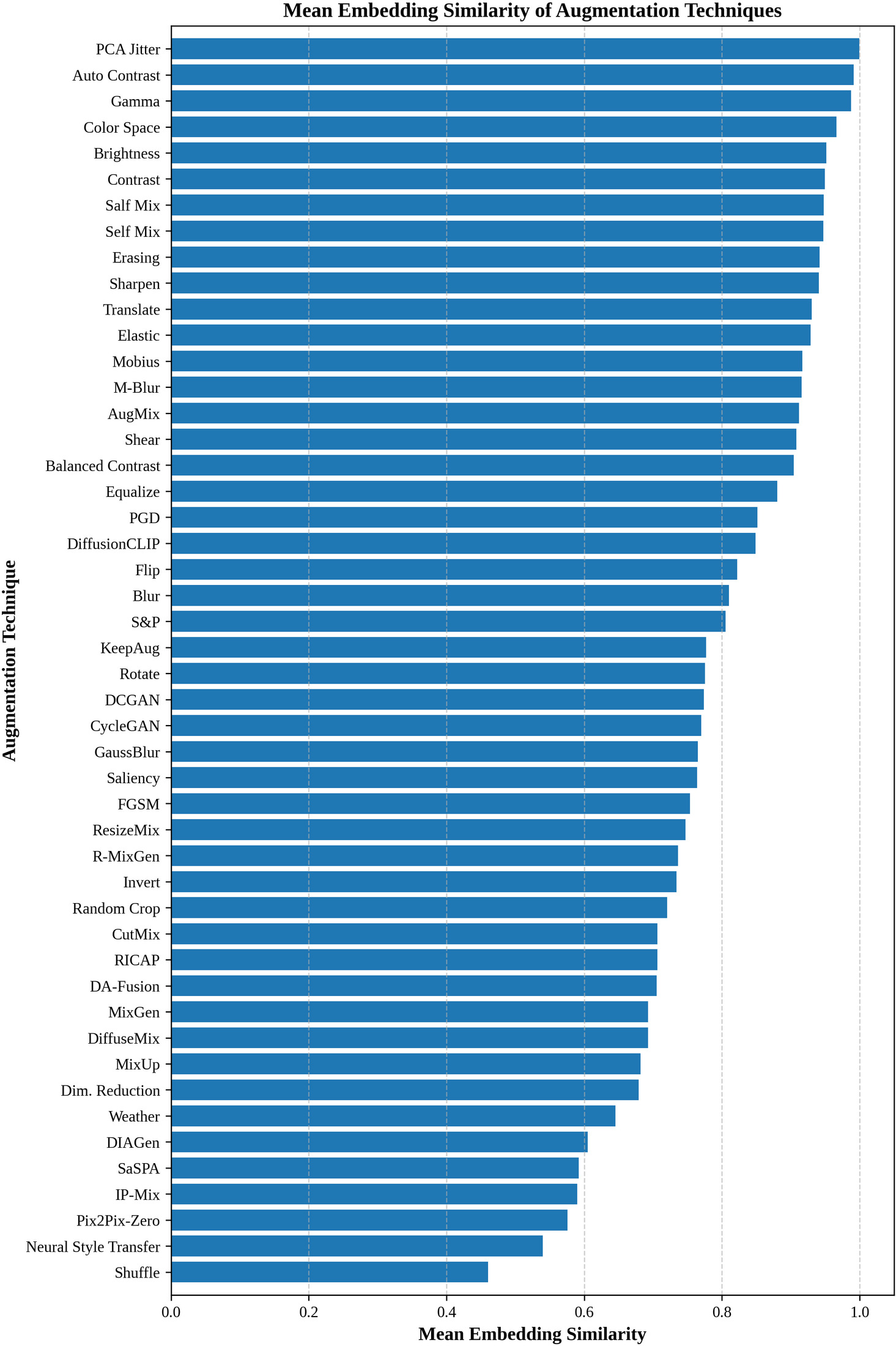}
\label{fig:similarity}
}
\caption{Comparison of mean failure {count} and embedding similarity for augmentation techniques across all datasets.}
\label{fig:FR_Sim_evaluation_plots}
\end{figure*}

\begin{tcolorbox}[title=RQ2a Summary, breakable]
Shuffle, Weather, and Dimension Reduction produce the largest embedding shifts, as indicated by their low cosine similarity values, and also exhibit high retrieval failure {counts}. In contrast, PCA Jitter, Auto Contrast, and Gamma Correction maintain high similarity with the original images and generally induce relatively few retrieval failures, indicating limited fault-detection capability under the evaluated configuration. However, embedding displacement alone is insufficient to predict retrieval failure. IP-Mix, KeepAug, and RICAP substantially alter the embedding representation while producing comparatively modest failure {counts}, suggesting that large representation changes can occur while sufficient class-relevant semantic information is preserved for successful retrieval. The standard-deviation results further indicate that Titan exhibits more stable embedding-similarity behavior than OpenCLIP across the evaluated conditions. Overall, the findings demonstrate that embedding similarity is useful for characterizing augmentation-induced representation changes, but should be considered together with retrieval failure behavior when assessing the effectiveness of an augmentation technique.
\end{tcolorbox}

\subsection{RQ2b: Embedding Model Uncertainty}

The uncertainty results are shown in \Cref{tab:dispersion_consistency,tab:mahalanobis_ensemble} {provided in Appendix,} utilizing four complementary metrics: embedding dispersion, pairwise distance, Mahalanobis distance, and ensemble-based uncertainty. 

\begin{figure}[!htbp]
\centering
\includegraphics[width=\columnwidth]{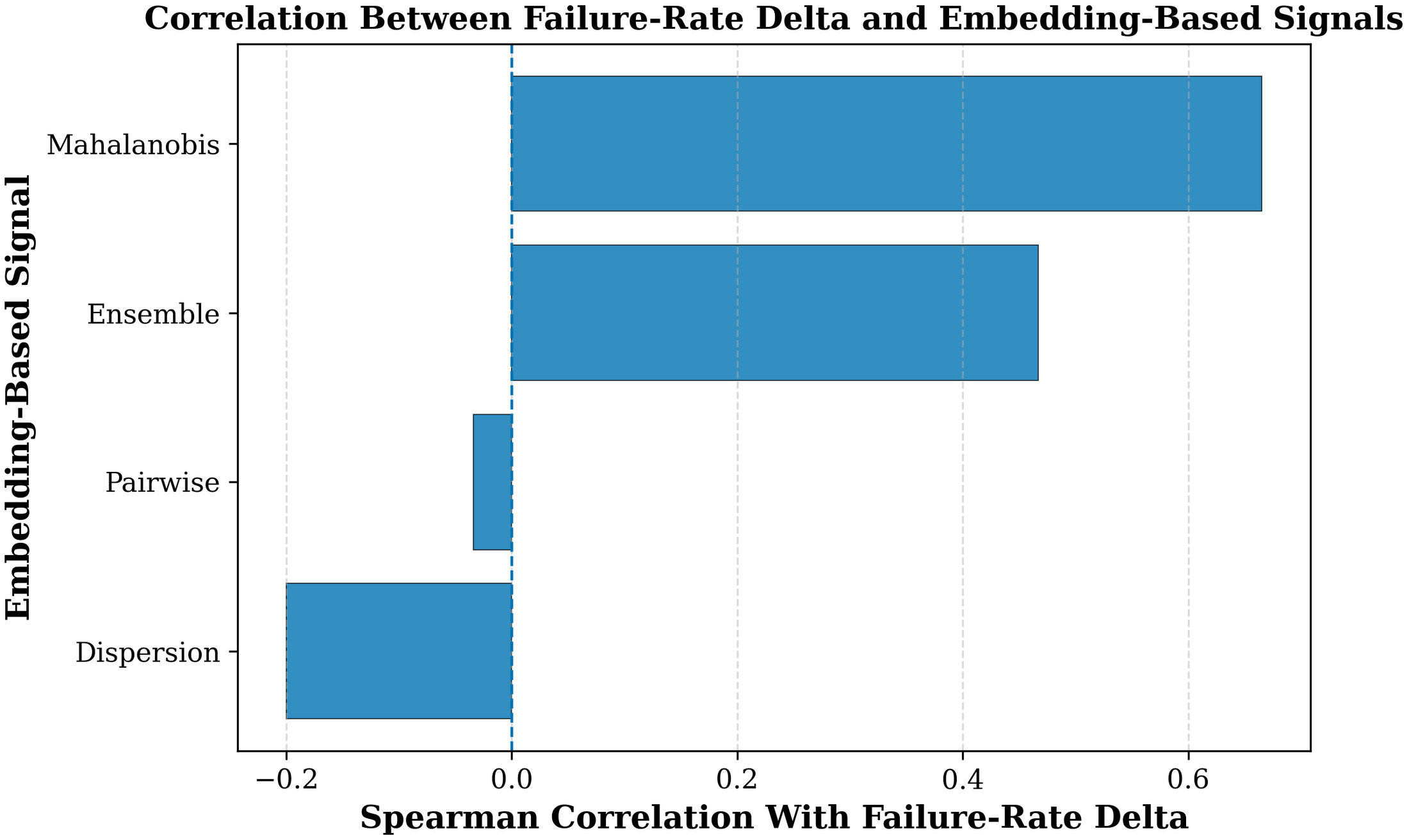}
\caption{Correlation Between failure {count} and Uncertainty Metrics}
\label{fig:spearman_test}
\end{figure}

\Cref{fig:spearman_test} illustrates the Spearman rank correlations between the failure count and the evaluated uncertainty-related metrics, while \Cref{tab:spearman_uncertainty} reports the corresponding correlation coefficients and p-values. Statistical significance was assessed using a two-sided Spearman correlation test with a significance level of $\alpha=0.05$. The results reveal statistically significant positive associations between the failure count and both Mahalanobis distance and ensemble-based uncertainty. Specifically, the corresponding correlation coefficients are $\rho=0.67$ and $\rho=0.47$, respectively, indicating strong and moderate monotonic relationships. Therefore, augmentation techniques associated with larger Mahalanobis distances or higher ensemble uncertainty generally produce a greater number of retrieval failures. 
{The stronger relationship observed for Mahalanobis distance can be attributed to its covariance-aware formulation. Unlike conventional unweighted distance measures, Mahalanobis distance accounts for both the scale and correlation structure of the embedding dimensions. Consequently, deviations along low-variance or statistically informative directions receive greater importance than deviations along highly variable directions, resulting in a more stable characterization of distributional shift. Similarly, the ensemble-based uncertainty metric aggregates the outputs of multiple KNN configurations. Combining estimates obtained under different neighborhood settings reduces dependence on a single model configuration and provides a more robust representation of uncertainty. These properties explain why both measures exhibit reliable associations with augmentation-induced retrieval failures.}{In contrast, pairwise distance and embedding dispersion exhibit weak correlations with the failure count and statistically nonsignificant p-values. These results do not necessarily indicate that the metrics are dominated by noise; rather, they show that the available data provide insufficient evidence of a consistent monotonic relationship with retrieval failure. Embedding dispersion measures the deviation of individual embeddings from the corresponding category centroid. Because this centroid-based measure assigns equal importance to all embedding dimensions and may be influenced by atypical samples or naturally diverse category members, increased dispersion does not necessarily imply semantic degradation. Likewise, the near-zero correlation observed for pairwise distance indicates that this metric does not systematically increase or decrease with the failure count. Augmentation techniques with high pairwise distance may therefore exhibit either high or low failure counts, and the same behavior may occur for techniques with low pairwise distance.} {This behavior is particularly evident for sample-mixing techniques, which can increase embedding diversity by generating visually heterogeneous representations while preserving sufficient category-level semantic information for successful retrieval. Accordingly, an augmentation may move an embedding farther from its original position or increase within-category dispersion without shifting it toward a decision region associated with retrieval failure. These findings demonstrate that the magnitude of an embedding-space displacement alone is insufficient to characterize failure risk; the direction and statistical relevance of that displacement must also be considered.}

In this regard, Shuffle, Neural Style Transfer (NST), Pix2Pix-Zero, Weather, and IP-Mix exhibit the highest uncertainty among the evaluated augmentation techniques. However, high uncertainty does not necessarily correspond to a high failure {count}. For example, Pix2Pix-Zero, IP-Mix, and DIAGen produce substantial variations in the embedding space, yet these variations do not consistently result in retrieval failures. This finding indicates that both the magnitude and direction of the embedding displacement are important. Although these techniques generate large feature-space shifts, the resulting embeddings often remain within the same class-consistent region. In contrast, Shuffle and Weather perturb the images in ways that substantially damage task-relevant visual and semantic structures. Shuffle disrupts the spatial organization of image content, thereby weakening the structural cues required for correct semantic matching. Weather-based transformations may obscure or distort salient visual features through effects such as fog, rain, or illumination changes, which can move the corresponding embeddings toward less discriminative regions of the feature space. Pix2Pix-Zero, IP-Mix, and DIAGen, on the other hand, are semantic or generative augmentation techniques that tend to preserve object identity while introducing considerable stylistic or appearance-level diversity. Pix2Pix-Zero largely maintains the structural composition of the original image, whereas IP-Mix and DIAGen employ diffusion- or feature-based generation mechanisms that are conditioned to produce class-consistent samples. Consequently, these techniques may induce large embedding variations without crossing the decision boundaries associated with the original semantic category. Their embedding vectors therefore move within, or remain close to, the same class manifold, which explains why their high uncertainty does not necessarily translate into a correspondingly high failure {count}.\\
An more significant observation pertains to the adversarial augmentation techniques. In these techniques, OpenCLIP demonstrates a larger failure {count} compared to Titan, even though it maintains greater similarity between the original and altered embeddings. This outcome suggests that adversarial attacks do not inherently necessitate significant displacements inside the embedding space. They instead implement minor yet strategically focused perturbations that alter the embedding vectors towards directions likely to intersect semantic or retrieval decision boundaries, thereby misleading the model. The obtained results indicate that the efficacy of these attacks is mostly determined by the direction of the embedding displacement rather than its magnitude. This behavior occurs because adversarial perturbations are specifically designed to take advantage of the local geometry and weaknesses of the learned representation space. In contrast to destructive alterations like Shuffle or Weather, which significantly compromise image structure or look, adversarial techniques predominantly maintain visual content while effecting targeted modifications that diminish retrieval dependability. The important observation from this analysis is that a larger embedding dispersion does not necessarily correspond to higher retrieval uncertainty or failure. This finding further supports the need to employ multiple complementary metrics to assess model behavior from different perspectives, including failure {count}, embedding similarity, dispersion, Mahalanobis distance, and ensemble-based uncertainty. 

\begin{tcolorbox}[title=RQ2b Summary, breakable]
Shuffle, NST, Pix2Pix-Zero, Weather, and IP-Mix exhibit the highest uncertainty among the evaluated augmentation techniques. Mahalanobis distance and ensemble-based uncertainty show significant positive associations with retrieval failure {count}, whereas embedding dispersion and pairwise distance show weak, nonsignificant relationships. Overall, high uncertainty does not necessarily imply retrieval failure, indicating that both the magnitude and the nature of the embedding-space change are important.
\end{tcolorbox}

\begin{table}[!t]
\centering
\caption{Spearman Correlation Between Failure-Rate Delta and Embedding-Based Signals}
\label{tab:spearman_uncertainty}
\setlength{\tabcolsep}{7pt}
\renewcommand{\arraystretch}{1.05}
\small

\resizebox{\columnwidth}{!}{%
\begin{tabular}{lcc}
\hline
\textbf{Embedding-Based Signal}
& \textbf{Spearman $\rho$}
& \textbf{$p$-value}\\
\hline
Mahalanobis distance  & 0.67    & $1.91\times10^{-7}$ \\
Ensemble uncertainty & 0.47    & $7.50\times10^{-4}$  \\
pairwise distance & $-0.03$ & 0.840                \\
Embedding dispersion & $-0.20$ & 0.173                \\
\hline
\end{tabular}%
}
\end{table}

\subsection{RQ3: Semantic Realism}

Two validation investigations were performed to assess the reliability of the realism scores generated by the utilized VLM judge. The initial research examined prompt sensitivity by assessing the consistency of the resulting ratings across various prompt formulations. Three prompt configurations were specifically evaluated: the original prompt, a prompt with rearranged evaluation criteria, and a succinct prompt using a more direct instructional format. The first evaluation is conducted using all samples from the three datasets without applying any augmentation technique. The assessed prompts are displayed in Algorithm 2, and 3 presented in Appendix. For each category, 1000 samples are randomly selected, and the three prompts are sent separately to the VLM. The realism scores for all 3000 samples are stored, and the mean difference between the proposed prompt and each of the other two prompts is computed. \Cref{tab:realism_diff_summary} presents the differences between the realism scores obtained using the proposed prompt and the tested prompt variants. The results indicate that the VLM is robust to the ordering of the criteria (a mean deviation of $0.21$ of $18$, $\approx1\%$) but markedly more sensitive to their wording (a mean deviation of $2.82$ of $18$, $\approx16\%$). The absolute realism scores are therefore prompt-dependent, and we accordingly rely on the relative ordering of techniques rather than on absolute values; these results also highlight the importance of using consistent technical terminology and definitions across prompts. Consistent wording helps the model interpret the evaluation objective in the same way across different trials. {The first evaluated prompt produces results that are highly consistent with those obtained using the proposed prompt, whereas the second prompt exhibits a larger degree of deviation. 
The observed discrepancy motivates the second stage of the evaluation, which is described in the following paragraph. 

\begin{table}[!htbp]
\centering
\caption{Average Realism-Score Differences Across Prompt Variants Relative to the Proposed Prompt}
\label{tab:realism_diff_summary}
\setlength{\tabcolsep}{4pt}
\renewcommand{\arraystretch}{1.05}
\small

\begin{tabular}{lcccc}
\hline
\textbf{Method} 
& \textbf{CIFAR10} 
& \textbf{ImageNet} 
& \textbf{MN}
& \textbf{Mean}\\
\hline

Prompt1        & 0.293 & 0.241  & 0.094 & 0.209\\
Prompt2        & 2.881 & 2.659  & 2.915 & 2.817\\

\hline
\end{tabular}
\end{table}

Secondly, human evaluation and {Qwen2.5-VL initialized from the Qwen/Qwen2.5-VL-7B-Instruct pretrained and instruction-tuned checkpoint are utilized} to further analyze the dependability of the realism ratings and to compare the consistency of the assessed outcomes. Qwen2.5-VL is chosen due to its design for visual comprehension, image-based reasoning, document/image analysis, and dynamic-resolution image processing. To facilitate an equitable comparison, both VLM-based evaluators are chosen from an identical parameter scale, specifically 7 billion parameters. The identical realism prompt is presented to Qwen2.5-VL. For human review, the scoring criteria are simplified in order to enhance clarity and minimize ambiguity for raters. Four specific criteria are evaluated: overall semantic realism, object and scene preservation, artifact severity, and consistency of lighting and look. Each criterion is rated on a scale from 0 to 2, yielding a maximum human evaluation score of 8. This simplification is essential as human evaluators may not reliably evaluate intricate visual elements, such as sensor attributes, camera geometry, or depth-of-field authenticity. For the purpose of comparison with the VLM-based scores, the human evaluation scores are normalized to align with the score range of the model outputs. { Because manually evaluating every sample across all three datasets would require substantial time and human effort, a representative subset is selected for the human-validation stage. Specifically, 300 samples are randomly drawn from each dataset, resulting in a total of 900 images. The same sampled images are then independently evaluated by human assessors and the automated vision–language models to ensure a consistent basis for comparison. This sampling strategy reduces the annotation workload while maintaining coverage across the evaluated datasets. Furthermore, applying an identical sample set to all evaluators enables a direct assessment of the agreement between human judgments and model-generated realism scores.} \Cref{tab:realism_diff_models} presents the comparison of the final realism scores alongside the mean absolute disparities between each model and human judgment. The findings demonstrate that LLaVA yields more consistent scores compared to Qwen2.5-VL, especially on the CIFAR-10 dataset. LLaVA also aligns most closely with human ratings overall (mean absolute difference 1.15 versus 2.75 for Qwen2.5-VL), scoring slightly higher than humans on CIFAR-10 (12.41 vs.\ 11.74) and lower on ImageNet and MN. { Qwen2.5-VL yields lower realism scores compared to LLaVA on the CIFAR-10 dataset. This disagreement may stem from variations in the internal structures and training methodologies of the two models, encompassing their visual encoders, instruction-tuning approaches, tokenization processes, and image-processing techniques. LLaVA typically use a CLIP-based encoder for visual inputs at a specified resolution \cite{liu2023visual}, while Qwen2.5-VL utilizes an alternative visual representation and processing approach. The architectural and preprocessing variations can result in differing sensitivity levels when assessing low-resolution photos, consequently yielding consistently divergent absolute realism scores. However, because the LLaVA-based evaluation is sensitive to prompt formulation and exhibits discrepancies across prompt variants, the analysis relies on relative rankings rather than the absolute realism scores. This approach reduces the influence of prompt-dependent score shifts and provides a more robust basis for comparing the evaluated augmentation techniques.
}

\begin{table}[!htbp]
\centering
\caption{Average Realism-Score Differences Across Human, Qwen 2.5 and LLaVA judges}
\label{tab:realism_diff_models}
\setlength{\tabcolsep}{4pt}
\renewcommand{\arraystretch}{1.05}
\small

\begin{tabular}{lcccc}
\hline
\textbf{Method} 
& \textbf{CIFAR10} 
& \textbf{ImageNet} 
& \textbf{MN}
& \textbf{Mean Diff}\\
\hline

LLaVa        & 12.41 & 12.67  & 12.63 & 1.15\\
Qwen2.5      & 5.5   & 15.57  & 13.45 & 2.75\\
Human        & 11.74 & 14.09  & 13.99 & 0\\

\hline
\end{tabular}
\end{table}

\Cref{tab:realism} {provided in Appendix,} displays the LLaVA-derived realism scores for the assessed augmentation approaches across all datasets. This tables proves the relation between the realism score and the data characteristics such as image quality and visual details. ImageNet benefits higher realism score because of higher resolution and clear boundary and visual features while CIFAR10 because of lower resolution images achieves lower score. MN dataset is a mixture of different scores which present real-world situation and its inclusion of higher-resolution photographs taken under more advantageous and visually informative circumstances leading to lower realism score. 

According to \Cref{tab:realism}, Sharpen, Brightness, Contrast, Balanced Contrast, and SaSPA attain the highest observed realism scores, while Shuffle, Neural Style Transfer (NST), DCGAN, FGSM, and IP-Mix receive the lowest values. Photometric adjustments, such as Sharpen, Brightness, and Contrast, largely adjust pixel-level attributes including intensity, lighting, and local contrast, without significantly affecting the spatial arrangement, object identity, or semantic content of the image. As a result, the enhanced samples maintain visual coherence and align more closely with the natural image distribution. Likewise, SaSPA incorporates semantic differences while ensuring class consistency and structural plausibility, therefore upholding a significant level of perceptual realism.

Conversely, Shuffle, NST, DCGAN, FGSM, and IP-Mix impose significant structural, textural, or semantic distortions, potentially resulting in visually implausible samples. Shuffling alters the spatial configuration of image segments, potentially destroying globally significant object structures. NST substitutes the original texture and appearance with an alternate artistic depiction, thereby diminishing naturalness and visual coherence. Samples generated by DCGAN may exhibit synthesis artifacts, fuzzy features, or unrealistic object shapes due to constraints in the generative process. FGSM generates adversarial perturbations that can create artificial high-frequency patterns, especially with increased perturbation magnitudes. IP-Mix combines data from various pictures or feature regions, potentially resulting in uneven borders, improbable item compositions, and semantic discrepancies. These attributes collectively explain the diminished realism scores linked to these strategies.

 \begin{tcolorbox}[title=RQ3 Summary, breakable]{ Realism varies systematically by category. Photometric adjustments (Sharpen, Brightness, Contrast, Balanced Contrast) and SaSPA yield the highest realism, as they alter pixel-level appearance without disturbing scene structure or object identity. Shuffle, NST, DCGAN, FGSM, and IP-Mix yield the lowest, producing structural or textural artifacts with no counterpart in real deployment conditions. Fault-exposure power and realism are therefore partly opposed: several of the most failure-inducing techniques are also the least plausible as test inputs, which is precisely why both dimensions must be reported together.}\end{tcolorbox}

\subsection{RQ4:  Ranking and Most Effective Test Generators}
Across all four research questions --- retrieval failure {count} (RQ1), embedding-space similarity and diversity (RQ2a), embedding model uncertainty (RQ2b), and semantic realism (RQ3) --- the results converge on a consistent pattern as shown in \Cref{fig:failure_realism}. The efficacy of the implemented augmentation techniques is significantly affected by the quality and visual attributes of the datasets. Under the evaluated parameter settings, Neural Style Transfer, Weather simulation, Dimension Reduction, and Shuffle produced the highest observed failure {count}s and uncertainty scores across the two embedding models. However, these results should be interpreted as configuration-specific rather than severity-independent. Since each augmentation was evaluated using one predefined severity level, the relative ranking may change under milder or stronger perturbation settings. Therefore, the reported ranking identifies the most effective methods under the selected experimental configuration, rather than establishing a universal ranking across all possible augmentation severities. 
\Cref{fig:overall_aug_ranking} compares the failure {count}, realism scores, and stability of the implemented augmentation techniques. An effective augmentation technique should provide a suitable balance between these criteria by increasing the failure {count} while preserving semantic realism.  The overall performance of the evaluated augmentation techniques, analyzed at both the taxonomy and individual-technique levels, is reported in \Cref{tab:overall_category_weighted} and \Cref{fig:overall_aug_ranking}, respectively.  Accordingly, the resulting rankings reflect the ability of each taxonomy to induce retrieval failures, preserve perceptual realism, and maintain consistent performance across the evaluated datasets and embedding models.


The Diffusion, Photometric, and Self-Mix classifications identify the trade-off between realism and failure {count}, resulting in optimal performance with a moderate realism score, so determining the ideal candidate. Sample-mixing techniques, including {KeepAug, RICAP and IP-Mix,} do not introduce sufficiently targeted perturbations to consistently induce retrieval failures across the evaluated embedding models. In contrast, MixUp achieves a more better trade-off among failure {count}, semantic realism, and performance stability. 
Noise-based techniques show mixed behavior, with Dimension Reduction producing stronger perturbation effects than other techniques in the same category. The {Shuffle and NST} techniques achieve the highest failure {count}; however, it substantially disrupts the visual structure of the image by mixing image features and boundaries, which leads to a lower realism score. In contrast, Weather, Dimension Reduction, and random crop provide a more favorable trade-off by increasing the failure {count} while maintaining relatively higher realism. This behavior can be attributed to their ability to preserve the main visual features and structural content of the original images while still introducing meaningful perturbations. 

In general, the results indicate an inverse relationship, whereby augmentation techniques that induce larger changes in the embedding space, reflected by lower similarity scores, tend to produce higher failure {count}s. This trend suggests that stronger embedding displacement generally increases the likelihood of retrieval errors. However, several techniques, including IP-MIX, Robust MixGen, MixGen, RICAP, and KeepAug, deviate from this overall pattern. These techniques synthesize augmented samples by combining or selectively preserving content from existing images, thereby retaining substantial structural and semantic information from the original inputs. Although such operations may produce noticeable displacement in the embedding space, the induced changes are not necessarily aligned with the most failure-sensitive directions of the learned representation. Consequently, these techniques can exhibit relatively low embedding similarity without causing a corresponding increase in failure {count}. This observation indicates that the magnitude of the embedding shift alone is insufficient to characterize augmentation effectiveness; the direction of the displacement relative to the model’s decision and retrieval geometry is also a critical factor.

{\Cref{fig:weight_overal} provided in Appendix, illustrates the sensitivity of the overall ranking to different weight combinations. The results demonstrate that Weather, SaSPA, DiffuseMix, and Dimension Reduction maintain consistently high performance across the evaluated weighting configurations. This stability indicates that these augmentation techniques achieve an effective balance between inducing model failures and preserving semantic realism. Accordingly, they can introduce substantial yet plausible modifications that remain representative of realistic image variations and are therefore suitable for practical, real-world testing scenarios.}

The final comparison examines the effect of each augmentation technique on the two evaluated embedding models, as presented in \Cref{fig:models_failures} {provided in Appendix}. The horizontal axis represents the odds ratio obtained for the Titan model, whereas the vertical axis represents the corresponding odds ratio for OpenCLIP. Each marker denotes an individual augmentation technique. 
Shuffle, NST, weather simulation, and dimensionality reduction exhibit the strongest failure-inducing effects for both embedding models, as indicated by their odds ratios being the highest among the evaluated techniques along both axes.} The diagonal reference line represents equal augmentation effects on Titan and OpenCLIP. Techniques positioned above this line produce a larger increase in failure odds for OpenCLIP, whereas those below the line have a stronger effect on Titan. The larger number of techniques located above the diagonal indicates that OpenCLIP is generally more susceptible to augmentation-induced failures, suggesting greater robustness of the Titan model. This difference is particularly evident for techniques such as Rotation, Pix2Pixzero, NST, colour space, and saliency, which yield substantially higher odds ratios for OpenCLIP than for Titan.

\begin{tcolorbox}[title=RQ4 Summary, breakable]{The results indicate that \emph{challenge} and \emph{quality} are largely independent between augmentation categories, and selecting a test-generation method requires a careful trade-off between fault-exposure effectiveness and input plausibility. The reported rankings indicate the comparative performance of the assessed augmentation configurations, while subsequent research should broaden the analysis to various severity levels to determine if these trends persist throughout mild, moderate, and severe disturbances.  Under the weighted ranking, Diffusion, Photometric, and Self-Mix are the
highest-ranked augmentation categories, while Sample Mix, Adversarial, and
Selection rank lowest. At the individual-technique level, Weather simulation
and Dimension Reduction provide a strong balance between fault exposure and
realism, whereas Shuffle and NST expose more failures at the cost of
substantially lower realism. These rankings are specific to the evaluated
severity settings.
}\end{tcolorbox}

\begin{figure}[!htbp]
\centering
\includegraphics[width=\columnwidth]{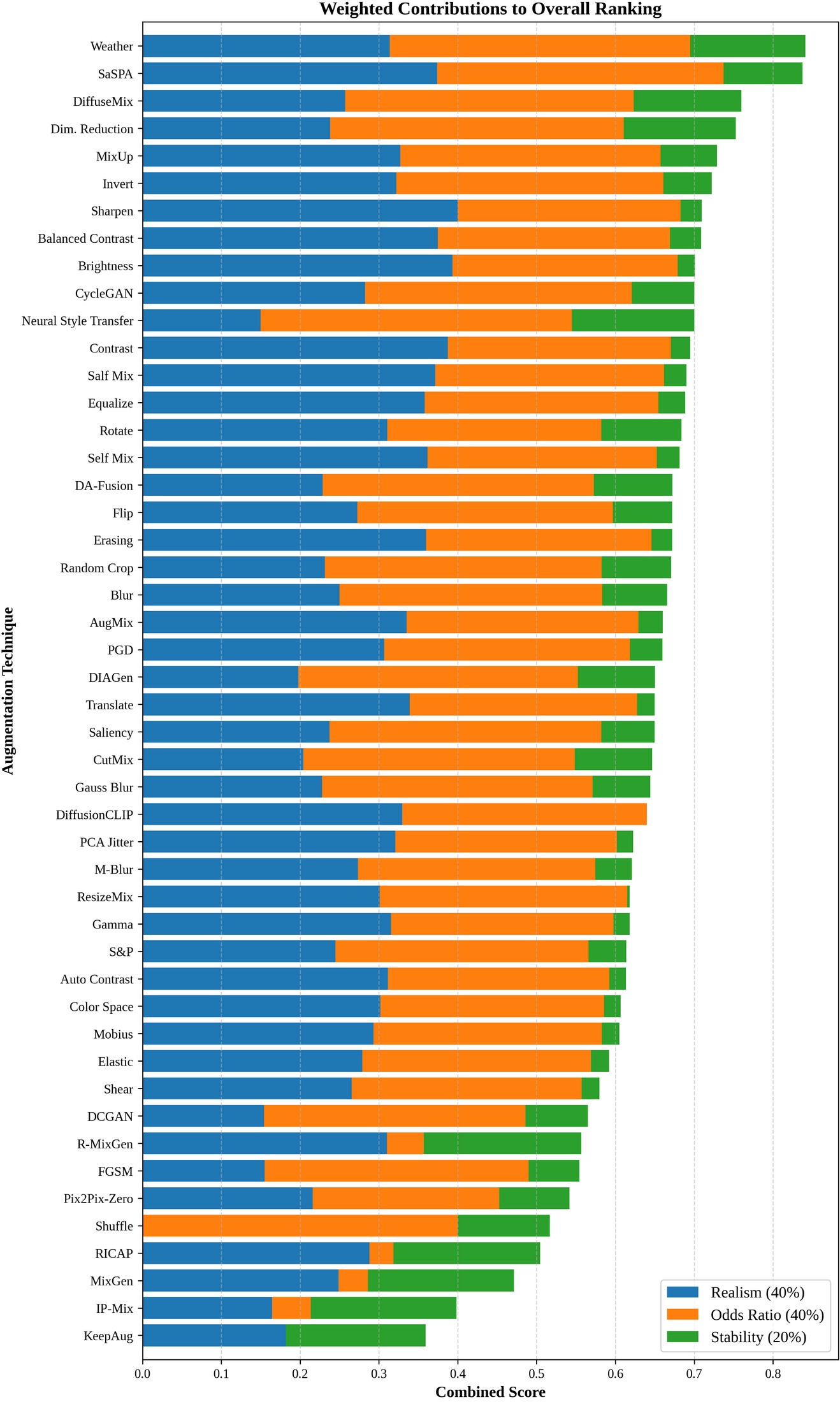}
\caption{Overall ranking of augmentation techniques based on mean performance across all datasets and models}
\label{fig:overall_aug_ranking}
\end{figure}

\begin{figure}[!htbp]
\centering
\includegraphics[width=\columnwidth]{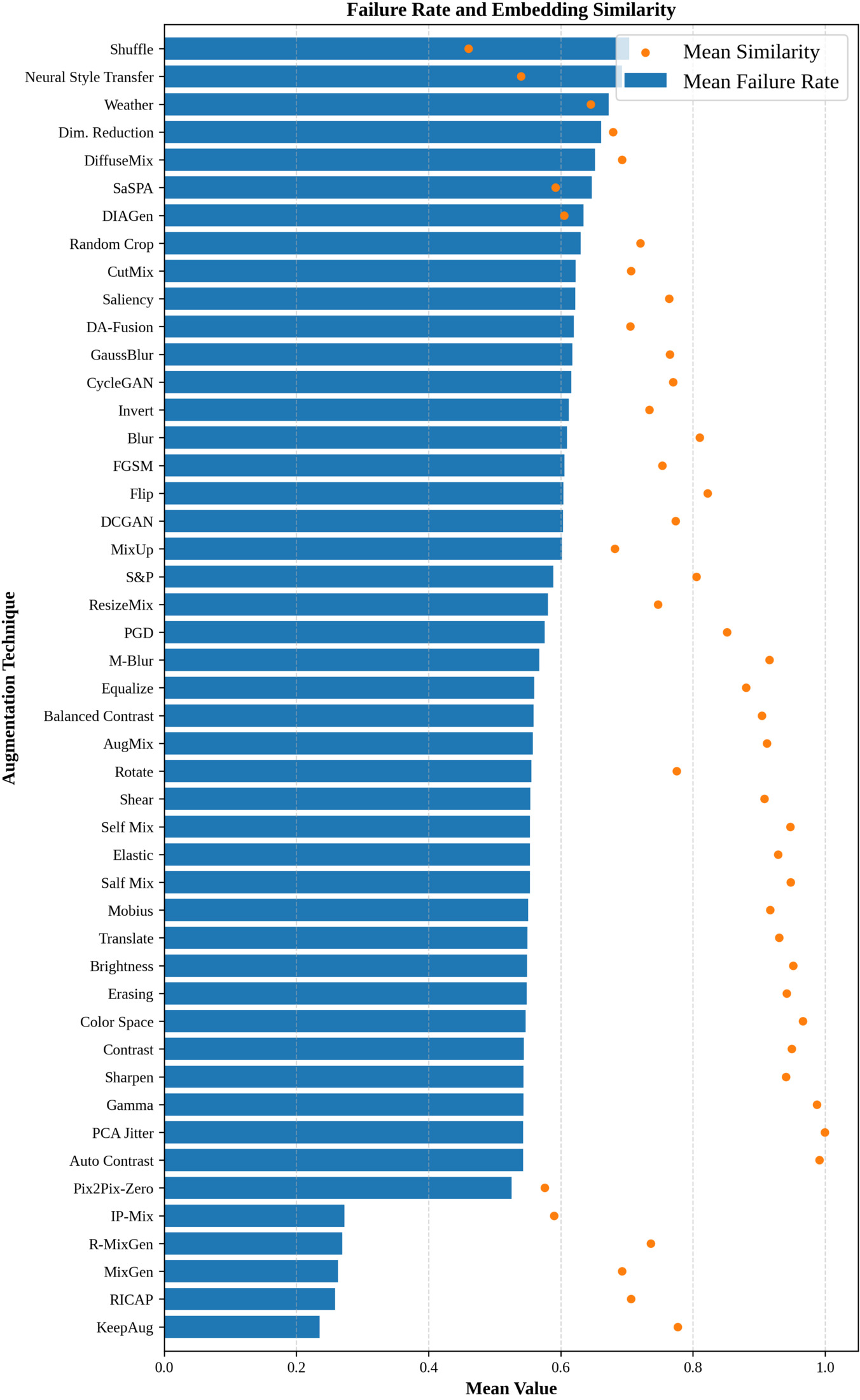}
\caption{Comparison of failure {count} and realism across augmentation techniques}
\label{fig:failure_realism}
\end{figure}


\begin{table}[!htbp]
\centering
\caption{OVERALL RANKING OF AUGMENTATION TAXONOMIES USING UNEQUAL WEIGHTS}
\label{tab:overall_category_weighted}
\setlength{\tabcolsep}{12pt}
\renewcommand{\arraystretch}{1}
\begin{tabular}{llc}
\hline
\textbf{Rank} & \textbf{Category}
& \textbf{Overall Score} \\
\hline
1  & Diffusion    & 0.6837 \\
2  & Photometric  & 0.6817 \\
3  & Self-Mix     & 0.6773 \\
4  & Filter       & 0.6602 \\
5  & GAN          & 0.6551 \\
6  & Geometric    & 0.6464 \\
7  & Noise        & 0.6409 \\
8  & Selection    & 0.6199 \\
9  & Adversarial  & 0.6070 \\
10 & Sample Mix   & 0.5482 \\
\hline
\end{tabular}
\end{table}

\section{Threats to Validity}
\label{sec:threats}

\textbf{Internal validity.} The realism scores produced by LLaVA depend on the wording of the evaluation criteria prompt. Our sensitivity analysis quantifies this: merely reordering the criteria while retaining the same terminology changes the mean realism score by only $0.21$ of $18$ ($\approx1\%$), whereas rephrasing the criteria with alternative wording shifts it by $2.82$ of $18$ ($\approx16\%$, \Cref{tab:realism_diff_summary}). The scores are thus robust to criterion order but sensitive to terminology. We reduced this risk by fixing the criteria and their wording prior to any experiment and applying them uniformly across all techniques and datasets, but the absolute realism values remain prompt-dependent; consequently, the analysis relies on the relative ordering of techniques rather than on absolute scores.

\textbf{Realism judge reliability at low resolution.} Both automated judges are 7-billion-parameter vision--language models, whose reliability degrades on very low-resolution inputs. This is visible in the cross-model comparison: Qwen2.5-VL scores CIFAR-10 at $5.5$ but ImageNet at $15.57$ (\Cref{tab:realism_diff_models}), a gap far larger than the corresponding difference for human raters. Realism values for CIFAR-10 ($32 \times 32$ pixels) should therefore be interpreted with caution and are not directly comparable across datasets of differing resolution.

\textbf{Construct validity.} Embedding-space distance is used as a proxy for semantic similarity. This is valid when the embedding model has been trained with contrastive objectives on diverse data, as is the case for both Titan and OpenCLIP. However, embedding distance may not perfectly capture human-perceived semantic similarity for all image categories.

\textbf{Metamorphic relation validity.} The metamorphic relation of Definition~\ref{def:mr} assumes that each augmentation preserves the semantic category of its source image. This assumption holds for mild transformations, but it is weaker for two groups of techniques. Balanced sample-mixing techniques (e.g., MixUp with $\lambda = 0.5$) genuinely combine two classes into a single image and are conventionally represented by soft labels rather than a single hard label; counting a retrieval from the mixed-in class as a fault therefore penalises the model for perceiving content that is in fact present. Similarly, strong transformations may alter the content itself: Weather reduces the original--augmented cosine similarity to $0.288$ on CIFAR-10 (\Cref{tab:embedding_similarity}), at which point a changed retrieval may be a correct response to a genuinely different image rather than a model fault. For these techniques the reported failure figures should be read as upper bounds, and the semantic-realism metric (RQ3) is used to distinguish augmentations that expose realistic hard cases from those that simply destroy class-relevant content.

\textbf{Severity of augmentation.} Each technique was evaluated at a single hyperparameter setting (\Cref{tab:hyperparameters}). The reported rankings are therefore configuration-specific rather than severity-independent properties of the techniques, and a milder or stronger setting may reorder them. Establishing whether the observed ordering holds across mild, moderate, and severe perturbations requires a severity sweep, which is left to future work.

\textbf{External validity.} The empirical study covers three datasets: CIFAR-10 ($32 \times 32$ pixels, 10 categories), ImageNet-1K (high-resolution, 1{,}000 categories), and the March Networks industry-partner dataset (20{,}000 images, 100 categories). CIFAR-10's low resolution may reduce realism scores for all augmentation categories relative to the higher-resolution datasets. Generalisability of findings is supported by the inclusion of both a standard academic benchmark and a real-world industrial dataset, though domain-specific datasets beyond surveillance and general object recognition may exhibit different patterns.

\textbf{Reliability.} The SLR was conducted by three reviewers covering non-overlapping publication-date ranges (2020--2022 and 2023--2025). Any conflicts in inclusion decisions were resolved through discussion among all three reviewers until consensus was reached. Inter-rater agreement on the inclusion decisions was not formally quantified using a statistical measure such as Cohen's $\kappa$; this is a limitation that future work should address. The Amazon Titan model is accessed via a cloud API, meaning exact reproducibility depends on the API version remaining available; the specific model version (Titan Multimodal Embeddings G1) is documented to support future replication.

\textbf{Conclusion validity.} No statistical significance testing was applied to the metric differences reported in the results tables. Comparisons between closely ranked augmentation techniques should therefore be interpreted cautiously. Future work should apply non-parametric pairwise tests (e.g., Wilcoxon signed-rank) with appropriate correction for multiple comparisons.


\section{Conclusion}
\label{sec:conclusion}

This paper presented a dual contribution toward improving the systematic testing of embedding-based image retrieval systems. First, a systematic literature review of 56 papers on image augmentation published between 2020 and 2025 was conducted, producing a ten-category taxonomy of augmentation techniques that organizes the landscape from simple geometric flips to text-guided diffusion models. Second, a large-scale empirical study evaluated each technique as a source of challenging test inputs for two embedding-based retrieval models, Amazon Titan and OpenCLIP, across three datasets (CIFAR-10, ImageNet-1K, and the March Networks industry-partner dataset), measuring the difficulty each augmentation poses to the model across four complementary metrics: retrieval failure {count}, embedding-space similarity and diversity, embedding model uncertainty (assessed through four estimators: dispersion, pairwise distance, Mahalanobis distance, and ensemble agreement), and semantic realism (LLaVA).

The key finding is that augmentation categories differ substantially and systematically across all evaluated metrics. At the tested severity level, weather simulation and Dimension Reduction produced the highest embedding uncertainty and failure {count}s across both models, making them the most effective test generators in terms of raw fault-detection power. Photometric techniques such as brightness adjustment, gamma correction, and color-space manipulation preserved embedding structure to the greatest degree, yielding the lowest failure {count}s, lowest uncertainty, and highest embedding similarity scores. GAN-based augmentation was among the lowest in realism, indicating that synthetic artifacts impair the visual plausibility of generated test inputs even when failure {count}s are moderate, highlighting that fault exposure and image quality are distinct and sometimes conflicting objectives. Mixing-based techniques showed a split: CutMix and IP-Mix shifted the embedding representation substantially but did not reliably induce retrieval failures, whereas MixUp offered the most useful middle ground, combining a moderate failure {count} with reasonable embedding diversity and structurally coherent augmented images. Because each technique was evaluated at a single severity level, these rankings should be read as configuration-specific rather than as severity-independent properties of the techniques; establishing whether they hold across mild, moderate, and severe settings is left to future work.

\section*{Acknowledgment}
This work was funded by March Networks. The authors gratefully acknowledge their valuable feedback and the resources they provided, including the industrial dataset used in the empirical evaluation.

\bibliographystyle{IEEEtran}
\bibliography{references}

\clearpage
\onecolumn
\raggedbottom
\appendices

\section{Additional Experimental Results} \label{app:additional_results}
This appendix presents supplementary results that are omitted from the main body to maintain the readability and structural consistency of the paper. The additional analyses and supporting materials corresponding to each research question (RQ) are presented separately in the following subsections.

\subsection{Hyperparameter Settings of the Evaluated Augmentation Techniques}
\noindent\textbf{\Cref{tab:hyperparameters}} presents the hyperparameter settings for all augmentation techniques evaluated in this study. All experiments and the corresponding results reported in this study are based on these parameter configurations. These parameter settings were determined empirically to produce meaningful and perceptible modifications to the original images.

\begin{table}[!htbp]
\centering
\caption{Summary of Hyperparameter Configurations for the Evaluated Augmentation Techniques}
\label{tab:hyperparameters}
\setlength{\tabcolsep}{2.5pt}
\renewcommand{\arraystretch}{1.08}
\scriptsize

\begin{tabular}{
@{}
>{\centering\arraybackslash}m{0.10\columnwidth}
>{\raggedright\arraybackslash}p{0.28\columnwidth}
>{\raggedright\arraybackslash}p{0.55\columnwidth}
@{}
}
\toprule
\textbf{Category} & \textbf{Augmentation} & \textbf{Hyperparameters} \\
\midrule

\multirow[c]{4}{=}{\centering
\rotatebox[origin=c]{90}{\textbf{Geometric}}}
& Rotation & $\theta = 30^\circ$ \\
& Flip & Vertical flipping \\
& Translation & $\Delta x = 20$, $\Delta y = 20$ \\
& Shearing & $s_x = 20$, $s_y = 20$ \\

\midrule
\multirow{7}{*}{\rotatebox[origin=c]{90}{\textbf{Photometric}}}
& Brightness & $\beta = 50$ \\
& Contrast & $\beta = 150$ \\
& Balanced Contrast & grid size $= 8$, threshold $=[1,4]$ \\
& Gamma Correction & $\gamma \in [30,120]$ \\
& Color Space & $\beta = 50$ \\
& PCA Jitter & $\sigma = 3$ \\
& Weather & $\sigma = 3.0$, blur $= 7$ \\

\midrule
\multirow{4}{*}{\rotatebox[origin=c]{90}{\textbf{Noise}}}
& Elastic Transform & $\alpha = 100$, $\sigma = 5$ \\
& Dimension Reduction & scale $= 20\%$ \\
& Mobius Transform & $(a,a_{in},b,b_{im},c,c_{im},d,d_{im}) = (1,0,0.3,0,0.15,0,1,0)$ \\
& Salt \& Pepper & ratio $= 5\%$, S/P $= 0.5$ \\

\midrule
\multirow{3}{*}{\rotatebox[origin=c]{90}{\textbf{Selection}}}
& Random Crop & $h_c = 50\%$, $w_c = 50\%$ \\
& Random Erasing & \makecell[l]{area $= 5\%$, \\aspect ratio $\in [0.3,3.0]$} \\
& Patch Shuffle & patch size $= 8$ \\

\midrule
\multirow{4}{*}{\rotatebox[origin=c]{90}{\textbf{Filter}}}
& Motion Blur & kernel $= 7$, $\theta \in [0^\circ,360^\circ]$ \\
& Blur & kernel $= 7$ \\
& Gaussian Blur & kernel $= 5$, $\sigma \in [10,20]$ \\
& Sharpen & $\alpha = 20$, lightness $= 50$ \\

\midrule
\multirow[c]{3}{=}{\centering
\rotatebox[origin=c]{90}{\textbf{Self-Mixing}}}
& AugMix & \makecell[l]{severity $= 3$,\\ strength $= 1.0$,\\ all ops $= \text{False}$} \\
& Self-Mix & patch ratio $= 10\%$ \\
& Salf-Mix & \makecell[l]{patch ratio $= 10\%$,\\ blend $= 0.9$} \\

\midrule
\multirow[c]{6}{=}{\centering
\rotatebox[origin=c]{90}{\textbf{Sample-Mix}}}
& MixGen & $\lambda = 0.5$ \\
& Robust MixGen & $\alpha = 0.2$ \\
& RICAP & vertical partitioning enabled \\
& IP-Mix & keep ratio $= 0.5$,
max patches $= 10$,
min fraction $= 0.01$ \\
& Saliency & $\alpha = 0.5$ \\
& KeepAug & keep ratio $= 0.5$, max patches $= 10$, min fraction $= 0.01$ \\

\midrule
\multirow{4}{*}{\rotatebox[origin=c]{90}{\textbf{GAN}}}
& DCGAN & \makecell[l]{strength $= 0.8$,\\ prompts = \texttt{scene of a ...}} \\
& CycleGAN (train) & \texttt{load\_size} $= 512$, \texttt{crop\_size} $= 256$ \\
& CycleGAN (test) & epoch $= 200$ \\
& Neural Style Transfer & num\_steps $= 300$, style\_weight $= 1{,}000{,}000$, content\_weight $= 1$ \\

\midrule
\multirow[c]{6}{=}{\centering
\rotatebox[origin=c]{90}{\textbf{Diffusion}}}
& Da-fusion & strength $= 0.8$ \\
& DIAGen & strength $= 0.8$ \\
& Pix2Pix-zero & \makecell[l]{strength $= 0.8$,\\ prompts = \texttt{scene of a ...}} \\
& DiffusionCLIP & prompts = \texttt{wet}, $t_0 = 500$, $n_{\text{inv}} = 40$, $n_{\text{test}} = 40$ \\
& DiffuseMix & prompts = \texttt{wet} \\
& SaSPA & meta class: object, prompt generator: scene of an object... \\

\midrule
\multirow{2}{*}{\rotatebox[origin=c]{90}{\textbf{Adv.}}}
& FGSM & $\epsilon = 20$ \\
& PGD & $\epsilon = 20$ \\

\bottomrule
\end{tabular}
\end{table}

\subsection{RQ1: Supplementary Failure Count Analysis}
\noindent\textbf{\Cref{tab:failure_count_std}} reports the mean failure count and corresponding STD across all evaluated datasets and embedding models. The results demonstrate that the effects of the evaluated augmentation techniques vary across datasets. Moreover, the STD values characterize the variability across categories within each dataset, indicating that the effect of a given augmentation technique is not necessarily uniform across different classes. 

\begin{table*}[!htbp]
\centering
\caption{Mean Failure Count and Standard Deviation of Augmentation Techniques\\
Across the Evaluated Embedding Models}
\label{tab:failure_count_std}
\setlength{\tabcolsep}{2.2pt}
\renewcommand{\arraystretch}{0.90}
\scriptsize

\begin{tabular}{ll@{\hspace{3pt}}ll@{\hspace{4pt}}ll@{\hspace{4pt}}ll}
\hline
\noalign{\vskip 2pt}
\textbf{Cat.} & \textbf{Aug.}
& \multicolumn{2}{c}{\textbf{CIFAR-10}}
& \multicolumn{2}{c}{\textbf{ImageNet}}
& \multicolumn{2}{c}{\textbf{MN}} \\
\cline{3-8}
\noalign{\vskip 2pt}
& & \textbf{Titan} & \textbf{OpenCLIP}
  & \textbf{Titan} & \textbf{OpenCLIP}
  & \textbf{Titan} & \textbf{OpenCLIP} \\
\hline
\noalign{\vskip 2pt}
\multicolumn{8}{c}{\textbf{\textit{Format: mean failure count (STD).}}} \\
\hline
\addlinespace[2pt]
\multirow{1}{*}{Base}
& Baseline & 572.5767 (243.1313) & 584.8544 (241.7174) & 237.0558 (58.6997) & 246.4794 (57.1076) & 205.9577 (50.6567) & 215.3052 (49.4382) \\
\addlinespace[2pt]
\hline
\addlinespace[2pt]
\multirow{4}{*}{Geom.}
& Rotate    & 990.2496 (196.4482) & 1009.9858 (180.0118)    & 252.4997 (129.3240) & 266.1494 (135.1272) & 405.0558 (99.1448) & 243.5085 (110.8359) \\
& Flip      & 836.2938 (250.9183) & 855.2781 (237.7982)     & 264.5521 (126.8911) & 288.3339 (128.8726) & 410.2310 (97.5343) & 425.8269 (97.3510) \\
& Translate & 536.9484 (334.6484) & 571.7898 (316.4504)     & 249.1792 (130.1832) & 269.8723 (134.9961) & 418.5611 (99.6760) & 433.7444 (98.5949) \\
& Shear     & 536.9465 (334.6868) & 571.9810 (316.4087)     & 258.6427 (128.2269) & 273.4162 (134.0809) & 417.7274 (99.9784) & 434.0201 (98.2263) \\
\addlinespace[2pt]
\hline
\addlinespace[2pt]
\multirow{10}{*}{Photo.}
& Auto Contrast     & 534.0475 (332.8658) & 570.5383 (315.8305) & 244.3727 (131.5337) & 265.2046 (136.7162) & 411.9820 (101.3265) & 430.6511 (98.8841) \\
& BCET              & 614.9601 (307.0521) & 678.0259 (284.2107) & 243.8222 (130.7278) & 264.2668 (135.7792) & 412.5621 (101.0370) & 431.8819 (98.4429) \\
& Brightness        & 536.9242 (334.7718) & 571.7426 (316.4592) & 251.3986 (130.1153) & 272.2311 (134.1626) & 414.4858 (101.1519) & 431.3794 (98.8966) \\
& Contrast          & 552.4230 (327.6226) & 573.3371 (312.2322) & 245.3082 (130.4486) & 261.5795 (136.2149) & 412.6204 (101.1724) & 432.0238 (98.5323) \\
& Color Space       & 535.1309 (336.0589) & 571.8434 (315.9304) & 250.4313 (130.5022) & 269.9329 (134.8916) & 412.4898 (101.2134) & 430.7034 (98.8555) \\
& Equalize          & 629.0842 (311.8683) & 620.8870 (300.0334) & 251.9932 (129.1087) & 268.0729 (134.4668) & 413.9265 (101.0213) & 433.7341 (98.2080) \\
& Gamma             & 537.1759 (334.6396) & 571.7588 (316.2336) & 244.0421 (131.4928) & 264.7652 (136.6969) & 413.1732 (101.0861) & 431.5680 (98.7651) \\
& PCA Jitter        & 537.0227 (334.7193) & 571.6666 (316.5139) & 244.4053 (131.5455) & 265.3919 (136.7087) & 411.9232 (101.2881) & 430.6062 (98.8717) \\
& Invert            & 764.0935 (272.8587) & 749.0798 (282.6006) & 280.1566 (118.4668) & 317.0129 (115.0333) & 422.8821 (97.9046)  & 437.9222 (97.6641) \\
& Weather           & 1118.7633 (136.1668)& 1131.4508 (129.5786)& 280.3793 (122.0618) & 314.5058 (121.7293) & 421.8739 (98.3696)  & 438.4340 (97.4984) \\
\addlinespace[2pt]\hline
\addlinespace[2pt]
\multirow{4}{*}{Noise}
& Elastic           & 536.8222 (333.6627)  & 571.0387 (314.0715)  & 255.4974 (130.1745) & 276.9368 (135.0306) & 416.7188 (100.0662) & 433.2899 (98.3539) \\
& S\&P              & 746.4121 (300.3190) & 730.1876 (274.6944) & 258.1973 (126.7785) & 281.6883 (130.4310) & 420.2749 (99.0105) & 436.3016 (97.9004) \\
& Mobius            & 536.8563 (334.7795)  & 571.7401 (316.4855)  & 255.0444 (129.2956) & 267.7460 (135.4345) & 417.9090 (100.0297) & 434.6715 (98.1629) \\
& Dim. Reduction    & 1117.4610 (137.9214) & 1132.5109 (129.4763) & 267.6607 (126.5183) & 293.1788 (129.7890) & 421.1242 (98.3525)  & 438.1548 (97.5480) \\
\addlinespace[2pt]
\hline
\addlinespace[2pt]
\multirow{3}{*}{Selection}
& Random Crop   & 869.2976 (247.7887) & 977.1768 (200.6151) & 271.2469 (124.6774) & 296.7524 (126.7032) & 422.6426 (98.1525)  & 437.5652 (97.7368) \\
& Erasing       & 556.4648 (330.5150) & 604.2227 (304.5863) & 243.6329 (131.2475) & 267.3103 (134.8751) & 413.1003 (100.6970) & 431.0297 (98.6176) \\
& Shuffle       & 911.9003 (244.9408) & 935.8021 (226.6450) & 386.0987 (66.3521)  & 399.5653 (69.5259)  & 424.1928 (97.5573)  & 438.6767 (97.3996) \\
\addlinespace[2pt]
\hline
\addlinespace[2pt]
\multirow{4}{*}{Filter}
& M-Blur    & 682.3415 (300.8933) & 689.5168 (267.2325) & 248.2191 (130.8450) & 267.3114 (136.1696) & 413.2589 (100.9786) & 431.2912 (98.7286) \\
& Blur      & 868.5595 (237.1563) & 894.9222 (221.5926) & 260.2230 (127.3530) & 277.1267 (130.7090) & 415.3001 (100.7656) & 434.2075 (98.2997) \\
& GaussBlur & 906.7924 (254.8747) & 947.0024 (230.6187) & 257.2303 (128.9027) & 272.7552 (134.9444) & 421.3524 (98.5897)  & 438.4071 (97.4938) \\
& Sharpen   & 544.2589 (324.6503) & 588.0881 (305.0086) & 242.7917 (131.1876) & 261.7336 (136.8349) & 412.0101 (101.3629) & 431.9279 (98.6398) \\
\addlinespace[2pt]
\hline
\addlinespace[2pt]
\multirow{3}{*}{Self-Mix}
& AugMix    & 581.0336 (321.7227) & 601.8362 (303.8199) & 254.7015 (128.2020) & 273.9938 (132.5360) & 415.4540 (100.3291) & 432.9607 (98.4398) \\
& Self Mix  & 577.9888 (324.2442) & 617.8929 (301.4522) & 245.9401 (131.3105) & 268.5979 (135.4932) & 415.0778 (100.3308) & 431.9650 (98.6850) \\
& Salf Mix  & 574.1259 (324.5167) & 615.7380 (302.3503) & 246.3753 (131.2037) & 269.1525 (135.4774) & 414.5802 (100.5206) & 431.7740 (98.6041) \\
\addlinespace[2pt]
\hline
\addlinespace[2pt]
\multirow{8}{*}{Sample Mix}
& MixGen    & 270.0935 (167.9003) & 275.3485 (168.1327) & 109.8253 (69.1138)    & 121.4188 (72.2896)    & 206.1801 (50.6284) & 215.3990 (49.4989) \\
& R-MixGen  & 275.8336 (148.9036) & 315.0668 (146.8323) & 113.1008 (64.3054)    & 124.0377 (66.7354)    & 207.2507 (51.1876) & 215.8706 (50.1460) \\
& MixUp     & 746.2413 (268.9037) & 829.6745 (231.6351) & 263.7011 (125.3427)   & 292.2706 (127.3774)   & 420.2654 (99.1349) & 435.4839 (98.2613) \\
& CutMix    & 904.9369 (231.5092) & 1013.3269 (174.9325)& 252.7744 (128.2238)   & 279.4574 (132.3353)   & 420.1678 (99.1332) & 435.0711 (98.2916) \\
& Saliency  & 752.2170 (286.3514) & 793.4339 (258.7763) & 311.4871 (104.1301)   & 309.6999 (117.8365)   & 421.6507 (98.7286) & 437.1413 (97.9787) \\
& RICAP     & 240.7691 (169.0666) & 247.7257 (164.3434) & 107.6193 (67.6911)    & 123.4640 (71.6772)    & 206.5546 (50.5455) & 215.3224 (49.5133) \\
& KeepAug   & 210.1471 (175.1638) & 216.1744 (169.1073) & 85.5239 (76.3481)     & 97.3807 (80.5415)     & 203.3264 (51.7076) & 213.2120 (49.9413) \\
& IP-Mix    & 303.4170 (174.6090) & 291.3822 (167.0008) & 118.8471 (66.4885)    & 128.3563 (69.5183)    & 206.0437 (50.6921) & 215.3127 (49.4754) \\
& ResizeMix & 766.7511 (288.2701) & 818.0704 (263.3811) & 233.7051 {(132.3889)}   & 259.6086 (137.7763)   & 418.4419 (99.8117) & 433.3584 (98.7064) \\
\addlinespace[2pt]
\hline
\addlinespace[2pt]
\multirow{3}{*}{GAN}
& DCGAN                 & 756.346 (287.1990)    & 805.195 (260.9617)    & 267.8198 (86.5070)  & 296.5628 (94.0993)  & 420.5732 (98.6338)  & 435.6908 (97.9313) \\
& CycleGAN              & 848.5462 (258.4762)   & 913.7942 (220.6773)   & 270.4465 (122.4845) & 287.6713 (125.4423) & 416.3703 (100.5847) & 433.3395 (98.5465) \\
& Neural Style Transfer & 1119.7751 (135.2682)  & 1133.0919 (130.0941)  & 316.0854 (106.5826) & 335.6744 (109.6855) & 424.2642 (97.6664)  & 438.8266 (97.3672) \\
\addlinespace[2pt]
\hline
\addlinespace[2pt]
\multirow{6}{*}{Diffusion}
& DiffusionCLIP & 659.5258 (302.4121) & 676.3180(292.0551)      & 255.9714 (129.4315) & 273.2729 (133.8264) & 419.9505 (99.2441)  & 437.2839 (97.5687) \\
& Pix2Pix-Zero  & 957.5956 (215.6848) & 1004.2383 (195.1902)    & 274.6313 (129.1365) & 299.1360 (133.0525) & 248.0051 (107.0901) & 265.4265 (112.2774) \\
& DiffuseMix    & 1116.1256 (137.1843)& 1102.2247 (145.7069)    & 254.7456 (128.4068) & 288.4453 (129.2172) & 421.2071 (98.7154)  & 436.5665 (97.7719) \\
& SaSPA         & 992.1392 (246.4158) & 1106.0767 (149.4098)    & 264.5491 (130.6308) & 290.8919 (135.7645) & 423.2177 (98.0231)  & 438.3398 (97.5262) \\
& DA-Fusion     & 958.7918 (212.1257) & 1004.1618 (191.2831)    & 242.3314 (131.9499) & 270.2326 (138.3204) & 421.7035 (98.3975)  & 436.5030 (97.9544) \\
& DIAGen        & 957.4367 (212.5426) & 999.9663 (195.1786)     & 261.3788 (128.3621) & 290.1284 (134.0517) & 423.6827 (97.6833)  & 438.4679 (97.4731) \\
\addlinespace[2pt]
\hline
\addlinespace[2pt]
\multirow{2}{*}{Adv.}
& FGSM      & 818.1260 (268.9167) & 845.4207 (255.6378) & 257.8032 (124.7683) & 281.3656 (130.2688) & 424.1927 (97.5822) & 436.7664 (97.6696) \\
& PGD       & 647.5483 (312.9077) & 781.5403 (276.4467) & 246.8483 (129.4301) & 266.4464 (136.1559) & 421.1881 (98.6994) & 435.5307 (97.9419) \\
\addlinespace[2pt]
\hline
\end{tabular}
\end{table*}

\noindent\textbf{\Cref{tab:glmm_ci_variance}} reports the 95\% confidence intervals (CIs) for the estimated effects. Green indicates augmentation techniques whose CIs lie entirely above one, whereas red indicates techniques whose CIs lie entirely below one. The CIs are consistent with the effect estimates reported in \Cref{tab:failure_rate_mixed_effects} and provide evidence that the observed effects are not attributable solely to sampling variability.

\begin{table*}[!htbp]
\centering
\caption{95\% Confidence Intervals and Category-Level Random-Effect Variances from Mixed-Effects Logistic Regression}
\label{tab:glmm_ci_variance}
\setlength{\tabcolsep}{1.7pt}
\renewcommand{\arraystretch}{0.92}
\scriptsize

\resizebox{\textwidth}{!}{%
\begin{tabular}{ll cc cc cc cc cc cc}
\hline
\noalign{\vskip 2pt}

\textbf{Cat.} & \textbf{Aug.}
& \multicolumn{4}{c}{\textbf{CIFAR-10}}
& \multicolumn{4}{c}{\textbf{ImageNet}}
& \multicolumn{4}{c}{\textbf{MN}} \\

\cline{3-6}
\cline{7-10}
\cline{11-14}

\noalign{\vskip 2pt}

&
& \multicolumn{2}{c}{\textbf{Titan}}
& \multicolumn{2}{c}{\textbf{OpenCLIP}}
& \multicolumn{2}{c}{\textbf{Titan}}
& \multicolumn{2}{c}{\textbf{OpenCLIP}}
& \multicolumn{2}{c}{\textbf{Titan}}
& \multicolumn{2}{c}{\textbf{OpenCLIP}} \\

\noalign{\vskip 2pt}

&
& \textbf{95\% CI} & \textbf{Var.}
& \textbf{95\% CI} & \textbf{Var.}
& \textbf{95\% CI} & \textbf{Var.}
& \textbf{95\% CI} & \textbf{Var.}
& \textbf{95\% CI} & \textbf{Var.}
& \textbf{95\% CI} & \textbf{Var.} \\

\hline
\noalign{\vskip 2pt}

\multirow{4}{*}{ \rotatebox[origin=c]{90}{ \scalebox{0.90}{\textbf{Geom.}}}} & Rotate & \negcell{$[0.7280,\,0.7303]$} & 0.0553 & \negcell{$[0.7206,\,0.7229]$} & 0.0444 & \negcell{$[0.1711,\,0.1718]$} & 0.3326 & \negcell{$[0.1476,\,0.1482]$} & 0.4307 & \negcell{$[0.8827,\,0.8883]$} & 0.2898 & \negcell{$[0.1014,\,0.1021]$} & 0.4452 \\

& Flip & \negcell{$[0.5287,\,0.5304]$} & 0.0862 & \negcell{$[0.5247,\,0.5263]$} & 0.0599 & \negcell{$[0.1872,\,0.1879]$} & 0.3287 & \negcell{$[0.1759,\,0.1766]$} & 0.3945 & \negcell{$[0.9669,\,0.9730]$} & 0.2754 & \negcell{$[0.8915,\,0.8980]$} & 0.2095 \\

& Translate & \negcell{$[0.2546,\,0.2554]$} & 0.2309 & \negcell{$[0.2713,\,0.2722]$} & 0.1478 & \negcell{$[0.1661,\,0.1667]$} & 0.3477 & \negcell{$[0.1513,\,0.1519]$} & 0.4387 & \poscell{$[1.1324,\,1.1397]$} & 0.2752 & \poscell{$[1.0773,\,1.0853]$} & 0.1692 \\

& Shear & \negcell{$[0.2546,\,0.2554]$} & 0.2309 & \negcell{$[0.2714,\,0.2723]$} & 0.1478 & \negcell{$[0.1792,\,0.1799]$} & 0.3301 & \negcell{$[0.1558,\,0.1564]$} & 0.4314 & \poscell{$[1.1139,\,1.1211]$} & 0.2863 & \poscell{$[1.0850,\,1.0930]$} & 0.1713 \\

\addlinespace[2pt]
\hline
\addlinespace[2pt]

\multirow{10}{*}{ \rotatebox[origin=c]{90}{ \scalebox{0.90}{\textbf{Photo.}}}} & Auto Contrast & \negcell{$[0.2528,\,0.2537]$} & 0.2279 & \negcell{$[0.2704,\,0.2713]$} & 0.1478 & \negcell{$[0.1602,\,0.1609]$} & 0.3475 & \negcell{$[0.1460,\,0.1467]$} & 0.4392 & $[0.9980,\,1.0044]$ & 0.2968 & $[0.9973,\,1.0047]$ & 0.1761 \\

& BCET & \negcell{$[0.3156,\,0.3166]$} & 0.1760 & \negcell{$[0.3530,\,0.3542]$} & 0.1193 & \negcell{$[0.1599,\,0.1605]$} & 0.3426 & \negcell{$[0.1454,\,0.1460]$} & 0.4322 & \poscell{$[1.0088,\,1.0153]$} & 0.2993 & \poscell{$[1.0279,\,1.0354]$} & 0.1720 \\

& Brightness & \negcell{$[0.2545,\,0.2554]$} & 0.2311 & \negcell{$[0.2713,\,0.2721]$} & 0.1478 & \negcell{$[0.1694,\,0.1701]$} & 0.3377 & \negcell{$[0.1550,\,0.1556]$} & 0.4200 & \poscell{$[1.0460,\,1.0527]$} & 0.2926 & \poscell{$[1.0152,\,1.0227]$} & 0.1787 \\

& Contrast & \negcell{$[0.2665,\,0.2674]$} & 0.2151 & \negcell{$[0.2727,\,0.2735]$} & 0.1440 & \negcell{$[0.1618,\,0.1624]$} & 0.3405 & \negcell{$[0.1422,\,0.1428]$} & 0.4402 & \poscell{$[1.0099,\,1.0164]$} & 0.2973 & \poscell{$[1.0315,\,1.0391]$} & 0.1689 \\

& Color Space & \negcell{$[0.2531,\,0.2539]$} & 0.2341 & \negcell{$[0.2714,\,0.2723]$} & 0.1470 & \negcell{$[0.1682,\,0.1689]$} & 0.3371 & \negcell{$[0.1522,\,0.1528]$} & 0.4231 & \poscell{$[1.0075,\,1.0139]$} & 0.2963 & $[0.9986,\,1.0059]$ & 0.1795 \\

& Equalize & \negcell{$[0.3260,\,0.3270]$} & 0.1848 & \negcell{$[0.3070,\,0.3080]$} & 0.1410 & \negcell{$[0.1706,\,0.1713]$} & 0.3311 & \negcell{$[0.1499,\,0.1506]$} & 0.4281 & \poscell{$[1.0350,\,1.0416]$} & 0.2968 & \poscell{$[1.0770,\,1.0850]$} & 0.1656 \\

& Gamma & \negcell{$[0.2547,\,0.2555]$} & 0.2311 & \negcell{$[0.2713,\,0.2722]$} & 0.1478 & \negcell{$[0.1598,\,0.1605]$} & 0.3477 & \negcell{$[0.1456,\,0.1462]$} & 0.4392 & \poscell{$[1.0204,\,1.0269]$} & 0.2959 & \poscell{$[1.0199,\,1.0274]$} & 0.1731 \\

& PCA Jitter & \negcell{$[0.2546,\,0.2554]$} & 0.2311 & \negcell{$[0.2712,\,0.2721]$} & 0.1479 & \negcell{$[0.1603,\,0.1609]$} & 0.3476 & \negcell{$[0.1463,\,0.1469]$} & 0.4389 & $[0.9970,\,1.0033]$ & 0.2964 & $[0.9962,\,1.0036]$ & 0.1763 \\

& Invert & \negcell{$[0.4507,\,0.4521]$} & 0.1165 & \negcell{$[0.4145,\,0.4158]$} & 0.1131 & \negcell{$[0.2134,\,0.2142]$} & 0.2697 & \negcell{$[0.2224,\,0.2233]$} & 0.3025 & \poscell{$[1.2374,\,1.2455]$} & 0.2651 & \poscell{$[1.2049,\,1.2140]$} & 0.1683 \\

& Weather & \negcell{$[0.9457,\,0.9486]$} & 0.0473 & \negcell{$[0.9227,\,0.9256]$} & 0.0335 & \negcell{$[0.2118,\,0.2126]$} & 0.2969 & \negcell{$[0.2149,\,0.2158]$} & 0.3566 & \poscell{$[1.2116,\,1.2195]$} & 0.2708 & \poscell{$[1.2224,\,1.2316]$} & 0.1686 \\

\addlinespace[2pt]
\hline
\addlinespace[2pt]

\multirow{4}{*}{ \rotatebox[origin=c]{90}{ \scalebox{0.90}{\textbf{Noise}}}} & Elastic & \negcell{$[0.2546,\,0.2554]$} & 0.2296 & \negcell{$[0.2708,\,0.2717]$} & 0.1468 & \negcell{$[0.1739,\,0.1746]$} & 0.3491 & \negcell{$[0.1593,\,0.1600]$} & 0.4406 & \poscell{$[1.0920,\,1.0990]$} & 0.2883 & \poscell{$[1.0649,\,1.0728]$} & 0.1743 \\

& S\&P & \negcell{$[0.4314,\,0.4328]$} & 0.1425 & \negcell{$[0.3985,\,0.3997]$} & 0.1041 & \negcell{$[0.1790,\,0.1797]$} & 0.3244 & \negcell{$[0.1673,\,0.1680]$} & 0.4030 & \poscell{$[1.1726,\,1.1802]$} & 0.2885 & \poscell{$[1.1525,\,1.1612]$} & 0.1786 \\

& Mobius & \negcell{$[0.2545,\,0.2553]$} & 0.2311 & \negcell{$[0.2713,\,0.2721]$} & 0.1478 & \negcell{$[0.1744,\,0.1751]$} & 0.3316 & \negcell{$[0.1491,\,0.1497]$} & 0.4360 & \poscell{$[1.1178,\,1.1250]$} & 0.2811 & \poscell{$[1.1035,\,1.1117]$} & 0.1694 \\

& Dim. Reduction & \negcell{$[0.9432,\,0.9461]$} & 0.0479 & \negcell{$[0.9247,\,0.9276]$} & 0.0334 & \negcell{$[0.1920,\,0.1927]$} & 0.3181 & \negcell{$[0.1813,\,0.1821]$} & 0.4060 & \poscell{$[1.1927,\,1.2005]$} & 0.2695 & \poscell{$[1.2128,\,1.2219]$} & 0.1682 \\

\addlinespace[2pt]
\hline
\addlinespace[2pt]

\multirow{3}{*}{ \rotatebox[origin=c]{90}{ \scalebox{0.86}{\textbf{Selection}}}} & Random Crop & \negcell{$[0.5646,\,0.5663]$} & 0.1109 & \negcell{$[0.6728,\,0.6748]$} & 0.0693 & \negcell{$[0.1963,\,0.1971]$} & 0.3335 & \negcell{$[0.1866,\,0.1874]$} & 0.4051 & \poscell{$[1.2316,\,1.2397]$} & 0.2710 & \poscell{$[1.1932,\,1.2022]$} & 0.1713 \\

& Random Erasing & \negcell{$[0.2690,\,0.2699]$} & 0.2191 & \negcell{$[0.2941,\,0.2950]$} & 0.1485 & \negcell{$[0.1593,\,0.1599]$} & 0.3492 & \negcell{$[0.1492,\,0.1498]$} & 0.4257 & \poscell{$[1.0190,\,1.0256]$} & 0.2902 & \poscell{$[1.0066,\,1.0140]$} & 0.1748 \\

& Shuffle & \negcell{$[0.6169,\,0.6188]$} & 0.1112 & \negcell{$[0.6174,\,0.6194]$} & 0.0829 & \negcell{$[0.4681,\,0.4699]$} & 0.1489 & \negcell{$[0.4193,\,0.4210]$} & 0.1937 & \poscell{$[1.2730,\,1.2814]$} & 0.2745 & \poscell{$[1.2309,\,1.2402]$} & 0.1705 \\

\addlinespace[2pt]
\hline
\addlinespace[2pt]

\multirow{4}{*}{ \rotatebox[origin=c]{90}{ \scalebox{0.90}{\textbf{Filter}}}} & Motion Blur & \negcell{$[0.3719,\,0.3731]$} & 0.1596 & \negcell{$[0.3624,\,0.3636]$} & 0.1194 & \negcell{$[0.1651,\,0.1658]$} & 0.3434 & \negcell{$[0.1485,\,0.1491]$} & 0.4360 & \poscell{$[1.0220,\,1.0286]$} & 0.2938 & \poscell{$[1.0130,\,1.0205]$} & 0.1754 \\

& Blur & \negcell{$[0.5642,\,0.5660]$} & 0.1035 & \negcell{$[0.5679,\,0.5696]$} & 0.0799 & \negcell{$[0.1820,\,0.1827]$} & 0.3170 & \negcell{$[0.1610,\,0.1617]$} & 0.4128 & \poscell{$[1.0624,\,1.0692]$} & 0.2928 & \poscell{$[1.0902,\,1.0983]$} & 0.1716 \\

& Gauss Blur & \negcell{$[0.6099,\,0.6118]$} & 0.1176 & \negcell{$[0.6310,\,0.6330]$} & 0.0924 & \negcell{$[0.1776,\,0.1782]$} & 0.3254 & \negcell{$[0.1551,\,0.1557]$} & 0.4285 & \poscell{$[1.1985,\,1.2064]$} & 0.2720 & \poscell{$[1.2214,\,1.2307]$} & 0.1685 \\

& Sharpen & \negcell{$[0.2610,\,0.2619]$} & 0.2125 & \negcell{$[0.2838,\,0.2847]$} & 0.1336 & \negcell{$[0.1586,\,0.1592]$} & 0.3438 & \negcell{$[0.1421,\,0.1427]$} & 0.4443 & $[0.9984,\,1.0048]$ & 0.2994 & \poscell{$[1.0290,\,1.0366]$} & 0.1716 \\

\addlinespace[2pt]
\hline
\addlinespace[2pt]

\multirow{5}{*}{ \rotatebox[origin=c]{90}{ \scalebox{0.90}{\textbf{Self-Mix}}}} & AugMix & \negcell{$[0.2880,\,0.2890]$} & 0.2017 & \negcell{$[0.2934,\,0.2943]$} & 0.1361 & \negcell{$[0.1743,\,0.1750]$} & 0.3259 & \negcell{$[0.1576,\,0.1583]$} & 0.4099 & \poscell{$[1.0655,\,1.0724]$} & 0.2902 & \poscell{$[1.0560,\,1.0638]$} & 0.1713 \\

& KeepAugment & \negcell{$[0.0813,\,0.0817]$} & 0.1463 & \negcell{$[0.0811,\,0.0814]$} & 0.1107 & \negcell{$[0.0350,\,0.0352]$} & 0.3876 & \negcell{$[0.0317,\,0.0319]$} & 0.4575 & \negcell{$[0.1144,\,0.1151]$} & 0.0564 & \negcell{$[0.0863,\,0.0869]$} & 0.0206 \\

& Robust MixGen & \negcell{$[0.1146,\,0.1150]$} & 0.0875 & \negcell{$[0.1298,\,0.1303]$} & 0.0456 & \negcell{$[0.0537,\,0.0539]$} & 0.2398 & \negcell{$[0.0481,\,0.0483]$} & 0.2718 & \negcell{$[0.1188,\,0.1195]$} & 0.0481 & \negcell{$[0.0884,\,0.0890]$} & 0.0185 \\

& SelfMix & \negcell{$[0.2855,\,0.2864]$} & 0.2053 & \negcell{$[0.3052,\,0.3062]$} & 0.1341 & \negcell{$[0.1620,\,0.1627]$} & 0.3497 & \negcell{$[0.1502,\,0.1508]$} & 0.4330 & \poscell{$[1.0579,\,1.0647]$} & 0.2837 & \poscell{$[1.0302,\,1.0378]$} & 0.1703 \\

& SALF-Mix & \negcell{$[0.2826,\,0.2835]$} & 0.2065 & \negcell{$[0.3036,\,0.3046]$} & 0.1345 & \negcell{$[0.1625,\,0.1632]$} & 0.3500 & \negcell{$[0.1506,\,0.1513]$} & 0.4352 & \poscell{$[1.0478,\,1.0545]$} & 0.2852 & \poscell{$[1.0251,\,1.0327]$} & 0.1716 \\

\addlinespace[2pt]
\hline
\addlinespace[2pt]

\multirow{7}{*}{ \rotatebox[origin=c]{90}{ \scalebox{0.90}{\textbf{Sample Mix}}}} & CutMix & \negcell{$[0.6091,\,0.6110]$} & 0.0958 & \negcell{$[0.7250,\,0.7272]$} & 0.0541 & \negcell{$[0.1711,\,0.1718]$} & 0.3416 & \negcell{$[0.1636,\,0.1643]$} & 0.4251 & \poscell{$[1.1698,\,1.1774]$} & 0.2820 & \poscell{$[1.1152,\,1.1235]$} & 0.1792 \\

& MixUp & \negcell{$[0.4327,\,0.4341]$} & 0.1258 & \negcell{$[0.4957,\,0.4973]$} & 0.0807 & \negcell{$[0.1865,\,0.1872]$} & 0.3233 & \negcell{$[0.1813,\,0.1821]$} & 0.3937 & \poscell{$[1.1720,\,1.1797]$} & 0.2780 & \poscell{$[1.1274,\,1.1358]$} & 0.1755 \\

& RICAP & \negcell{$[0.0955,\,0.0959]$} & 0.1357 & \negcell{$[0.0953,\,0.0957]$} & 0.0993 & \negcell{$[0.0495,\,0.0497]$} & 0.2745 & \negcell{$[0.0463,\,0.0465]$} & 0.3255 & \negcell{$[0.1181,\,0.1188]$} & 0.0488 & \negcell{$[0.0883,\,0.0888]$} & 0.0191 \\

& ResizeMix & \negcell{$[0.4498,\,0.4512]$} & 0.1595 & \negcell{$[0.4809,\,0.4824]$} & 0.1153 & \negcell{$[0.1474,\,0.1480]$} & 0.3577 & \negcell{$[0.1397,\,0.1403]$} & 0.4495 & \poscell{$[1.1298,\,1.1371]$} & 0.2824 & \poscell{$[1.0667,\,1.0746]$} & 0.1758 \\

& SaliencyMix & \negcell{$[0.4368,\,0.4381]$} & 0.1451 & \negcell{$[0.4576,\,0.4591]$} & 0.0990 & \negcell{$[0.2694,\,0.2704]$} & 0.2318 & \negcell{$[0.2108,\,0.2116]$} & 0.3189 & \poscell{$[1.2062,\,1.2141]$} & 0.2746 & \poscell{$[1.1795,\,1.1884]$} & 0.1656 \\

& IPMix & \negcell{$[0.1268,\,0.1272]$} & 0.1149 & \negcell{$[0.1161,\,0.1165]$} & 0.0856 & \negcell{$[0.0567,\,0.0569]$} & 0.2561 & \negcell{$[0.0494,\,0.0496]$} & 0.3036 & \negcell{$[0.1174,\,0.1181]$} & 0.0516 & \negcell{$[0.0882,\,0.0888]$} & 0.0194 \\

& MixGen & \negcell{$[0.1112,\,0.1116]$} & 0.0991 & \negcell{$[0.1094,\,0.1098]$} & 0.0701 & \negcell{$[0.0503,\,0.0505]$} & 0.2907 & \negcell{$[0.0451,\,0.0453]$} & 0.3355 & \negcell{$[0.1176,\,0.1183]$} & 0.0503 & \negcell{$[0.0883,\,0.0889]$} & 0.0189 \\

\addlinespace[2pt]
\hline
\addlinespace[2pt]

\multirow{3}{*}{ \rotatebox[origin=c]{90}{ \scalebox{0.90}{\textbf{GAN}}}} & DCGAN & \negcell{$[0.4413,\,0.4427]$} & 0.1400 & \negcell{$[0.4692,\,0.4707]$} & 0.0992 & \negcell{$[0.1938,\,0.1945]$} & 0.3048 & \negcell{$[0.1895,\,0.1903]$} & 0.3686 & \poscell{$[1.1793,\,1.1870]$} & 0.2717 & \poscell{$[1.1335,\,1.1419]$} & 0.1700 \\

& CycleGAN & \negcell{$[0.5397,\,0.5414]$} & 0.1218 & \negcell{$[0.5904,\,0.5922]$} & 0.0797 & \negcell{$[0.1984,\,0.1992]$} & 0.2804 & \negcell{$[0.1779,\,0.1786]$} & 0.3480 & \poscell{$[1.0845,\,1.0916]$} & 0.2842 & \poscell{$[1.0662,\,1.0741]$} & 0.1687 \\

& NST & \negcell{$[0.9477,\,0.9506]$} & 0.0472 & \negcell{$[0.9258,\,0.9287]$} & 0.0334 & \negcell{$[0.2777,\,0.2788]$} & 0.2376 & \negcell{$[0.2550,\,0.2560]$} & 0.3005 & \poscell{$[1.2751,\,1.2835]$} & 0.2750 & \poscell{$[1.2362,\,1.2456]$} & 0.1697 \\

\addlinespace[2pt]
\hline
\addlinespace[2pt]

\multirow{6}{*}{ \rotatebox[origin=c]{90}{ \scalebox{0.90}{\textbf{Diffusion}}}} & DiffuseCLIP & \negcell{$[0.3512,\,0.3524]$} & 0.1782 & \negcell{$[0.3511,\,0.3522]$} & 0.1262 & \negcell{$[0.1760,\,0.1767]$} & 0.3262 & \negcell{$[0.1565,\,0.1572]$} & 0.4157 & \poscell{$[1.1648,\,1.1724]$} & 0.2924 & \poscell{$[1.1837,\,1.1926]$} & 0.1699 \\

& Pix2Pix-Zero & \negcell{$[0.6793,\,0.6814]$} & 0.0885 & \negcell{$[0.7112,\,0.7134]$} & 0.0631 & \negcell{$[0.1976,\,0.1983]$} & 0.3930 & \negcell{$[0.1857,\,0.1865]$} & 0.4823 & \negcell{$[0.1575,\,0.1584]$} & 0.4444 & \negcell{$[0.1215,\,0.1223]$} & 0.4701 \\

& DiffuseMix & \negcell{$[0.9406,\,0.9435]$} & 0.0474 & \negcell{$[0.8691,\,0.8718]$} & 0.0388 & \negcell{$[0.1740,\,0.1747]$} & 0.3315 & \negcell{$[0.1762,\,0.1769]$} & 0.3956 & \poscell{$[1.1953,\,1.2031]$} & 0.2872 & \poscell{$[1.1608,\,1.1695]$} & 0.1764 \\

& SaSPA & \negcell{$[0.9214,\,0.9243]$} & 0.0475 & \negcell{$[0.6945,\,0.6967]$} & 0.0529 & \negcell{$[0.1864,\,0.1871]$} & 0.3441 & \negcell{$[0.1758,\,0.1765]$} & 0.4572 & \poscell{$[1.2466,\,1.2548]$} & 0.2792 & \poscell{$[1.2192,\,1.2284]$} & 0.1723 \\

& DA-Fusion & \negcell{$[0.6811,\,0.6832]$} & 0.0861 & \negcell{$[0.7113,\,0.7135]$} & 0.0594 & \negcell{$[0.1563,\,0.1569]$} & 0.3800 & \negcell{$[0.1496,\,0.1502]$} & 0.4848 & \poscell{$[1.2073,\,1.2152]$} & 0.2746 & \poscell{$[1.1587,\,1.1674]$} & 0.1742 \\

& DIAGen & \negcell{$[0.6791,\,0.6812]$} & 0.0872 & \negcell{$[0.7051,\,0.7072]$} & 0.0628 & \negcell{$[0.1802,\,0.1809]$} & 0.3772 & \negcell{$[0.1746,\,0.1754]$} & 0.4625 & \poscell{$[1.2589,\,1.2672]$} & 0.2733 & \poscell{$[1.2236,\,1.2329]$} & 0.1704 \\

\addlinespace[2pt]
\hline
\addlinespace[2pt]

\multirow{2}{*}{ \rotatebox[origin=c]{90}{ \scalebox{0.90}{\textbf{Adv.}}}} & FGSM & \negcell{$[0.5050,\,0.5065]$} & 0.1336 & \negcell{$[0.5103,\,0.5119]$} & 0.1064 & \negcell{$[0.1787,\,0.1794]$} & 0.3199 & \negcell{$[0.1666,\,0.1673]$} & 0.4094 & \poscell{$[1.2730,\,1.2814]$} & 0.2771 & \poscell{$[1.1670,\,1.1757]$} & 0.1663 \\

& PGD & \negcell{$[0.3405,\,0.3416]$} & 0.1870 & \negcell{$[0.4439,\,0.4453]$} & 0.1246 & \negcell{$[0.1637,\,0.1644]$} & 0.3396 & \negcell{$[0.1472,\,0.1479]$} & 0.4431 & \poscell{$[1.1945,\,1.2023]$} & 0.2771 & \poscell{$[1.1287,\,1.1371]$} & 0.1631 \\

\addlinespace[2pt]
\hline

\end{tabular}%
}
\end{table*}

\begin{figure*}[!htbp]
\centering

\subfloat[MN dataset.]{%
    \hspace*{-24mm}%
    \includegraphics[width=0.72\textwidth]{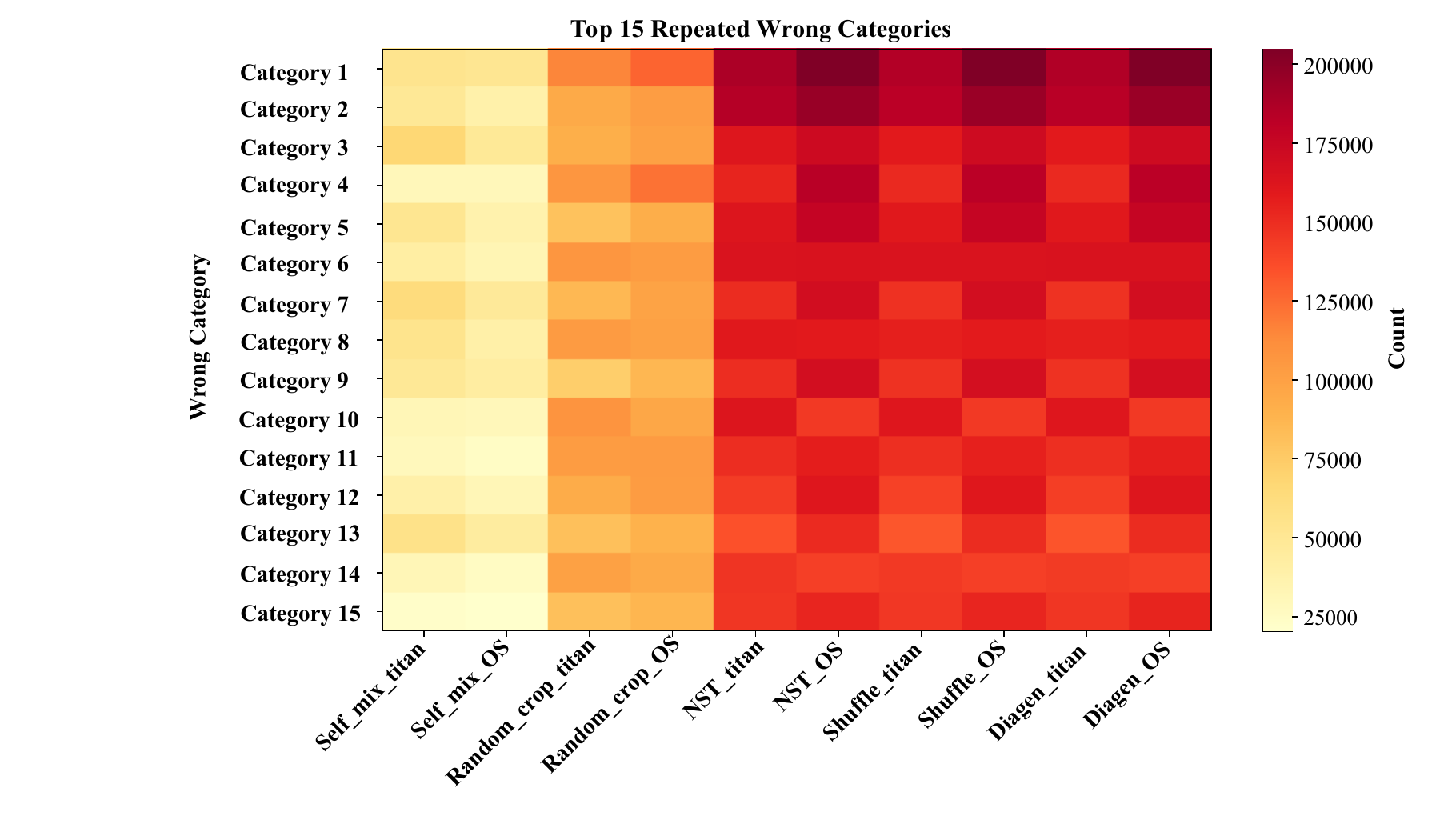}%
    \label{fig:heatmap_mn}
}

\vspace{-2mm}

\subfloat[CIFAR-10 dataset.]{%
    \includegraphics[width=0.62\textwidth]{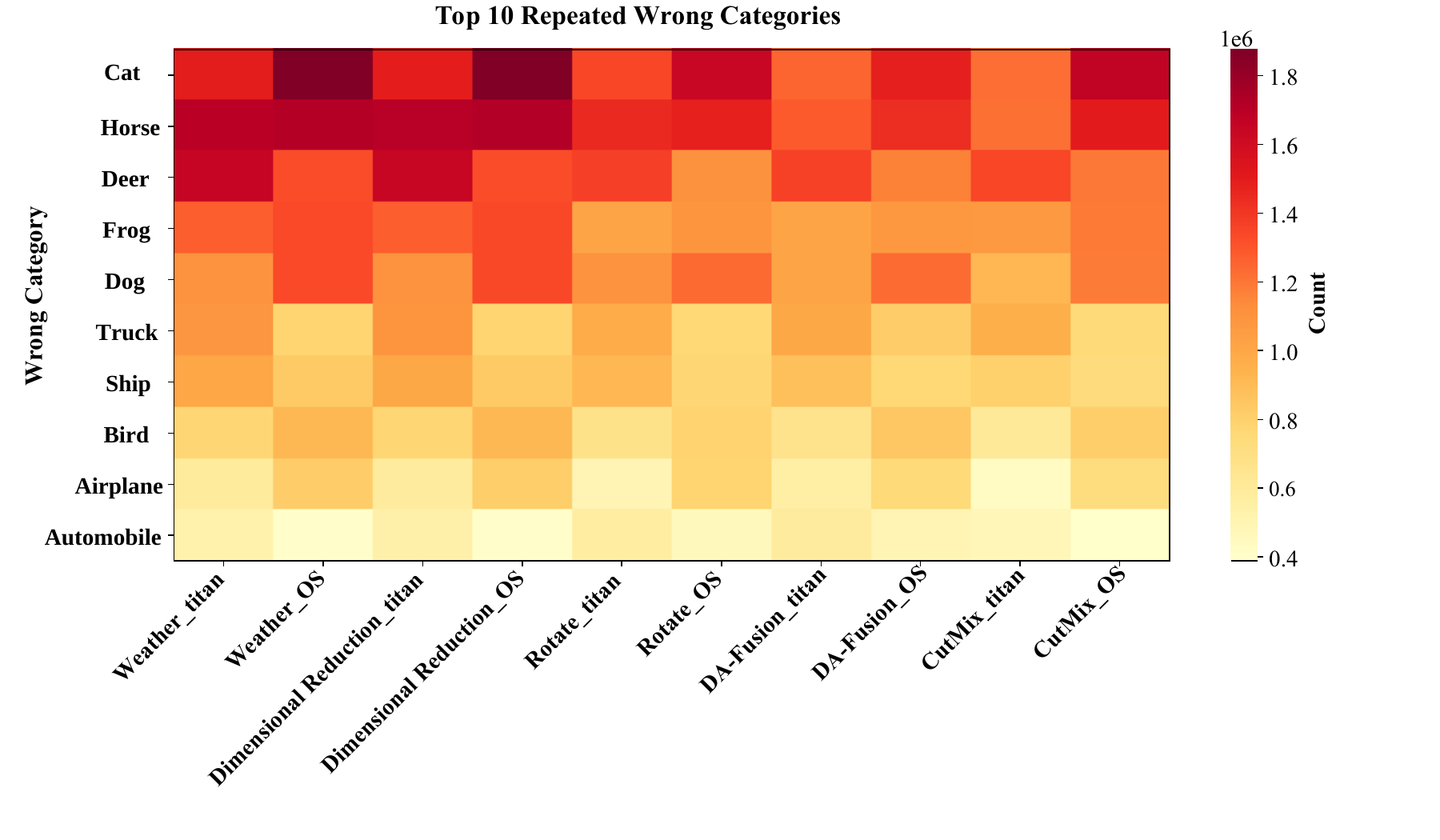}%
    \label{fig:heatmap_cifar10}
}

\vspace{-2mm}

\subfloat[ImageNet dataset.]{%
    \includegraphics[width=0.62\textwidth]{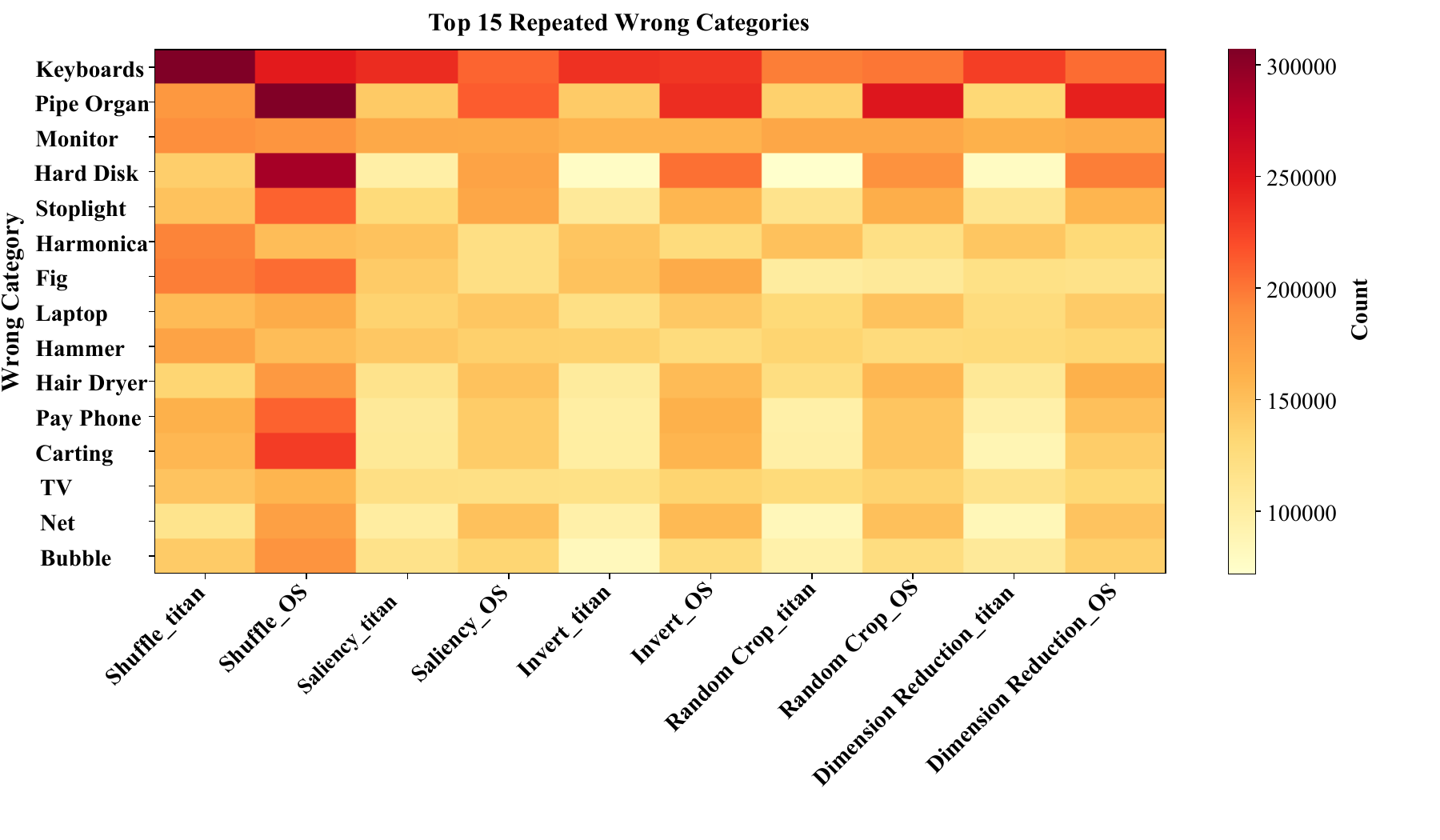}%
    \label{fig:heatmap_imagenet}
}

\caption{Comparison of the most frequently repeated incorrect categories for the selected augmentation techniques across (a) the MN, (b) CIFAR-10, and (c) ImageNet datasets.}
\label{fig:heatmap}
\end{figure*}

\noindent\textbf{\Cref{fig:heatmap}} presents the 15 categories with the highest failure counts across the three evaluated datasets. A common observation across all datasets is that categories containing images with multiple objects tend to exhibit higher failure counts under different augmentation techniques. This finding suggests that visually complex scenes containing multiple objects present a more challenging retrieval scenario for the evaluated embedding models. 

\begin{figure}[!htbp]
\centering

\subfloat[Original image]{%
    \includegraphics[
        width=0.55\columnwidth,
        height=60mm,
        keepaspectratio
    ]{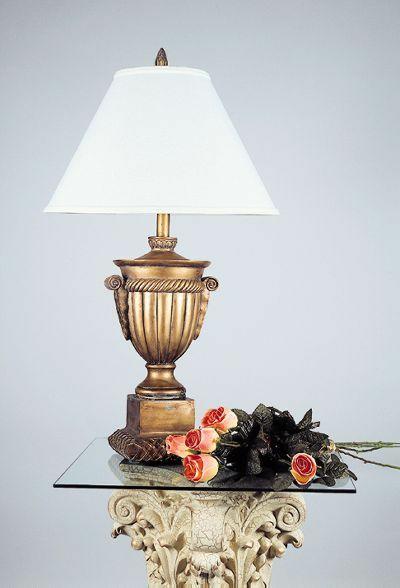}%
    \label{fig:misclass_original}
}
\hspace{0.1\textwidth}
\subfloat[Augmented image]{%
    \includegraphics[
       width=0.55\columnwidth,
        height=60mm,
        keepaspectratio
    ]{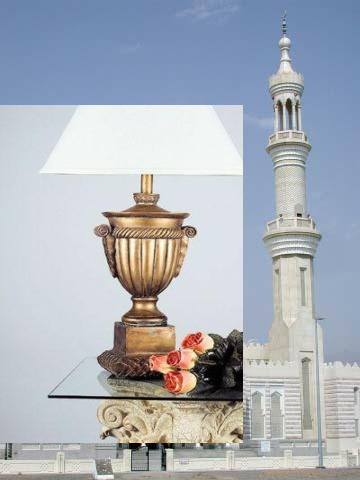}%
    \label{fig:misclass_augmented}
}

\caption{Comparison between (a) the original image and
(b) its corresponding SaliencyMix-augmented counterpart.}
\label{fig:misclass_sample}
\end{figure}

\noindent\textbf{\Cref{fig:misclass_sample}} presents an example from the ImageNet dataset in which the original image is labeled as a table lamp. After applying the SaliencyMix augmentation technique, the resulting image is incorrectly associated with the mosque category, indicating that the introduced salient-region composition substantially alters the dominant semantic cues used by the embedding model for category recognition.

\subsection{RQ2a: Supplementary Embedding-Space Similarity Results}
\noindent\textbf{\Cref{tab:embedding_similarity}} presents the pairwise embedding similarity between each original image and its corresponding augmented image across the evaluated datasets and augmentation techniques. The similarity is quantified using cosine similarity.

\Cref{tab:embedding_similarity} reports the standard deviation of embedding similarity across all evaluated datasets and embedding models. Overall, Titan exhibits greater stability than OpenCLIP, as reflected by its generally lower STD values across different image categories and augmentation conditions. Nevertheless, both models demonstrate relatively stable similarity behavior between the original and augmented images for most augmentation techniques. Among the evaluated techniques, Robust MixGen, MixUp, and MixGen produce the highest STD values. These techniques generate augmented samples by combining visual content from multiple source images, which can introduce substantial variations in semantic structure, object composition, and local appearance. As a result, the degree of similarity between the original and augmented images varies considerably across samples, thereby creating more challenging conditions for the embedding models. In contrast, Blur, PCA Jitter, Translate, and Brightness generally produce smaller STD values. These techniques apply more uniform transformations to individual images while preserving most of their semantic content and global structure. Consequently, the resulting embeddings remain more consistent across samples, leading to lower variability in similarity scores.

\begin{table*}[]
\centering
\caption{Mean Embedding Similarity and Standard Deviation of Augmentation Techniques for the Evaluated Embedding Models}
\label{tab:embedding_similarity}
\setlength{\tabcolsep}{1.5pt}
\renewcommand{\arraystretch}{0.90}
\scriptsize

\resizebox{\textwidth}{!}{%
\begin{tabular}{
ll
cc cc
cc cc
cc cc}
\hline
\noalign{\vskip 2pt}

\textbf{Cat.} & \textbf{Aug.}
& \multicolumn{4}{c}{\textbf{CIFAR-10}}
& \multicolumn{4}{c}{\textbf{ImageNet}}
& \multicolumn{4}{c}{\textbf{MN}} \\

\cline{3-6}
\cline{7-10}
\cline{11-14}
\addlinespace[4pt]
&
& \multicolumn{2}{c}{\textbf{Titan}}
& \multicolumn{2}{c}{\textbf{OpenCLIP}}
& \multicolumn{2}{c}{\textbf{Titan}}
& \multicolumn{2}{c}{\textbf{OpenCLIP}}
& \multicolumn{2}{c}{\textbf{Titan}}
& \multicolumn{2}{c}{\textbf{OpenCLIP}} \\

\cline{3-4}
\cline{5-6}
\cline{7-8}
\cline{9-10}
\cline{11-12}
\cline{13-14}
\addlinespace[3pt]
&
& \textbf{Mean} & \textbf{STD}
& \textbf{Mean} & \textbf{STD}
& \textbf{Mean} & \textbf{STD}
& \textbf{Mean} & \textbf{STD}
& \textbf{Mean} & \textbf{STD}
& \textbf{Mean} & \textbf{STD} \\

\hline

\addlinespace[3pt]
\multirow{4}{*}{\rotatebox[origin=c]{90}{\scalebox{0.90}{\textbf{Geom.}}}}
& Rotate    & 0.6720 & 0.0871 & 0.5256 & 0.1239 & 0.8646 & 0.0507 & 0.8530 & 0.0616 & 0.8898 & 0.0873 & 0.8469 & 0.0622 \\
& Flip      & 0.8056 & 0.0796 & 0.7350 & 0.1162 & 0.8680 & 0.0839 & 0.8429 & 0.1019 & 0.8523 & 0.0628 & 0.8287 & 0.0868 \\
& Translate & 0.9999 & 0.0001 & 0.9999 & 0.0001 & 0.9092 & 0.0473 & 0.8726 & 0.0630 & 0.9014 & 0.0349 & 0.9001 & 0.0388 \\
& Shear     & 0.9999 & 0.0001 & 0.9999 & 0.0001 & 0.8439 & 0.0568 & 0.8283 & 0.0698 & 0.8943 & 0.0366 & 0.8813 & 0.0474 \\

\noalign{\vskip 2pt}
\hline
\noalign{\vskip 2pt}

\multirow{10}{*}{\rotatebox[origin=c]{90}{\scalebox{0.90}{\textbf{Photo.}}}}
& Auto Contrast     & 0.9873 & 0.0169 & 0.9748 & 0.0296 & 0.9966 & 0.0150 & 0.9952 & 0.0174 & 0.9982 & 0.0130 & 0.9970 & 0.0184 \\
& BCET              & 0.8662 & 0.0580 & 0.7701 & 0.0799 & 0.9525 & 0.0305 & 0.9173 & 0.0421 & 0.9704 & 0.0185 & 0.9488 & 0.0277 \\
& Brightness        & 0.9999 & 0.0001 & 0.9999 & 0.0001 & 0.9163 & 0.0323 & 0.8962 & 0.0452 & 0.9526 & 0.0197 & 0.9440 & 0.0271 \\
& Contrast          & 0.9679 & 0.0228 & 0.9375 & 0.0369 & 0.9372 & 0.0359 & 0.9241 & 0.0360 & 0.9735 & 0.0133 & 0.9549 & 0.0185 \\
& Color Space       & 0.9971 & 0.0016 & 0.9950 & 0.0030 & 0.9348 & 0.0311 & 0.9210 & 0.0416 & 0.9794 & 0.0114 & 0.9697 & 0.0166 \\
& Equalize          & 0.8705 & 0.0770 & 0.8074 & 0.0994 & 0.8737 & 0.0794 & 0.8662 & 0.0809 & 0.9463 & 0.0395 & 0.9175 & 0.0473 \\
& Gamma             & 0.9983 & 0.0016 & 0.9973 & 0.0029 & 0.9949 & 0.0095 & 0.9901 & 0.0174 & 0.9749 & 0.0268 & 0.9678 & 0.0318 \\
& PCA Jitter        & 0.9997 & 0.0006 & 0.9996 & 0.0013 & 0.9994 & 0.0017 & 0.9993 & 0.0027 & 0.9995 & 0.0008 & 0.9996 & 0.0009 \\
& Invert            & 0.7586 & 0.0696 & 0.6872 & 0.0981 & 0.7629 & 0.0932 & 0.7188 & 0.1263 & 0.7656 & 0.0820 & 0.7106 & 0.1072 \\
& Weather           & 0.5045 & 0.0504 & 0.2880 & 0.0895 & 0.7786 & 0.0747 & 0.7383 & 0.0870 & 0.8251 & 0.0530 & 0.7386 & 0.0592 \\

\noalign{\vskip 2pt}
\hline
\noalign{\vskip 2pt}

\multirow{4}{*}{\rotatebox[origin=c]{90}{\scalebox{0.90}{\textbf{Noise}}}}
& Elastic        & 0.9954 & 0.0056 & 0.9904 & 0.0117 & 0.8925 & 0.0656 & 0.8679 & 0.0776 & 0.9174 & 0.0374 & 0.9095 & 0.0496 \\
& S\&P           & 0.8144 & 0.0421 & 0.7047 & 0.0744 & 0.8398 & 0.0717 & 0.8128 & 0.0901 & 0.8450 & 0.0698 & 0.8162 & 0.0920 \\
& Mobius         & 0.9999 & 0.0001 & 0.9999 & 0.0001 & 0.8744 & 0.0447 & 0.8682 & 0.0558 & 0.8949 & 0.0369 & 0.8648 & 0.0499 \\
& Dim. Reduction & 0.5191 & 0.0542 & 0.2734 & 0.0860 & 0.8213 & 0.0691 & 0.7986 & 0.0776 & 0.8594 & 0.0369 & 0.8028 & 0.0427 \\

\noalign{\vskip 2pt}
\hline
\noalign{\vskip 2pt}

\multirow{3}{*}{\rotatebox[origin=c]{90}{\scalebox{0.80}{\textbf{Selection}}}}
& Random Crop & 0.7239 & 0.0783 & 0.5169 & 0.1241 & 0.7800 & 0.0887 & 0.7266 & 0.1111 & 0.7923 & 0.0595 & 0.7825 & 0.0696 \\
& Erasing     & 0.9455 & 0.0300 & 0.8814 & 0.1241 & 0.9500 & 0.0341 & 0.9362 & 0.0465 & 0.9726 & 0.0224 & 0.9654 & 0.0284 \\
& Shuffle     & 0.7094 & 0.0529 & 0.5944 & 0.0818 & 0.3803 & 0.0562 & 0.2790 & 0.0831 & 0.4220 & 0.0452 & 0.3776 & 0.0647 \\

\noalign{\vskip 2pt}
\hline
\noalign{\vskip 2pt}

\multirow{4}{*}{\rotatebox[origin=c]{90}{\scalebox{0.93}{\textbf{Filter}}}}
& M-Blur    & 0.8785 & 0.0833 & 0.8189 & 0.1224 & 0.9403 & 0.0463 & 0.9270 & 0.0530 & 0.9692 & 0.0240 & 0.9610 & 0.0292 \\
& Blur      & 0.7256 & 0.1211 & 0.5774 & 0.1862 & 0.8721 & 0.0652 & 0.8612 & 0.0767 & 0.9302 & 0.0514 & 0.8959 & 0.6981 \\
& GaussBlur & 0.7069 & 0.0683 & 0.5415 & 0.1274 & 0.8781 & 0.0464 & 0.8668 & 0.0585 & 0.8383 & 0.0490 & 0.7578 & 0.0466 \\
& Sharpen   & 0.9442 & 0.0211 & 0.8902 & 0.0367 & 0.9550 & 0.0267 & 0.9285 & 0.0354 & 0.9769 & 0.0083 & 0.9509 & 0.0215 \\

\noalign{\vskip 2pt}
\hline
\noalign{\vskip 2pt}

\multirow{3}{*}{\rotatebox[origin=c]{90}{\scalebox{0.82}{\textbf{Self-Mix}}}}
& AugMix   & 0.9454 & 0.1181 & 0.9074 & 0.1500 & 0.8993 & 0.0885 & 0.8829 & 0.0981 & 0.9260 & 0.0760 & 0.9114 & 0.0891 \\
& Self Mix & 0.9457 & 0.0386 & 0.9072 & 0.0584 & 0.9616 & 0.0340 & 0.9580 & 0.0407 & 0.9530 & 0.0302 & 0.9586 & 0.0272 \\
& Salf Mix & 0.9492 & 0.0351 & 0.9164 & 0.0514 & 0.9487 & 0.0336 & 0.9422 & 0.0420 & 0.9626 & 0.0230 & 0.9660 & 0.0216 \\

\noalign{\vskip 2pt}
\hline
\noalign{\vskip 2pt}

\multirow{9}{*}{\rotatebox[origin=c]{90}{\scalebox{0.90}{\textbf{Sample Mix}}}}
& MixGen    & 0.7612 & 0.0978 & 0.6488 & 0.1264 & 0.6127 & 0.1539 & 0.5520 & 0.1732 & 0.7819 & 0.0988 & 0.7996 & 0.1031 \\
& R-MixGen  & 0.7835 & 0.2035 & 0.7135 & 0.2606 & 0.6576 & 0.3038 & 0.6174 & 0.3323 & 0.8177 & 0.1686 & 0.8271 & 0.1657 \\
& MixUp     & 0.7338 & 0.1116 & 0.5559 & 0.1770 & 0.6593 & 0.1568 & 0.6005 & 0.1791 & 0.7728 & 0.0955 & 0.7685 & 0.1044 \\
& CutMix    & 0.6702 & 0.1138 & 0.5649 & 0.1689 & 0.7213 & 0.1229 & 0.6694 & 0.1449 & 0.8011 & 0.0761 & 0.8106 & 0.0767 \\
& Saliency  & 0.8215 & 0.0987 & 0.7183 & 0.1342 & 0.7002 & 0.1424 & 0.6419 & 0.1590 & 0.8396 & 0.0945 & 0.8626 & 0.0909 \\
& RICAP     & 0.7308 & 0.0737 & 0.6262 & 0.0820 & 0.6578 & 0.0831 & 0.5887 & 0.0959 & 0.8102 & 0.0773 & 0.8229 & 0.0789 \\
& KeepAug   & 0.7634 & 0.0819 & 0.6362 & 0.1140 & 0.7699 & 0.0958 & 0.7361 & 0.1126 & 0.8745 & 0.0720 & 0.8822 & 0.0730 \\
& IP-Mix    & 0.6522 & 0.0800 & 0.5249 & 0.0985 & 0.5356 & 0.1236 & 0.4740 & 0.1469 & 0.6928 & 0.0890 & 0.6582 & 0.1128 \\
& ResizeMix & 0.7395 & 0.0946 & 0.6052 & 0.1362 & 0.7607 & 0.1500 & 0.7302 & 0.1724 & 0.8312 & 0.0910 & 0.8154 & 0.0284 \\

\noalign{\vskip 2pt}
\hline
\noalign{\vskip 2pt}

\multirow{3}{*}{\rotatebox[origin=c]{90}{\scalebox{0.90}{\textbf{GAN}}}}
& DCGAN                 & 0.7883 & 0.1145 & 0.6637 & 0.1670 & 0.7919 & 0.1116 & 0.7440 & 0.1378 & 0.8321 & 0.0913 & 0.8227 & 0.1031 \\
& CycleGAN              & 0.7111 & 0.1033 & 0.5511 & 0.1355 & 0.7799 & 0.0701 & 0.7559 & 0.0869 & 0.9168 & 0.0341 & 0.9055 & 0.0336 \\
& NST                   & 0.4255 & 0.0604 & 0.3234 & 0.0860 & 0.6340 & 0.1068 & 0.5952 & 0.1252 & 0.6790 & 0.0830 & 0.5826 & 0.1031 \\

\noalign{\vskip 2pt}
\hline
\noalign{\vskip 2pt}

\multirow{6}{*}{\rotatebox[origin=c]{90}{\scalebox{0.90}{\textbf{Diffusion}}}}
& DiffusionCLIP & 0.9110 & 0.0934 & 0.8549 & 0.1222 & 0.8767 & 0.0603 & 0.8556 & 0.0678 & 0.8336 & 0.0644 & 0.7959 & 0.0836 \\
& Pix2Pix-Zero  & 0.6757 & 0.0962 & 0.5068 & 0.1378 & 0.6027 & 0.1261 & 0.5341 & 0.1550 & 0.6027 & 0.1264 & 0.5327 & 0.1546 \\
& DiffuseMix    & 0.5254 & 0.0939 & 0.4290 & 0.1138 & 0.8406 & 0.0804 & 0.7883 & 0.1028 & 0.7968 & 0.1015 & 0.7759 & 0.1298 \\
& SaSPA         & 0.4866 & 0.0708 & 0.4253 & 0.1032 & 0.6438 & 0.1046 & 0.5977 & 0.1358 & 0.7251 & 0.0807 & 0.6732 & 0.1059 \\
& DA-Fusion     & 0.6708 & 0.1007 & 0.5033 & 0.1411 & 0.7731 & 0.1150 & 0.7271 & 0.1414 & 0.7862 & 0.0809 & 0.7692 & 0.1069 \\
& DIAGen        & 0.6688 & 0.1027 & 0.5030 & 0.1423 & 0.6558 & 0.1280 & 0.5924 & 0.1567 & 0.6306 & 0.1049 & 0.5788 & 0.1275 \\

\noalign{\vskip 2pt}
\hline
\noalign{\vskip 2pt}

\multirow{2}{*}{\rotatebox[origin=c]{90}{\scalebox{0.90}{\textbf{Adv.}}}}
& FGSM & 0.7512 & 0.0808 & 0.6316 & 0.1107 & 0.8159 & 0.0712 & 0.8345 & 0.0891 & 0.6419 & 0.0947 & 0.8449 & 0.0666 \\
& PGD  & 0.8726 & 0.0513 & 0.6921 & 0.1045 & 0.9177 & 0.0436 & 0.9434 & 0.0359 & 0.7931 & 0.0816 & 0.8902 & 0.0430 \\

\hline
\end{tabular}%
}
\end{table*}
\subsection{RQ2b: Supplementary Embedding-Uncertainty Results}
\noindent\textbf{\Cref{tab:dispersion_consistency,tab:mahalanobis_ensemble}} present the uncertainty of the evaluated embedding models under different augmentation techniques using four estimators: embedding dispersion, pairwise distance, Mahalanobis distance, and ensemble-based uncertainty. Embedding dispersion measures the deviation of each image embedding from the mean embedding of its corresponding class. Pairwise distance quantifies the difference between the embeddings of an original image and its augmented counterpart. Mahalanobis distance measures this deviation while accounting for the covariance structure of the embedding space. For the ensemble-based estimator, ten KNN models are employed, and uncertainty is quantified using the difference between the maximum and minimum predicted probabilities for the category corresponding to each original image.
\begin{table}[!t]
\centering
\caption{Embedding Dispersion and Pairwise Distance Across Augmentation Techniques}
\label{tab:dispersion_consistency}
\setlength{\tabcolsep}{2pt}
\renewcommand{\arraystretch}{1.0}
\scriptsize

\resizebox{\textwidth}{!}{%
\begin{tabular}{@{}p{0.02\textwidth}p{0.12\textwidth}cccccc@{}}
\hline
\textbf{Cat.} & \textbf{Aug.}
& \multicolumn{2}{c}{\textbf{CIFAR-10}}
& \multicolumn{2}{c}{\textbf{ImageNet}}
& \multicolumn{2}{c}{\textbf{MN}} \\
\cline{3-8}
& & \textbf{Titan} & \textbf{OpenCLIP}
  & \textbf{Titan} & \textbf{OpenCLIP}
  & \textbf{Titan} & \textbf{OpenCLIP} \\
\hline
\multicolumn{8}{c}{\textit{Values are reported as dispersion / pairwise distance.}} \\
\hline
\noalign{\vskip 3pt}

\multirow{4}{*}{\rotatebox[origin=c]{90}{\scalebox{0.90}{\textbf{Geom.}}}}
& Rotate    & 0.5953/0.3279 & 0.7143/0.4744 & 0.6775/0.1353 & 0.7041/0.1469 & 0.5107/0.1101 & 0.5313/0.1530 \\
& Flip      & 0.5500/0.1944 & 0.6540/0.2650 & 0.6708/0.1319 & 0.7005/0.1570 & 0.5205/0.1476 & 0.5364/0.1712 \\
& Translate & 0.5177/0.0001 & 0.6210/0.0001 & 0.6634/0.0907 & 0.6963/0.1273 & 0.4967/0.0985 & 0.4983/0.0998 \\
& Shear     & 0.5177/0.0001 & 0.6210/0.0001 & 0.6793/0.1560 & 0.7062/0.1716 & 0.5035/0.1056 & 0.5065/0.1186 \\

\noalign{\vskip 2pt}
\hline
\noalign{\vskip 2pt}

\multirow{10}{*}{\rotatebox[origin=c]{90}{\scalebox{0.90}{\textbf{Photo.}}}}
& Auto Contrast     & 0.5183/0.0127 & 0.6222/0.0252 & 0.6584/0.0033 & 0.6888/0.0047 & 0.4868/0.0017 & 0.4910/0.0029 \\
& Balanced Contrast & 0.5365/0.1338 & 0.6491/0.2299 & 0.6629/0.0474 & 0.6970/0.0826 & 0.4897/0.0295 & 0.4980/0.0511 \\
& Brightness        & 0.5177/0.0001 & 0.6210/0.0001 & 0.6722/0.0836 & 0.7039/0.1037 & 0.4977/0.0473 & 0.5011/0.0559 \\
& Contrast          & 0.5216/0.0321 & 0.6269/0.0625 & 0.6667/0.0627 & 0.6978/0.0756 & 0.4906/0.0264 & 0.4988/0.0450 \\
& Color Space       & 0.5179/0.0029 & 0.6213/0.0050 & 0.6689/0.0651 & 0.6997/0.0789 & 0.4901/0.0205 & 0.4948/0.0302 \\
& Equalize          & 0.5373/0.1295 & 0.4190/0.1926 & 0.6738/0.1262 & 0.7018/0.1337 & 0.4950/0.0536 & 0.5059/0.0824 \\
& Gamma             & 0.5177/0.0016 & 0.6210/0.0027 & 0.6588/0.0050 & 0.6891/0.0098 & 0.4883/0.0250 & 0.4934/0.0321 \\
& PCA Jitter        & 0.5177/0.0002 & 0.6210/0.0003 & 0.6583/0.0005 & 0.6888/0.0006 & 0.4868/0.0004 & 0.4909/0.0003 \\
& Invert            & 0.5533/0.2414 & 0.6556/0.3128 & 0.6869/0.2370 & 0.7222/0.2811 & 0.5447/0.2343 & 0.5606/0.2893 \\
& Weather           & 0.6576/0.4954 & 0.7799/0.7119 & 0.6965/0.2213 & 0.7296/0.2616 & 0.5325/0.1748 & 0.5636/0.2613 \\

\noalign{\vskip 2pt}
\hline
\noalign{\vskip 2pt}

\multirow{4}{*}{\rotatebox[origin=c]{90}{\scalebox{0.90}{\textbf{Noise}}}}
& Elastic        & 0.5179/0.0046 & 0.6214/0.0095 & 0.6707/0.1074 & 0.7017/0.1320 & 0.5024/0.0825 & 0.5055/0.0904 \\
& S\&P           & 0.5686/0.1856 & 0.6789/0.2953 & 0.6827/0.1601 & 0.7108/0.1871 & 0.5229/0.1549 & 0.5288/0.1837 \\
& Mobius         & 0.5177/0.0001 & 0.6210/0.0001 & 0.6753/0.1255 & 0.7001/0.1317 & 0.5002/0.1050 & 0.5058/0.1351 \\
& Dim. Reduction & 0.6514/0.4809 & 0.7870/0.7266 & 0.6961/0.1786 & 0.7182/0.2013 & 0.5167/0.1405 & 0.5407/0.1971 \\

\noalign{\vskip 2pt}
\hline
\noalign{\vskip 2pt}

\multirow{3}{*}{\rotatebox[origin=c]{90}{\scalebox{0.86}{\textbf{Selection}}}}
& Random Crop & 0.5736/0.2761 & 0.7108/0.4830 & 0.6725/0.2199 & 0.7058/0.2733 & 0.5126/0.2076 & 0.5177/0.2174 \\
& Erasing     & 0.5303/0.0545 & 0.6411/0.1186 & 0.6669/0.0499 & 0.6983/0.0637 & 0.4916/0.0273 & 0.4968/0.0345 \\
& Shuffle     & 0.5833/0.2906 & 0.6939/0.4055 & 0.7803/0.6196 & 0.8253/0.7209 & 0.6708/0.5779 & 0.7184/0.6223 \\

\noalign{\vskip 2pt}
\hline
\noalign{\vskip 2pt}

\multirow{4}{*}{\rotatebox[origin=c]{90}{\scalebox{0.93}{\textbf{Filter}}}}
& M-Blur     & 0.5343/0.1215 & 0.6384/0.1811 & 0.6643/0.0596 & 0.6945/0.0729 & 0.4903/0.0307 & 0.4942/0.0389 \\
& Blur       & 0.5289/0.2743 & 0.6329/0.4223 & 0.6774/0.1278 & 0.7046/0.1387 & 0.4983/0.0697 & 0.5069/0.1040 \\
& Gauss Blur & 0.5854/0.2931 & 0.7048/0.4585 & 0.6788/0.1218 & 0.7064/0.1331 & 0.5278/0.1616 & 0.5567/0.2421 \\
& Sharpen    & 0.5266/0.0558 & 0.6356/0.1098 & 0.6644/0.0449 & 0.6977/0.0714 & 0.4907/0.0230 & 0.5006/0.0490 \\

\noalign{\vskip 2pt}
\hline
\noalign{\vskip 2pt}

\multirow{3}{*}{\rotatebox[origin=c]{90}{\scalebox{0.90}{\textbf{Self-Mix}}}}
& AugMix   & 0.5261/0.0546 & 0.6255/0.0925 & 0.6675/0.1006 & 0.6975/0.1170 & 0.4949/0.0739 & 0.4994/0.0885 \\
& Self Mix & 0.5230/0.0543 & 0.6285/0.0927 & 0.6615/0.0383 & 0.6910/0.0419 & 0.4934/0.0413 & 0.4945/0.0469 \\
& Salf Mix & 0.5228/0.0507 & 0.6276/0.0835 & 0.6633/0.0512 & 0.6925/0.0577 & 0.4916/0.0373 & 0.4936/0.0339 \\

\noalign{\vskip 2pt}
\hline
\noalign{\vskip 2pt}

\multirow{9}{*}{\rotatebox[origin=c]{90}{\scalebox{0.90}{\textbf{Sample-Mix}}}}
& MixGen   & 0.5447/0.2387 & 0.6559/0.3512 & 0.6977/0.3872 & 0.7303/0.4479 & 0.5188/0.2180 & 0.5112/0.2003 \\
& R-MixGen & 0.5295/0.2165 & 0.6337/0.2865 & 0.6743/0.3423 & 0.7058/0.3825 & 0.4968/0.1822 & 0.4930/0.1728 \\
& MixUp    & 0.5513/0.2661 & 0.6764/0.4440 & 0.6770/0.3406 & 0.7091/0.3994 & 0.5067/0.2271 & 0.5074/0.2314 \\
& CutMix   & 0.5843/0.3297 & 0.7219/0.5649 & 0.6789/0.2786 & 0.7093/0.3305 & 0.5102/0.1988 & 0.5075/0.1893 \\
& Saliency & 0.5364/0.1784 & 0.6495/0.2816 & 0.6794/0.2997 & 0.7123/0.3580 & 0.5045/0.1603 & 0.4873/0.1373 \\
& RICAP    & 0.5594/0.2691 & 0.6691/0.3738 & 0.6856/0.3421 & 0.7217/0.4112 & 0.5078/0.1897 & 0.5061/0.1770 \\
& KeepAug  & 0.5479/0.2365 & 0.6647/0.3637 & 0.6786/0.2300 & 0.7088/0.2638 & 0.5017/0.1236 & 0.5037/0.1254 \\
& IP-Mix   & 0.5767/0.3477 & 0.6871/0.4751 & 0.7122/0.4643 & 0.7430/0.5259 & 0.5471/0.3071 & 0.5539/0.3417 \\
& ResizeMix& 0.5558/0.2604 & 0.6760/0.3947 & 0.6665/0.2392 & 0.6971/0.2697 & 0.5082/0.1687  & 0.5117/0.1845 \\

\noalign{\vskip 2pt}
\hline
\noalign{\vskip 2pt}

\multirow{3}{*}{\rotatebox[origin=c]{90}{\scalebox{0.90}{\textbf{GAN}}}}
& DCGAN                 & 0.5502/0.2116 & 0.6663/0.3362 & 0.6876/0.2080 & 0.7217/0.2559 & 0.5231/0.1678 & 0.5261/0.1772 \\
& CycleGAN              & 0.5736/0.2888 & 0.6937/0.4488 & 0.6837/0.2200 & 0.7129/0.2440 & 0.5009/0.0831 & 0.5047/0.0944 \\
& NST                   & 0.6881/0.5744 & 0.7785/0.6765 & 0.7167/0.3615 & 0.7503/0.4047 & 0.5683/0.3209 & 0.5990/0.4173 \\

\noalign{\vskip 2pt}
\hline
\noalign{\vskip 2pt}

\multirow{6}{*}{\rotatebox[origin=c]{90}{\scalebox{0.90}{\textbf{Diffusion}}}}
& DiffusionCLIP         & 0.5265/0.0889 & 0.6321/0.1450 & 0.6707/0.1232 & 0.6999/0.1443 & 0.5202/0.1663 & 0.5270/0.2040 \\
& Pix2Pix-Zero          & 0.5783/0.3242 & 0.7075/0.4931 & 0.6872/0.3972 & 0.7234/0.4658 & 0.6888/0.3972 & 0.7258/0.4672 \\
& DiffuseMix            & 0.6332/0.4745 & 0.7230/0.5709 & 0.6745/0.1593 & 0.7086/0.2116 & 0.5241/0.2031 & 0.5258/0.2240 \\
& SaSPA                 & 0.6017/0.5133 & 0.6857/0.5746 & 0.6846/0.3561 & 0.7144/0.4022 & 0.5295/0.2748 & 0.5370/0.3267 \\
& DA-Fusion             & 0.5785/0.3292 & 0.7078/0.4967 & {0.6754/0.2268} & {0.7063/0.2728} & 0.5274/0.2137 & 0.5316/0.2307 \\
& DIAGen                & 0.5784/0.3311 & 0.7079/0.4969 & 0.6900/0.3441 & 0.7235/0.4075 & 0.5724/0.3693 & 0.5875/0.4211 \\

\noalign{\vskip 2pt}
\hline
\noalign{\vskip 2pt}

\multirow{2}{*}{\rotatebox[origin=c]{90}{\scalebox{0.90}{\textbf{Adv.}}}}
& FGSM & 0.5676/0.2487 & 0.6823/0.3683 & 0.6860/0.1840 & 0.7059/0.1654 & 0.5833/0.3580 & 0.5249/0.1550 \\
& PGD  & 0.5407/0.1273 & 0.6669/0.3078 & 0.6680/0.0822 & 0.6926/0.0565 & 0.5255/0.2068 & 0.5175/0.1097 \\

\hline
\end{tabular}%
}
\end{table}

\begin{table}[!t]
\centering
\caption{MAHALANOBIS DISTANCE AND ENSEMBLE-BASED UNCERTAINTY UNDER DIFFERENT AUGMENTATION TECHNIQUES}
\label{tab:mahalanobis_ensemble}
\setlength{\tabcolsep}{2pt}
\renewcommand{\arraystretch}{1.0}
\scriptsize

\resizebox{\textwidth}{!}{%
\begin{tabular}{@{}p{0.02\textwidth}p{0.12\textwidth}cccccc@{}}
\hline
\textbf{Cat.} & \textbf{Aug.}
& \multicolumn{2}{c}{\textbf{CIFAR-10}}
& \multicolumn{2}{c}{\textbf{ImageNet}}
& \multicolumn{2}{c}{\textbf{MN}} \\
\cline{3-8}
& & \textbf{Titan} & \textbf{OpenCLIP}
  & \textbf{Titan} & \textbf{OpenCLIP}
  & \textbf{Titan} & \textbf{OpenCLIP} \\
\hline
\noalign{\vskip 2pt}
\multicolumn{8}{c}{\textbf{\textit{Format: Mahalanobis distance / ensemble uncertainty}}} \\
\noalign{\vskip 2pt}
\hline
\noalign{\vskip 3pt}

\multirow{4}{*}{\rotatebox[origin=c]{90}{\scalebox{0.90}{\textbf{Geom.}}}}
& Rotate    & 24.9045/0.4432 & 26.7502/0.4563 & 23.0270/0.0981 & 23.5328/0.1060 & 20.1802/0.4838 & 20.6768/0.5228 \\
& Flip      & 23.6117/0.2858 & 24.7922/0.2912 & 22.9514/0.1599 & 23.4486/0.1581 & 20.1885/0.4987 & 20.6482/0.5270 \\
& Translate & 23.5884/0.0215 & 24.8577/0.0203 & 22.9361/0.1027 & 23.5131/0.1122 & 20.0824/0.6152 & 20.5720/0.6585 \\
& Shear     & 23.5883/0.0212 & 24.8596/0.0204 & 23.0853/0.1169 & 23.5684/0.1211 & 20.0853/0.6307 & 20.5787/0.6546 \\

\noalign{\vskip 2pt}
\hline
\noalign{\vskip 2pt}

\multirow{10}{*}{\rotatebox[origin=c]{90}{\scalebox{0.90}{\textbf{Photo.}}}}
& Auto Contrast     & 23.8025/0.0229 & 25.0589/0.0215 & 18.1559/0.0794 & 18.3387/0.0868 & 15.8607/0.6145 & 16.0209/0.6617 \\
& BCET              & 24.4839/0.1037 & 25.4160/0.1152 & 22.3195/0.0819 & 23.1243/0.0868 & 19.5581/0.5991 & 20.2105/0.6446 \\
& Brightness        & 23.5884/0.0212 & 24.8577/0.0204 & 22.4830/0.0815 & 23.1414/0.0877 & 19.6842/0.6013 & 20.2743/0.6451 \\
& Contrast          & 23.9861/0.0286 & 25.2336/0.0297 & 22.3978/0.0823 & 23.0826/0.0873 & 19.4931/0.0823 & 20.1596/0.0872 \\
& Color Space       & 23.7175/0.0214 & 24.9705/0.0206 & 22.1959/0.0784 & 22.8684/0.0847 & 19.1879/0.6165 & 19.8484/0.6627 \\
& Equalize          & 24.3299/0.1042 & 25.6920/0.1181 & 22.8227/0.1019 & 23.3655/0.1023 & 19.7491/0.5987 & 20.3537/0.6417 \\
& Gamma             & 23.6268/0.0215 & 24.8905/0.0207 & 19.8542/0.0791 & 20.4839/0.0862 & 18.9726/0.6184 & 19.5484/0.6628 \\
& PCA Jitter        & 23.5993/0.0212 & 24.8606/0.0203 & 18.1999/0.0795 & 18.1698/0.0870 & 16.1857/0.6143 & 16.0377/0.6614 \\
& Invert            & 23.9941/0.2005 & 25.2668/0.2145 & 23.1437/0.2080 & 23.5931/0.2152 & 20.1719/0.6410 & 20.6222/0.6664 \\
& Weather           & 27.1433/0.5908 & 27.3580/0.6014 & 23.1988/0.1385 & 23.6139/0.1297 & 20.2022/0.6289 & 20.6563/0.6720 \\

\noalign{\vskip 2pt}
\hline
\noalign{\vskip 2pt}

\multirow{4}{*}{\rotatebox[origin=c]{90}{\scalebox{0.90}{\textbf{Noise}}}}
& Elastic        & 23.7711/0.0230 & 25.0004/0.0232 & 22.8417/0.1126 & 23.3815/0.1181 & 19.9301/0.6280 & 20.4344/0.6399 \\
& S\&P           & 25.2115/0.0772 & 26.9034/0.1169 & 23.0698/0.1204 & 23.5478/0.1268 & 20.0793/0.6412 & 20.5707/0.6673 \\
& Mobius         & 23.5883/0.0212 & 24.8577/0.0204 & 22.9636/0.0933 & 23.4720/0.1005 & 20.0644/0.6272 & 20.5776/0.6495 \\
& Dim. Reduction & 26.5958/0.6010 & 27.4063/0.6086 & 23.0846/0.1244 & 23.5109/0.1177 & 20.1519/0.6381 & 20.5995/0.6582 \\

\noalign{\vskip 2pt}
\hline
\noalign{\vskip 2pt}

\multirow{3}{*}{\rotatebox[origin=c]{90}{\scalebox{0.86}{\textbf{Selection}}}}
& Random Crop & 24.7469/0.3641 & 25.7322/0.3912 & 23.1681/0.2011 & 23.6773/0.1954 & 20.0320/0.6375 & 20.5601/0.6627 \\
& Erasing     & 24.1263/0.0335 & 25.1700/0.0409 & 22.0154/0.0858 & 22.7574/0.0944 & 19.0967/0.5980 & 19.7032/0.6439 \\
& Shuffle     & 25.5331/0.3180 & 26.5754/0.3722 & 24.8202/0.4587 & 24.9681/0.5771 & 20.5699/0.6272 & 20.8349/0.6586 \\

\noalign{\vskip 2pt}
\hline
\noalign{\vskip 2pt}

\multirow{4}{*}{\rotatebox[origin=c]{90}{\scalebox{0.93}{\textbf{Filter}}}}
& M-Blur     & 23.9423/0.1397 & 24.9319/0.1489 & 22.2551/0.0840 & 22.9095/0.0907 & 19.4776/0.6174 & 20.0385/0.7112 \\
& Blur       & 24.7152/0.3389 & 25.4755/0.3701 & 22.7881/0.0932 & 23.2968/0.0963 & 19.8005/0.6209 & 20.3484/0.6669 \\
& Gauss Blur & 25.1381/0.3637 & 25.7486/0.4079 & 22.8418/0.0904 & 23.3556/0.0946 & 20.1942/0.6421 & 20.6505/0.6867 \\
& Sharpen    & 24.0512/0.0298 & 25.3907/0.0331 & {22.1910/0.0808} & {22.9968/0.0855} & 19.4674/0.5967 & 20.1724/0.6421 \\

\noalign{\vskip 2pt}
\hline
\noalign{\vskip 2pt}

\multirow{3}{*}{\rotatebox[origin=c]{90}{\scalebox{0.90}{\textbf{Self-Mix}}}}
& AugMix   & 23.6908/0.0501 & 24.8752/0.0579 & 22.2869/0.1019 & 22.8987/0.1068 & 19.4678/0.6113 & 20.0207/0.6508 \\
& Self Mix & 23.3603/0.0512 & 24.3348/0.0547 & 21.9323/0.0870 & 22.4998/0.0949 & 19.5697/0.6060 & 20.1027/0.6458 \\
& Salf Mix & 23.3964/0.0513 & 24.4323/0.0577 & 22.3564/0.0894 & 22.9504/0.0969 & 19.5495/0.6052 & 20.0813/0.6444 \\

\noalign{\vskip 2pt}
\hline
\noalign{\vskip 2pt}

\multirow{9}{*}{\rotatebox[origin=c]{90}{\scalebox{0.90}{\textbf{Sample-Mix}}}}
& MixGen   & 23.5033/0.2246 & 24.8742/0.2603 & 23.1895/0.2269 & 23.6652/0.2267 & 20.0449/0.6717 & 20.6451/0.6953 \\
& R-MixGen & 22.4181/0.0519 & 24.1003/0.0601 & 22.0586/0.1017 & 22.4383/0.1075 & 19.1901/0.6252 & 19.5836/0.6693 \\
& MixUp    & 24.4333/0.2560 & 25.5893/0.3484 & 23.1531/0.1565 & 23.6335/0.1623 & 20.0167/0.6297 & 20.5349/0.6665 \\
& CutMix   & 24.3311/0.3676 & 25.2264/0.4087 & 23.2039/0.1479 & 23.6644/0.1497 & 20.0168/0.6383 & 20.5444/0.6791 \\
& Saliency & 23.0936/0.1507 & 24.1300/0.1852 & 23.1218/0.1814 & 23.6192/0.1816 & 19.9373/0.6752 & 20.6026/0.6760 \\
& RICAP    & 23.5251/0.2471 & 24.6733/0.2439 & 23.1813/0.2300 & 23.6656/0.2271 & 19.9946/0.6488 & 20.6110/0.6929 \\
& KeepAug  & 23.2847/0.2270 & 24.4306/0.2875 & 23.1164/0.1611 & 23.5988/0.1615 & 19.9456/0.6406 & 20.4806/0.6921 \\
& IP-Mix   & 23.3979/0.3369 & 25.0817/0.4058 & 23.2588/0.3212 & 23.7024/0.3084 & 20.0854/0.6953 & 20.6891/0.6940 \\
& ResizeMix& 23.7149/0.2270 & 25.4456/0.3224 & 22.9897/0.0725 & 23.4908/0.767  & 20.0086/0.6293 & 20.5190/0.6745 \\
\noalign{\vskip 2pt}
\hline
\noalign{\vskip 2pt}

\multirow{3}{*}{\rotatebox[origin=c]{90}{\scalebox{0.90}{\textbf{GAN}}}}
& DCGAN    & 25.6870/0.2509 & 26.8840/0.2702 & 23.1109/0.1575 & 23.5907/0.1899 & 20.0914/0.6514 & 20.5932/0.6908 \\
& CycleGAN & 24.8292/0.3025 & 25.7590/0.3307 & 23.1717/0.1245 & 23.6380/0.1287 & 20.0633/0.6286 & 20.5434/0.6725 \\
& NST & 27.6003/0.4513 & 29.4089/0.5085 & 23.3719/0.2880 & 23.7429/0.2979 & 20.1947/0.7006 & 20.6599/0.7159 \\

\noalign{\vskip 2pt}
\hline
\noalign{\vskip 2pt}

\multirow{6}{*}{\rotatebox[origin=c]{90}{\scalebox{0.90}{\textbf{Diffusion}}}}
& DiffusionCLIP         & 23.1160/0.0374 & 24.6488/0.0435  & 22.9477/0.0876 & 23.4870/0.0914 & 20.1367/0.6395 & 20.5999/0.6825 \\
& Pix2Pix-Zero          & 24.4627/0.4236 & 26.3514/0.5298  & 23.3693/0.2749 & 23.7550/0.2908 & 21.1728/0.2749 & 21.3399/0.2760 \\
& DiffuseMix            & 22.9349/0.2922 & 24.8757/0.3603  & 23.0830/0.1140 & 23.5801/0.1245 & {20.0839}/0.6538  & 20.5658/0.6906 \\
& SaSPA                 & 22.5517/0.0141 & 25.0285/0.0109  & 23.2714/0.1187 & 23.7025/0.1192 & 20.0628/0.6511 & 20.5682/0.7070 \\
& DA-Fusion             & 24.4651/0.4111 & 26.3819/0.5239  & 23.2155/0.1306 & 23.6470/0.1310 & 20.0499/0.6370 & 20.5608/0.6785 \\
& DIAGen                & 24.4571/0.4190 & 26.4270/0.5279  & 23.3754/0.2432 & 23.7445/0.2188 & 20.0972/0.6801 & 20.5985/0.6820 \\

\noalign{\vskip 2pt}
\hline
\noalign{\vskip 2pt}

\multirow{2}{*}{\rotatebox[origin=c]{90}{\scalebox{0.90}{\textbf{Adv.}}}}
& FGSM & 24.7308/0.2308 & 26.3674/0.2308 & 23.1864/0.1491 & 23.4648/0.1464 & 20.2082/0.6958 & 20.5610/0.6750 \\
& PGD  & 23.9363/0.0934 & 25.8994/0.2627 & 22.7470/0.0951 & 23.0228/0.0973 & 20.048/0.6343  & 20.4306/0.6647 \\

\hline
\end{tabular}%
}
\end{table}

\subsection{RQ3: Supplementary Semantic-Realism Results}
\noindent\textbf{Algorithm 2, and 3} present the modified prompts compared with the original prompt used in this study. The primary objective is to evaluate the sensitivity of the final realism score to variations in keyword selection and keyword ordering.

\begin{table}[H]
\centering
\refstepcounter{myalgorithm}
\label{alg:prompt_variant_2}
\textbf{Algorithm~\themyalgorithm: Pseudocode for Prompt Variant 2 in Realism Sensitivity Analysis}

\vspace{2pt}
\setlength{\tabcolsep}{3pt}
\renewcommand{\arraystretch}{1.05}
\footnotesize

\begin{tabular}{p{0.07\columnwidth} p{0.86\columnwidth}}
\hline
\textbf{Step} & \textbf{Operation} \\
\hline
\addlinespace[2pt]
1 & Provide the augmented image $I_{\mathrm{aug}}$ to the VLM $M$. \\

2 & Define $M$ as an image realism evaluator and instruct it to assess the visual realism of $I_{\mathrm{aug}}$ using only visible evidence in the image. \\

3 & Evaluate $I_{\mathrm{aug}}$ according to nine realism criteria: semantic coherence of the scene, physical plausibility, perspective and camera geometry, lighting consistency, depth-of-field and focus realism, texture and material realism, edge integrity and object boundaries, noise, compression, and sensor characteristics, and local visual artifacts. \\

4 & Assign $r_i \in \{0,1,2\}$ to each criterion, where 0, 1, and 2 denote unrealistic, partially realistic, and realistic conditions, respectively. \\

5 & Compute the total realism score as $R_s=\sum_{i=1}^{9} r_i$. \\

6 & Ensure that the final score satisfies $0 \leq R_s \leq 18$. \\

7 & Suppress all explanations and comments in the model response. \\

8 & Return only the final score using the predefined format: \texttt{Total Score: <0--18>/}. \\
\addlinespace[2pt]
\hline
\end{tabular}
\end{table}

\begin{table}[H]
\centering
\refstepcounter{myalgorithm}
\label{alg:prompt_variant_3}
\textbf{Algorithm~\themyalgorithm: Pseudocode for Prompt Variant 3 in Realism Sensitivity Analysis}

\vspace{2pt}
\setlength{\tabcolsep}{3pt}
\renewcommand{\arraystretch}{1.05}
\footnotesize

\begin{tabular}{p{0.07\columnwidth} p{0.86\columnwidth}}
\hline
\textbf{Step} & \textbf{Operation} \\
\hline
\addlinespace[2pt]
1 & Provide the augmented image $I_{\mathrm{aug}}$ to the VLM $M$. \\

2 & Instruct $M$ to evaluate the realism of $I_{\mathrm{aug}}$ using only the visible visual content of the image. \\

3 & Define nine realism factors for assessment: physical realism, lighting realism, camera perspective and geometry, focus and depth-of-field realism, texture and material quality, object boundaries and edge quality, noise, compression, or sensor effects, scene-level semantic consistency, and small local artifacts. \\

4 & Assign a score $r_i \in \{0,1,2\}$ to each realism factor, where 0, 1, and 2 denote not realistic, somewhat realistic, and realistic conditions, respectively. \\

5 & Compute the total realism score as $R_s = \sum_{i=1}^{9} r_i$. \\

6 & Ensure that the final score satisfies $0 \leq R_s \leq 18$. \\

7 & Suppress all explanations, comments, and intermediate outputs in the response. \\

8 & Return only the final score using the predefined output format: \texttt{Total Score: <0--18>/}. \\
\addlinespace[2pt]
\hline
\end{tabular}
\end{table}

\noindent\textbf{\Cref{tab:realism}} presents the model-based realism scores obtained for all evaluated augmentation techniques across the three datasets. These scores quantify the perceptual realism of the generated images, with higher values indicating that the augmented samples retain more realistic visual characteristics according to the employed evaluation model. The results enable a systematic comparison of how different augmentation techniques affect image realism across datasets with different visual characteristics and complexities.

\begin{table}[!t]
\centering
\caption{LLAVA REALISM SCORES FOR AUGMENTED IMAGES}
\label{tab:realism}
\setlength{\tabcolsep}{2pt}
\renewcommand{\arraystretch}{0.90}
\small

\begin{tabular}{llccc}
\hline
\textbf{Cat.} & \textbf{Aug.}
& \textbf{CIFAR-10} & \textbf{ImageNet} & \textbf{MN} \\
\hline
\noalign{\vskip 2pt}

\multirow{4}{*}{Geom.}
& Rotate    & 11.36 & 11.73 & 7.36 \\
& Flip      & 8.79  & 10.94 & 8.63 \\
& Translate & 9.97  & 11.47 & 10.60 \\
& Shear     & 10.06 & 10.59 & 7.32 \\

\hline
\noalign{\vskip 2pt}

\multirow{10}{*}{Photo.}
& Auto Contrast     & 9.89  & 12.61 & 8.01 \\
& Balanced Contrast & 8.24  & 12.65 & 13.12 \\
& Brightness        & 10.01 & 12.05 & 12.97 \\
& Contrast          & 9.42  & 12.18 & 13.10 \\
& Color Space       & 9.93  & 12.16 & 7.89 \\
& Equalize          & 8.18  & 11.65 & 13.25 \\
& Gamma             & 9.94  & 12.56 & 8.22 \\
& PCA Jitter        & 9.98  & 12.59 & 8.46 \\
& Invert            & 7.65  & 10.70 & 12.74 \\
& Weather           & 7.75  & 12.00 & 10.89 \\

\hline
\noalign{\vskip 2pt}

\multirow{4}{*}{Noise}
& Elastic        & 9.69 & 11.61 & 7.41 \\
& S\&P           & 9.49 & 10.83 & 6.50 \\
& Mobius         & 9.97 & 11.66 & 7.86 \\
& Dim. Reduction & 7.83 & 11.04 & 7.57 \\

\hline
\noalign{\vskip 2pt}

\multirow{3}{*}{Selection}
& Random Crop & 7.61 & 11.04 & 7.42 \\
& Erasing     & 8.61 & 11.84 & 12.73 \\
& Shuffle     & 6.75 & 4.52  & 2.00 \\

\hline
\noalign{\vskip 2pt}

\multirow{4}{*}{Filter}
& M-Blur     & 8.34 & 11.97 & 8.09 \\
& Blur       & 7.96 & 11.21 & 7.94 \\
& Gauss Blur & 6.94 & 11.27 & 7.66 \\
& Sharpen    & 9.62 & 12.56 & 13.23 \\

\hline
\noalign{\vskip 2pt}

\multirow{3}{*}{Self-Mix}
& AugMix   & 9.03 & 10.56 & 12.22 \\
& Self Mix & 8.90 & 12.15 & 12.24 \\
& Salf Mix & 9.20 & 12.34 & 12.29 \\

\hline
\noalign{\vskip 2pt}

\multirow{8}{*}{Sample Mix}
& MixGen   & 7.93 & 9.36  & 9.75 \\
& R-MixGen & 9.46 & 11.29 & 9.69 \\
& MixUp    & 8.02 & 10.75 & 12.62\\
& CutMix   & 6.96 & 10.60 & 7.01 \\
& Saliency & 7.81 & 10.43 & 8.17 \\
& RICAP    & 9.21 & 10.98 & 9.01 \\
& KeepAug  & 6.99 & 9.017 & 7.33 \\
& IP-Mix   & 6.91 & 7.81  & 7.66 \\
& ResizeMix& 8.98 & 11.23 & 9.72 \\

\hline
\noalign{\vskip 2pt}

\multirow{3}{*}{GAN}
& DCGAN                 & 6.81  & 8.18  & 6.81  \\
& CycleGAN              & 6.76  & 10.12 & 12.02 \\
& Neural Style Transfer & 6.314 & 8.83  & 6.42  \\

\hline
\addlinespace[2pt]
\multirow{9}{*}{Diffusion}
& DiffusionCLIP         & 9.06  & 12.13 & 11.95\\
& Pix2Pix-Zero          & 7.831 & 8.82  & 8.57 \\
& Diffusemix            & 7.18  & 10.50 & 9.82 \\
& SaSPA                 & 11.59 & 10.72 & 11.65\\
& DA-Fusion             & 7.58  & 10.97 & 7.38 \\
& DIAGen                & 7.64  & 8.19  & 8.39 \\

\hline
\addlinespace[2pt]
\multirow{2}{*}{Adv.}
& FGSM & 8.23 & 7.88  & 5.74 \\
& PGD  & 9.50 & 10.32 & 10.43 \\
\hline
\end{tabular}
\end{table}

\subsection{RQ4: Overall-Ranking Sensitivity Analysis}

\begin{figure}[!htbp]
\centering
\includegraphics[width=0.9\textwidth]{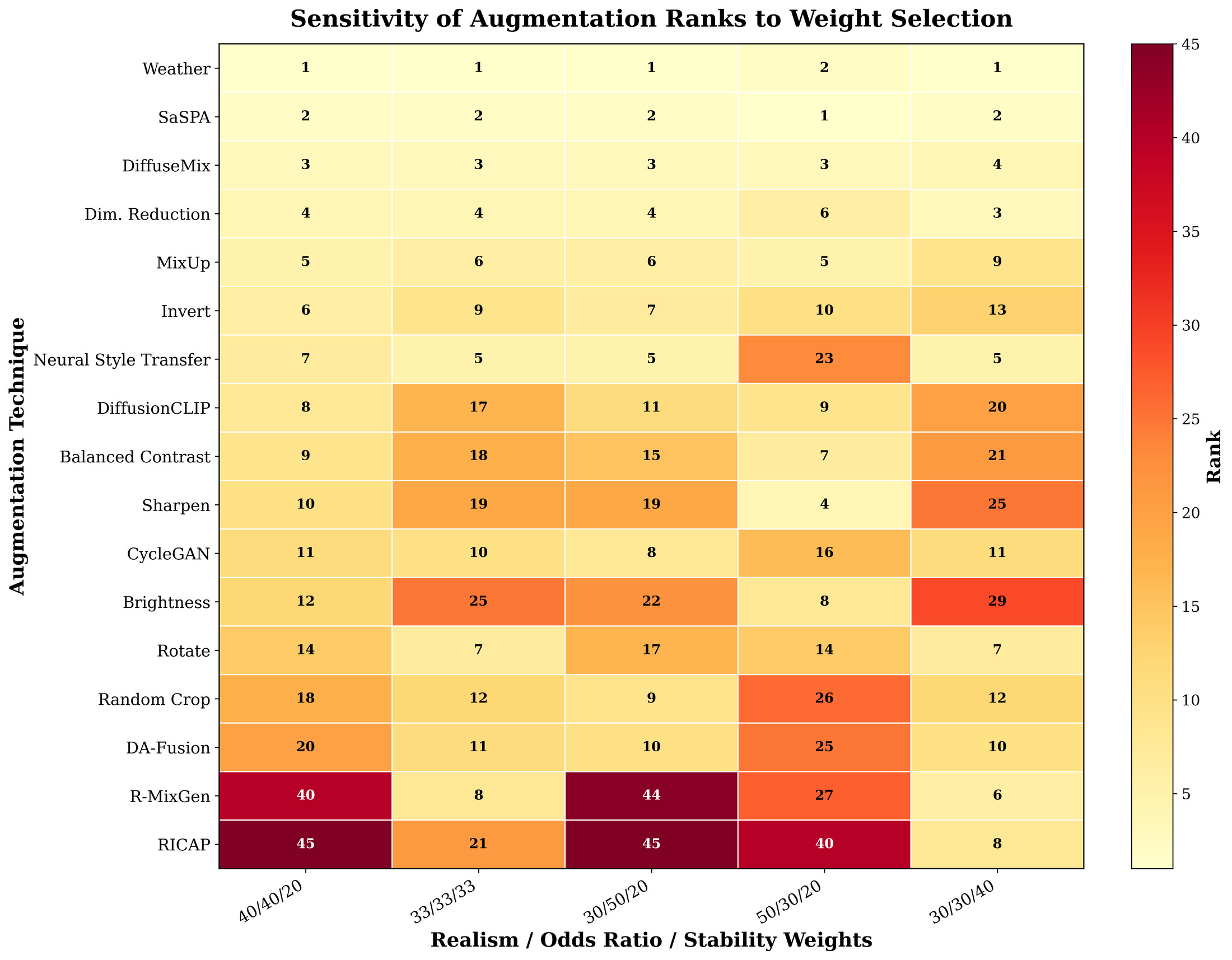}
\caption{Sensitivity of the overall augmentation ranking to alternative
weighting configurations.}
\label{fig:weight_overal}
\end{figure}

\noindent\textbf{\Cref{fig:weight_overal}} compares the effect of different weighting configurations across all considered evaluation metrics. This sensitivity analysis assesses the extent to which variations in the assigned weights influence the overall ranking of augmentation techniques. In particular, it evaluates whether the techniques identified as the most effective remain consistently ranked when the relative importance of failure rate, realism, and stability is varied.

\begin{figure}[!htbp]
\centering
\includegraphics[width=0.9\textwidth]{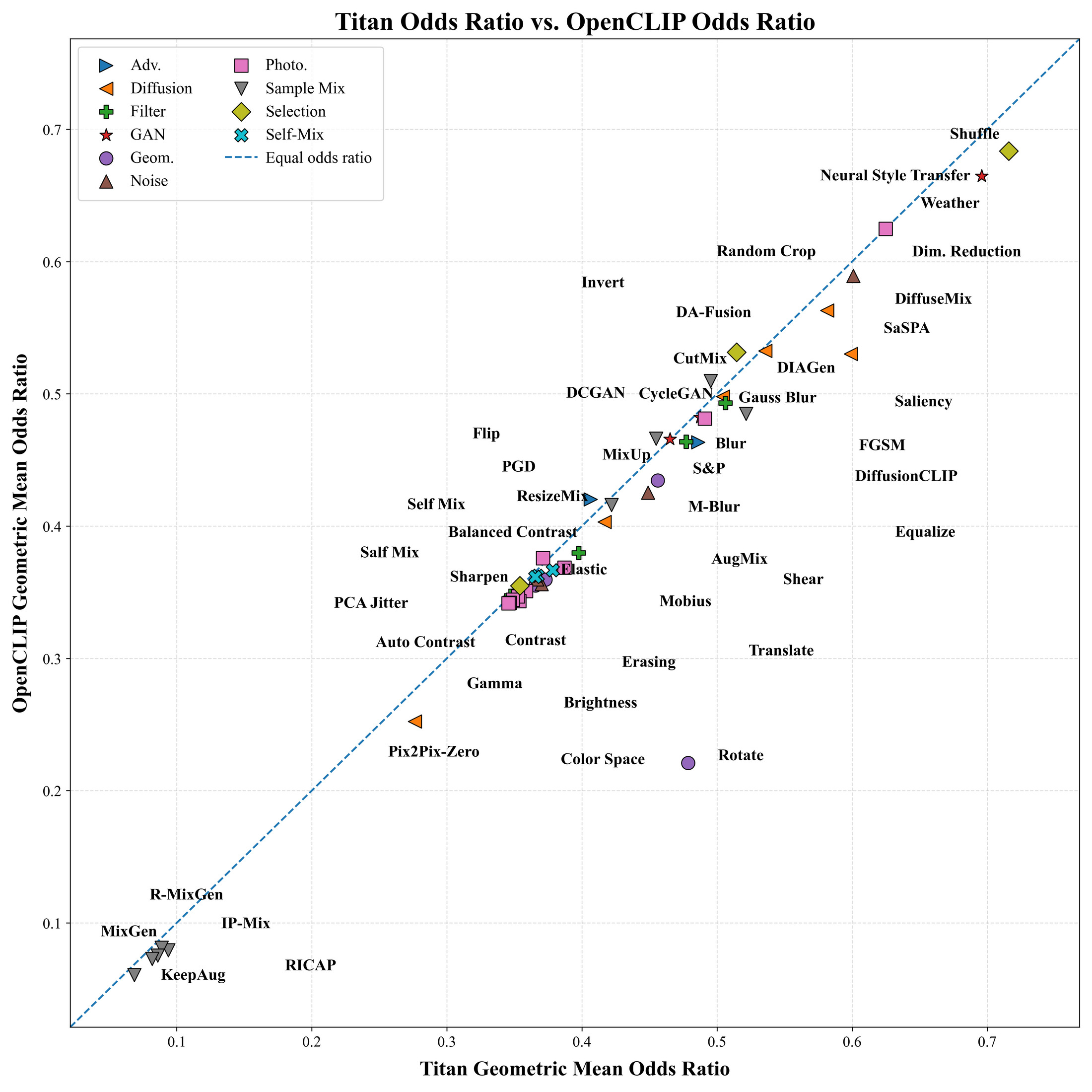}
\caption{Failure-rate comparison between Titan and OpenCLIP across augmentation techniques.
}
\label{fig:models_failures}
\end{figure}

\noindent\textbf{\Cref{fig:models_failures}} compares the effect of all evaluated augmentation techniques on the failure counts of the two embedding models. Techniques located closer to the diagonal line exhibit comparable effects on both models, whereas a greater distance from the diagonal indicates a larger difference in model sensitivity to the corresponding augmentation. Furthermore, techniques positioned toward the upper-right corner produce higher failure counts for both models, indicating a stronger and more consistent effect on retrieval performance.
\newpage

\end{document}